\documentclass[12pt,oneside,onecolumn]{report}
\usepackage{graphicx,bm,floatflt,amssymb,epsf,epsfig,rotate,multirow,axodraw,cite,subeqn}
\newsavebox{\fmbox}

\newcommand{\be}{\begin{eqnarray}}
\newcommand{\ee}{\end{eqnarray}}

\newcommand{\br}{$^8$B~}
\newcommand{\li}{$^8$Li~}
\newcommand{\he}{$^6$He~}
\newcommand{\neon}{$^{18}$Ne~}
\newcommand{\chr}{\mbox{$\breve{\rm C}$erenkov~}}

\newcommand{\ms}{\Delta m^2_{21}}
\newcommand{\ma}{\Delta m^2_{31}}

\newcommand{\sss}{\sin^2 \theta_{12}}
\newcommand{\sch}{\sin^2 \theta_{13}}
\newcommand{\stch}{\sin^2 2\theta_{13}}

\newcommand{\sta}{\sin^22 \theta_{23}}
\newcommand{\mst}{\Delta m^2_{21}{\mbox {(true)}}}
\newcommand{\mat}{\Delta m^2_{31}{\mbox {(true)}}}
\newcommand{\sgnma}{$\mathrm{sgn}(\Delta m^2_{31})$}

\newcommand{\ssst}{\sin^2 \theta_{12}{\mbox {(true)}}}

\newcommand{\stcht}{\sin^2 2\theta_{13}{\mbox {(true)}}}

\newcommand{\stat}{\sin^22 \theta_{23}{\mbox {(true)}}}
\newcommand{\dcpt}{\delta_{CP}{\mbox {(true)}}}
\newcommand{\tmt}{$\theta_{23}$}
\newcommand{\tet}{$\theta_{13}$}
\newcommand{\tem}{$\theta_{12}$}

\newcommand{\sig}{$3\sigma$}
\newcommand{\Rsl}{{\not \! \!{R}}}
\newcommand{\ldm}{\Delta m_{31}^2}

\def\petau{{${P_{e \tau}}$}}

\def\pemu{{${P_{e \mu }}$ }}

\def\pee{{${P_{ee}}$ }}
\def\nue{{\nu_e}}
\def\anue{{\bar\nu_e}}
\def\numu{{\nu_{\mu}}}
\def\anumu{{\bar\nu_{\mu}}}
\def\nutau{{\nu_{\tau}}}
\def\anutau{{\bar\nu_{\tau}}}

\def\n{\nu}
\def\lp {\lambda'}

\newcommand{\ie}{{\it i.e.}}

\def\ltap{\ \raisebox{-.4ex}{\rlap{$\sim$}} \raisebox{.4ex}{$<$}\ }
\def\gtap{\ \raisebox{-.4ex}{\rlap{$\sim$}} \raisebox{.4ex}{$>$}\ }
\def\lsim{\:\raisebox{-0.5ex}{$\stackrel{\textstyle<}{\sim}$}\:}
\def\gsim{\:\raisebox{-0.5ex}{$\stackrel{\textstyle>}{\sim}$}\:}
\begin{document}
\thispagestyle{empty}


\begin{center}

{\LARGE \bf SOME ASPECTS OF \\
\vskip0.4cm
NEUTRINO MIXING AND OSCILLATIONS}
\end{center}

\vskip 3cm
\begin{center}
{\bf {
THESIS SUBMITTED TO \\
\vskip0.2cm
THE UNIVERSITY OF CALCUTTA \\
\vskip0.2cm
FOR THE DEGREE OF \\
\vskip0.2cm
DOCTOR OF PHILOSOPHY (SCIENCE) 
}}
\end{center}

\vskip 3cm
\begin{center}
{\bf {
By \\
\vskip0.5cm
{\large{\bf SANJIB KUMAR AGARWALLA}} \\
\vskip0.5cm
DEPARTMENT OF PHYSICS, \\
\vskip0.2cm
UNIVERSITY OF CALCUTTA
}}

\vskip0.5cm

{\bf {SEPTEMBER 2008}}

\end{center}


\pagenumbering{roman}
\newpage
\mbox{}
\newpage
\begin{center}
{\large\bf -- To My Parents --}
\end{center}

\newpage
\mbox{}
\newpage
\begin{center}
{\underline{\Large\bf Acknowledgements}}
\end{center}
\vskip 15pt

First and foremost, I would like to express my deepest respect and most 
sincere gratitude to my supervisor, Professor Amitava Raychaudhuri, for
introducing me into the fascinating world of neutrinos and for being an
excellent teacher and a perfect guide. I convey my regards to him 
not only for his continuous support and patience throughout this study, 
which was a great comfort, but also for the new dimensions he brought 
into my life. Without his inspiring discussions, invaluable guidance and 
exact encouragement, this accomplishment would have never been fulfilled.
For almost everything I know about neutrino phenomenology, I owe to him.
I consider myself very fortunate to have him as my supervisor who gave
me immense freedom to pursue my interests and always encouraged me to 
take part in various summer schools, workshops, conferences in India 
and abroad from the very starting point of my research career. 
This was quite helpful in building my confidence to discuss physics
issues with other scientists which most of the times culminated into
fruitful collaborations. I am greatly indebted to his assistance and 
understanding in matters of non-academic concern which have helped me 
endure some difficult times during my study period. His suggestions
will always help me in moving the things ahead in my future research
career.

There are no words to describe such a great honour that I have the
chance to convey my sincere gratefulness to my senior collaborator
Dr. Sandhya Choubey with whom I have done most of my work.
She has been my constant source of inspiration for gaining grounds 
in research work and I strongly feel indebted to her as nothing would 
have been possible without her support. I would like to thank her
for her close daily guidance, her patience in assisting me during 
the course of this work, her kindness, and her opportune encouragements.
Her commitment to research and dedication to physics, and even her 
unique style of language and her sense of humour have all greatly 
influenced me. 
 
I am very much thankful to Dr. Srubabati Goswami, my senior collaborator, 
with whom a part of the research work contained in this thesis was done
and whose extraordinary energy and enthusiasm for physics has set a 
great example for me.

I would like to thank Dr. Abhijit Samanta for my first successful
collaboration and for helping me regarding the use of computers. 
He also shared his expertise on various issues related to the
ICAL@INO detector simulation with me.

Mention must be made of my senior research colleague Dr. Rathin Adhikari,
who extended a helping hand both in academic and non-academic issues
whenever it was necessary.

I would like to thank Dr. Subhendu Rakshit for successful
collaboration and providing a friendly environment for discussing
physics.

I express my gratitude to Dr. Walter Winter and Dr. Patrick Huber for
providing me useful informations regarding the simulations related to
long baseline neutrino experiments at various stages of this work. 

I would like to express my sincere gratitude and appreciation to
Professor Mina Ketan Parida who introduced me to the topic of
renormalization group evolution analysis during his stay at
Harish-Chandra Research Institute and provided me an opportunity
to collaborate with two big names in the field of neutrino
physics - Professor R. N. Mohapatra and Professor G. Rajasekaran.

The constructive suggestions and useful comments from 
Professor Manfred Lindner, Professor A. Smirnov, Professor G. Raffelt,
Professor Andre de Gouvea, Professor Mats Lindroos, 
Professor Mauro Mezzetto, Professor J. J. Gomez Cadenas, 
Dr. Andrea Donini, Dr. Pilar Hernandez, Dr. Osamu Yasuda 
and Dr. Francesco Terranova helped me a lot at different 
stages of this work.

I gratefully acknowledge generous helps offered by Professor 
Naba Kumar Mondal who is the spokesperson of the INO collaboration.
At this stage I would like to extend my thanks to Professor Kamales Kar,
Professor Sudeb Bhattacharya, Professor G. Rajasekaran, 
Professor H. S. Mani, Professor M. V. N. Murthy, 
Professor Raj Gandhi, Dr. Amol Dighe, Dr. D. Indumathi,
Dr. Brajesh Choudhary, Dr. Gobinda Majumder, Dr. Debasish Majumdar, 
Dr. Ambar Ghosal, Dr. Satyajit Saha, Dr. Subhasish Chattopadhyay, 
Dr. Abhijit Bandhopadhyay for having useful discussions on INO. 
I also acknowledge the moral support from my friends Saikat Biswas, Tapasi Ghosh, 
Satyajit Jena who are working for the INO collaboration. 

I benefitted from discussing physics with Professor Amitava Datta,
Professor Ashoke Sen, Professor Rajesh Gopakumar, Professor Dileep Jatkar, 
Professor Biswarup Mukhopadhyaya, Professor Soumitra Sengupta, 
Professor Anjan S. Joshipura, Professor Sreerup Raychaudhuri, 
Professor Gautam Bhattacharyya, Dr. Subhendra Mohanty, 
Dr. C. S. Aulakh, Dr. Anindya Datta, Dr. Asesh Krishna Datta 
and Dr. Raghavan Rangarajan at various stages.

I am much obliged to Professor Tapan Kumar Das, Professor Subinay Dasgupta, 
Dr. Anirban Kundu, Dr. Parongama Sen and Dr. Gautam Gangopadhyay from 
Calcutta University who have helped me in various ways during this work.

I wish to thank Dr. Arunansu Sil who helped me at the initial stage
of my research work to grab the basics of neutrino oscillations.
I would also like to thank my friend Shashank Shalgar for sharing
his insights in neutrino physics and computer related issues.
I admire the moral support that I received from Swarup Majee.

My heartiest thanks to Soumitra Nandi, Biplob Bhattacharjee, 
Kamalika Basu Hazra, Anasuya Kundu, Pratap K. Das, Anjan Kumar Chandra 
from Calcutta University for providing me the moral support 
during the initial stage of this work. I also would like to
thank Professor S. D. Adhikari, Dr. R. Thangadurai, 
Anamitra Mukherjee, Arijit Saha, Pomita Ghosal, Atashi Chatterjee,
Kalpataru Pradhan, Sudhir Kumar Gupta, Subhaditya Bhattacharya,
Priyotosh Bandyopadhyay, R. Srikanth H., Shailesh Lal, 
Mahender Singh for sharing light moments with me and for
their constant encouragement during my stay at Harish-Chandra 
Research Institute. 

I gratefully acknowledge the cluster facilities of 
Harish-Chandra Research Institute for computational work.
I received help from Professor Jasjeet Singh Bagla regarding
the use of this cluster facility. I would like to thank 
Mr. Sanjai Verma, Mr. Shahid Ali Farooqui and Mr. Chandan Kanaojia
for providing help regarding computers at various stages during
this work.

I would like to acknowledge support from the project (SP/S2/K-10/2001)
of the Department of Science and Technology, India for providing the 
financial support during the initial part of this work.
I also acknowledge support from the XIth Plan Neutrino Project
of the Harish-Chandra Research Institute.

Last but not least with all my heart, my deepest thanks are devoted 
to my beloved family~:~my father, my mother and my brother, Rakesh, 
for their ever patience and constant support, their daily encouragements
and boundless love. All these things I could do, just for them.

\vskip 1.5cm\noindent
{\rm Department of Physics}, \hfill {\rm Sanjib Kumar Agarwalla}\\
{\rm University of Calcutta}. \hfill {\rm September 2008}.

\newpage
\mbox{}
\newpage
\begin{center}
{\underline{\Large\bf ABSTRACT OF THE THESIS}}
\end{center}

\vskip 10pt

Neutrino physics is a very intense field of research having implications 
in different branches of physics, such as high energy physics, 
quantum field theory, cosmology, astrophysics, nuclear physics and geophysics. 
Spectacular results on neutrino oscillations in the last 
several years have triggered a lot of interest in neutrinos, from experimental
as well as theoretical point of view, and many future neutrino experiments are in 
preparation or under discussion to sharpen our understanding about these tiny
particles. This thesis addresses several aspects of these issues.

We have studied the physics reach of an experiment where 
neutrinos produced in a beta-beam facility at CERN are observed in a large 
magnetized iron calorimeter (ICAL) at the India-based Neutrino Observatory 
(INO). The idea of beta-beam is based on the production of a pure, intense,
collimated beam of electron neutrinos or their antiparticles via the beta decay
of accelerated radioactive ions circulating in a storage ring. Interestingly, 
the CERN-INO distance of 7152 km happens to be tantalizingly close to the 
so-called ``magic'' baseline where the sensitivity to the neutrino mass ordering 
(sign of $\Delta m^2_{31} \equiv m_3^2 - m_1^2$)
and more importantly, $\theta_{13}$, goes up significantly, 
while the sensitivity to the unknown CP phase is absent. 
This permits such an experiment 
involving the golden $P_{e\mu}$ channel to make precise measurements of the 
mixing angle $\theta_{13}$ and neutrino mass hierarchy avoiding the issues of 
intrinsic degeneracies and correlations which plague other baselines. 

We propound the possibility of using
large matter effects in the survival channel, $P_{ee}$, at long baselines 
for determination of the neutrino mass ordering and the mixing angle 
$\theta_{13}$. Matter effects in the 
transition probabilites $P_{e \mu}$ and $P_{e \tau}$ act in consonance
to give an almost two-fold effect in the survival channel. In addition,
the problem of spurious solutions due to the leptonic CP phase and the atmospheric 
mixing angle $\theta_{23}$ does not crop up. Thus a beta-beam enables one to 
exclusively study the $P_{ee}$ survival probability with the help of proposed megaton class water detectors like UNO, HyperKamiokande, MEMPHYS.

We have also explored the possibility of detecting new physics signals
in dedicated experiments using a near and a far detector and a beta-beam source. 
We focus on the possible impact of 
flavor-changing and flavor-diagonal neutral current interactions that might 
crop up at the production point, in the oscillation stage, or at the detection 
point when one will deal with these upcoming facilities. 
For an example, the R-parity violating Supersymmetric model allows these kind 
of interactions and long baseline neutrino oscillation experiments may well 
emerge as test beds for this kind of models.

\newpage
\mbox{}
\tableofcontents
\newpage
\addcontentsline{toc}{chapter}{~~~~List of Figures}
\listoffigures
\newpage
\mbox{}
\listoftables
\addcontentsline{toc}{chapter}{~~~~List of Tables}
\newpage
\mbox{}
\newpage
\begin{center}

{\underline{\Large\bf Neutrino of Love}}

\vskip 5pt

I go undetected \\
In all my interactions \\
I cannot be seen \\
From any point of view \\
You won't know if I'm here \\
Except when I'm gone \\
I'm the neutrino of love \\
And I'm coming over you
\vskip 2pt
You cannot keep me in a cage \\
No matter how thick the walls \\
I will escape \\
You cannot hold me in a box \\
Cannot bind me with a lock \\
Cannot keep me anyway \\
I'm not afraid of the dark \\
I'm the neutrino \\
Neutrino of love \\
I'm the neutrino \\
Neutrino of love
\vskip 2pt
I go undetected \\
In all my interactions \\
I cannot be seen \\
From any point of view \\
You won't know if I'm here \\
Except when I'm gone \\
I'm the neutrino of love \\
And I'm coming over you
\vskip 2pt
I'm the neutrino baby \\
Neutrino of love
\vskip 2pt
Cannot inhibit my infiltration \\
Neutrino
\vskip 2pt
Cannot prevent my penetration \\
Neutrino
\vskip 2pt
I am the neutrino

\end{center}

\begin{flushright}
{\bf Dylan Casey: guitar and vocals, 2001}
\end{flushright}

\newpage
\mbox{}
\newpage
\chapter{A Preamble to Neutrino Physics}
\pagenumbering{arabic}
The musical description of neutrinos by Dylan Casey in his 
song ``Neutrino of Love'' is really fantastic. Yes indeed,
neutrinos are elusive, mysterious, yet abundant. Despite that 
(or because of that!), even after fifty years of it's discovery, 
it still poses many mysteries and creates challenges to the 
physicists who want to detect it. 
Like electrons, they are elementary particles. F. Reines would narrate 
neutrino as, it is ``...the most tiny quantity of reality ever imagined by a
human being''.

Neutrino physics is a very intense and exciting field of research 
having wide range of implications in high energy physics, quantum field theory, 
cosmology, astrophysics, nuclear physics, and geophysics. 
Marvellous results on neutrino oscillations in the last several years have 
triggered a lot of enthusiasm and interest in neutrinos, from experimental 
as well as theoretical point of view. One of the most important facts is that 
neutrino physics is a data driven field - for several years now, new data are 
pouring at an outstanding rate. Our understanding of neutrinos has improved
dramatically in the past ten years and there is no doubt that neutrino oscillation 
is an exclusive example of experimental evidence for physics beyond the 
Standard Model of particle physics. This success sets a fantastic example 
of a road-map in which both theoretical understanding and experimental achievements
have walked hand in hand to provide us with the first evidence of physics beyond the
Standard Model. These developments culminated in the Nobel prize 
for physics in the year 2002, which was awarded to two pioneers in neutrino 
physics. Masatoshi Koshiba was awarded the prize for the
detection of neutrinos from a supernova and Ray Davis Jr. for his detection 
of solar neutrinos. 

Neutrino physics is now poised to move into the precision regime.
Active attempts are under way to commence the era of precision neutrino
measurement science which will surely widen the horizon of our knowledge
about neutrinos. A number of high-precision neutrino oscillation experiments 
have been contrived to sharpen our understanding about these tiny particles.
This is the right time to ask how different planned/proposed next generation 
experiments in the coming decades would perform to explicate the nature of 
neutrinos and our thrust for new physics. This thesis is an effort to have a 
look on several aspects of these issues.

\section{\fbox{Neutrino In a Nutshell}}

Neutrinos are electrically neutral particles of spin ${1 \over 2}$
with a very tiny mass, almost 500 000 times smaller than the mass of 
the electron, which itself is 2000 times smaller than the proton mass. 
There are at least three species (or flavours) of very light neutrinos,
$\nue$, $\numu$ and $\nutau$, which are left handed, and their antiparticles,
$\anue$, $\anumu$ and $\anutau$, which are right handed. 
After the photon, the neutrino is the most abundant particle 
in the Universe~: each cubic meter of the Universe contains about 
30 million neutrinos, which are remnant from the Big Bang, similar 
to the well known cosmic microwave background. It also arrives ``unscathed'' 
from the farthest reaches of the Universe, carrying information about its source.
The interactions of neutrinos are mediated by heavy 
$W^{\pm}$ and $Z^{0}$ bosons and therefore at low energies they talk 
feebly with ordinary matter and pass through the Earth very much like 
light through a crystal. If a matter target as big as Earth is placed 
in front of 100 billion neutrinos, only one of them is likely to interact 
with it. The mean free path of a 1 MeV neutrino in lead is about 1 light year! 
Therefore neutrino detection requires very large detectors and/or
very intense neutrino beams.

\section{\fbox{Neutrino Odyssey}}

Let us have a look at the incredible journey of discovery into one of
Nature's most elusive particles. In a letter to colleagues on 4th December,
1930, Wolfgang Pauli \cite{pauli} postulated the existence of neutrinos to 
guarantee the energy conservation in radioactive beta-decay. After the discovery 
of the neutron by James Chadwick two years later, it was first speculated that 
the particle predicted by Pauli could be the neutron. However, soon it was 
realized that Pauli's particle had to be much lighter than the neutron. 
In 1933, Enrico Fermi introduced the name neutrino, where he used the Italian 
syllable ``-ino'' to indicate ``small neutron''. More than two decades after 
Pauli's letter proposing the neutrino, in 1956, Clyde Cowan and Frederick Reines
\cite{cowan_reines} observed the antineutrinos (the antimatter partners of 
neutrinos) emitted by a nuclear reactor.
This neutrino is later determined to be the partner of the electron. 
In 1969, neutrinos produced by the Sun's burning were detected 
by Ray Davis with a detector based on Chlorine in an underground laboratory 
in the Homestake mine in USA. This experiment reported that less than half 
the expected neutrinos were detected. This originated the long-standing 
``solar neutrino problem''. The scope that the missing electron neutrinos 
may have transformed into another type (undetectable to this experiment) 
was soon suggested, but the lack of our knowledge of the solar model on 
which the expected neutrino rates were based was initially considered a more 
likely explanation. 

In 1987, neutrinos from a supernova in the Large 
Magellanic Cloud were also detected. Only 19 events were observed 
\cite{sup1,sup2,sup3} and they established the standard picture of core-collapse
supernovae. In recent years, several experiments could confirm the existence 
of neutrino oscillations. In 1998, the Super-Kamiokande experiment \cite{sk1} 
reported the evidence for oscillations of atmospheric neutrinos. It was a
crucial juncture for neutrino physics. Neutrino oscillation demands that
neutrinos do have a mass and the observation of large mixing angle was
completely beyond the range of our imagination because in analogy to the 
quark mixing, it was the common belief that if neutrinos mixed at all then
the mixing should be small.

The year 2002 was a spectacular year for neutrino physics.
The neutral current (NC) data of the SNO \cite{sno1} solar neutrino experiment 
provided an independent determination of the total flux of active 
neutrinos from the Sun. The combined SNO and other solar neutrino 
data finally could establish an explanation of the longstanding 
solar neutrino problem in terms of neutrino oscillations.
The KamLAND \cite{kl,kltalk} reactor neutrino experiment confirmed 
the oscillation hypothesis observing disappearance of $\anue$ and 
constrained the mixing parameters to the so-called 
LMA-MSW \cite{lma_msw} solution.
The K2K \cite{k2k_ahn} experiment is the first long baseline
experiment which uses novel man-made accelerator beams.  
This experiment provided an independent confirmation of the 
explanation for the observed atmospheric neutrino anomaly by supplying the 
data which is consistent with it's oscillation interpretation.  

All the results from these dedicated experiments forced us to recall
the pioneering work by Gribov and Pontecorvo \cite{p1,p2}. In 1968, they
showed that flavour conversions can arise if neutrinos are massive and mixed.
Neutrino oscillation is the only phenomenon which can describe the outcome of
all these experiments and it can easily explain the disappearance of both 
atmospheric $\numu$'s and solar $\nue$'s. There are two recent very good 
reviews \cite{maltoni_garcia,strumia_vissani} on this topic in general which
can give us more insight.

\section{\fbox{What are the Main Sources of Neutrinos?}}

Neutrinos are the most common matter particles in the universe.
Neutrinos are produced via weak interactions (like beta-decays in atomic nuclei). 
In number, they exceed the constituents of ordinary matter 
(electrons, protons, neutrons) by a factor of ten billion. 
We have used neutrinos from many different origins with different 
energy ranges to study neutrino oscillations and the properties 
of neutrino sources. Mainly we can classify the neutrino sources into
two categories~: 

\begin{enumerate}

\item
{\bf Natural sources of neutrinos.}
\item
{\bf Artificial man-made sources of neutrinos.}

\end{enumerate}  

\subsection{Natural Sources of Neutrinos}

\subsubsection{\fbox{The Sun}}

One of the strongest neutrino sources is our Sun. The Sun shines not only 
in light but also in electron neutrinos produced in the thermonuclear
reactions which generate the solar energy. These reactions take place via
two main chains, the $pp$ chain and the CNO cycle. There are five reactions 
which produce $\nu_e$ in the $pp$ chain and three in the CNO cycle.
Both chains result in the overall fusion of protons into $^4$He~:
\begin{equation}
    4p \to ^4\rm{He} + 2 e^+ + 2\nu_e + \gamma,
\end{equation}
where the energy released in the reaction, 
$Q = 4m_p - m_{^{4}{He}} - 2 m_e \simeq 26$ MeV, is mostly radiated through 
the photons and the neutrinos carry only a small fraction of it, 
$\langle E_{2\nu_e}\rangle = 0.59$ MeV. Hence, the observation of solar
neutrinos provides direct evidence for the nuclear process in the 
center of the sun. Moreover, current solar neutrino data allow a 
quantitative test of the Standard Solar Model (SSM) \cite{bahcall}. 
By the measurement of the solar neutrino flux the temperature in the 
center of the sun can be determined with the impressive accuracy of 1\%.

The Sun emits about $2\times 10^{38}$ electron neutrinos per second, 
leading to the neutrino flux at the surface of the earth of 
$\sim 6\times 10^{10}$ $cm^{-2} s^{-1}$ in the energy range 
$E \le 0.42$ MeV and $\sim 5\times 10^{6}$ $cm^{-2} s^{-1}$ 
in the energy range 0.8 MeV $\lsim E \le 15$ MeV. The detection
mechanism of solar neutrinos is very sophisticated and involves  
mainly the radiochemical processes (Homestake, Gallex, 
Sage, GNO detectors) \cite{solar,borex} and the water \chr techniques (Kamiokande,
SuperKamiokande, SNO detectors) \cite{sk1,sno1,sk2,sno2}.

\subsubsection{\fbox{The Earth's Atmosphere}}

Earth's atmosphere is another crucial source of electron and muon neutrinos and 
their antiparticles which are created in the hadronic showers induced by primary
cosmic rays. Atmospheric neutrinos were first detected in the 1960's by the
Kolar Gold Field experiment in India \cite{kgf} and the underground experiments 
in South Africa \cite{africa}. The following chain of reactions depicts the 
main mechanism of production of the atmospheric neutrinos~:
\begin{equation}
\begin{array}{lllll}
p(\alpha, ...)+Air&\rightarrow &\pi^{\pm}(K^{\pm})&+ & X  \\
&   &\pi^{\pm}(K^{\pm})&\rightarrow
&\mu^{\pm}+\nu_{\mu}(\bar{\nu}_{\mu})\\
& & & &\mu^{\pm}\rightarrow e^{\pm}+\nu_{e}(\bar{\nu}_{e})+
\bar{\nu}_{\mu}(\nu_{\mu})
\end{array}
\label{nuprod}
\end{equation}

Atmospheric neutrinos can be directly detected in large mass underground
detectors predominantly by means of their charged current (CC) interactions~:
\begin{eqnarray}
&  \nu_{e}(\bar{\nu}_{e})+A\rightarrow e^{-}(e^{+})+X\,,
\nonumber \\
&  \nu_{\mu}(\bar{\nu}_{\mu})+A\rightarrow \mu^{-}(\mu^{+})+X\,.
\label{nudet}
\end{eqnarray}

Atmospheric neutrinos cover a wide range of energy starting from few
MeV to hundreds of GeV. The typical flux of atmospheric neutrinos at 
the earth's surface is $\sim 10^{-1}$ $cm^{-2} s^{-1}$.
These neutrinos are observed in underground experiments with bigger and 
better detectors using different detection techniques and leading to different 
type of events depending on their energy. In the last ten years, the high 
precision and large statistics data from the Super-Kamiokande \cite{sk1} 
experiment (using water \chr detectors) has played an important role to 
solve the atmospheric neutrino puzzle through the concept of neutrino oscillation.  
It has received important confirmation from the iron calorimeter detectors 
Soudan2 \cite{soudan} and MACRO \cite{macro}. In June 1998, in the Neutrino98 
conference, Super-Kamiokande collaboration presented evidence of $\nu_\mu$ 
oscillations based on the angular distribution for their contained event data sample.

Recently, in India, to observe atmospheric neutrinos, the proposal for a 
large magnetized iron calorimeter detector (ICAL) with charge identification 
capability is being evaluated by the INO \cite{ino} collaboration.
We will discuss about this detector in detail later. 

\subsubsection{\fbox{The Earth's Crust}}

The Earth contains a certain amount of natural radioactivity,
and the decay of these radioactive elements is an important and perhaps 
main source of geothermal heat. These same decays also generate particles 
known as geoneutrinos. ``Geoneutrinos'' are electron antineutrinos
produced by beta-decays of the unstable, radioactive nuclei in the decay 
chains of $^{238}$U, $^{232}$Th, $^{40}$K. The amount of neutrinos coming 
from this natural radioactivity is huge~:~about 6 millions per second and per 
cm$^2$ with the energy E $\lsim 1$ MeV.

A careful observation of the arrival directions of neutrinos generated
in the decay of natural radioactive elements in the Earth's interior
can give us a three-dimensional view of the Earth's composition 
and shell structure. This will provide a new and detailed understanding 
of the origin of the Earth's geothermal heat, and will finally answer the 
question of how much heat comes from radioactive decays, and how much is 
``primordial'' heat leftover from the birth of the Earth. 
The mapping of the Earth's interior might also help give answers to such questions 
as ``What powers the magnetic field of the Earth?'' and 
``What dominates the geodynamo?''. To actually take a neutrino picture of the Earth 
is quite challenging technically, but not impossible.

Recently the KamLAND experiment, which was primarily designed to
measure anti-neutrinos from nuclear reactors, reported 9 events 
\cite{geoneutrino} due to geoneutrinos. This marks the first detection 
of neutrinos from the Earth's interior, and already demonstrates that 
radioactivity is an important heat source for the Earth.

\subsubsection{\fbox{The Supernovae within our galaxy}}

The life of a star ends often with a huge explosion called Supernova, which
can be even brighter than a whole galaxy. However, only a tiny amount of
the total released energy is emitted as light. About 99\% of the energy 
is released in the form of neutrinos having energies in the range 10-30 MeV.
Roughly, they emit $\sim 6\times 10^{58}$ neutrinos and antineutrinos 
of all flavours over the time interval of about ten seconds. 
The neutrino luminosity of a gravitational collapse-driven supernova 
is typically 100 times its optical luminosity.
The neutrino signal emerges from the core of a star promptly after
core collapse, whereas the photon signal may take hours or days to
emerge from the stellar envelope. The neutrino signal can therefore
give information about the very early stages of core collapse,
which is inaccessible to other kinds of astronomy.

In 1987, there was a Supernova in the Large Magellanic Cloud within 
our galaxy. Indeed, about 19 neutrinos have been observed, which 
confirmed our basic understanding of the Supernova explosion mechanism. 
Neutrinos from gravitational collapse can be detected in various ways. 
For water \chr detectors, such as Super-Kamiokande, the most important 
detection reaction is the absorption of electron antineutrinos on protons~:
\begin{equation}
\anue + p \to n + e^{+}.
\label{eq:sup_detec}
\end{equation}
The positron from this reaction, which retains most of the energy of 
the incoming neutrino, is detected from it's \chr light.

\subsubsection{\fbox{The Big-Bang}}

The most extraordinary explosion of all, the Big Bang, created
more neutrinos than any other source which has existed since.
The ``standard'' model of the Big-Bang predicts, like for the 
photons, a cosmic background of neutrinos. These relic neutrinos 
still exist, have a number density of about 110 $cm^{-3}$ for each
neutrino species and a black-body spectrum with the average energy
of about 0.0005 eV. The energy of these neutrinos is too small so that
no experiment, even very huge, has been able to detect them.

\subsubsection{\fbox{The Ultra-High Energetic Cosmic Neutrino Sources}}

Astrophysical neutrinos can also be produced with remarkably high
energies. Neutrinos born in the center of an active galaxy can arrive 
on Earth with more energy than we will ever be able to create with a 
terrestrial accelerator. Gamma-ray bursts (GRB) and active galactic 
nuclei (AGN) jets have been suggested as sources of high-energy, 
$> 10^{14}$ eV, neutrinos, with fluxes that may be detectable 
in a kilometer-squared effective area telescope \cite{tele}.

\subsection{Artificial Man-made Sources of Neutrinos}

\subsubsection{\fbox{The Nuclear Reactors (Power Plants)}}

Electron antineutrinos with $E \sim$ MeV are produced copiously 
in the process of generating electrical power in nuclear power plants 
using controlled fission technique. A 3 GW plant releases about 
$7.7\times 10^{20}$ $\anue$ per second and creates a flux of 
$\sim 6\times 10^{11}$ $cm^{-2} s^{-1}$ at 100 $m$. Due to the low 
energy, $e$'s are the only charged leptons which can be produced in the 
neutrino CC interaction. If the $\anue$ oscillated to another flavour, 
its CC interaction would not be observed. Therefore only disappearance 
experiments can be performed with reactors.

The KamLAND experiment \cite{kl}, a 1000 ton liquid
scintillation detector, is currently in operation in the Kamioka mine
in Japan. This underground site is located at an average distance of
150-210 km from several Japanese nuclear power stations.
This experiment has played a crucial role to establish the fact that the
solar neutrino puzzle can be explained by the so-called 
LMA-MSW \cite{lma_msw} solution. Gosgen \cite{gosgen}, Krasnoyarsk \cite{krasnoyarsk},
Bugey \cite{bugey}, CHOOZ \cite{CHOOZ} and Palo Verde \cite{paloverde}
are the examples of reactor experiments which are performed at 
relatively short or intermediate baselines. It is worth-while to 
mention here that none of these experiments find a positive evidence
of flavour mixing.

\subsubsection{\fbox{The Particle Accelerators}}

Now neutrino physics has entered into precision age from discovery era. 
Only in a well tuned, fully optimized environment will it be possible
to perform precision measurements of neutrino oscillation parameters.
A high intensity neutrino source with known spectrum is most desirable 
for precision measurements, the consensus direction for the future.
Man-made accelerator based neutrino beams of the energy ranging typically 
between 30 MeV to 30 GeV are the novel, intense sources of neutrinos,
an important tool for studying neutrino properties. Beam shape parameters 
play a very key role for the measurement of oscillation length, while the
absolute normalization is crucial for the determination of the mixing angle.
One of the most important features of the neutrino beam is that   
one can control the flux of the produced neutrinos and can tune
the main parameters that govern the systematic uncertainties on the
neutrino fluxes. Currently, there are three widely different schemes for 
producing neutrino beams and they mainly differ from each other on the issue 
of what particle is decaying (pion decay, muon decay and radioactive ion decay) 
to give rise to the neutrinos. 
For a detailed discussion of future beams and
their comparison see \cite{report1,report2} and references therein.

\begin{enumerate}

\item
\fbox{\bf Conventional Neutrino Beams}

Conventional neutrino beams are produced by shooting a target with as 
many protons as can be provided, and then focusing the produced mesons 
(mostly pions and some kaons) into an evacuated decay pipe where they 
are allowed to decay. The mesons will decay primarily to muons and
muon neutrinos. This decay chain is quite similar in nature to the decay
process through which the atmospheric neutrinos are produced. But one
should keep in mind that the energy spectrum of the conventional neutrino 
beam is quite different compared to that of atmospheric one. 
From meson decay (two-body decay) kinematics it follows that the neutrino 
energy is given by
\begin{equation}
E_\nu = {\frac{m^2_{\pi(K)}-m^2_\mu}{m^2_{\pi(K)}}}\frac{E_{\pi(K)}}{(1+\gamma^2
\theta^2)},
\label{eq:pion_decay}
\end{equation}
where $\gamma$ is the Lorentz boost of the parent meson, $E_{\pi(K)}$
it's energy and $\theta$ the angle of the neutrino with respect to the
meson flight direction. The polarity of the focusing device has to be 
reversed to produce a beam of $\anumu$. In both the cases, there is
always some contamination of $\nue$ or $\anue$ due to the three-body
decays of the kaons and daughter muons.

We can classify the conventional neutrino beams into three categories~:
the Wide Band Beams (WBB), the Narrow Band Beams (NBB) and the Off-Axis 
Beams (OAB). 

The main feature of WBB is that they have wide energy 
spectrum with high neutrino flux. They are perfectly suited to
make discoveries. But they have some limitations. In a WBB, the
irreducible fraction of $\nue$ originating from the meson decays
is quite significant and these $\nue$ produce electrons inside the detector
which play the role of intrinsic beam related background when we try to
extract interesting results using the $\numu \rightarrow \nue$
appearance channel. Another problem is that WBB comes with a tail
of high energy neutrinos and often these neutrinos produce $\pi^0$
inside the detector via NC process. Now, the early showering of gamma's from 
the $\pi^0$ decay can be misidentified for a $\nu_e$ CC interaction.
If the signal corresponds to a small part of the energy spectrum, 
it could be overwhelmed by the beam induced background coming from 
the region which is outside the signal.

The NBB are quite opposite in nature. This facility can provide us almost 
monochromatic energy spectra by judiciously choosing a small momentum 
bite of the parent $\pi$ and $K$. However, the neutrino yield is quite
suppressed which causes problem for oscillation searches.

The OAB \cite{offaxis} is a classic example of a neutrino beam with high 
flux and a narrow energy spectrum. This technique requires designing a
beam-line which can produce and focus a wide range of mesons in a
given direction (as in the WBB case), but then placing the detectors
at an angle with respect to that direction. Since the pion decay is a
two-body process (see Eq. \ref{eq:pion_decay}), we can obtain neutrinos 
of a given energy at a given angle between the pion direction 
and the detector location. But the most important feature to be noted
that at this given angle, the energy of the neutrino produced in the
decay of the pion becomes practically independent of the boosted
pion energy. Now, if we place a detector at this particular angle
with respect to the decay pipe then it will see a neutrino beam with
a very narrow energy spread compared to the on-axis beam.
Furthermore, this off-axis technology helps us to reduce the 
background coming from the intrinsic $\nue$ contamination  
in the beam and a smaller fraction of high energy tails reduces the 
background from NC events.
As a result, it improves the signal-to-background ratio a lot.

But it is important to mention that, independent of the adopted solution, 
all conventional neutrino beams have some common problems. 
The first major drawback is that the hadron production in the proton-target 
interaction has large uncertainties due to lack of data and theoretical 
difficulties in describing hadronic processes. It creates obstacles in 
predicting the neutrino flux and spectrum with good accuracy.
Secondly, in addition to the dominant flavour in the beam (typically $\numu$) 
there is a contamination (at the few percent level) from
other flavours ($\anumu$, $\nue$ and $\anue$) resulting into a 
``multiflavour'' neutrino beam. 

A recently terminated conventional beam experiment is the K2K
experiment \cite{k2k_ahn} where a neutrino beam was directed towards
the Super-Kamiokande detector from the KEK accelerator. This 
experiment has already confirmed the disappearance of $\numu$ as 
predicted by atmospheric neutrino data. The MINOS \cite{minos}
experiment in US, and the CERN to Gran Sasso (CNGS) experiment
OPERA \cite{opera} are the two conventional beam experiments which
are collecting data now. Another CNGS experiment ICARUS \cite{icarus}
will start taking data soon.

\item
\fbox{\bf SuperBeams}

The technology of conventional beam experiments with some 
technical upgrades is known as superbeam. These experiments 
are `super' in the sense that they will use proton beams of
unprecedented strength around 1 - 4 MW and detectors with 
large fiducial mass. All superbeams use a near detector for a 
better control of the systematics. The most advanced superbeam 
proposals are the J-PARC to Super-Kamiokande experiment (T2K)
\cite{t2k} in Japan, and the NuMI off-axis experiment (NO$\nu$A)
\cite{nova}, using a neutrino beam produced at Fermilab in US.

\item
\fbox{\bf Neutrino Factory}

\begin{figure}[t]
\begin{center}
\includegraphics[width=12.0cm, height=8.0cm]{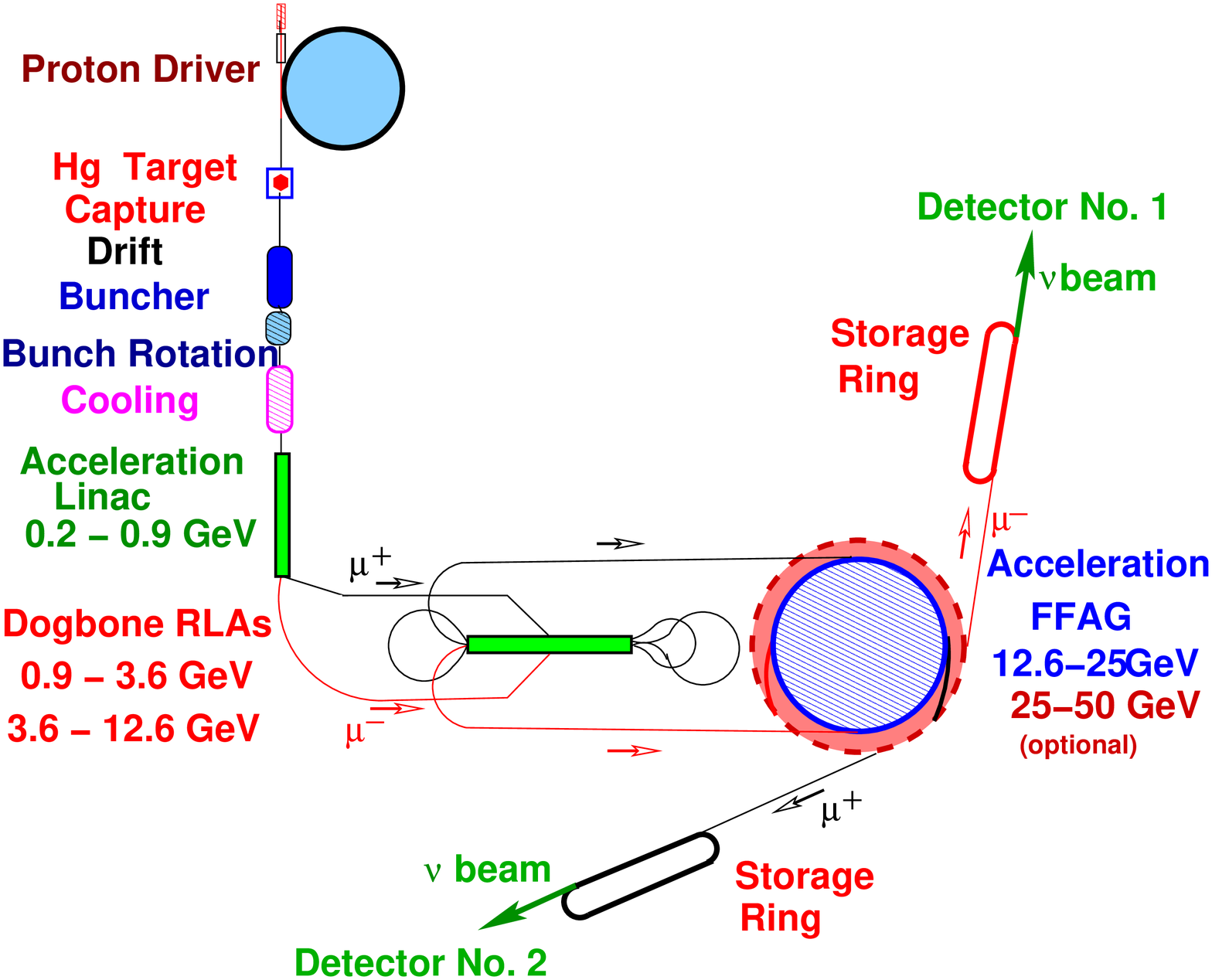}
\caption{\label{fig:nf_layout}
The schematic layout of the neutrino factory set-up. The web 
address http://www.cap.bnl.gov/mumu/project/ISS/ is the source of this figure.}
\end{center}
\end{figure}

The term ``neutrino factory'' \cite{geer} has been associated to 
describe neutrino beams created by the decays of high energy muons 
which are circulated in a storage ring with long straight sections. 
The decay of muons in these straight sections produces an intense,
well known and pure beam of $\numu$ and $\anue$. If $\mu^+$ are stored, 
$\mu^+\rightarrow e^+\nue\anumu$ decays generate a beam consisting 
of equal numbers of $\nue$ and $\anumu$. The overall layout is shown
in Fig. \ref{fig:nf_layout}. The muons are obtained via 
pion decay. To make a muon beam which can be accelerated, first the 
muons have to be cooled in phase space with the help of ionization 
cooling technique which is being studied by the MICE experiment 
\cite{mice}. There are other ways also being examined to tackle 
this problem. 
Several design studies have been performed in this direction 
in Europe, the United States and Japan \cite{report1,others}.  
Typical neutrino factories are being considered with
muon energies ranging from 20 GeV to 50 GeV with $\sim 10^{21}$
useful muon decays per year inside the storage ring.

In a neutrino factory experiment with stored $\mu^-$, the appearance 
channel, $\anue \to \anumu$ give rise to $\mu^+$ in the detector via 
CC deep inelastic scattering. These anti-muons are called wrong sign
muon events, since they have the opposite charge relative to the muons
produced by the $\numu$ in the beam itself. These wrong sign muons have 
to be cleanly separated inside the detector from the muons created by the 
surviving $\numu$. This charge identification can be achieved with a 
magnetized iron calorimeter detector like the proposed ICAL@INO \cite{ino}.

Muon decay is well known and it will provide the neutrino flux and
spectrum with minimal systematic uncertainties, compared to 
conventional neutrino beams. Radiative effects on the muon decay 
is negligible. Ultimately, the flux from a neutrino factory is expected 
to be known with a precision of the order of $10^{-3}$. Another vital
point to be noted is that a neutrino factory beam has a sharp cut-off at
the energy of the stored muons which reduces the background from NC
events. In a neutrino factory, the neutrinos with higher energy opens
up the possibility to study the oscillation channels like
$\numu \to \nutau$ and $\nue \to \nutau$ because one can produce 
tau leptons inside the detector \cite{silver_nufact}.

\item
\fbox{\bf Beta-beam}

\begin{table}[t]
\begin{center}
\begin{tabular}{||c||c||c||c||c||c||c||} \hline \hline
   Ion & $\tau$ (s) &
$E_0$ (MeV)
   & $f$& Decay fraction & Beam \\
\hline
  $^{18} _{10}$Ne &   2.41 & 3.92&820.37&92.1\%& $\nu_{e}$    \\
  $^6 _2$He   &   1.17 & 4.02&934.53&100\% &$\bar\nu_{e}$    \\
\hline
 $^{8} _5$B& 1.11 & 14.43&600872.07&100\%&$\nu_{e}$    \\
 $^8 _3$Li& 1.20 &13.47 &425355.16& 100\% & $\bar\nu_{e}$    \\
\hline \hline
\end{tabular}
\caption{\label{tab:ions}
Beta decay parameters: lifetime $\tau$,
electron total end-point energy
$E_0$, $f$-value or Fermi integral
and decay fraction for various ions~\cite{beta}. }
\end{center}
\end{table}

\begin{figure}[t]
\includegraphics[width=8.0cm, height=7.0cm]{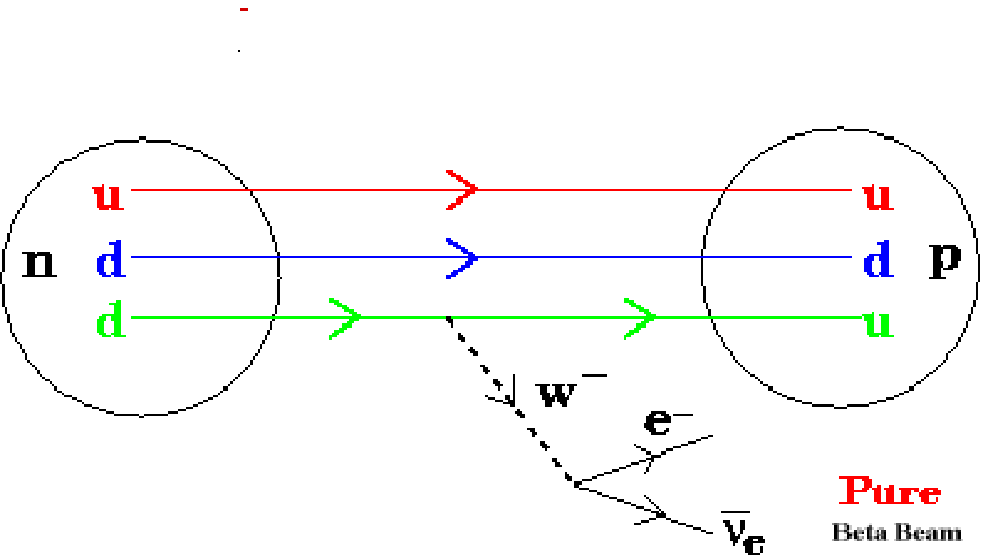}
\vglue -7.0cm \hglue 8.8cm
\includegraphics[width=8.0cm, height=7.0cm]{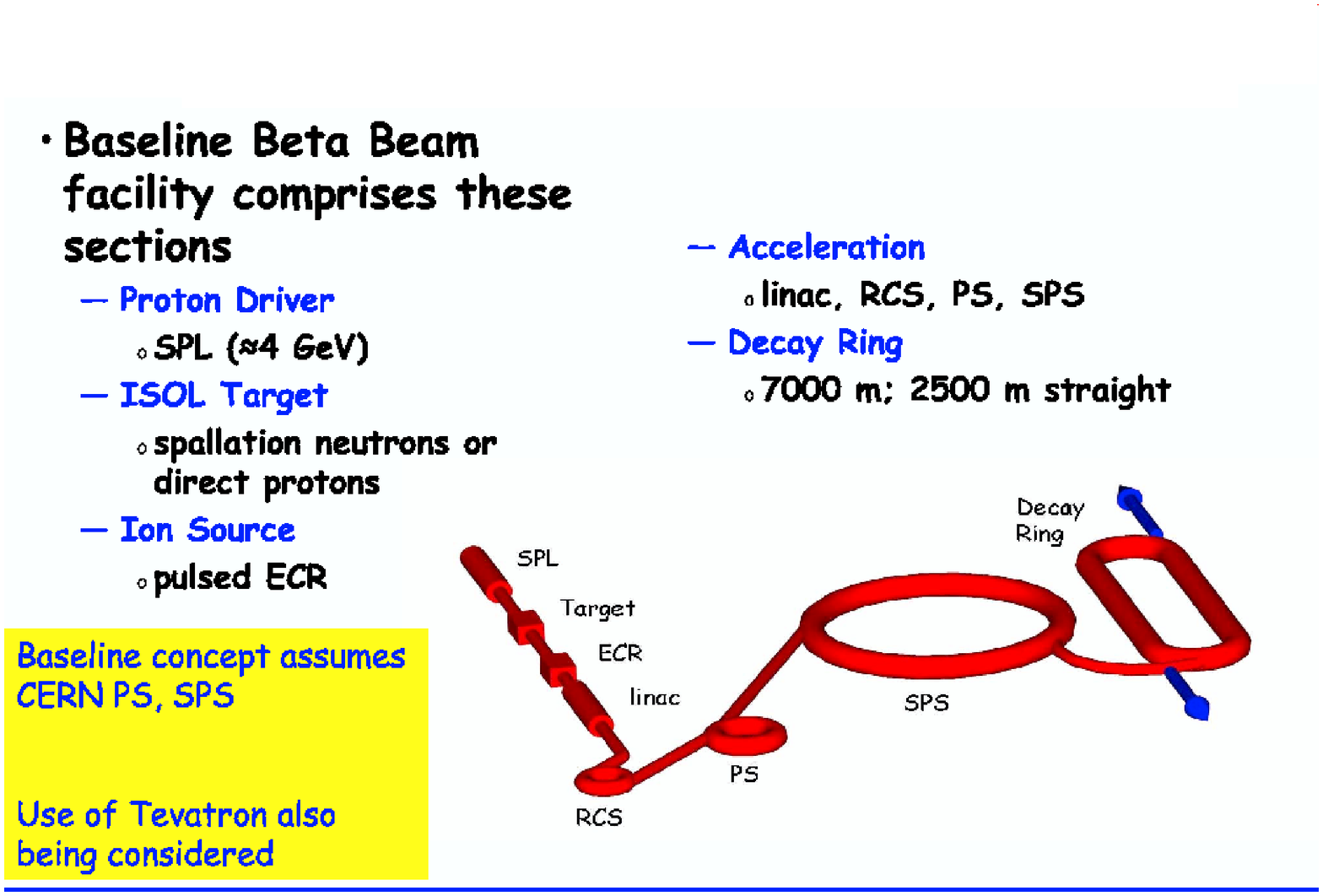}
\caption{\label{fig:beta_layout}
Left panel depicts the beta-decay process which is the source of
pure beta-beam. The proposed schematic layout of the beta-beam set-up at 
CERN (http://beta-beam.web.cern.ch/beta-beam/task/index.asp) has been 
displayed in the right panel.}
\end{figure}

Zucchelli \cite{zucc} put forward the novel idea of a beta-beam
\cite{volpe,cernmemphys,
paper1,betaino1,betaino2,pee,bboptim,twobaseline,rparity1,rparity2,
conf_talks,sanjib_talks,oldpapers,donini130,doninibeta,newdonini,bc,bc2,fnal,betaoptim,
doninialter,boulby}, which is based on the 
concept of creating a pure, well understood, intense, collimated beam 
of $\nu_{e}$ or $\bar\nu_{e}$ through the beta-decay 
(see left panel of Fig. \ref{fig:beta_layout}) of completely 
ionized radioactive ions. Firstly, radioactive nuclides are created by 
impinging a target by accelerated protons. These unstable nuclides are
collected, fully ionized, bunched, accelerated and then stored in a
decay ring (see for e.g. \cite{lindroos,betabeampage}).
The decay of these highly boosted ions in the straight sections of the 
decay ring \cite{jacques} produces the so-called beta-beam.
Feasibility of this proposal and its physics potential is being studied 
in depth \cite{iss}, and will take full advantage of the existing
accelerator complex at CERN and/or FNAL (Fermi National Accelerator Laboratory). 
The proposed schematic layout of the beta-beam set-up at 
CERN\footnote{A detailed R\&D on this issue is
being pursued by Mats Lindroos and his collaborators.} is given in the 
right panel of Fig. \ref{fig:beta_layout}. 
The main future challenge lies in building an intense proton driver and 
the hippodrome-shaped decay ring which are essential for this programme. 

It has been proposed to produce $\nu_e$ beams through the decay of 
highly accelerated $^{18}$Ne ions 
({$^{18} _{10}$Ne}  $\rightarrow$  $^{18} _{9}$F  $+$  $e^+ +  {\nue}$)
and ${\anue}$ from $^6$He 
({$^6 _2$He}  $\rightarrow$  $^6 _3$Li  $+$  $e^- +  {\anue}$)
\cite{zucc,jacques}. More recently, $^{8}$B 
({$^{8} _{5}$B}  $\rightarrow$  $^{8} _{4}$Be  $+$  $e^+ +  {\nue}$)
and $^{8}$Li ({$^8 _3$Li}  $\rightarrow$  $^8 _4$Be  $+$  $e^- +  {\anue}$)
\cite{rubbia,mori} with much larger end-point energy have been suggested as 
alternate sources since these ions can yield higher energy $\nue$ and 
$\anue$ respectively, with lower values of the Lorentz boost 
$\gamma$ \cite{betaino1,betaino2,bboptim,twobaseline,rparity1,rparity2,
newdonini,doninialter}. 
Details of the four beta-beam candidate ions can be found in 
Table~\ref{tab:ions}. $^8$B and $^8$Li decay to the broad $2^+$ 
first excited state of $^8$Be which has an energy width of $\sim 1.5$ MeV. 
It may be possible to store radioactive ions producing 
beams with both polarities in the same ring. This will enable running the 
experiment in the $\nue$ and $\anue$ modes simultaneously.

In the low $\gamma$ design of beta-beams, the standard luminosity
taken for the \neon and \he are $1.1 \times 10^{18}$ ($\nu_e$) and 
$2.9\times 10^{18}$ ($\bar{\nu}_e$) useful decays per year,
respectively \cite{beamnorm}. We will discuss about this newly 
proposed facility later in great detail. 

\end{enumerate}

\section{\fbox{Neutrino~: ``NU'' Horizons}}

We live in an exciting time when the light of new discoveries is breaking 
apart our long-held picture of the Standard Model. This revolution began in part
with the widely confirmed assertion that neutrinos have mass, and it 
will continue to be waged by upcoming neutrino experiments. Spectacular 
results from a series of experiments over the last four decades 
\cite{sk1,kl,k2k_ahn,solar,borex,sk2,CHOOZ,minos,recent_chooz,k2k,limits,thomas} 
have firmly established the phenomenon of neutrino oscillation and paved 
the way for the ``golden'' age of neutrino physics. Since neutrino 
oscillations can occur only if there is a mass difference between at least
two neutrinos, an observation of this effect proves that at least one non-zero
neutrino mass exists.

Neutrinos are strictly massless in the Standard Model of particle physics and 
the finite neutrino masses required by the experimental data provide 
the first hint for 
physics beyond the Standard Model, and make an extension of the theory necessary. 
No doubt that this has put the Standard Model in a paradoxical situation. 
Moreover, the fact that 
neutrino masses are so tiny (very much smaller than that of any other known fermion) 
should find an explanation in the new theory.

Recent discoveries on neutrinos might provide unique information on a more
complete theory of elementary particles. The sensitivity of neutrino experiments
to very tiny mass scales might provide the scope to learn something about
physics at very high energy scales (i.e., at very small distances), which
will never be accessible in particle accelerator experiments. Therefore,
information from neutrinos is complementary to the one from accelerator
experiments, and it may provide a key to a so-called Grand Unified Theory,
in which the electromagnetic, the weak and the strong interactions are
unified to one fundamental force.

Another puzzle of modern physics is the origin of matter. In the
so-called Leptogenesis mechanism the origin of matter in the very
first moments after the Big Bang is related to neutrinos. In that
theory the small asymmetry between matter and anti-matter is generated
by processes involving neutrinos in the early stage of the Universe.
In this way a theory of neutrino may even provide the reason for our
existence. Neutrinos have played a key role in shaping the Universe
as we see today. We have just started our journey in the mysterious 
world of neutrinos, a tiny creature of Nature. A long journey is 
waiting for us ahead and many experimental approaches are required
to get the full view. In the near future, the Large Hadron Collider 
(LHC) will start it's quest for Higgs and it is expected that the LHC
will explore the mechanism of electroweak symmetry breaking and provide
clues of new heavy degrees of freedom. This will certainly boost up the
future road map of the neutrino physics programme and it is for sure that
neutrino physics, a bit player on the physics stage in yesteryears, has now 
donned a central role and will play a crucial part in the high energy physics
programme.

\section{\fbox{Layout of the Doctoral Work}}

We organize the description of the doctoral work in the following way. 
The first part of chapter 2 deals with the basic introduction to the 
quantum mechanics of neutrino oscillation in vacuum under both two and 
three flavour frameworks. Then we discuss the importance of matter effects
in neutrino oscillations. In the second half of chapter 2 we take a look
at our present understanding of neutrino parameters and we identify the
major unknowns in the neutrino sector. Finally we close chapter 2 by 
giving a brief note on the future neutrino road-map based on 
long baseline experiments. In chapter 3 we underscore in detail the
physics advantage of an experimental set-up where neutrinos produced
in a beta-beam facility at CERN would be observed in the proposed
large magnetized iron calorimeter detector (ICAL) at the India-based
Neutrino Observatory (INO). The CERN-INO beta-beam set-up offers an
excellent avenue to use the ``Golden'' channel ($\nue \rightarrow \numu$) 
oscillation probability for a simultaneous determination of the neutrino
mass ordering and $\theta_{13}$. The merit of the earth matter effects in the
${\rm {\nu_e \to \nu_e}}$ survival probability at long baselines
in order to cleanly determine the third leptonic mixing angle 
$\theta_{13}$ and the sign of the atmospheric neutrino mass squared difference,
$\Delta m^2_{31}$, using a beta-beam as a $\nue$ source has been
discussed in chapter 4. In chapter 5 we study the possibility of detecting 
new physics signals in a dedicated neutrino beta-beam experiment with the
source at CERN and the detector at the proposed INO. These new physics 
signals arise in the R-parity violating supersymmetric models (RPVSM)
\cite{rparity} due to the flavor-changing neutral current (FCNC) and 
flavor-diagonal neutral current (FDNC) interactions of neutrinos with 
matter in the oscillation stage. In chapter 6 we show that 
a detector placed near a beta-beam storage ring can probe lepton number 
violating interactions that might crop up at the production and the detection 
point, as predicted by supersymmetric theories with R-parity non-conservation.
Finally we end with the conclusion and an outlook in chapter 7.

\newpage
\mbox{}
\chapter{Neutrino Oscillations Revisited}
This chapter is organized as follows. In the first section 
we discuss the role of neutrinos in the Standard Model.
An introduction to the basic formalism of neutrino oscillations 
in vacuum has been given in the subsequent section. We consider
both two and three flavour frameworks. In the following section we
concentrate on the importance of matter effects in neutrino 
oscillations. Then we turn our focus on the present global
understanding of the neutrino mass-mixing parameters. This is
to set the stage for the next section where we take a drive to the
unknown territories of the neutrino sector. Finally we take a look
at the future neutrino road-map. 

\section{\fbox{Neutrinos in the Standard Model}}

\begin{table}[t]
\begin{center}
$$
\boldmath
\begin{array}{||c||c||c|c|c||c||}
\hline\hline
{\rm Particles} & {\rm Notation} & {SU(3)_C}& {SU(2)_L} & {U(1)_Y} & {U(1)_{em}}\\
\hline
&&&&&\\
{\rm Leptons} & l^{I}_L \equiv \pmatrix{\nu^{I} \cr e^{I}}_L & 1 & 2 & -1/2 &
{\begin{array}{c} ~0 \\ -1\end{array}}\\
 & e^{I}_R & 1 & 1 & -1 & -1\\
&&&&&\\
\hline
&&&&&\\
{\rm Quarks} & q^{I}_L \equiv \pmatrix{u^{I} \cr d^{I}}_L & 3 & 2 & ~~1/6 &
{\begin{array}{c} ~~2/3 \\ -1/3\end{array}}\\
 & u^{I}_R & {3} & 1 & ~~2/3 & ~~2/3\\
 & d^{I}_R & {3} & 1 & -1/3 & -1/3\\
&&&&&\\
\hline\hline
\end{array}
$$
\end{center}
\caption{\label{tab:contents}
Matter contents of the Standard Model with their corresponding 
gauge quantum numbers. $I = 1,2,3$ is the generation
index. The electromagnetic charge listed in the last column
is defined as $Q_{em} = T_3 + Y$.}
\end{table}

The Standard Model is a quantum field theory and it unifies 
strong, weak and electromagnetic interactions. The Standard Model 
is based on the gauge group 
$SU(3)_C \times SU(2)_L \times U(1)_Y$, broken spontaneously 
by the Higgs mechanism to $SU(3)_C \times U(1)_{em}$. 
The transformation groups $SU(3)_C$, $SU(2)_L$ and $U(1)_Y$ correspond 
to quantum numbers called colour, weak isospin and weak hypercharge~($Y$) 
respectively. Experimental observations suggest that the matter fields 
are fermionic, consisting of leptons and quarks.
In the Standard Model, these leptons and quarks are chiral fermions~:~
\begin{equation}
\psi_{R, L} = \frac{1}{2} (1 \pm \gamma_5) \psi,
\label{eq:chiral}
\end{equation}
where $\psi = \psi_L + \psi_R$. The left-handed components transform 
as doublets under $SU(2)_L$, whereas the right-handed components are 
$SU(2)_L$ singlets. Within the framework of the Standard Model there are no 
right-handed neutrinos or left-handed antineutrinos. The particle content 
of the Standard Model is depicted in Table~\ref{tab:contents}.
It can be readily seen from Table~\ref{tab:contents} that neutrinos 
do not take part in strong and electromagnetic interactions and they
only undergo weak interactions, that is, they are singlets of 
$SU(3)_C \times U(1)_{em}$. Precision data of the $Z$-decay width
at the $e^+e^-$ collider at LEP indicate that there are three
neutrinos which take part in weak interactions. These are called the
{\em active} neutrinos. Often a scenario with four neutrinos is
discussed. So this fourth neutrino, if it exists, should be {\em
sterile} to weak interactions. This sterile neutrino does not have
any Standard Model interactions and passes undetected through the 
experimental set up. 

In weak interactions, parity is maximally violated and only the 
left-handed states do couple with the gauge bosons $W^{\pm}$ while
the right-handed states do not, unlike in the case of electromagnetism
where both the states couple to the photon with the same strength.  
The unbroken electromagnetic gauge symmetry ensures the masslessness
of photon. But no such fundamental principle predicts the neutrino to
be massless. In the mass-term, the left- and right-handed chiral states
are coupled to each other in the following way~:
\begin{equation}
m \bar\psi \psi = m (\bar\psi_{L} \psi_{R} + \bar\psi_{R} \psi_{L}),
\label{eq:mass_term}
\end{equation}
which can be easily obtained with the help of Eq. \ref{eq:chiral} and using
the properties of $\gamma_{5}$. Therefore the existence of both 
left-handed and right-handed chiral components is mandatory to have a nonzero mass 
of a fermion. But within the Standard Model there is no
right-handed neutrino and as a consequence the neutrino does not have any mass.    
So non-zero neutrino mass, as required by the neutrino oscillation data,
indicates the existence of physics beyond the Standard Model.

As we have already mentioned, the Standard Model has three active 
neutrinos accompanying the charged lepton mass eigenstates, $e$, $\mu$ and $\tau$. 
Therefore there is a possibility of weak CC interactions between the neutrinos 
and their corresponding charged leptons which is given by~:
\begin{equation}
    -{\mathcal{L}}_{CC} =
    \frac{g}{\sqrt{2}} \sum_I \bar{\nu}_{LI}
    \gamma^\mu
    \ell^-_{LI} W_\mu^+ + h.c..
\label{eq:leptons_cc}
\end{equation}
The Standard Model neutrinos also have NC interactions~:
\begin{equation}
    - {\mathcal{L}}_{NC}
    = \frac{g}{2\cos\theta_W} \sum_I \bar{\nu}_{LI}
    \gamma^\mu\nu_{LI} Z_\mu^0.
\label{eq:neutrino_cc}
\end{equation}
Thus, within the Standard Model, all the neutrino interactions 
are described by Eqs. \ref{eq:leptons_cc} and \ref{eq:neutrino_cc}.

\section{\fbox{Oscillations in Vacuum}}

\begin{figure}[t]
\begin{center}
\includegraphics[width=10.0cm, height=8.0cm]{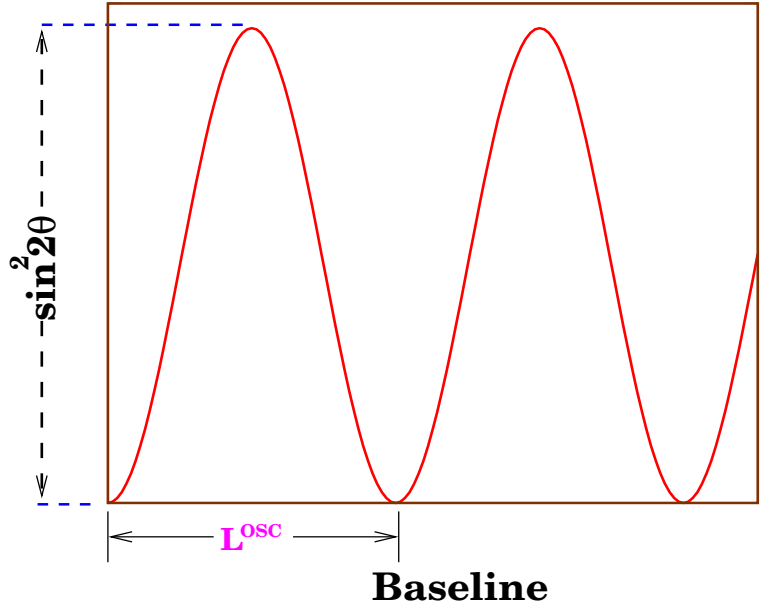}
\caption{\label{fig:osc}
The two flavour $\nu_e$ to $\nu_{\mu}$ oscillation 
probability (Eq. \ref{eq:2flavosc}). The mixing angle $\theta$
governs the oscillation amplitude and the frequency depends on 
$\Delta m^2_{12}=m_{2}^2 - m_{1}^2$, where $m_1$ and
$m_2$ are the masses of the mass eigenstates $\nu_1$ 
and $\nu_2$ respectively.}
\end{center}
\end{figure}

Neutrino oscillation is a simple quantum mechanical phenomenon
in which neutrino changes flavour as it propagates. 
In 1957, Pontecorvo \cite{pontecorvo} gave the concept of 
neutrino oscillation based on a two-level quantum system. 
This phenomenon arises if neutrinos have masses and there is mixing.

In the course of time evolution of a stationary state $|\Psi_k\rangle$
(which is one of the eigenstates of the hamiltonian of a system with the
energy eigenvalue $E_k$), it will remain in the same state and will be 
associated with a phase in the form ($\hbar = c = 1$)~:
\begin{equation}
|\Psi_k(t)\rangle=e^{-i E_k t} |\Psi_k(0)\rangle.
\label{eq:phase}
\end{equation}
Now if we deal with an arbitrary non-stationary state of the system 
which is a superposition of say two eigenstates (with eigenenergies $E_1$ and $E_2$) 
of the hamiltonian of the system, then it will not be in the same state 
after time evolution. The chances that the system will remain in it's initial state 
will be an oscillatory function of time with frequency $(E_2-E_1)$.

In order to implement this mechanism in case of neutrinos, 
we consider the fact that neutrinos $(\nue, \numu, \nutau)$
are produced or detected via weak interactions and therefore
they are referred to as weak-eigenstate neutrinos (denoted as $\nu_\alpha$) 
that means they are the weak doublet-partners of $e^{-}, \mu^{-}, \tau^{-}$ 
respectively. In such a scenario, if we assume that neutrinos are massive, 
then in general, it is not mandatory that the mass-matrix of neutrinos 
written in this weak (flavour) basis will have to be diagonal. 
So, it follows that the mass eigenstate neutrinos $\nu_i, i = 1, 2, 3$ 
(the basis in which the neutrino mass matrix is diagonal) are
not identical to the weak or flavour basis\footnote{The charged lepton 
mass-matrix is diagonal in this basis.} and we have for $n$ number
of light neutrino species  
\begin{equation}
    |\nu_\alpha\rangle = \sum_{i=1}^{n} U^*_{\alpha i} |\nu_i\rangle,
\label{eq:formula1}
\end{equation}
where $U$ is the unitary leptonic mixing matrix known as the 
Pontecorvo-Maki-Nakagawa-Sakata (PMNS) matrix \cite{pmns}.
This matrix is analogous to the CKM matrix in the quark sector.
Because of Eq. \ref{eq:formula1}, the probability of finding a neutrino 
created in a given flavour state to be in the same state 
(or any other flavour state) oscillates with time. Suppose a given source is producing a neutrino 
flux of given flavour $|\nu_{\alpha}\rangle$ at $t=x=0$, the neutrino 
state at a later time $t$ is then
\begin{equation}
    |\nu_\alpha(t)\rangle = \sum_{i=1}^{n} U^*_{\alpha i} |\nu_i(t)\rangle
= \sum_{i=1}^{n} U^*_{\alpha i} e^{- i E_i t} |\nu_i(0)\rangle,
\label{eq:formula2}
\end{equation}
and propagates as an admixture of the mass eigenstates.  
The neutrino oscillation probability, i.e. the transformation probability
of a flavour eigenstate neutrino $|\nu_{\alpha} \rangle$ into another
one $| \nu_{\beta} \rangle$ is then
\begin{equation}
    \label{eq:palbe}
    P_{\alpha\beta} = |\langle\nu_\beta|\nu_\alpha(t)\rangle|^2 =
    |\sum_{i=1}^n \sum_{j=1}^n U^*_{\alpha i} U_{\beta j}
    \langle\nu_j|\nu_i(t)\rangle|^2 \,.
\label{eq:formula3}
\end{equation}
To obtain the oscillation probability for antineutrinos one has to 
replace $U$ by $U^*$. For ultrarelativistic neutrinos 
with small mass one can assume $p_i\simeq p_j\equiv p\simeq E$ and we
have
\begin{equation}
    E_i = \sqrt{p_i^2 + m_i^2} \simeq
    p + \frac{{m_i^2}}{2E},
\label{eq:formula4}
\end{equation}
where $E_i$ and $m_i$ are, respectively, the energy and the mass of
the neutrino mass eigenstate $\nu_i$.

Now using the orthogonality relation 
$\langle\nu_j|\nu_i\rangle = \delta_{ij}$ and with the help of
Eq. \ref{eq:formula4}, we obtain the transition probability  
that an initial $\nu_{\alpha}$ of energy $E$ gets converted to a 
$\nu_{\beta}$ after traveling a distance $L (= t)$ in vacuum as
\be
P_{\alpha\beta} = \delta_{\alpha\beta} 
&-&4\sum_{i< j}^n\mbox{Re}[U^*_{\alpha i}U^*_{\beta j} U_{\beta i} U_{\alpha j}]
\sin^2 X_{ij}\nonumber\\
&+&2\sum_{i<j}^n\mbox{Im}[U^*_{\alpha i}U^*_{\beta j} U_{\beta i} U_{\alpha j}]
\sin 2 X_{ij} \,,
\label{eq:formula5}
\ee
where
\begin{equation}
    X_{ij} = \frac{(m_i^2-m_j^2) L}{4 E} = 1.27 \,
    \frac{\Delta m^2_{ij}}{eV^2} \, \frac{L/E}{m/MeV} \,.
\label{eq:formula6}
\end{equation}
$\Delta m^2_{ij} = m_i^2 - m_j^2$ is known as the mass splitting and neutrino
oscillations are only sensitive to this mass squared difference but not 
to the absolute neutrino mass scale. The transition probability
(depicted by Eq. \ref{eq:formula5}) has an oscillatory behaviour with
oscillation lengths
\begin{equation}
L^{osc}_{ij}=\frac{4\pi E}{\Delta m^2_{ij}}\,\simeq \,2.48\;m\,\frac{E\,\mbox{(MeV)}}
{\Delta m^2_{ij}\, (\mbox{eV}^2)}=\,2.48\;km\,\frac{E\,\mbox{(GeV)}}
{\Delta m^2_{ij}\, (\mbox{eV}^2)}
\label{eq:formula7}
\end{equation}
and the amplitudes are proportional to the elements in the mixing matrix.  
Since neutrino oscillations can occur only if there is a mass difference 
between at least two neutrinos, an observation of this effect proves that 
at least one non-zero neutrino mass exists.

In general, to construct a unitary $n \times n$ matrix we need $n(n-1)/2$ 
angles and $n(n+1)/2$ phases. If neutrinos are Dirac type in nature then
$2n-1$ phases can be absorbed by a proper re-phasing of the left-handed fields, 
leaving $(n-1)(n-2)/2$ physical phases. Therefore CP violation is only 
possible in the case of $n \ge 3$ generations. In the Majorana case there 
is no freedom to re-phase the neutrino fields; only $n$ phases can be 
removed using the charged lepton fields, leaving $n(n-1)/2$ physical phases. 
Out of these $(n-1)(n-2)/2$ are the
usual Dirac phases, while $n-1$ are specific to the Majorana case and are
called Majorana phases. The latter do not lead to any observable effects
in neutrino oscillations and are not discussed further in this thesis.

\subsection{Two-Flavour case}

In a simple case with only two families of neutrinos, the mixing matrix 
depends on a single parameter $\theta$ (known as the mixing angle),
\begin{equation}
U=\left(\begin{array}{cc}
\cos\theta & \sin\theta \\
-\sin\theta & \cos\theta \end{array} \right)\,
\label{eq:formula8}
\end{equation}
and there is a single mass-squared difference $\Delta m^2$.
Using this form of $U$ in Eq. \ref{eq:formula5}, one obtains
\begin{equation}
P_{\nu_{\alpha} \rightarrow \nu_{\beta}} = \sin^2 2\theta~
\sin^2({1.27 \Delta m^2 {L\over E}} ),
\label{eq:2flavosc}
\end{equation}

\begin{equation}
P_{\nu_{\alpha} \rightarrow \nu_{\alpha}} = 1 - \sin^2 2\theta~
\sin^2({1.27 \Delta m^2 {L\over E}} ),
\label{eq:2flavsurv}
\end{equation}
where $\Delta m^2$ is in $\mbox{eV}^2$, $L$ is in m (km) and $E$ in MeV (GeV).
In the above equations, $\theta$ and $\Delta m^2$ (these are fundamental constants
like the electron mass or the Cabibbo-angle) determine the oscillation amplitude
and frequency respectively (see figure \ref{fig:osc}). The various neutrino
sources like the Sun, the atmosphere, reactors and accelerators 
(discussed in the previous chapter) provide neutrinos with average energy 
varying in a wide range and the distances between the sources and the 
detectors are also quite different which offer the possibility to probe
various oscillation frequencies. In order to penetrate a given value of 
$\Delta m^2$, the experiment has to be performed with 
$E/L\approx\Delta m^2$ which is equivalent to $L\sim L^{osc}$.

In an oscillation experiment, if we search for a new flavour then we call 
it an appearance experiment. On the other hand if we look for a reduction in the 
neutrino flux from the source then it is known as a disappearance experiment.  
In both the cases, the measurement of $\Delta m^2$ and $\theta$ are 
correlated with each other. Therefore an error in measuring one of the
parameters causes an additional uncertainty on the other one. To locate
the position of the peak clearly, we need an experiment with good enough
energy resolution ($\Delta E\ll L \Delta m^2$) otherwise the information 
on $\Delta m^2$ gets blurred and the experimental signal becomes energy
independent ($\sim \frac{1}{2}\sin^22\theta$). The proper energy calibration 
of the detector (it provides the absolute energy scale) is also a very crucial 
issue while measuring the mass splitting. We can probe very small values of
$\theta$ with an appearance experiment because the measurement is done
relative to zero. On the contrary, a disappearance experiment measures 
relative to unity. Therefore the impact of certain systematical errors
is quite different in both the cases. In an appearance experiment, the 
signal is proportional to $\theta$ and therefore an accurate understanding 
of the level of background is very vital while measuring small values
of $\theta$. But in case of a disappearance measurement the issue of total 
normalization is crucial because a large normalization error makes it tough
to detect deviations from unity.

\subsection{Three-Flavour case}

Let us now consider the case of three neutrino flavours. In a three neutrino 
framework, the $3\times3$ unitary mixing matrix $U$ depends on three mixing
angles \tem, \tet ~and \tmt ~and one CP-violating phase $\delta_{CP}$
(ignoring Majorana phases). The mixing matrix $U$ can be parameterized as 
   
\begin{equation}
U = V_{23}W_{13}V_{12},
\label{eq:ckmrot}
\end{equation}
where
\begin{equation}
V_{12} = \left(\matrix{c_{12} & s_{12} & 0 \cr -s_{12} & c_{12} &
0 \cr 0 & 0 & 1}\right), W_{13} = \left(\matrix{c_{13} & 0 &
s_{13}e^{-i\delta_{CP}} \cr 0 & 1 & 0 \cr -s_{13}e^{i\delta_{CP}} & 0 &
c_{13}}\right), V_{23} = \left(\matrix{1 & 0 & 0 \cr 0 & c_{23} &
s_{23} \cr 0 & -s_{23} & c_{23}}\right).
\label{eq:euler_rotations}
\end{equation}
$c_{ij}$ = $\cos\theta_{ij}$ and $s_{ij}$ = $\sin\theta_{ij}$.
The neutrino mixing matrix takes the form
\be
U = \pmatrix
{c_{12} c_{13} & s_{12} c_{13} & s_{13} e^{-i \delta_{CP}}
\cr
-s_{12} c_{23} - c_{12} s_{23} s_{13} e^{i \delta_{CP}}
& c_{12} c_{23} - s_{12} s_{23} s_{13} e^{i \delta_{CP}}
& s_{23} c_{13} \cr
s_{12} s_{23} - c_{12} c_{23} s_{13} e^{i \delta_{CP}}
&
-c_{12} s_{23} - s_{12} c_{23} s_{13} e^{i \delta_{CP}}
& c_{23} c_{13}
}.
\label{eq:upmns}
\ee

The probabilities of oscillations between various flavour states
are given by Eq. \ref{eq:formula5}.   
For three neutrino species, one can identify 
$\Delta m^2_{\odot} = \Delta m^2_{21} > 0$ as the neutrino mass squared
difference responsible for the solar neutrino oscillations and the dominant
atmospheric neutrino oscillations are caused by the 
$|\Delta m^2_{A}| = |\Delta m^2_{31}| \cong |\Delta m^2_{32}| \gg
\Delta m^2_{21}$. $\theta_{12} = \theta_{\odot}$ and
$\theta_{23} = \theta_{A}$ are the solar and atmospheric 
neutrino mixing angles, respectively. The angle \tet ~is the 
so-called ``CHOOZ mixing angle'' which connects the solar sector with 
the atmospheric one and determines the impact of the three flavour 
effects. Three flavour effects leave their imprints in an experiment which
is sensitive to both the mass splittings. It is true that for the precision
measurements of the mass squared differences or the two large mixing angles,
these effects are not that vital but they can play a key role in the
determination of the small mixing angle and in many cases, they decide
the fate of an experiment. Like in the quark sector, in the case of three 
generations there can be CP-violation in the neutral lepton sector provided 
that $\theta_{13}$ and the CP phase $\delta_{CP}$ are non-zero. If CP 
is not conserved, the oscillation probabilities for neutrinos are
different from those for antineutrinos. The CP-odd asymmetries are 
defined as

\begin{equation}
\Delta P_{\alpha\beta}\equiv P(\nu_{\alpha} 
\to \nu_{\beta};\,L)-P(\bar{\nu}_{\alpha} \to \bar{\nu}_{\beta};\,L)\,.
\label{eq:cp_odd1}
\end{equation}

CPT invariance ensures that $\Delta P_{\alpha\beta}=-\Delta P_{\beta\alpha}$.    
With the help of Eq. \ref{eq:upmns} we have

\begin{eqnarray}
\Delta P_{e\mu}=\Delta P_{\mu \tau}=\Delta P_{\tau e}=4 J_{CP}
\times \left[\sin\left(\frac{\Delta m_{21}^2}{2E}L
\right)+ \sin\left(\frac{\Delta m_{32}^2}{2E}L\right)+ \sin\left(
\frac{\Delta m_{13}^2}{2E}L\right)\right]\,,
\label{eq:cp_odd2}
\end{eqnarray}
where
\begin{equation}
J_{CP} = 
\frac{1}{8}\cos\theta_{13}\,\sin 2\theta_{13}\,\sin 
2\theta_{23}\,\sin 2\theta_{12}\,\sin\delta_{CP}\,
\label{eq:jcp}
\end{equation} 
which is known as Jarlskog CP-odd invariant \cite{jarlskog}. 
One can easily draw the following conclusions by observing Eq. \ref{eq:cp_odd2}.

\begin{enumerate}

\item
The CP-odd asymmetry vanishes if $\delta_{CP}$ is zero or $180^\circ$ and
it comes with full strength if $\delta_{CP}$ equals to $90^\circ$ or $270^\circ$.

\item
It vanishes if any of the mixing angles \tem, \tet ~or \tmt ~is 
zero or $90^\circ$. This indicates that we need at least three 
generations to observe CP-violation which is suppressed 
by the smallest of the mixing angles, \tet.

\item
The mass squared differences satisfy the
relation $\Delta m_{21}^2+\Delta m_{32}^2+\Delta m_{13}^2=0$. 
Therefore, if any one of the $\Delta m_{ij}^2$ is zero then the 
CP-odd asymmetry vanishes immediately. It ensures the fact that
the three flavour effects are essential for probing the 
leptonic CP-violation.  
 
\end{enumerate}

There is no doubt that it would be of great importance if we can establish the 
CP-nonconservation in the leptonic sector and future long baseline neutrino
oscillation experiments will provide a crucial hint along this direction.

\section{\fbox{Oscillations in Matter}}

\begin{figure}[h]
\begin{center}
\begin{picture}(359,80)(-5,-40)
\ArrowLine(40,50)(70,25)
\ArrowLine(40,-50)(70,-25)
\Photon(70,25)(70, -25)3 4
\Text(95,0)[r]{\small$W^\pm$}
\Text(95,35)[r]{\small$e$}
\Text(100,-35)[r]{\small$\nu_e$}
\Text(40,-35)[l]{\small$e$}
\Text(40,35)[l]{\small$\nu_e$}
\ArrowLine(70,25)(100,50)
\ArrowLine(70,-25)(100,-50)
\ArrowLine(240,50)(270,25)
\ArrowLine(240,-50)(270,-25)
\Photon(270,25)(270, -25)3 4
\Text(290,0)[r]{\small$Z^0$}
\Text(325,35)[r]{\small$\nu_{e, \mu,\tau}$}
\Text(325,-35)[r]{\small$p, n, e$}
\Text(220,-35)[l]{\small$p,n,e$}
\Text(220,35)[l]{\small$\nu_{e,\mu,\tau}$}
\ArrowLine(270,25)(300,50)
\ArrowLine(270,-25)(300,-50)
\end{picture}
\end{center}
\caption{\label{fig:scattering}
Feynman diagrams showing neutrino scattering inside the matter.
Left panel depicts CC interactions whereas right panel describes
NC processes.}
\end{figure}
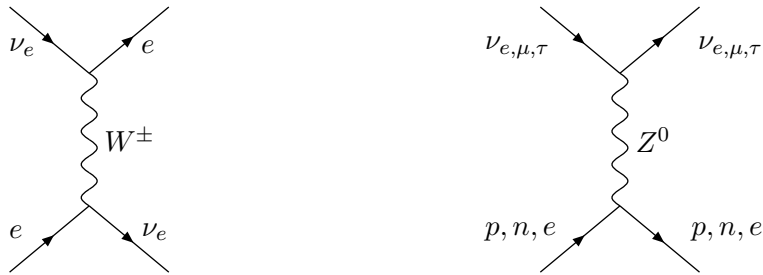

\begin{table}[t]
\begin{center}

\begin{tabular}{||c||c||c||} \hline\hline

\multicolumn{1}{||c||}{\bf Medium}
& \multicolumn{1}{|c||}{\bf Matter Density}
& \multicolumn{1}{|c||}{\bf $V_{CC}$}
\cr
\hline\hline

Solar core & $\sim$ 100 g/cm$^3$ & $\sim$ 10$^{-12}$ eV \cr
\hline
Earth core & $\sim$ 10 g/cm$^3$ & $\sim$ 10$^{-13}$ eV \cr
\hline
Supernova & $\sim$ 10$^{14}$ g/cm$^3$ & $\sim$ eV \cr

\hline\hline
\end{tabular}
\caption{\label{tab:density}
The impact of matter potential in various mediums.
}
\end{center}
\end{table}

Neutrino propagation through matter can modify oscillations significantly.
When neutrinos travel through matter ({\it e.g.} in the Sun, Earth or a
supernova), the weak interaction couples the neutrinos to matter and
besides few hard scattering events there is also coherent forward
elastic scattering of neutrinos with matter particles they encounter
along the way. This motion of neutrinos can be compared with the visible
light traveling through glass. The crucial point to be noted here is that
the coherent forward elastic scattering amplitudes are not the same
for all neutrino flavours. The ordinary matter consists of electrons,
protons and neutrons but it does not contain any muons or tau-leptons.
One can readily see from the right panel of Fig. \ref{fig:scattering} that 
neutrinos of all three flavours ($\nu_e$, $\nu_\mu$ and $\nu_\tau$)
interact with the electrons, protons and neutrons of matter through 
flavour independent NC interaction mediated by $Z^0$ bosons and these
contributions are same for neutrinos of all three flavours and 
therefore these interactions do not have any impact on neutrino 
oscillation probabilities. Interestingly, the electron neutrinos have 
an additional contribution due to their CC interactions 
(see left panel of Fig. \ref{fig:scattering}) with the ambient electrons of 
the medium which are mediated by the $W^\pm$ exchange. This extra matter
potential comes in the form

\begin{equation}
A = \pm 2\sqrt{2} G_F N_e E\,,
\label{eq:matter_potential} 
\end{equation}
where $G_F$ is the Fermi coupling constant, $N_e$ is the electron 
number density inside the Earth and $E$ is the neutrino energy. 
The $+$ sign refers to neutrinos while the $-$ to antineutrinos.
The electron density ($N_e$) is connected to the matter density ($\rho$) in the 
following way  
\begin{equation}
V_{CC} = \sqrt{2} G_F N_e \simeq 7.6 Y_e \frac{\rho}{10^{14}
\mbox{g}/\mbox{cm}^3} \mbox{eV}\,,
\label{eq:matter_density}
\end{equation}
where $Y_e = \frac{N_e}{N_p + N_n}$ is the relative number density.
$N_p$, $N_n$ are the proton and neutron densities in Earth matter
respectively. In an electrically neutral, isoscalar medium, we have 
$N_e = N_p = N_n$ and $Y_e$ comes out to be 0.5. 
The strength of $V_{CC}$ inside the Sun, Earth and 
a supernova is given in Table \ref{tab:density}. If we compare $V_{CC}$ with 
$\Delta m^2/2E$ then we can evaluate the importance of matter effects on
neutrino oscillations. If we consider a neutrino of 5 GeV passing through 
the core of the Earth then $V_{CC}$ is comparable with $\Delta m^2/2E$ 
($= 2.5 \times 10^{-13}$ eV if $\Delta m^2 = 2.5 \times 10^{-3}$ eV$^2$).     

Staying in the two-flavour formalism, the time evolution of the
flavour eigenstates in matter is given by the following Schr\"{o}dinger
equation

\be
i\frac{d}{dt}\left(\matrix{\nu_{\alpha} \cr \nu_{\beta}}\right) 
= \frac{1}{2E}\left[U\left(\matrix{m_{1}^{2} & 0 \cr 0 & m_{2}^{2}}\right)U^{\dagger}
+\left(\matrix{A(t) & 0 \cr 0 & 0}\right)\right]\left(\matrix{\nu_{\alpha} \cr 
\nu_{\beta}}
\right),
\label{eq:evo_matter}
\ee
where $U$ is the mixing matrix defined by Eq. \ref{eq:formula8}. 
In the constant matter density case the problem boils down to a 
stationary one and a trivial diagonalization of the Hamiltonian 
is required to obtain the solution. In the presence of matter, 
the vacuum oscillation parameters are mapped to the new 
parameters\footnote{The new parameters in matter carry a superscript $m$.}
in the following way

\be
{{
{(\Delta m^2)^m} }} &=&
{{
\sqrt{(\Delta m^2 \cos 2 \theta - A)^2 +
(\Delta m^2 \sin 2 \theta)^2} }},\nonumber \\
{{\sin 2 \theta^m}}
&=& {{\sin 2 \theta ~\Delta m^2/
(\Delta m^2)^m }}.
\label{eq:matter_param}
\ee
The so-called MSW-resonance \cite{lma_msw} condition is met at
\begin{equation}  
\Delta m^2 \cos 2 \theta = A.
\label{eq:resonance}
\end{equation}
At MSW-resonance, $\sin 2 \theta^m = 1$ (from Eq. \ref{eq:matter_param}
and \ref{eq:resonance}) which immediately signifies that independent of 
the value of the vacuum mixing angle $\theta$, the mixing in matter 
is maximal {\it i.e.} $\theta^m = \pi/4$.
This resonance occurs for neutrinos (antineutrinos) if $\Delta m^2$ is
positive (negetive). So the sign of $\Delta m^2$ determines the 
oscillation probability. From Eq. \ref{eq:resonance}, the resonance 
energy can be expressed as
\begin{equation} 
E_{res} = 11.16 \mbox{GeV} \left[\frac{|\Delta m^{2}|}{2.5 \times 10^{-3} 
\mbox{eV}^2}\right] \cdot \left[\frac{\cos 2 \theta}{0.95}\right] 
\cdot \left[\frac{2.8~\mbox{g}/\mbox{cm}^3}{\rho}\right].
\label{eq:resonance_energy}
\end{equation}
In course of neutrino propagation through the upper Earth mantle
with $\rho = 2.8~\mbox{g}/\mbox{cm}^3$, the resonance occurs at
roughly 11.2 GeV provided that $|\Delta m^{2}| = 2.5 \times 10^{-3} 
\mbox{eV}^2$ and $\cos 2 \theta = 0.95$. The matter effect arises 
due to matter and not anti-matter and this fact is responsible 
for the observed asymmetry between neutrino and antineutrino 
oscillation probabilities even in the two neutrino case.

In the presence of three neutrinos the time evolution of
flavour eigenstates in matter can be written as
\be
i\frac{d}{dt}\left(\matrix{\nu_{e} \cr \nu_{\mu} \cr
\nu_{\tau}}\right) = \frac{1}{2E}\left[U\left(\matrix{m_{1}^{2} & 0 &
0 \cr 0 & m_{2}^{2} & 0 \cr 0 & 0 &
m_{3}^{2}}\right)U^{\dagger}+\left(\matrix{A(t)
& 0 & 0 \cr 0 & 0 & 0 \cr 0 & 0 &
0}\right)\right]\left(\matrix{\nu_{e} \cr \nu_{\mu} \cr \nu_{\tau}}
\right),
\label{eq:evo_matter_3flav}
\ee
where $U$ is the mixing matrix defined by Eq. \ref{eq:upmns}.
Even in the constant matter density case, it is very tough to
obtain the analytical solution of the above equation. 
Under certain approximations, it is possible to obtain some
expressions which are quite useful in explaining the underlying 
physics. In the next chapter, we will discuss a few such cases
in detail. In the case of three flavours, besides the genuine 
CP asymmetry caused by the CP phase $\delta_{CP}$, we also have
fake CP asymmetry induced by matter which causes obstacles in
extracting the information on $\delta_{CP}$. Now let us see
what is our current understanding of these various neutrino
mass-mixing parameters?
    
\section{\fbox{Present Status}}

\begin{table}
\begin{center}
    \begin{tabular}{lcc}
        \hline \hline
         Parameter &
         Best fit  &
         ~~3$\sigma$~(1 d.o.f)
        \\
        \hline
        $\Delta m^2_{21}$~[$10^{-5}~\mbox{eV}^{2}$]
        & 7.6 & 7.1--8.3 \\
        $|\Delta m^2_{31}|$~[$10^{-3}~\mbox{eV}^{2}$]
        & 2.4 & 2.0--2.8 \\
        $\sin^2\theta_{12}$
        & 0.32 & 0.26--0.40\\
        $\sin^2\theta_{23}$
        & 0.50 & 0.34--0.67\\
        $\sin^2\theta_{13}$
        & 0.007 & $\leq$ 0.050 \\
        \hline \hline
\end{tabular}
\caption{\label{tab:best-fit}
Best-fit values and \sig (1 d.o.f) constraints on the oscillation
parameters under three-flavour scheme from global data including solar, atmospheric,
reactor (KamLAND and CHOOZ) and accelerator (K2K and MINOS)
experiments. This table has been taken from \cite{thomas}.}
\end{center}
\end{table}

Neutrino physics has entered the precision era, with the thrust
now shifting to detailed understanding of the structure of the
neutrino mass matrix, accurate reconstruction of which
would shed light on the underlying new physics that gives
rise to neutrino mass and mixing. The full mass matrix is given
in terms of nine parameters, the three neutrino masses,
the three mixing angles and the three CP violating phases.
Neutrino oscillation experiments are sensitive to only
two mass squared differences, all the three mixing angles and
the so-called Dirac CP phase. The remaining parameters,
comprising of the absolute neutrino mass scale and the
two so-called Majorana phases, have to be determined elsewhere.
We already have very good knowledge on the two mass squared 
differences and two of the three mixing angles.
Results from solar neutrino experiments \cite{solar,sk2}
which have been collecting data for more than four decades have 
now culminated in choosing the Large Mixing Angle (LMA) solution. 
The latest addition to this huge repertoire of experimental data is the
result from the on-going Borexino experiment \cite{borex}, and this
result is consistent with the LMA solution. This conclusion from 
solar neutrino experiments has been corroborated independently
by the KamLAND reactor antineutrino experiment \cite{kl,kltalk},
and a combined analysis of the solar and KamLAND data
gives as best-fit $\ms=7.6\times 10^{-5}$ eV$^2$ and $\sss=0.32$ 
\cite{kltalk,limits,thomas}. The other mass squared difference $\ma$ and
mixing angle $\theta_{23}$ are now pretty well determined
by the zenith angle dependent atmospheric $\numu$ data in
SuperKamiokande \cite{sk1,sk2,sk3} and the long baseline experiments
K2K \cite{k2k_ahn,k2k} and MINOS \cite{minos}. The combined data from
the atmospheric and long baseline experiments have pinned
down $|\ma| = 2.4\times 10^{-3}$ eV$^2$ and $\sin^22\theta_{23}=1$.
Our knowledge on the third mixing angle $\theta_{13}$ is restricted 
to an upper bound of $\sch < 0.033$ from the global analysis of all 
solar, atmospheric, long baseline and reactor data, including the 
CHOOZ \cite{CHOOZ,recent_chooz} results in particular.
We do not have any information on $\delta_{CP}$ and it is fully
unconstrained yet. All the best-fit values and \sig (1 d.o.f) constraints 
on the oscillation parameters under a three-flavour scheme from global data 
including solar, atmospheric, reactor (KamLAND and CHOOZ) and accelerator 
(K2K and MINOS) experiments have been summarized in Table \ref{tab:best-fit}. 
Recent cosmological observations (WMAP 5-year and other data) predict that 
the sum of three neutrino masses would be less than 0.19 eV \cite{mass}.
In the next section, we will focus on the major unknowns in the 
neutrino sector that we would like to resolve in the next 
ten to fifteen years down the line.

\section{\fbox{Missing Links}}

Despite the spectacular achievements in the last ten years or so, a lot of
information is still required to complete our understanding of the neutrino
sector. In the following, we have to tried to list them up :

\begin{itemize}

\item
Are the neutrinos Dirac or Majorana particles?
\item
How many neutrino species are there? Do sterile neutrinos exist?
Are three-flavour oscillations enough?
\item
What is the mass scale of the neutrinos? Why are neutrino masses so small?
\item
Does the neutrino have a non-zero magnetic moment?
\item
Why is the pattern of the neutrino mixing so different
from that of the quarks? Is there any connection between
quarks and leptons?
\item
What are the precise values of $\ms$ and $\sss$?
\item
What is the precise value of $|\ldm|$?
What is the sign of $\ldm$ or the character of the neutrino mass hierarchy?
\item
Is the ordering of the neutrino mass states hierarchical or quasi-degenerate?
\item
Is $\sta$ exactly maximal (= 1)?
If $\sta$ $\neq$ 1, what is its octant?
\item
How tiny is $\stch$? Is it zero? If it is non-zero then what is the
precise value of $\theta_{13}$?
\item
Does the behaviour of neutrinos violate CP? Can we probe the existence of the three
CP odd phases?
\item
Is there any link between the low energy CP violation in the
neutral lepton sector and the observed matter anti-matter asymmetry
(baryon asymmetry) in the universe ?
\item
Can we probe the sub dominant effects due to possible new physics in the
next generation high precision neutrino oscillation experiments?
\item
What is the importance of neutrino physics in our understanding of
dark matter and dark energy?
\item
Is there any connection between neutrino mass and leptogenesis
and galaxy-cluster formation?
\item
What is the role of neutrinos in connecting the predictions at the
grand-unification scale with low energy phenomena in the framework
of see-saw mechanism and supersymmetric extensions of the Standard
Model?

\end{itemize}

\begin{figure}[t]
\begin{center}
\includegraphics[width=12.0cm, height=8.0cm]{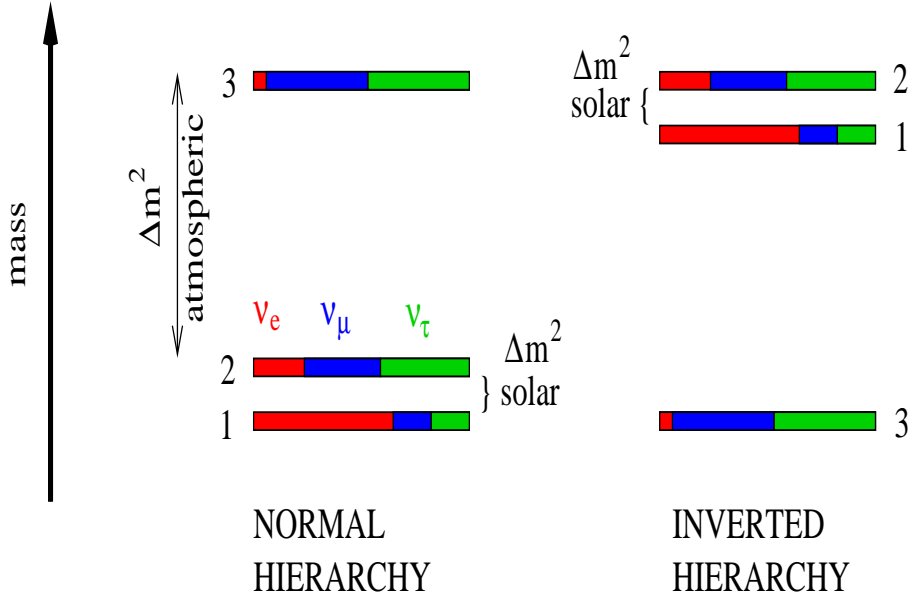}
\caption{\label{fig:hierarchy}
The sign of $\Delta m^2_{31}$ = $m_3^2 - m_1^2$ is not known.
The neutrino mass spectrum can be normal or inverted hierarchical.}
\end{center}
\end{figure}

Out of these several unanswered questions (listed above), there are several 
issues which can be answered in the future dedicated long baseline neutrino
oscillation experiments. Let us discuss a few of them.
While the solar neutrino data have confirmed 
that $\ms >0$ at a C.L. of more than $6\sigma$, we still do not know what 
is the sign of $\ma$, \ie, the neutrino mass hierarchy\footnote{Though we call 
this the neutrino mass {\it hierarchy}, what we mean is basically the 
{\it ordering} of the neutrino mass states (see Fig. \ref{fig:hierarchy}). 
Therefore, our discussions are valid for both hierarchical as well as 
quasi-degenerate mass spectra. We define $\Delta m_{ij}^2 = m_i^2 - m_j^2$
and refer to $sgn(\ma)$ as the neutrino mass hierarchy --
$sgn(\ma) >0$ is called ``normal hierarchy''(NH) while $sgn(\ma) <0$ 
is called ``inverted hierarchy''(IH).}.
Knowing the ordering of the neutrino masses is of prime importance, 
because it dictates the structure of the neutrino mass matrix, and hence 
could give vital clues towards the underlying theory of neutrino masses
and mixing. Knowing the $sgn(\ma)$ could have other far-reaching
phenomenological consequences. For instance, if it turns out the
$\ma < 0$ and yet neutrino-less double beta-decay is not observed even in 
the very far future experiments, that would be a strong hint that the 
neutrinos are not Majorana particles (see for {\it e.g.} \cite{0vbbus} and
references therein). A vital task to be undertaken in the immediate future
is the determination of the hitherto unknown mixing angle $\theta_{13}$.
Discovery of a non-zero value for $\theta_{13}$ would open up the
possibility of observing CP-violation in the lepton sector.  
Non-zero $\theta_{13}$ also brings in the possibility of large Earth matter 
effects \cite{lma_msw,barger_msw} for GeV energy accelerator neutrinos 
traveling over long distances. Effect of matter on neutrino oscillations
depends on the \sgnma{} and is opposite for neutrinos and antineutrinos. 
For a given \sgnma{} it enhances the oscillation probability in
one of the channels and suppresses it in the other. Therefore, comparing 
the neutrino signal against the antineutrino signal in very long baseline 
experiments gives us a powerful tool to determine \sgnma. A way to
judge the merit of an experiment is by observing it's limiting
reach in the small mixing angle $\theta_{13}$ where it is still 
sensitive to the mass ordering and the leptonic CP-violation.    

\section{\fbox{Future Neutrino Road-Map}}

Determining the unknown mixing angle $\theta_{13}$,
the CP phase $\delta_{CP}$, and the neutrino mass ordering have emerged 
as the next frontiers in this field. All these three quantities can be 
probed by experimentally measuring the so-called ``golden'' 
channel~\cite{golden} oscillation probability $P_{e\mu}$ 
($\nue \rightarrow \numu$) or its T-conjugate channel $P_{\mu e}$.
A series of ambitious projects are under discussion which plan to use 
this oscillation channel. The on-going and near future experiments include
the MINOS experiment in the US~\cite{minos}, and the CNGS experiments
ICARUS~\cite{icarus} and OPERA~\cite{opera} in Europe.
Next experiments in line will be T2K in Japan~\cite{t2k}
and NO$\nu$A in US \cite{nova}. All these experiments
will use muon neutrino beams from conventional accelerator
sources in order to observe $P_{\mu e}$.
Collectively and in combination with short-baseline reactor
experiments, such as Double Chooz \cite{chooz2}, these experiments
are expected to improve the bound on $\theta_{13}$ to about
$\stch < 0.01$ ($90\%$ CL)~\cite{huber10}.
The mass hierarchy and CP violation,
though in principle accessible using the combined
data from the T2K and NO$\nu$A experiments, can be determined
only for values\footnote{We distinguish between the
``true'' values of the oscillation parameters, which are
the values chosen by Nature, and their fitted values.
Throughout this work we denote the true value of a parameter
by putting ``(true)'' after the symbol for the parameter.}
of $\stcht$ close to the current bound and for some fraction of the possible 
values of the CP phase $\dcpt$.
The sensitivity of these experiments is mainly restricted
by statistics, while for larger luminosity set-ups,
the intrinsic $\nue$ background poses a natural limitation for
experiments sensitive to $\numu$ oscillations into $\nue$.
Therefore, if Nature has not been very kind we will need larger 
sophisticated experiments to complete our understanding of the neutrinos, 
possibly using an alternate smart technology with powerful, 
well measured neutrino beams from upgraded accelerator facilities and
improved detector technologies. The beta-beam and neutrino factory 
can play a crucial role in this direction in future. A summary of the 
potential of selected neutrino factory and beta-beam set-ups have been 
compiled by the physics  working group of the International Scoping Study 
for a future neutrino Factory, superbeam and beta-beam, 
in their report~\cite{iss}. In the next chapter, we will expound in detail 
the physics reach of an experimental set-up in which the proposed large 
magnetized iron detector at the India-based Neutrino Observatory (INO) 
\cite{ino} would serve as the far detector for a CERN-based 
monoflavour beta-beam \cite{zucc}. 

\chapter{Neutrino mass ordering and $\theta_{13}$ with a magical
Beta-beam experiment at INO}
\section{\fbox{Introduction}}

The most promising avenue for the determination of $\theta_{13}$, 
the CP phase $\delta_{CP}$ and \sgnma{} is the $\nue \rightarrow \numu$ 
oscillation channel $P_{e\mu}$ (or its T-conjugate $P_{\mu e}$), often 
referred to in the literature as the ``golden channel'' \cite{golden}.
The $P_{e\mu}$ channel can be studied in experiments which use an initial 
$\nue$ (or $\anue$) beam and a detector which can efficiently 
see muons\footnote{This is complementary to the standard accelerator beams 
where the initial (anti)neutrino flux consists of $\numu$ (or $\anumu$) and
where the relevant channel is the probability $P_{\mu e}$. Produced from
decay of accelerated pions, these conventional (anti)neutrino beams suffer 
from an additional hurdle of an intrinsic $\nue$ ($\anue$) contamination, 
which poses a serious problem of backgrounds. A beta-beam is comprised of 
pure $\nue$ (or $\anue$).}. Also in order to probe small values of $\stch$, one
needs the following~: 

\begin{enumerate}

\item
\fbox{\bf Low backgrounds}

As far as the low background requirement is concerned, it is an advantage 
to use a pure flavor neutrino beam without any intrinsic beam contamination.
A beta-beam \cite{zucc} is well suited for this purpose.
An alternative approach is the so-called neutrino factory \cite{geer}.
The detector backgrounds coming mainly from
neutral current interactions and mis-identification of particles,
can be reduced by imposing intelligent cuts. The atmospheric neutrino backgrounds,
which can be important for beta-beams at lower energies, can be suppressed
using timing and directional information.

\item
\fbox{\bf Large statistics}

Statistics can be increased by a higher beam power and the size and
efficiency of the detector.

\item
\fbox{\bf Reduced systematical uncertainties}

Beam-related systematic uncertainties can be
reduced to a large extent by working with a two detector set-up, one very
close to the beam line and another serving as the far detector.
The systematic uncertainties coming from the lack of knowledge of the
neutrino-nucleus interaction cross-sections are another important source of error.
These can be controlled to some degree by the near-far two detector
set-up, but they cannot be canceled completely~\cite{crossthomas}.

\end{enumerate}

For a beta-beam, a variety of plausible set-ups have been proposed 
in the literature~\cite{volpe,cernmemphys,
paper1,betaino1,betaino2,pee,bboptim,twobaseline,rparity1,rparity2,
conf_talks,sanjib_talks,oldpapers,donini130,doninibeta,newdonini,bc,bc2,fnal,betaoptim,
doninialter,boulby}. The proposal which poses minimal challenge
for the beta-beam design, is the commonly called 
CERN-MEMPHYS project~\cite{cernmemphys,oldpapers,donini130}.
It proposes to use the EURISOL ion source to produce the radioactive
source ions $^{18}$Ne and $^6$He, and demands a Lorentz boost factor 
$\gamma \simeq 100$ for them,
which can be produced using the existing
accelerator facilities at CERN. The far detector MEMPHYS,
a megaton water detector with fiducial
mass of 440~kton, will have to be built in the
Fr\'ejus tunnel, at a distance of 130 km from CERN.
Another possible beta-beam set-up using water detector but 
higher boost factors and an intermediate baseline option was put forth in
\cite{bc,bc2} (see also \cite{doninialter}). In these papers
authors have used a high $\gamma$ $^{18}$Ne and $^6$He
beta-beam option at CERN and 440 kton fiducial volume water detector at
GranSasso or Canfranc, which corresponds to $L=730$ and 650 km respectively.
Excellent sensitivity to $\theta_{13}$ and CP violation is expected 
\cite{bc,bc2} from this proposal. Set-ups with a neutrino beam from 
CERN to GranSasso or CanFranc~\cite{doninibeta}, from CERN to Boulby mine
\cite{boulby} and from Fermilab~\cite{fnal} ($L \sim 300$ km) have also 
been proposed, and their sensitivity reach has been explored.
Set-up with two sets of source ions
with different boost factor for
each set but with the same baseline was proposed in
\cite{doninialter}. In \cite{newdonini} the authors
consider the complementary situation where they take
only one set of source ions, \br and $^8$Li, with $\gamma=350$
and two different baselines, $L=2000$ km and 7000 km.
Very high gamma beta-beam options have been studied in
\cite{paper1,bc2,betaoptim}. The physics potential
of the low energy beta-beam option was probed in \cite{volpelow}.
A comparison of the physics reach among different
beta-beam experimental proposals can be found in \cite{volpe,conf_talks}.
Here we would like to discuss the the possibility of measuring to a 
very high degree of accuracy the mixing angle $\theta_{13}$ and 
\sgnma{} {\it aka}, the neutrino mass ordering, in an experimental set-up
\cite{paper1,betaino1,betaino2,bboptim,twobaseline} where a pure and 
intense multi-GeV $\nue$ (using $^8$B ion) and/or $\anue$ (using $^8$Li ion) 
beta-beam with the Lorentz boost $\gamma$ between 250 and 650 is shot
from CERN to the India-based Neutrino Observatory (INO) \cite{ino}
where a large magnetized iron calorimeter (ICAL) detector is expected 
to come-up soon.

Even if both neutrinos and antineutrinos are used, a serious complication 
with all long baseline experiments involving the golden $P_{e\mu}$ channel 
arises from discrete degeneracies which manifest in three forms:

\begin{enumerate}

\item 
\fbox{\bf the ($\theta_{13}, \delta_{CP}$) intrinsic degeneracy~\cite{intrinsic},}

\item 
\fbox{\bf the (\sgnma{}, $\delta_{CP}$) degeneracy~\cite{minadeg},}

\item 
\fbox{\bf the ($\theta_{23}, \pi/2-\theta_{23}$) degeneracy~\cite{th23octant}.}

\end{enumerate}
This leads to an eight-fold degeneracy \cite{eight}, with several
spurious or ``clone'' solutions in addition to the true one and
severely deteriorates the sensitivity of any experiment.
It has been shown \cite{eight,magic} that the problem of clone solutions
due the first two types of degeneracies can be evaded by choosing
the baseline of the experiment equal to the characteristic
refraction length due to the matter inside Earth
\cite{lma_msw,eight,magic,magic2,petcov}. This special value goes 
by the colloquial name ``magic baseline'' \cite{magic}.
As we will discuss in detail later, at this baseline the sensitivity to the
mass hierarchy and, more importantly, $\theta_{13}$, goes up
significantly \cite{bboptim,magic,nufactoptim}, while the sensitivity to
$\delta_{CP}$ is absent.

Interestingly, the CERN-INO distance of 7152 km happens to be
tantalizingly close to the magic baseline ($\sim$ 7500 km). This large 
baseline also enhances the matter effect and requires traversal through
denser regions of the Earth. Thus, for neutrinos (antineutrinos)
with energies in the range 3-8 GeV sizable matter effects are
induced if the mass hierarchy is normal (inverted). A unique
aspect of this set-up is the possibility of observing
near-resonant matter effects in the $\nue\rightarrow \numu$
channel. In fact, to our knowledge, what we propose here is the
only experimental situation where near-resonant matter effects
can be effectively used in a long baseline experiment to study
the neutrino mass matrix.  We show in this work that the
presence of this near-maximal Earth matter effect not only
maximizes the sensitivity to the neutrino mass hierarchy, it also
gives the experiment an edge in the determination of the mixing
angle $\theta_{13}$. The increase in the probability $P_{e\mu}$
due to near-resonant matter effects, compensates for the fall in
the beta-beam flux due to the very long baseline, so that one can
achieve sensitivity to $\theta_{13}$ and mass hierarchy which is
comparable, even better, than most other proposed experimental
set-ups.  Therefore a beta-beam experiment with its source at CERN and
the detector at INO could emerge as a powerful tool for a
simultaneous determination of the neutrino mass hierarchy and
$\theta_{13}$.

In \cite{paper1} such an experimental set-up was considered for
the first time and the physics potential explored. It was
demonstrated that both the neutrino mass hierarchy as well as
$\theta_{13}$ may be probed through such a set-up.
The ions considered for the beta-beam in that work were the
most commonly used $^{18}$Ne (for $\nue$) and $^6$He (for $\anue$). 
One requires very high values of the Lorentz boost for these ions 
($\gamma \sim 10^3$) because the energy $E$ of the beam has to be 
in the few GeV range for achieving near-resonant matter effects and 
to enable detection in the ICAL detector, which is expected to have a 
threshold of about 1 GeV. Such high values of $\gamma$, although possible
in principle, might turn out to be very difficult to realize.
Subsequently, two other ions, $^8$B (for $\nue$) and $^8$Li (for $\anue$),  
have been projected as viable options for the beta-beam 
source \cite{rubbia,mori}.  
The main advantage that these ions offer is their substantially higher 
end-point energy, $E_0$. This allows one to access $E\sim$ few GeV very 
easily with medium values of $\gamma$, that could be possible to achieve with 
either the existing CERN technology, or with the projected upgrades.

In this chapter we will mainly present our results with $^8$B and $^8$Li as
the candidate ions for the beta-beam source at CERN and ICAL@INO as the 
far detector. We perform a detailed $\chi^2$ analysis of the projected data 
in a future CERN-INO beta-beam experiment and show our results with
$\gamma$ ranging between 250 and 650. We begin in section 3.2 with a brief 
description of the beta-beam produced at the source. In section 3.3 we
discuss the oscillation probability and highlight the importance of the 
magic baseline and the near-resonant matter effects in the $P_{e\mu}$ channel. 
Section 3.4 contains the results for the event rates expected in the ICAL detector
and a discussion addressing the issue of background rejection capability of ICAL.    
In section 3.5 we present the details of the statistical analysis
used. In section 3.6 we focus on the sensitivity to the neutrino mass hierarchy.
The sensitivity of the results on the mixing angle $\theta_{13}$
are presented in section 3.7. We end in section 3.8 with our conclusions.

\section{\fbox{The Beta-beam Fluxes}}

Very pure and intense $\nue$ and/or $\anue$ beams can be produced by 
the decay of highly accelerated radioactive beta unstable ions, 
circulating in a storage ring \cite{jacques}. This is what is called a ``beta-beam'' 
(see section 1.3.2) and was first proposed by Piero Zucchelli \cite{zucc}.
The selection of the beta unstable parent ion is determined by a variety of 
factors essential for efficiently producing, bunching, accelerating and 
storing these ions in the storage ring. A beta-beam comes with the following
novel features~:

\begin{enumerate}

\item
{\bf This neutrino beam would be very suitable for precision experiments
because it is pure and mono-flavor and hence beam related backgrounds are 
almost absent.}

\item
{\bf The neutrino spectrum depends only on the beta decay total
end-point energy $E_0$ and the Lorentz boost of the radioactive 
ions $\gamma$. The spectral shape can therefore be very well 
determined which makes it almost free of systematic uncertainties.}

\item
{\bf The flux normalization is determined by the number of useful ion 
decays in the straight section of the beam.}

\item
{\bf The source is very intense.}

\item
{\bf The beam divergence is controlled by the Lorentz boost $\gamma$.
Hence by increasing $\gamma$, we can produce a higher beam collimation 
and increase the beam intensity along the forward direction 
$\propto \gamma^2$.}

\item
{\bf The $\nue$ and $\anue$ beta-beam fluxes could be produced 
simultaneously in distinct bunches from the storage ring.
Therefore, in principle, it should be straight-forward to
distinguish the neutrino from the antineutrino events at
any detector using the nano-second time resolution that most
detectors possess and the presence of magnetic field is not
mandatory.}

\item
{\bf It can be built using the existing CERN facilities.}

\end{enumerate}

\subsection{The Beta-decay Spectrum}

In the rest frame of the radioactive ion the beta-decay spectrum is given by
\begin{subequations}
\begin{equation}
\frac{d^2 \Gamma^*}{d\Omega^* d E^*}
= \frac {1} {4\pi\,m_e^5 \,f
\,\tau} (E_0 - E^*)E^*
\sqrt{ (E_0-E^*)^2-m_e^2}
\label{eq:rspectra} ,
\end{equation}

where $m_e$ is the electron mass, $E_0$ the electron
total end-point energy, $E^*$ and $\tau$ are the neutrino
energy and lifetime of the decaying ion respectively
in the latter's rest frame and $f$ is defined in the fashion

\begin{equation}
f(y_e)\equiv {1\over 60 y_e^5} \left\{ \sqrt{1-y_e^2} (2-9 y_e^2 - 8
y_e^4) + 15 y_e^4 \mathrm{Log} \left[{y_e \over
1-\sqrt{1-y_e^2}}\right]\right\}
\end{equation}
\end{subequations}

where $y_e=m_e/E_0$.

In the lab frame,
the flux of the unoscillated beta-beam at the detector is given by
\begin{equation}
\phi_{\nue}(E,\theta)
 =\frac{1}{4\pi L^2}\frac {N_\beta} {m_e^5 \,f}
\frac{1}{\gamma(1-\beta \cos\theta)}
 (E_0 - E^*) E^{*2} \sqrt{ (E_0-E^*)^2-m_e^2},
\label{eq:flux}
\end{equation}
where $L$ is the distance
between the source and detector, $N_\beta$ are
the number of useful decays in the storage ring
per unit time,
$\theta$ is the angle between
the neutrino flight direction and the direction in which
the ions are boosted\footnote{We work with the on-axis flux for which
$\theta=0$.}
and $\gamma$
is the Lorentz boost such that
$E^* =  \gamma E(1-\beta \cos\theta)$, $E$ being the neutrino
energy in the lab frame.
The maximum energy of neutrinos produced by a beta-beam with a Lorentz
factor $\gamma$ is given by
\begin{subequations}
\begin{equation}
E_{\rm max}=\frac{(E_0-m_e)}{\gamma(1-\beta\cos\theta)}.
\label{eq:enumax}
\end{equation}
For the on-axis ($\theta = 0$) neutrino beam, the above expression 
looks like the following
\begin{equation}
E_{\rm max}=\gamma(E_0-m_e)(1+\beta)~~~~~{\rm{(on-axis)}}.
\label{eq:on-axis}
\end{equation}
\end{subequations}
\subsection{Candidate Ions for the Beta-beam}

\begin{figure}
\includegraphics[width=0.51\textwidth]{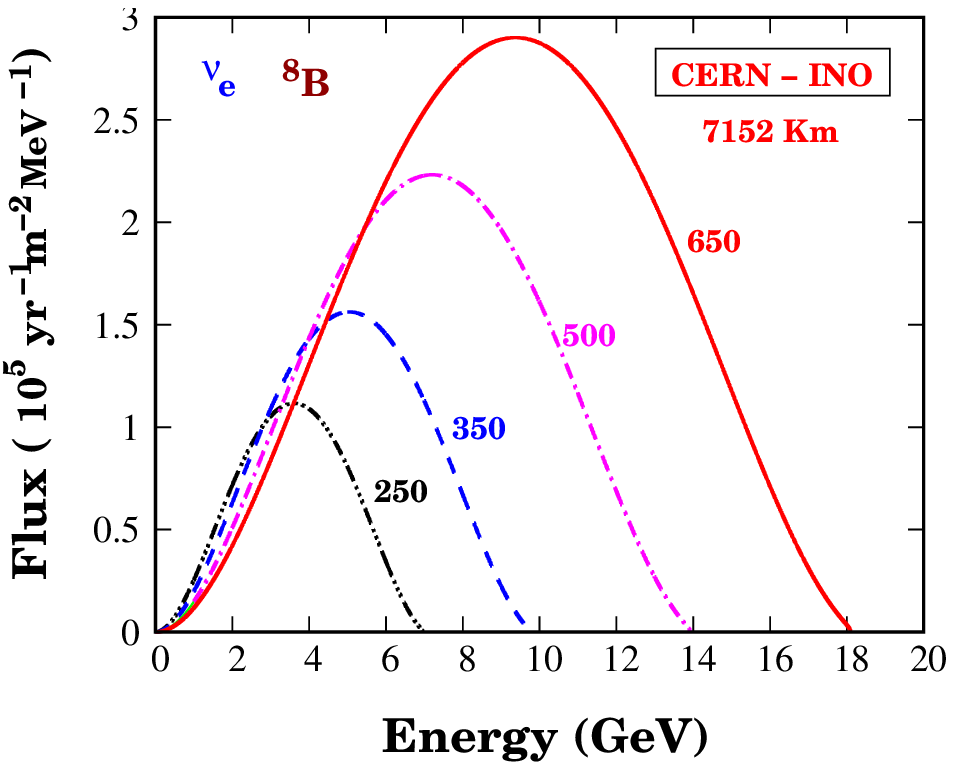}
\includegraphics[width=0.51\textwidth]{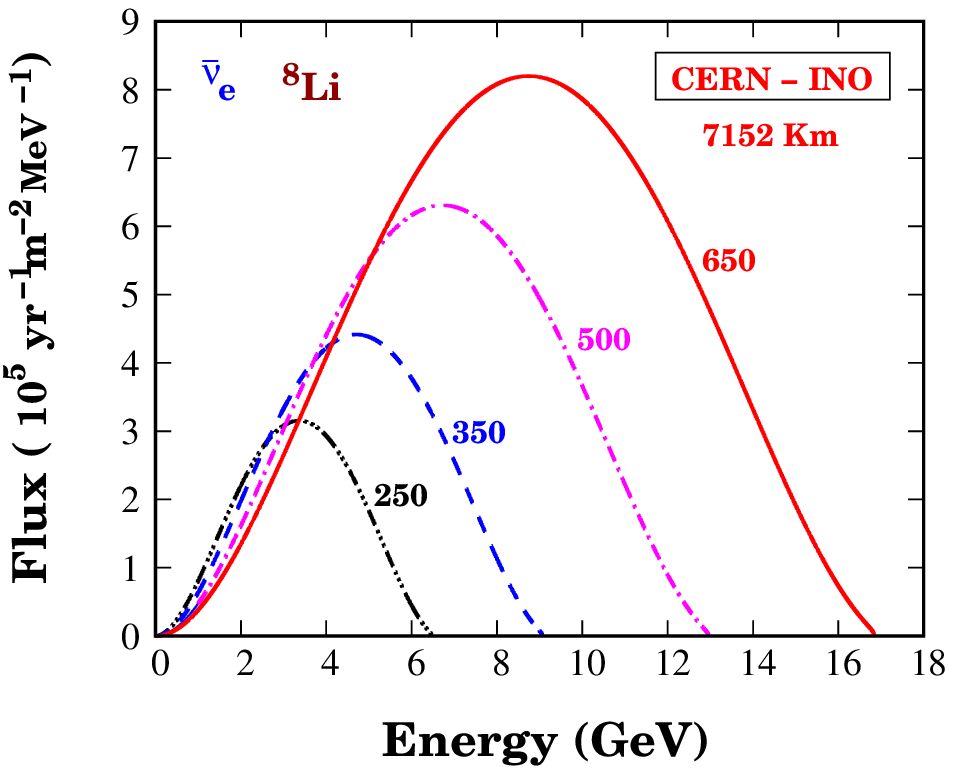}
\vskip1.0cm
\includegraphics[width=0.51\textwidth]{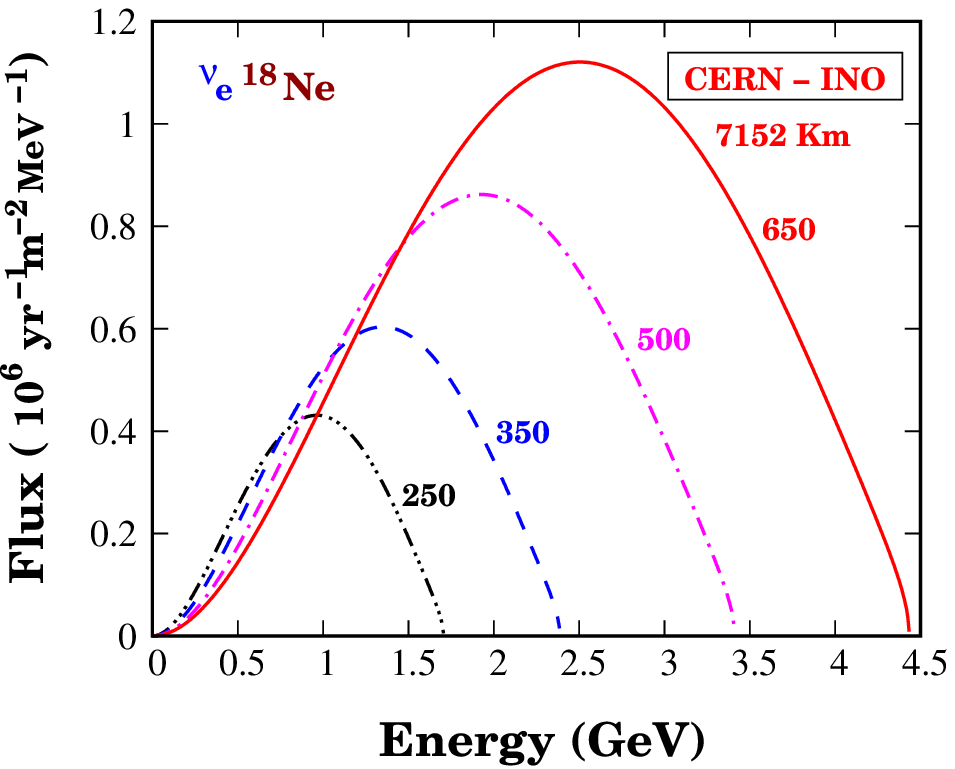}
\includegraphics[width=0.51\textwidth]{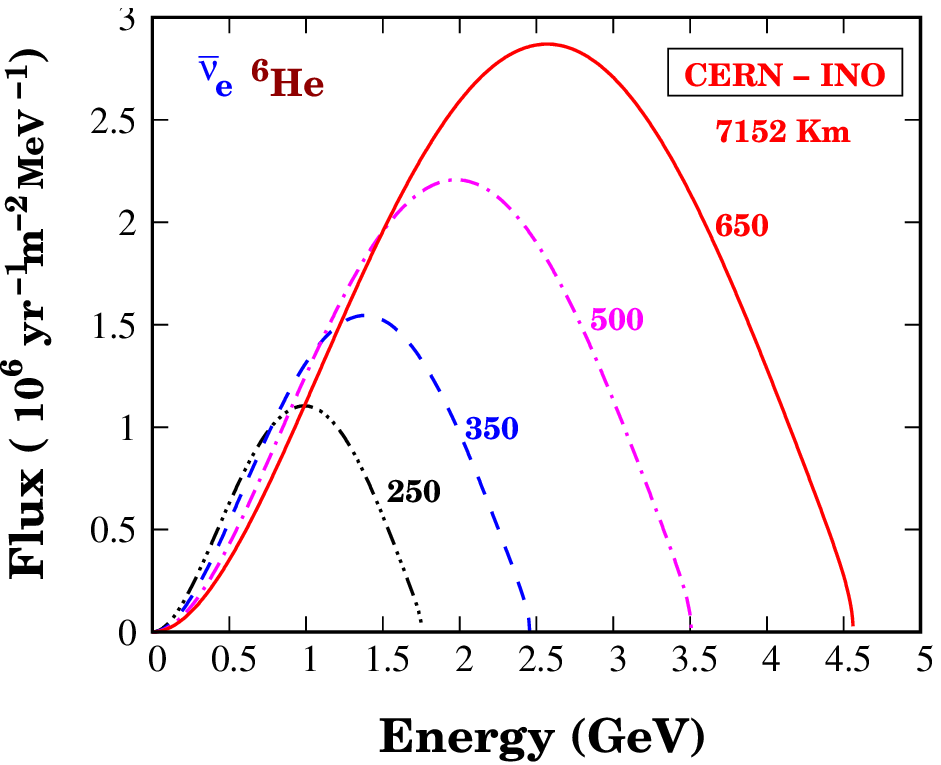}
\caption{\label{fig:flux}
The unoscillated beta-beam flux spectrum arriving at ICAL@INO.
The upper panels are for \br (left panel) and \li (right panel),
while lower panels are for \neon (left panel) and \he (right panel).
Due to the finite energy width ($\sim 1.5$ MeV) of the $2^+$ first 
excited state of $^8$Be, the maximum neutrino energy available in the 
decay of $^8$Li and $^8$B may be increased by 5 - 6\%.
}
\end{figure}

\begin{figure}[t]
\includegraphics[width=8.0cm, height=7.0cm]{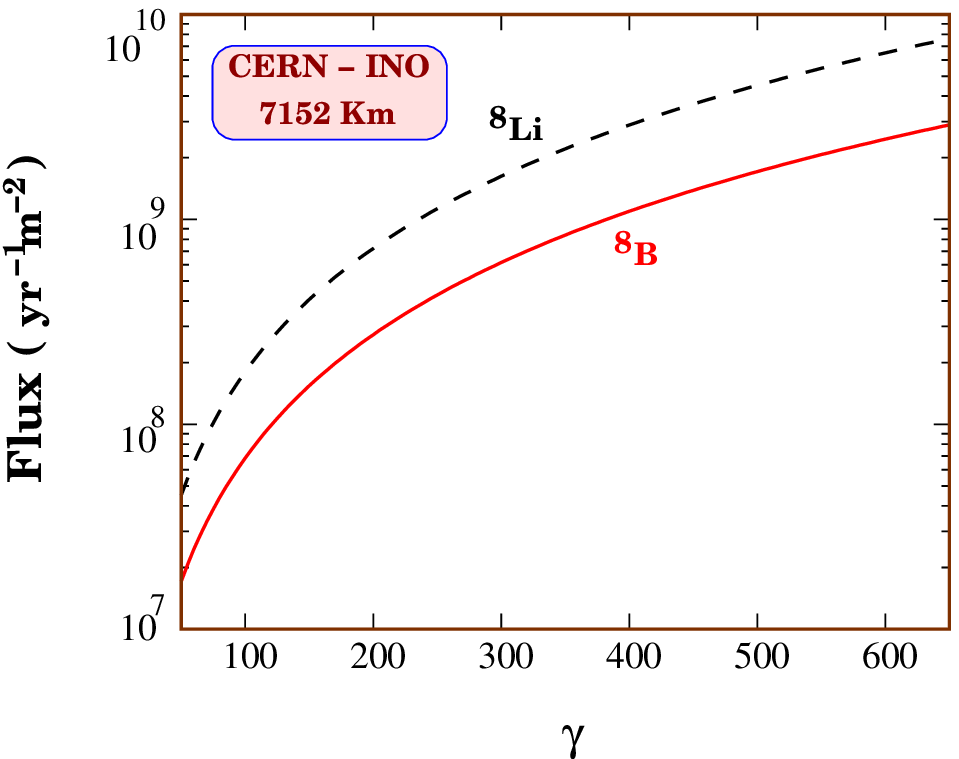}
\vglue -7.0cm \hglue 8.8cm
\includegraphics[width=8.0cm, height=7.0cm]{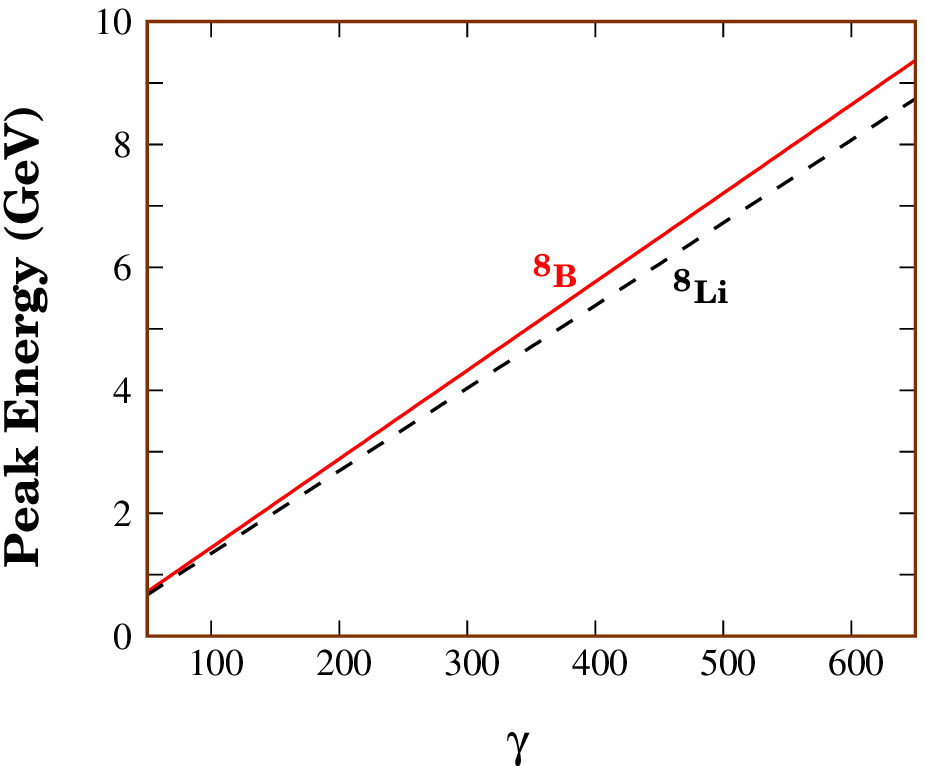}
\caption{\label{fig:flux1}
Left panel shows
the total unoscillated flux in $yr^{-1}m^{-2}$ expected at INO, as a function of
the Lorentz factor $\gamma$. The solid (dashed)
line corresponds to $^8$B ($^8$Li) and we have assumed
$1.1\times 10^{18}$ ($2.9\times 10^{18}$) useful ions
decays per year. Right panel shows the energy at which the
unoscillated flux peaks, as a function of $\gamma$.
}
\end{figure}

From Eq. (\ref{eq:flux}) it is seen that the total flux and
energy of the beta-beam at the far detector depends mainly on the end
point energy $E_0$ of the beta-decay ion and the Lorentz boost
factor $\gamma$. The flux increases as both $E_0$ and $\gamma$
increase. Larger $\gamma$ results in better collimation of the
impinging flux, thereby increasing the statistics.  The spectrum
also shifts to larger energies as $E_0$ and $\gamma$ increase.
The neutrino cross section in the detector increases with energy, 
so for the same total flux a harder spectrum further enhances the statistics.  
Since the flux falls as $1/L^2$ as the source-detector distance increases, 
high values of either $E_0$ or $\gamma$ or both are needed to have sizable
number of events at the far detector. $E_0$ is an intrinsic property of the 
decaying ion while $\gamma$ is restricted by the nature of the accelerators 
and the machine design. The other desired properties which the ion should have 
include high production yield, large decay fraction, and a lifetime that is
long enough to allow the ions to be adequately accelerated. It is also easier 
to store larger number of lower-$Z$ isotopes in the storage ring \cite{autin}.
The characteristic features of the four different ions which have been discussed 
in the literature as possible candidates for the beta-beam have already been 
shown in Table \ref{tab:ions} (section 1.3.2). While $^{18}$Ne and $^8$B are 
$\beta^+$ emitters (producing a $\nue$ beam), $^6$He and $^8$Li are $\beta^-$ 
emitters (producing a $\anue$ beam). They all have comparable lifetimes, 
conducive to the requirements necessary for the beta-beam, very high (or maximal) 
decay fraction and very low $A/Z$ ratio. The point to be noted that the end-point 
energies, $E_0$, for $^8$B and $^8$Li are much larger than those for $^{18}$Ne 
and $^6$He.

In this work we are interested in the physics potential of an 
experimental set-up with a beta-beam at CERN and a large magnetized
iron detector in India. For such a large baseline, one needs a
very high value of $\gamma$, both to cross the
detector energy threshold as well as to get reasonable statistics
in the detector. In particular, with $^6$He (or $^{18}$Ne) as the
source ion, one needs $\gamma \gtap 1000$.  Such high values of
$\gamma$ can only be achieved by using the LHC itself. As noted
above, $E_0$ for the alternative options for the radioactive ion source 
($^8$B and $^8$Li) is higher by a factor of more than 3.  This means that 
it should be possible to produce high intensity, high energy beams with 
$^8$B and $^8$Li for a much lower boost factor\footnote{The loss in 
collimation is not significant as we show later.}.
In the low $\gamma$ design, the standard numbers taken for the 
\neon and \he are $1.1 \times 10^{18}$ ($\nu_e$) and $2.9\times 10^{18}$
($\bar{\nu}_e$) useful decays per year, respectively \cite{beamnorm}.
Earlier, only these ions were considered because it was believed that
$^8$B could not be produced with the standard ISOLDE techniques.
Since most exercises focused on observation of CP-violation, it
was necessary to have both $\nue$ and $\anue$ beams with similar
spectra, so $^8$Li (though considered in \cite{autin}) was also
generally ignored. Interest in both these ions have been
rekindled recently \cite{rubbia,mori,doninialter}, as it appears that
having intense $^8$B and $^8$Li fluxes should be feasible using
the ionization cooling technique \cite{rubbia,mori}. In what follows,
we will vary the value of $\gamma$ to test the physics potential
of the CERN beta-beam INO-ICAL set-up and, following the current
practice, assume that it is possible to get $2.9\times 10^{18}$ 
useful decays per year\footnote{Wherever not explicitly mentioned, 
these reference numbers of useful ion decays for 
$\nu_e$ and $\bar{\nu}_e$ are chosen.} for $^8$Li and $1.1\times 10^{18}$ 
for $^8$B for all values of $\gamma$. Note, however, that new ideas suggest
luminosities higher even by a factor of ten or so, depending on
the isotopes used, by using a recirculating ring to improve the performance 
of the ion source \cite{rubbia,mori}. Studies have shown that it is possible 
to accelerate $^6$He to $\gamma \ltap 250$ with the existing facilities at CERN, 
while $\gamma=250-600$ should be accessible with the ``Super-SPS'', an
upgraded version of the SPS with super-conducting magnets
\cite{doninibeta,bc,bc2}. The Tevatron at FNAL could in principle
also be used to produce a beta-beam with $\gamma < 600$.

The upper panels of Fig. \ref{fig:flux} depict the unoscillated 
$^8$B (left-hand panel) and $^8$Li (right-hand panel) 
beta-beam flux expected at ICAL@INO, for four different benchmark 
values of $\gamma$. The lower panels of Fig. \ref{fig:flux} show the 
corresponding spectra for the $^{18}$Ne (left-hand panel) and 
$^{6}$He (right-hand panel) beta-beam. Note that
even though apparently it might seem from the figures that the
$\nue$ ($\anue$) flux is larger for $^{18}$Ne
($^6$He), in reality, for a given $\gamma$, the total flux is
given by the area under the respective curves. One can easily
check that for a fixed $\gamma$ this is same for both the ions as
expected, since we have assumed equal number of decays for both
$^{18}$Ne and $^8$B for $\nue$ ($^6$He and $^8$Li for
$\anue$)\footnote{Of course, the on-axis flux increases with
$\gamma$ because of better collimation of the beam.}. Also
note that even though the total $\nue$ ($\anue$) flux remains the
same for both the ions for a given $\gamma$, we would expect that 
the number of events produced in the detector is much enhanced for 
the $^8$B ($^8$Li) beam since the energy of the beam is larger 
and the CC cross sections increase with the neutrino energy.
In the left panel of Fig. \ref{fig:flux1} we show the energy integrated
total number of unoscillated neutrinos in units of $yr^{-1}m^{-2}$
arriving at ICAL@INO, as a function of the Lorentz boost $\gamma$.
The solid (dashed) line corresponds to $^8$B ($^8$Li).
The figure shows that the energy integrated flux arriving at the detector 
increases almost quadratically with $\gamma$. Note that with the same
accelerator, the Lorentz boost acquired by $^8$B is 1.67 times larger 
than that by $^8$Li, determined by the charge to mass ratios of the ions.
The right panel of Fig. \ref{fig:flux1} depicts the energy at which 
the unoscillated flux peaks, as a function of the Lorentz boost $\gamma$. 
It turns out that this peak energy is roughly half the maximum energy of
the beam, which is given as 
$E_{\rm max} \simeq 2(E_0-m_e)\gamma$ (see Eq. \ref{eq:enumax}).

\section{\fbox{Neutrino Propagation and the ``Golden Channel''}}

In this section we will very briefly review some issues 
related to the neutrino oscillation probability.

\subsection{The ``Golden Channel'' ($\nue \rightarrow \numu$)}

The expression for $P_{e\mu}$ in matter~\cite{lma_msw,barger_msw},
up to second order terms in the small quantities $\theta_{13}$ and
$\alpha \equiv \ms/\ma$, is given by\footnote{This particular low order 
expansion of the transition probability is valid only in the range
of $L$ and $E$ where the resonance condition $\hat{A}=1$ is never reached. 
For the $E$ and $L$ range that we consider in this work, this condition 
is satisfied and Eq. (\ref{eq:pemu}) fails, as we discuss later 
in this section. Nonetheless, we present this expression to illustrate
the effect of the magic baseline.} \cite{golden,freund}
\be
P_{e\mu} & \simeq & \sin^2 2\theta_{13} \, \sin^2 \theta_{23}
\frac{\sin^2[(1- \hat{A}){\Delta}]}{(1-\hat{A})^2}
\nonumber \\
&\pm&   \alpha  \sin 2\theta_{13} \,  \xi \sin \delta
\sin({\Delta})  \frac{\sin(\hat{A}{\Delta})}{\hat{A}}
\frac{\sin[(1-\hat{A}){\Delta}]}{(1-\hat{A})}
\nonumber  \\
&+&   \alpha  \sin 2\theta_{13} \,  \xi \cos \delta
\cos({\Delta})  \frac{\sin(\hat{A}{\Delta})}{\hat{A}}
\frac{\sin[(1-\hat{A}){\Delta}]} {(1-\hat{A})}
\nonumber  \\
&+&  \alpha^2 \, \cos^2 \theta_{23}  \sin^2 2\theta_{12}
\frac{\sin^2(\hat{A}{\Delta})}{\hat{A}^2},
\label{eq:pemu}
\ee
where
\begin{equation}
\Delta\equiv \frac{\ma L}{4E},
\end{equation}

\begin{equation}
\xi \equiv \cos\theta_{13} \,
\sin 2\theta_{12} \, \sin 2\theta_{23},
\end{equation}

\begin{equation}
\hat{A} \equiv \frac{A}{\ma}.
\label{eq:matt}
\end{equation}
Above, $A$ is the matter potential (see Eq. \ref{eq:matter_potential}).
The Eq. \ref{eq:pemu} has been derived under the 
constant matter density approximation. 
The first term of Eq. (\ref{eq:pemu}) can be used to extract 
information about the value of $\theta_{13}$. This is also the 
term which has the largest Earth effect and this effect of matter 
can be used to determine the sign of $\ma$. In Eq. (\ref{eq:pemu}) 
the second term has the CP violating part. This term is 
positive for neutrinos and negative for antineutrinos.
The third term, though $\delta_{CP}$ dependent, is CP conserving.
The last term is independent of both $\theta_{13}$ and $\delta_{CP}$ 
and depends mainly on the solar parameters $\ms$ and $\theta_{12}$. 
In very long baseline experiments, this term has sizable matter effects and
if the true value of $\theta_{13}$ turns out to be zero (or nearly zero),
this would be the only surviving term in $P_{e\mu}$ which could still be 
used to study matter enhanced oscillations \cite{d21msw}.
While we will use this formula to discuss our results in some cases, 
our simulation is based on the exact probabilities.

\subsection{Eight-fold Degeneracy}

Despite its advantage in determining the most interesting
oscillation parameters, this channel is, however, rife
with the problem of ``degeneracies'' --
the ($\theta_{13}$ - $\delta_{CP}$) intrinsic degeneracy \cite{intrinsic},
the ($sgn(\ma)$ - $\delta_{CP}$) degeneracy \cite{minadeg}
and the octant of $\theta_{23}$ degeneracy \cite{th23octant} --
leading to an overall eight-fold degeneracy in the parameter 
values \cite{eight}, of which, obviously, only one is true. 
These degeneracies always deteriorate the performance of an
experiment. 

\subsection{Remedy with ``Magic'' Baseline}

To tackle these problems, there are various 
schemes suggested in the literature which can be broadly categorized in the 
following fashion~:

\begin{enumerate} 

\item
combine data from several experiments observing
the golden channel, but with different baselines $L$
and neutrino energies $E$~\cite{intrinsic,diffLnE,t2ksimulation},

\item
combine data from accelerator experiments observing
different oscillation channels~\cite{silver,dissappear,pee},

\item
combine the golden channel data with those
from atmospheric neutrino experiments~\cite{addatm,cernmemphys},

\item
combine the golden channel data with those 
from reactor antineutrino experiments \cite{addreact},

\item
kill the spurious clone solutions at the ``magic baseline'' \cite{magic}.

\end{enumerate}

\begin{figure}[t]
\begin{center}
\includegraphics[width=12.0cm, height=8.0cm]{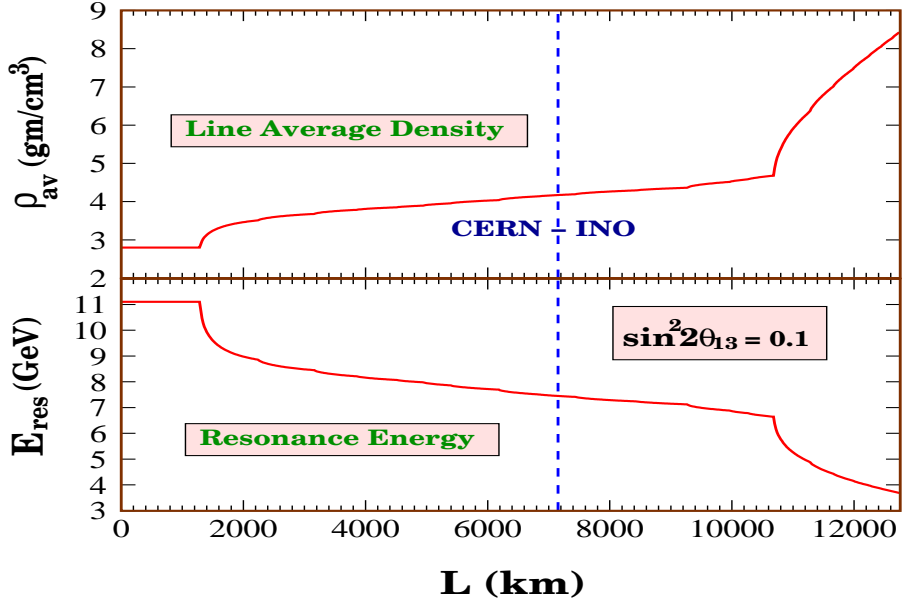}
\caption{\label{fig:reso_omsd}
The upper panel shows the constant line average density of the Earth
for various baselines calculated using the Preliminary Reference 
Earth Model (PREM) \cite{prem} and the lower panel depicts the 
corresponding resonance energy (where $\theta^m_{13} = \pi/4$)
at those baselines. The blue dashed vertical line is drawn
at the CERN-INO baseline.
}
\end{center}
\end{figure}

\begin{table}
\begin{center}

\begin{tabular}{||c||c||c||c||} \hline\hline

\multicolumn{1}{||c||}{\bf Source Location}
& \multicolumn{1}{|c||}{\bf Detector Location}
& \multicolumn{1}{|c||}{\bf Distance}
& \multicolumn{1}{|c||}{\bf Dip Angle}
\cr
\hline\hline

& Fr\'{e}jus & 132 km & 0.6$^\circ$ \cr
& Canfranc  &  657 km &  3.0$^\circ$ \cr
& Gran Sasso & 730 km & 3.2$^\circ$ \cr
& Boulby Mine & 1050 km & 4.7$^\circ$ \cr
& Pyh\"{a}salmi & 2283 km & 10.3$^\circ$ \cr
& Canary Islands & 2763 km & 12.5$^\circ$ \cr
{\bf CERN} & Baksan & 2910 km & 13.2$^\circ$ \cr
& Soudan Mine & 6557 km & 31.0$^\circ$ \cr
& Ash-River & 6575 km & 31.0$^\circ$ \cr
& {\bf INO} & {\bf 7152 km} & {\bf 34.0$^\circ$} \cr
& Homestake & 7346 km & 35.2$^\circ$ \cr
& Henderson & 7743 km & 37.4$^\circ$ \cr
& Super-K   & 8732 km & 43.2$^\circ$ \cr

\hline\hline
\end{tabular}

\caption{\label{tab:cern_distance}
Possibilities for long baselines from CERN.
In the third column, the distances (in km) from CERN to various 
existing/proposed underground neutrino observatories spread over 
the entire world are given. The respective dip angles 
are also mentioned in the last column. The distance of INO from 
CERN is 7152 km with a dip angle of 34.0$^\circ$.}

\end{center}
\end{table}

Now let us see how the concept of ``magic'' baseline works.
A particularly interesting scenario arises when the condition
\begin{equation}
\sin(\hat{A}\Delta)=0
\label{eq:condmagic}
\end{equation}
is satisfied. In such an event, the last three terms in Eq.
(\ref{eq:pemu}) drop out and the $P_{e\mu}$ channel enables 
a clean determination of $\theta_{13}$ and $sgn(\ma)$.

Since $\hat{A}\Delta = A L/4E$ by definition, the 
first non-trivial solution of Eq. \ref{eq:condmagic}
comes in the form $\rho L = \sqrt{2}\pi/G_F Y_e$.
This gives
\begin{equation}
\frac{\rho}{[{\rm gm/cm^3}]}\frac{L}{[\rm km]} \simeq 32725~.
\end{equation}
From the upper panel of Fig. \ref{fig:reso_omsd}, one can immediately observe 
that the above condition is satisfied for the ``magic baseline'' \cite{magic}
\begin{equation}
L_{\rm magic} \simeq 7690 ~{\rm km},
\end{equation}
where the constant line average density of the Earth\footnote{We use the 
full PREM profile
in our simulation without any approximations.} estimated from the 
PREM profile \cite{prem} comes out to be 4.25 gm/cm$^3$. 
An enlightening discussion on the physical meaning of the magic baseline 
is given in \cite{magic2}.

Table~\ref{tab:cern_distance} depicts the distances 
(in km) and the corresponding dip angles (in degree)
from CERN to various existing/proposed underground neutrino 
observatories spread over the entire world.
The CERN-INO distance corresponds to $L=7152$ km, which is 
tantalizingly close to the magic baseline. We therefore expect 
that the beta-beam experiment we consider here should give
an essentially degeneracy-free measurement of both 
$\theta_{13}$ and $sgn(\ma)$.

\subsection{Near-Resonant Matter Effects}

The large CERN-INO baseline, of course, results in very significant
Earth matter effects in the $P_{e\mu}$ channel. In fact one can readily see
from the upper panel of Fig. \ref{fig:reso_omsd} that for the baseline 
of 7152 km, the line average Earth matter density calculated using the 
PREM profile is $\rho_{av}=4.17$ gm/cm$^3$, for which the 
resonance occurs at (see the lower panel of Fig. \ref{fig:reso_omsd})
\be
E_{res} &\equiv& {|\ma| \cos 2\theta_{13} \over
2\sqrt{2} G_F N_e}\label{eq:eres1}~\\
&=& 7.45~{\rm GeV}~,
\label{eq:eres}
\ee
for $|\ma|=2.5\times 10^{-3}$ eV$^2$ and $\stch=0.1$.
One can check (cf. Fig. \ref{fig:flux})
that this is roughly in the ballpark where we expect good enough
beta-beam flux using $^8$B and $^8$Li ions with the
$\gamma$ in the range 350-650.
The resonance energy is $\propto$ $\cos 2\theta_{13}$ and therefore
it depends on $\theta_{13}$ mildly. At the CERN-INO baseline, the
resonance energy can vary within a very small range of 
7.1 to 7.85 GeV depending on the present allowed span of $\theta_{13}$. 

\subsection{The Benchmark Oscillation Parameters}

\begin{table}[t]
\begin{center}
\begin{tabular}{||c||c||} \hline\hline
\multicolumn{1}{||c||}{{\rule[0mm]{0mm}{6mm}{Benchmark Values}}}
& \multicolumn{1}{|c||}{\rule[-3mm]{0mm}{6mm}{$1\sigma$ estimated error}}
\cr
\hline \hline
$|\Delta m^2_{31}{\rm (true)}| = 2.5 \times 10^{-3} \ {\rm eV}^2$
& $\sigma(|\Delta m^2_{31}|)=1.5\%$
\cr \hline
$\sin^2 2 \theta_{23}{\rm (true)}| = 1.0$
& $\sigma(\sta)=1\%$
\cr \hline
$\Delta m^2_{21}{\rm (true)} = 8.0 \times 10^{-5} \ {\rm eV}^2$
& $\sigma(\Delta m^2_{21})=2\%$
\cr \hline
$\sin^2\theta_{12}{\rm (true)} = 0.31$
& $\sigma(\sss)=6\%$
\cr \hline
$\rho{\rm (true)} = 1~{\rm (PREM)}$ & $\sigma(\rho)=5\%$
\cr
\hline \hline
\end{tabular}
\caption{\label{tab:bench}
Chosen benchmark values of oscillation
parameters and their $1\sigma$ estimated errors. The last row
gives the corresponding values for the Earth matter density.}
\end{center}
\end{table}

Note that the low order expansion of the probability $P_{e\mu}$
given by Eq. (\ref{eq:pemu}) is valid only for values of
$E$ and Earth matter density $\rho$ (and hence $L$)
where flavour oscillations are far from resonance, {\it i.e.},
$ \hat{A}\ll 1$. In the limit $\hat{A} \sim 1$,
one can check that even though the analytic expression for $P_{e\mu}$
given by Eq. (\ref{eq:pemu}) remains finite, the resultant probability 
obtained is incorrect \cite{Takamura:2005df}.
We reiterate that Eq. (\ref{eq:pemu}) was presented only
in order to elucidate the importance of the magic baseline. 
For all the numerical results presented in this thesis, we calculate the 
exact three generation oscillation probability using the full
realistic PREM \cite{prem} profile for the Earth matter density.
Unless stated otherwise, we have generated our simulated
data for the benchmark values in the first column of Table~\ref{tab:bench}.
These values have been chosen in conformity with the status of
the oscillation parameters in the light of the current neutrino
data~\cite{limits}. The values of $\stcht$, $\dcpt$ and
mass hierarchy which are allowed to vary in our study, will be
mentioned wherever applicable. When we fit this simulated data, we
marginalize over all the oscillation parameters, the Earth matter density,
as well as the neutrino mass ordering, as applicable.
We expect to have a better knowledge of all the parameters mentioned in 
Table~\ref{tab:bench} when the Beta-beam facility comes up. In particular,
we assume that the $1\sigma$ error on them will be reduced
to the values shown in the second column of
Table~\ref{tab:bench}~\cite{huber10,solarprecision,tomography}.
Therefore, we impose ``priors'' on these quantities, with the
corresponding $1\sigma$ error. In the fit, we allow for a 5\% uncertainty 
in the PREM profile and take it into account by inserting a prior and
marginalizing over the density normalization.
We analyze the full spectral data from {\rm both} the
neutrino ($^8$B) and antineutrino ($^8$Li) run expected in the
CERN-INO beta-beam set-up.

\subsection{Phenomenology with $P_{e\mu}$}

\begin{figure}[t]
\begin{center}
\includegraphics[width=16.0cm, height=9.5cm]{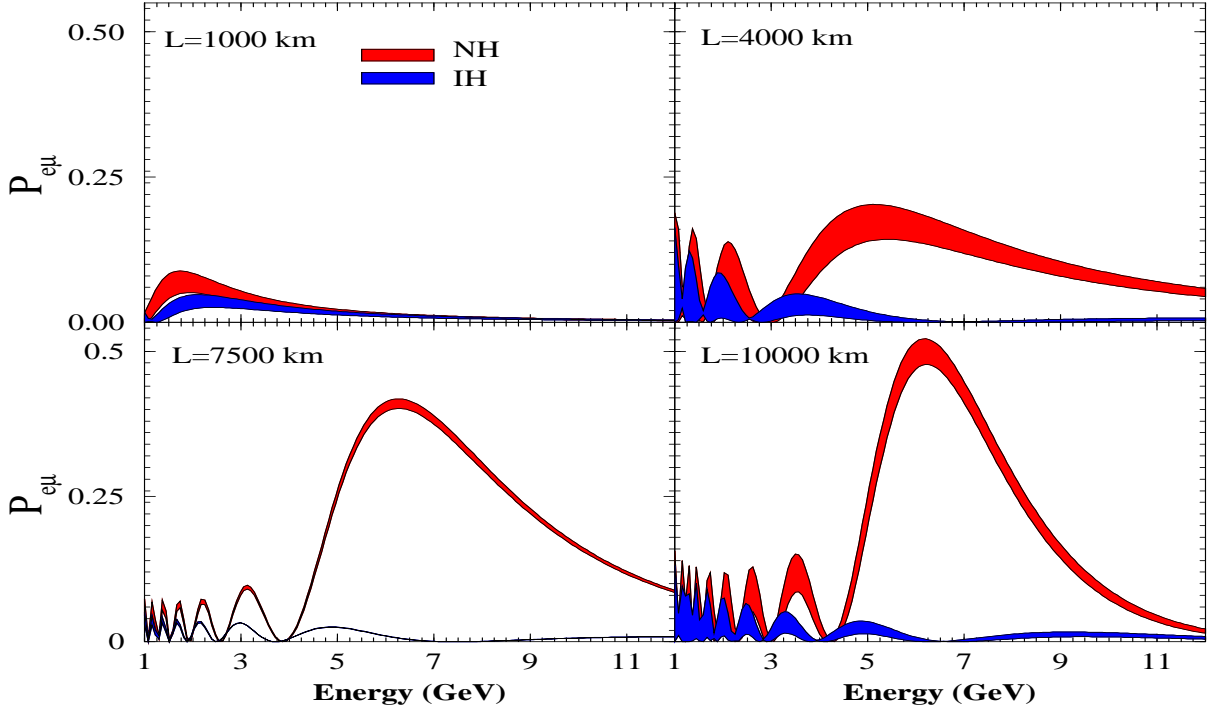}
\caption{\label{fig:p12Efixedcp}
The transition probability $P_{e\mu}$ as a function of $E$
for four values of the baseline $L$. The band reflects the effect
of the unknown $\delta_{CP}$. The dark (red) shaded band is for
the NH while the light (cyan) shaded
band is for the IH. We have taken $\stch=0.1$
and for all other oscillation parameters we assume the
benchmark values given in Table \ref{tab:bench}.}
\end{center}
\end{figure}

\begin{figure}[t]
\begin{center}
\includegraphics[width=16.0cm, height=9.5cm]{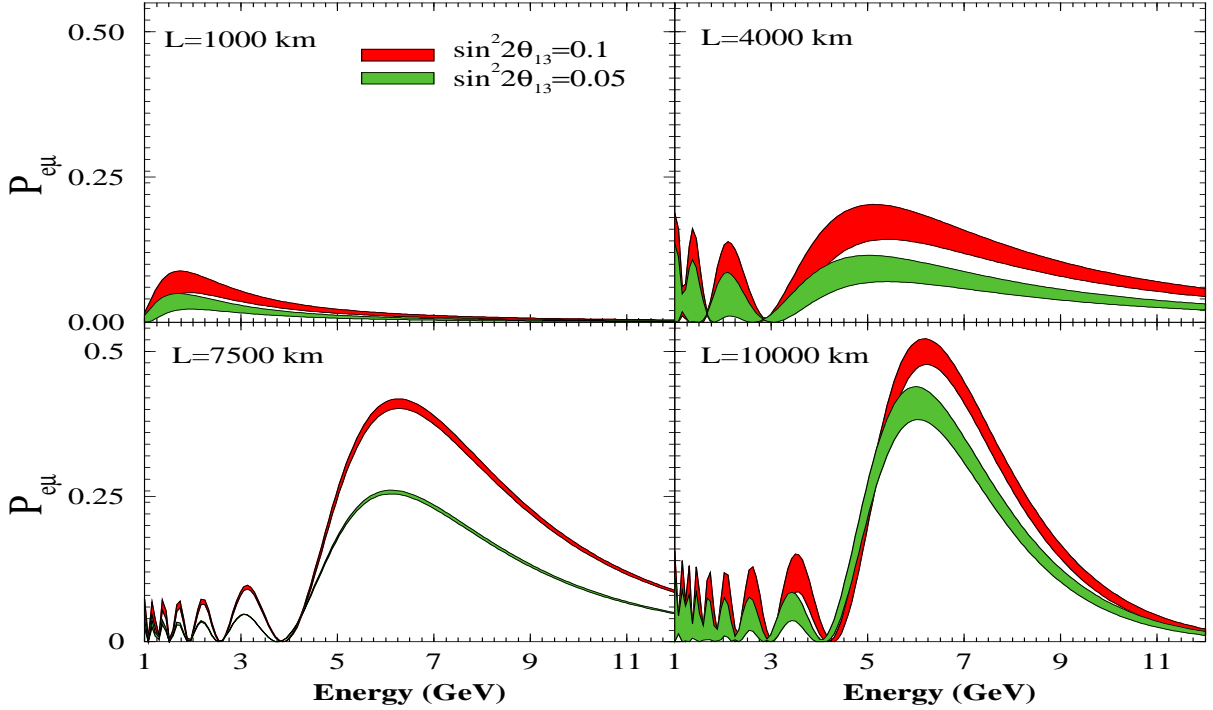}
\caption{\label{fig:p12Efixedt13}
The dark (red) shaded band is the same as in Fig. \ref{fig:p12Efixedcp}.
The light (green) shaded band shows the corresponding $P_{e\mu}$
for $\stch=0.05$. Values of all the other oscillation parameters
are same as in Fig. \ref{fig:p12Efixedcp} and the hierarchy is assumed
to be normal.}
\end{center}
\end{figure}

The exact neutrino transition probability using the full PREM density profile is
given in Fig. \ref{fig:p12Efixedcp} as a function of the
neutrino energy, for four different baselines (four panels).
We allow $\delta_{CP}$ to take on all possible values between 
0-360 degrees and the resultant probability is shown as a band, 
with the thickness of the band reflecting the effect of 
$\delta_{CP}$ on $P_{e\mu}$. The figure is drawn assuming the
benchmark values of the oscillation parameters
given in Table \ref{tab:bench} and $\stch=0.1$.
We show the probability for both the NH (dark band) 
as well as the IH (light band). As discussed in detail above, 
for $L=7500$ km, which is close to the magic baseline, the
effect of the CP phase is seen to be almost negligible,
while for all other cases the impact of $\delta_{CP}$ on $P_{e\mu}$ 
is seen to be appreciable. In fact, for $L=1000$ km, the probability
corresponding to the NH and IH become almost
indistinguishable due to the uncertainty arising from the unknown value
of $\delta_{CP}$. As the baseline is increased, Earth matter
density increases, enhancing the impact of matter effects.
The probability for NH is hugely enhanced for
the neutrinos, while for the IH matter effects
do not bring any significant change.
This difference in the predicted probability, evident
in the panels corresponding to $L$ = 4000, 7500 and 10000 km
of Fig. \ref{fig:p12Efixedcp}, can be used to
determine the neutrino mass ordering.

In Fig. \ref{fig:p12Efixedt13} we display the dependence of the
neutrino probability $P_{e\mu}$ on the mixing angle $\theta_{13}$
for four different baselines. The dark bands, as in Fig.
\ref{fig:p12Efixedcp}, are for NH and $\stch=0.1$
with full variation of $\delta_{CP}$, while the light bands are
for NH and $\stch=0.05$.  The impact of matter
effect in increasing the $\theta_{13}$ sensitivity of a given
experimental set-up is evident from the figure. The $\theta_{13}$
sensitivity for $L=1000$ km can be seen to be much weaker than
for the other cases, since matter effects are smaller.
However, the most striking feature seen in Fig.
\ref{fig:p12Efixedt13} is the effect of the
magic baseline in enhancing the sensitivity of the
experiment to $\theta_{13}$. The figure
clearly shows that the difference in
the predicted $P_{e\mu}$ for the two values of $\stch$
is largest for $L=7500$ km (since effect of
$\delta_{CP}$ is the least) and thus an experiment at this
baseline is most suitable for probing $\theta_{13}$.
Figs. \ref{fig:p12Efixedcp} and \ref{fig:p12Efixedt13}
therefore reinforce our choice of the near-magic baseline
as one of the best options for determining the neutrino mass hierarchy
and $\theta_{13}$, since both these parameters
are directly related to large matter effects and
the uncertainty of $\delta_{CP}$ could prove to be a hindrance in 
their measurement at non-magic baselines.

\subsection{One Mass Scale Dominance}

\begin{figure}[t]
\begin{center}
\includegraphics[width=12.0cm, height=8.0cm]{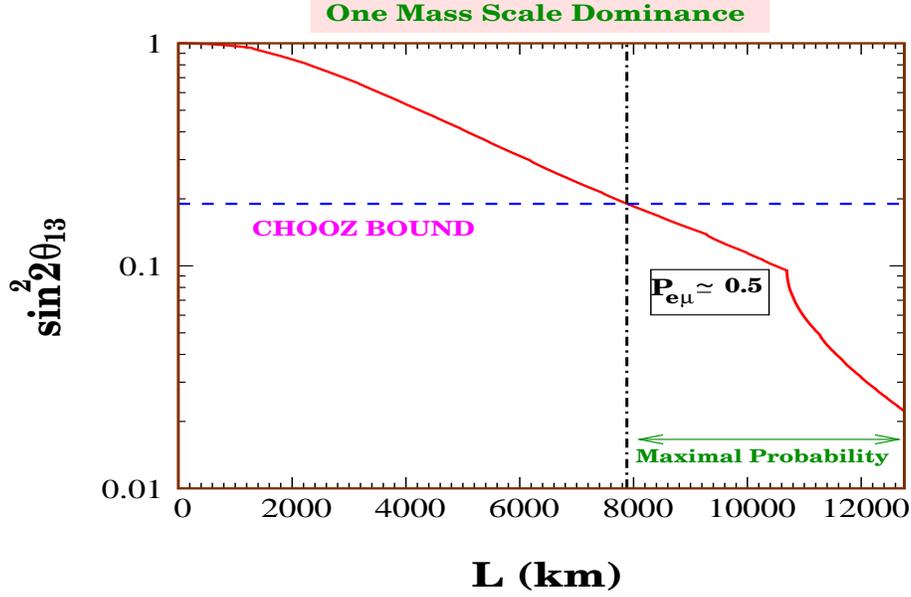}
\caption{\label{fig:omsd}
The values of $\stch$ needed for the maximal matter effect
(from Eq. \ref{eq:eecondtn} with $p$ = 0) at different baselines 
using the constant line average density of the Earth. The blue dashed 
horizontal line shows the present upper bound on $\stch$ which 
predicts that the maximal probability (= 0.5) can only be achieved for 
$L \ge$ 7880 km (see the black dot-dashed vertical line).}
\end{center}
\end{figure}

Now let us discuss the ``Golden'' channel oscillation probability 
in the approximation where the solar mass squared difference $\ms$ can be neglected.
This is known as one mass scale dominance (OMSD) approximation.  
Under this condition with the constant matter density approximation, 
the probability $P_{e\mu}$ can be expressed as  
\begin{equation}
P_{e \mu} =
\sin^2\theta_{23} \sin^22\theta_{13}^{m} 
\sin^2 \left[1.27 (\Delta m^2_{31})^{m} {L}/{E} \right],
\label{eq:omsd_p_emu}
\end{equation}
where $\sin^22\theta_{13}^{m}$ and $(\Delta m^2_{31})^{m}$ are given 
by Eq. \ref{eq:matter_param}.
The probability $P_{e\mu}$ would be largest 
(= 0.5, assuming $\theta_{23} = \pi/4$) if the conditions
\begin{equation}
\sin^2 2 \theta^m_{13} = 1
\label{eq:cond1} 
\end{equation}
and
\begin{equation}
\sin^2 \left[1.27 (\Delta m^2_{31})^m L/E \right]=1
\label{eq:cond2}
\end{equation}
are satisfied simultaneously.
The first condition is achieved at the resonance energy.
The second condition gives the energy where the $(\ma)^m$
driven oscillatory term is maximal,
\begin{equation}
E^{m}_{max} = \frac{1.27 (\Delta m^2_{31})^m L}{(2p+1) \pi/2}, \, p=0,1,2..
\label{eq:max_osc}
\end{equation}
Maximum matter effect is obtained when \cite{anuls:2001zn,gandhi1,gandhi2} 
\begin{equation}
E_{res} = E^m_{max}, 
\label{eq:max_cond}
\end{equation}
which gives (using Eqs. \ref{eq:matter_param} \& \ref{eq:eres1}),
\begin{equation}
(\rho L)^{max} =
\frac{(2p+1) \pi \times 5.18 \times 10^3} { \tan 2\theta_{13}} ~{\rm km ~gm/cm^3}.
\label{eq:eecondtn}
\end{equation}
Although both $E_{res}$ and $E^{m}_{max}$ depend on the value of 
$\Delta m^2_{31}$, the distance at which we get the maximum matter 
effect is independent of $\Delta m^2_{31}$. However, it is controlled 
by $\theta_{13}$ very sensitively. In Fig. \ref{fig:omsd}, we have 
plotted the values of $\stch$ required for the maximal matter effect 
at different baselines using the constant line average density of the 
Earth as predicted by the PREM profile. For $p=0$, we can have maximal
matter effect only if $L \ge 7880$ km because of the upper bound 
of 0.19 on $\stch$ \cite{CHOOZ,limits}. At $L=7152$ km, for 
maximal matter effect $\stch\simeq 0.23$, which is already 
ruled out by the present constraint. Therefore at the CERN-INO baseline,
maximal matter effect cannot be achieved and the oscillation probability
peaks roughly at 6 GeV (see the Figs. \ref{fig:p12Efixedcp} 
\& \ref{fig:p12Efixedt13}) instead of 7.45 GeV

\section{\fbox{Event Rates in ICAL@INO}}

\subsection{The ICAL Detector at INO}

\begin{figure}[t]
\includegraphics[width=8.0cm, height=7.0cm]{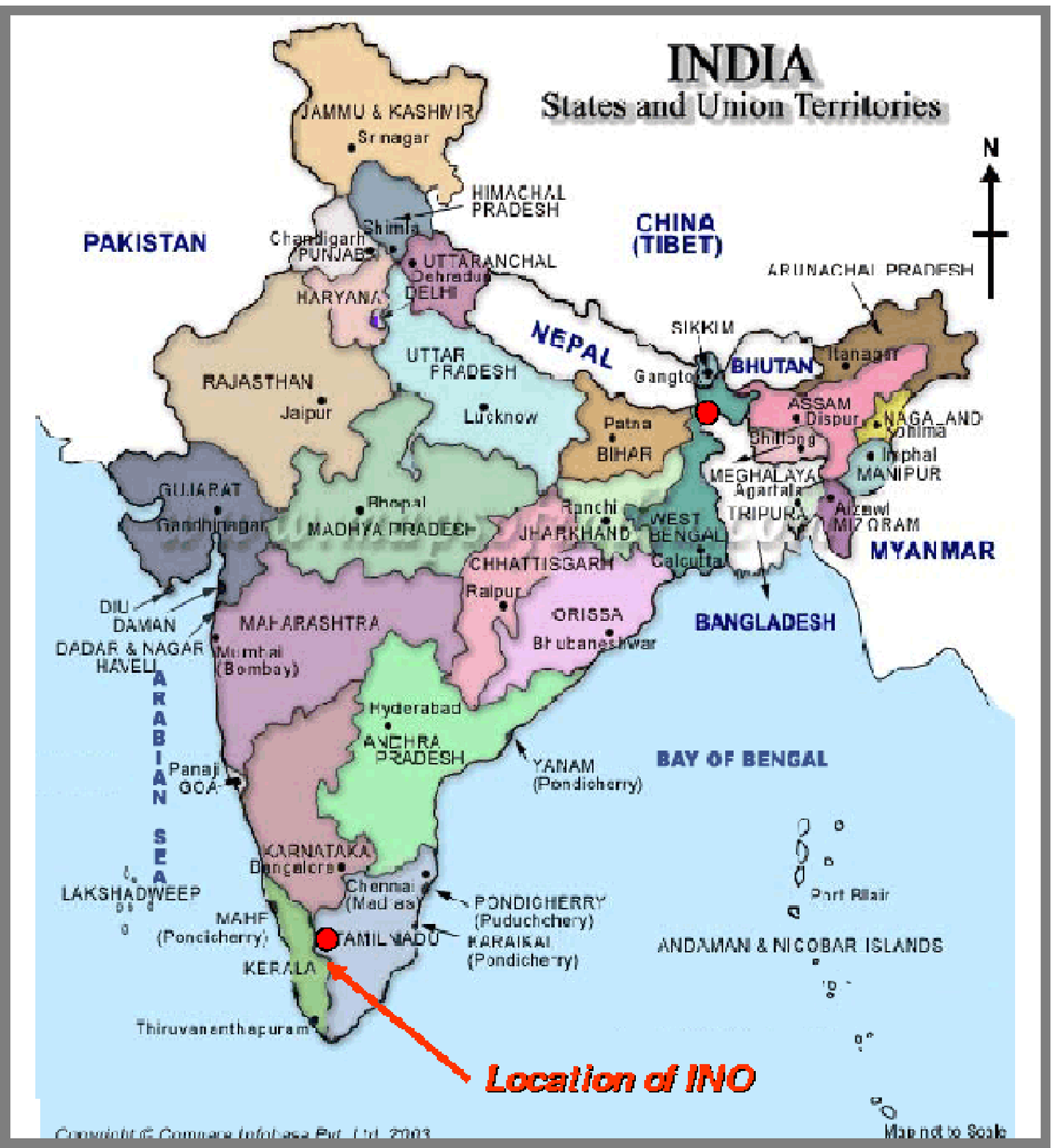}
\vglue -7.05cm \hglue 8.4cm
\includegraphics[width=8.4cm, height=7.0cm]{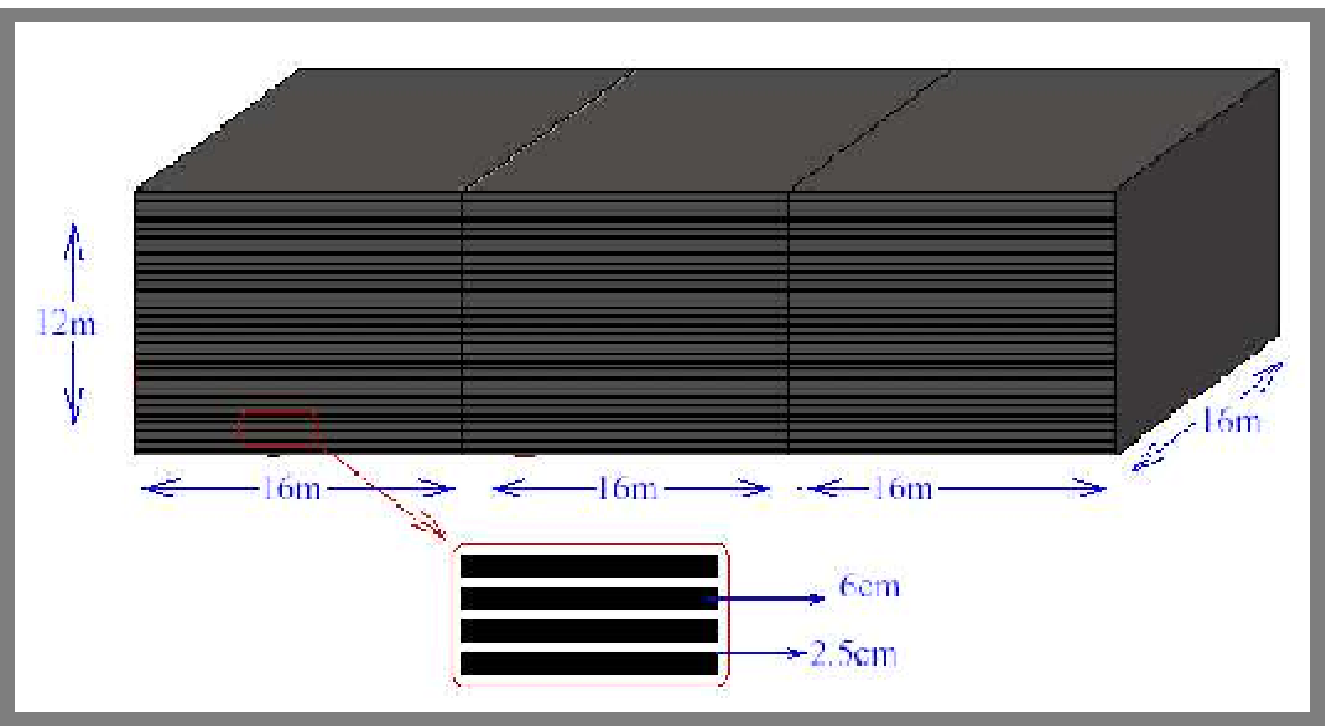}
\caption{\label{fig:ino}
Left panel shows the location of the India-based Neutrino Observatory.
The schematic view of the 50 kton iron calorimeter detector
consisting of three modules each having 140 layers of iron plates
is given in the right panel. These figures are obtained from the
official website of INO (http://www.imsc.res.in/$\sim$ino/).
}
\end{figure}

The proposed large magnetized iron calorimeter at the 
India-based Neutrino Observatory \cite{ino} is planned 
to have a total mass of 50 kton at startup, which might be
later upgraded to 100 kton. The left panel of Fig. 
\ref{fig:ino} shows the location of the INO facility  
which is expected to come up at PUSHEP (lat. North 11.5$^\circ$,
long. East 76.6$^\circ$), situated close to Bangalore
in southern India. The distances and the corresponding 
dip angles from INO to various existing accelerator based 
laboratories are given in Table \ref{tab:ino_distance}.
The JAERI (Japan Atomic Energy Research Institute) to INO
distance is 6477 km with a dip angle of 30.5$^\circ$.
The distances from INO to CERN and RAL (Rutherford Appleton Laboratory) 
are 7152, and 7653 km, respectively. These baselines are quite close
to the magic baseline. In this sense, the location of INO is 
ideal for futuristic long baseline experiments. The BNL 
(Brookhaven National Laboratory) and the FNAL are the two major 
accelerator based facilities in US and they constitute the 
baselines of 11081, and 11306 km, respectively from INO with
the dip angles $> 60^\circ$. The first phase of this detector
(starting around 2012) will focus on atmospheric neutrino
measurements. The ICAL detector will have a modular structure with 
a total lateral size of $48{\rm m} \times 16 {\rm m}$, divided into 
three modules of $ 16{\rm m}\times  16 {\rm m}$ each (see the right panel of 
Fig. \ref{fig:ino}). Each of these modules will have 140 horizontal
layers of $\sim 6$ cm thick iron plates, separated from
each other by a gap of $\sim 2.5$ cm to hold the active detector material, 
giving a total height of 12 m for the full detector. 
The active detector elements will be $2$ cm thick glass resistive plate
chambers (RPC) and will be filled with a suitable gas mixture, 
which will be recycled with approximately one volume change per day. 
An internal magnetic field of $\sim 1.3$ Tesla would be applied over 
the entire detector. The detector will be surrounded by an external layer of
scintillator or proportional gas counters which will act both
as veto to identify external muon backgrounds as well as to identify
partially contained events.

\begin{table}
\begin{center}

\begin{tabular}{||c||c||c||c||} \hline\hline

\multicolumn{1}{||c||}{\bf Detector Location}
& \multicolumn{1}{|c||}{\bf Source Location}
& \multicolumn{1}{|c||}{\bf Distance}
& \multicolumn{1}{|c||}{\bf Dip Angle}
\cr
\hline\hline

& JAERI & 6477 km & 30.5$^\circ$ \cr
& {\bf CERN} & 7152 km & 34.0$^\circ$ \cr
{\bf INO} & RAL & 7653 km & 36.8$^\circ$ \cr
& BNL & 11081 km & 60.3$^\circ$ \cr
& FNAL  & 11306 km &  62.4$^\circ$ \cr

\hline\hline
\end{tabular}

\caption{\label{tab:ino_distance}
The distances from INO to various accelerator based laboratories
spread over the entire world. The corresponding dip angles are
also noted.}

\end{center}
\end{table}

\begin{table}[t]
\begin{center}

\begin{tabular}{||c||c||} \hline\hline

Total Mass & 50 kton \cr
\hline
Energy threshold & 1 GeV \cr
\hline
Detection Efficiency ($\epsilon$) & 80\% \cr
\hline
Charge Identification Efficiency ($f_{ID}$)& 95\% \cr
\hline
Energy Resolution ($\sigma$) (GeV) & 0.15E(GeV) \cr
\hline
Bin Size & 1 GeV \cr
\hline
Background Rejection & 0.0001 \cr
\hline
Signal error & 2.5\% \cr
\hline
Background error & 5\% \cr
\hline\hline
\end{tabular}
\caption{\label{tab:detector}
Detector characteristics of ICAL@INO used in the
simulations \cite{mind}. The bin size is kept fixed, 
while the number of bins is varied according to the maximum energy.
}
\end{center}
\end{table}

We are interested in measuring the golden channel probability $P_{e\mu}$. 
Since we have a $\nue$ ($\anue$) flux in the beam, we need a detector which 
is sensitive to muons (antimuons). The detector should have a suitable
energy threshold, depending on the energy spectrum of the
beta-beam. In addition, it should have a good energy resolution and 
low backgrounds. According to the ongoing detector simulation study performed 
by the INO collaboration, the detector energy 
threshold for $\mu^\pm$ is expected to be around $\sim 1$ GeV and charge 
identification efficiency\footnote{In case of a beta-beam, the neutrino
and antineutrino events produced in distinct bunches can be 
distinguished from each other using the nano-second time resolution 
of the detector. The charge identification efficiency is incorporated since 
that helps in reducing the neutral current backgrounds.} will be about 95\%.
The detection efficiency of ICAL after cuts is expected to be about 80\%.  
In what follows, we will present our numerical results assuming an energy 
threshold of 1 GeV, detector charge identification efficiency
as 95\% and detection efficiency as 80\% for $\mu^\pm$ 
(cf. Table \ref{tab:detector}). However, we will also show the impact 
of changing the threshold. We have seen that our results remain unaffected 
if the energy threshold is raised to 2 GeV for the entire range of assumed 
Lorentz boost factor $\gamma=250-650$. For $\gamma > 350$ the threshold 
can be even 3 GeV, while for $\gamma > 500$ one can work with an energy 
threshold of 4 GeV, without changing the final results. 
This can be seen from Fig. \ref{fig:flux}; for $\gamma=350(500)$, 
the majority of neutrinos arriving at INO-ICAL would have $E>3(4)$ GeV.
The energy resolution of the detector is expected to be reasonable and 
we assume that the neutrino energy will be reconstructed with an uncertainty 
parameterized by the Gaussian energy resolution function with $\sigma_E=0.15E$,
where $E$ is the energy of the neutrino and $\sigma_E$
is related with the half width at half maximum (HWHM) in the fashion
HWHM = 1.17$\sigma$. We will present and compare the sensitivity of this 
experimental set-up with and without the full spectral analysis using {\rm both} 
the neutrino ($^8$B) and antineutrino ($^8$Li) run.

\subsection{Oscillation Signal at ICAL@INO}

The $\numu$ induced $\mu^-$ event spectrum at INO is estimated using
\be
N_{i} = T\, n_n\, f_{ID}\,\epsilon~  \int_0^{E_{\rm max}} dE
\int_{E_{A_i}^{\rm min}}^{E_{A_i}^{\rm max}}
dE_A \,\phi_{\nue}(E) \,\sigma_\numu(E) \,R(E,E_A)\, P_{e\mu}(E) \, ,
\label{eq:events}
\ee
where $T$ is the total running time (taken as five years),
$\phi_{\nue}(E)$ (given by Eq. \ref{eq:flux}) is the unoscillated 
beta-beam flux at the detector, $\epsilon$ is the detector efficiency,
$n_n$ are the number of target nucleons in the detector,
$f_{ID}$ is the charge identification efficiency and
$R(E,E_A)$ is the energy resolution function of the detector,
for which we assume a Gaussian function. The quantities $E$ and $E_A$ 
are the true and reconstructed neutrino energy respectively.
$E^{\rm max}$ is the maximum energy of the neutrinos for a given 
Lorentz factor $\gamma$ (given by Eq. (\ref{eq:enumax})) and $P_{e\mu}$ 
is the $\nue\rightarrow \numu$ oscillation probability. For muon events, 
$\sigma_\numu$ is the neutrino interaction cross-section. 
The expression for the $\mu^+$ signal in the detector from a 
$\anue$ beta-beam flux is given by replacing $\phi_\nue$ by 
$\phi_\anue$, $\sigma_\numu$ by $\sigma_\anumu$ and $P_{e\mu}$ 
by $P_{\bar e{\bar\mu}}$. For the neutrino-nucleon interaction 
we consider quasi-elastic scattering, single-pion production and 
deep inelastic scattering and use the cross sections given in the 
Globes package \cite{globes} which are taken from 
\cite{Messier:1999kj,Paschos:2001np}.

\begin{enumerate}

\item
\fbox{\bf Event Rates vs. $\stch$}

\begin{figure}
\includegraphics[width=8.0cm, height=7.0cm]{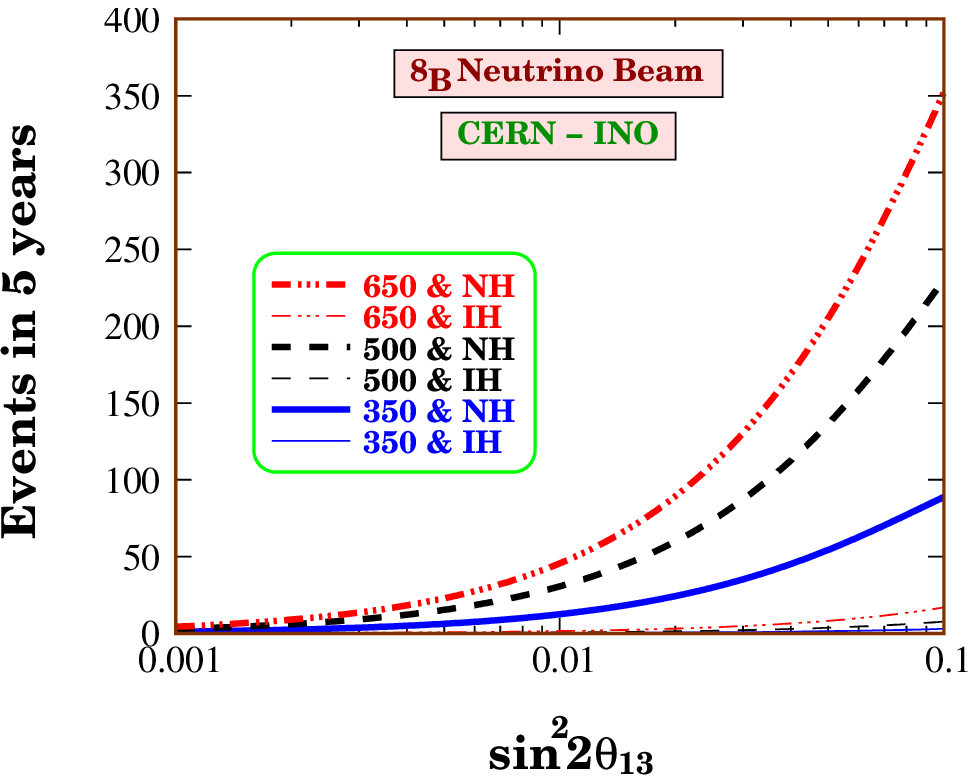}
\vglue -7.0cm \hglue 8.5cm
\includegraphics[width=8.0cm, height=7.0cm]{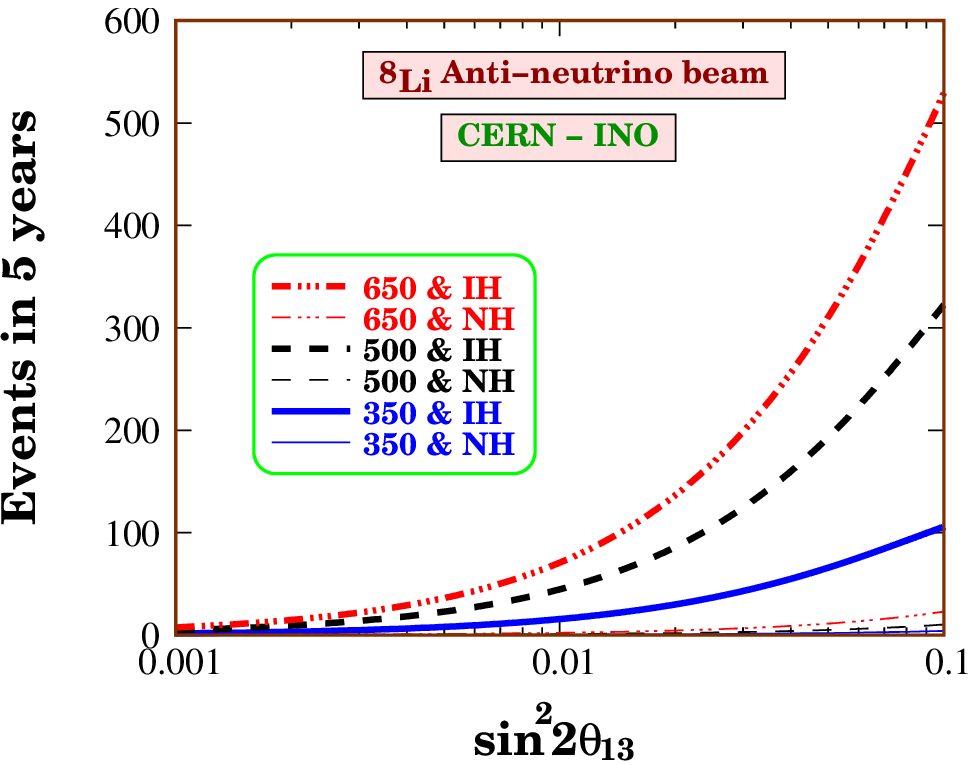}
\caption{\label{fig:rates}
The expected number of events in
5 years running time, as a function of $\stch$. The value of $\gamma$ and
the hierarchy chosen corresponding to each curve is shown in the
figure legend.}
\end{figure}

Fig. \ref{fig:rates} portrays the number of events expected
in INO-ICAL from 5 years exposure of an $^8$B (for $\nue$, shown
in the left-hand panel) or $^8$Li (for $\anue$, shown in the right-hand panel)
beta-beam from CERN. The expected number of events is presented as a
function of $\stch$ for both NH and IH
for three benchmark values of the boost factor $\gamma$.
Large resonant matter effects in the neutrino channel for
NH drives the number of expected events to very
large values, compared to what would be expected for
IH. Similarly, in the antineutrino channel
we have resonant matter effects for IH and 
the number of predicted events is many times larger than for 
NH\footnote{In fact, matter effects are seen to suppress the 
oscillation probability for neutrinos (antineutrinos) compared 
to that in vacuum when the hierarchy is inverted (normal).}. 
This difference in the number of events is seen to increase with $\stch$.

From the left-hand panel of Fig. \ref{fig:rates} it can be noted that 
for $\gamma=500$ and $\stch=0.05$, the predicted number of neutrino
events for NH is 137, while that for IH is only 4. 
This implies that if the NH was true, we could
comprehensively rule out the wrong IH using the neutrino run.
If IH is true then antineutrino run will be very useful
(see the right panel of Fig. \ref{fig:rates}) to rule out the
wrong NH. Note that while the interaction cross section
for the $\anue$s are smaller, the flux itself is larger 
owing to the (assumed) larger number of
decays per year for $^8$Li. Thus the statistics
expected in both the neutrino as well as the
antineutrino channel is comparable.
As discussed above, resonant matter effects in the
neutrino (antineutrino) channel for the normal (inverted)
hierarchy, results in substantial enhancement in the
observed number of events. In particular, we note from
Fig. \ref{fig:rates} that the number of events in the
near-resonant channels depends strongly on the value of 
$\stch$, since the extent of matter effects
is dictated directly by $\theta_{13}$. We can see
from the figure that the dependence of the event rate on
$\stch$ is much enhanced due to matter effects.

\item
\fbox{\bf Iso-event Curves in the $\stch$-$\delta_{CP}$ plane}

\begin{figure}
\includegraphics[width=8.0cm, height=7.0cm]{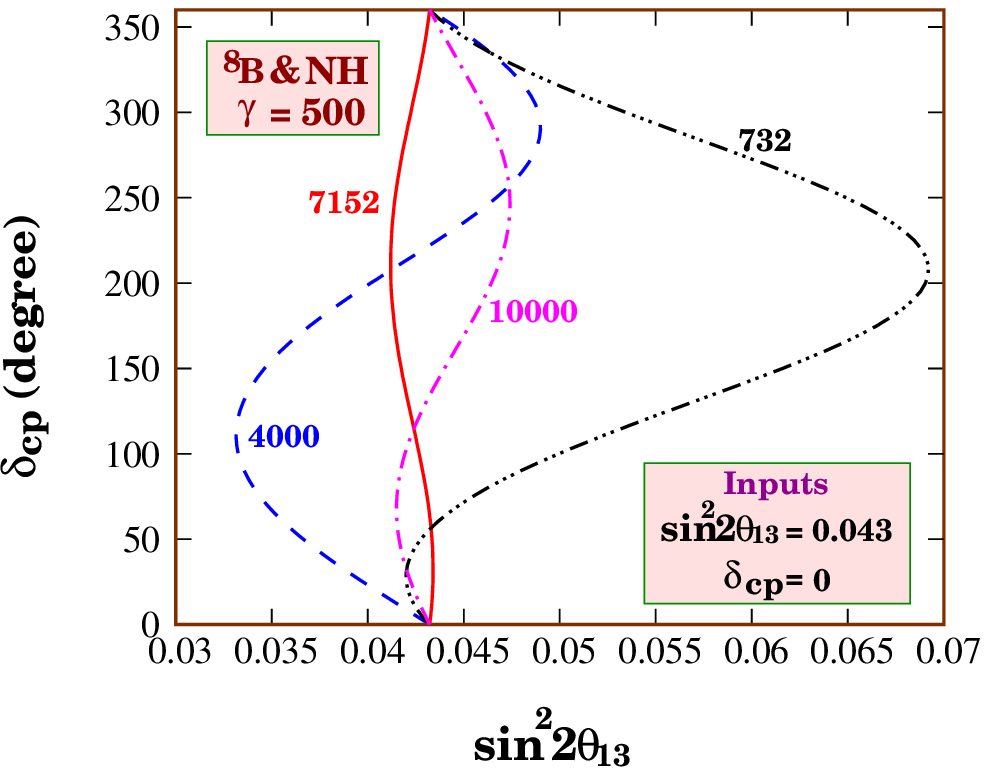}
\vglue -7.0cm \hglue 8.5cm
\includegraphics[width=8.0cm, height=7.0cm]{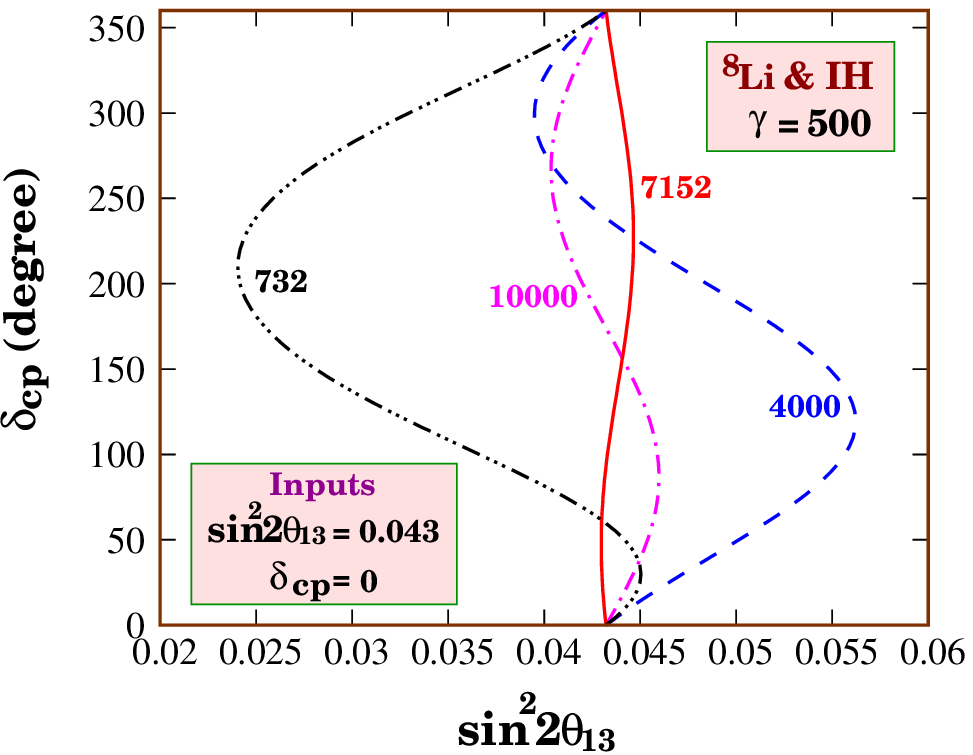}
\caption{\label{fig:deg}
The iso-event curves in the $\delta_{CP}$-$\stch$ plane
for four baselines are shown in the figure. The true
values of $\delta_{CP}$ and $\stch$ are assumed to
be $0^\circ$ and $0.043$ respectively.
Left- (right-) hand panel
is for the $\nue$ ($\anue$) beta-beam. The assumed
hierarchy is mentioned in the figure.
}
\end{figure}

Fig. \ref{fig:deg} shows the effect of $\delta_{CP}$
on the variation of the measured rate with the value of the mixing angle $\theta_{13}$,
for four different baselines 732 km, 4000 km, 7152 km and 10000 km.
The left-hand panel shows the results for the $\nue$ beta-beam
assuming a normal mass hierarchy,
while the right-hand panel gives the
same for the $\anue$ beta-beam flux with IH.
For all cases in any panel, a 50 kton magnetized iron calorimeter 
as the far detector and the same flux created at the
source have been considered.
Each curve gives the sets of values of
$\{\stch,\delta_{CP}\}$ which give the same observed rate
in the detector as the set $\{\stch=0.04,\delta_{CP}=0^\circ\}$.
In other words, if the true value of $\stch$ and
$\delta_{CP}$ were 0.04 and $0^\circ$ respectively, then every
point on a given curve would also be a solution for that
experiment. It is clear from this figure that combining
results from experiments at different baselines helps
solve/reduce the problem. However,
the most important issue exemplified here is the fact that
for the baseline 7152 km, which is the CERN-INO distance,
the effect of the unknown value of $\delta_{CP}$ on
the measurement of $\theta_{13}$ and the mass hierarchy,
is very small. This happens because this
distance corresponds to a near-magic baseline for which, as
noted earlier, the $\delta_{CP}$ dependent terms are
almost vanishing.

\item
\fbox{\bf Event Rates vs. $L$}

\begin{figure}[t]
\begin{center}
\includegraphics[width=10.0cm]{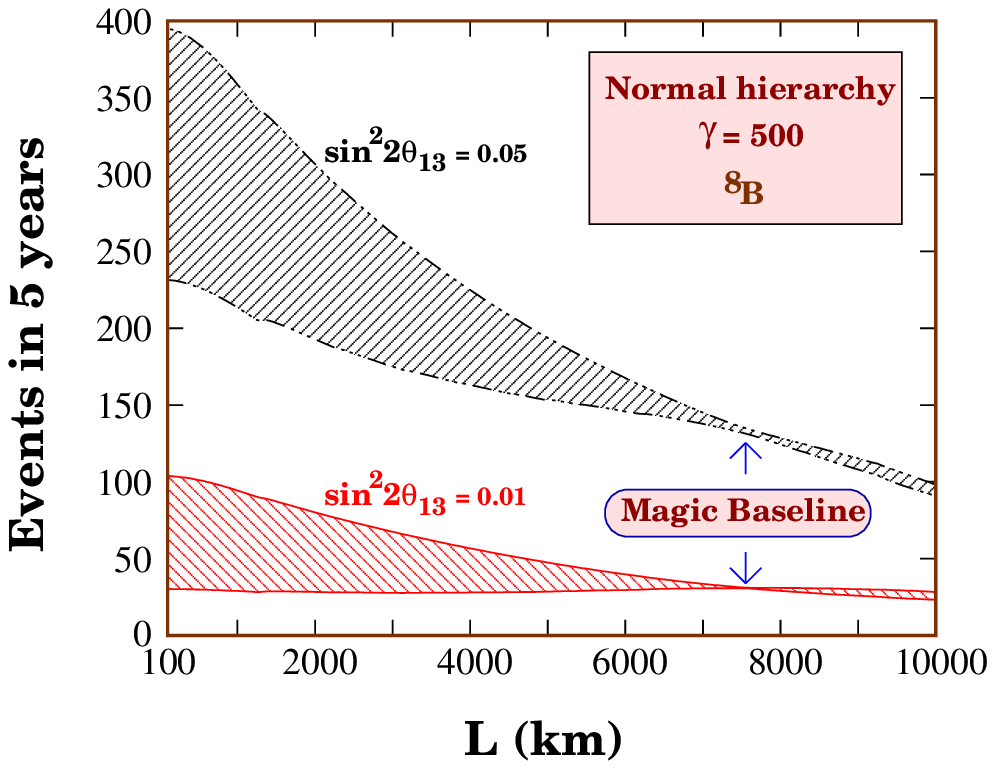}
\caption{\label{fig:magic}
Total number of expected
events in five years as a function of the
baseline $L$ for the $^8$B source with $\gamma=500$ and for
two values of $\stch$ and assuming that the
NH is true.
The hatched areas show the
expected uncertainty due to the CP phase.
}
\end{center}
\end{figure}

\begin{figure}[t]
\begin{center}
\includegraphics[width=15.0cm]{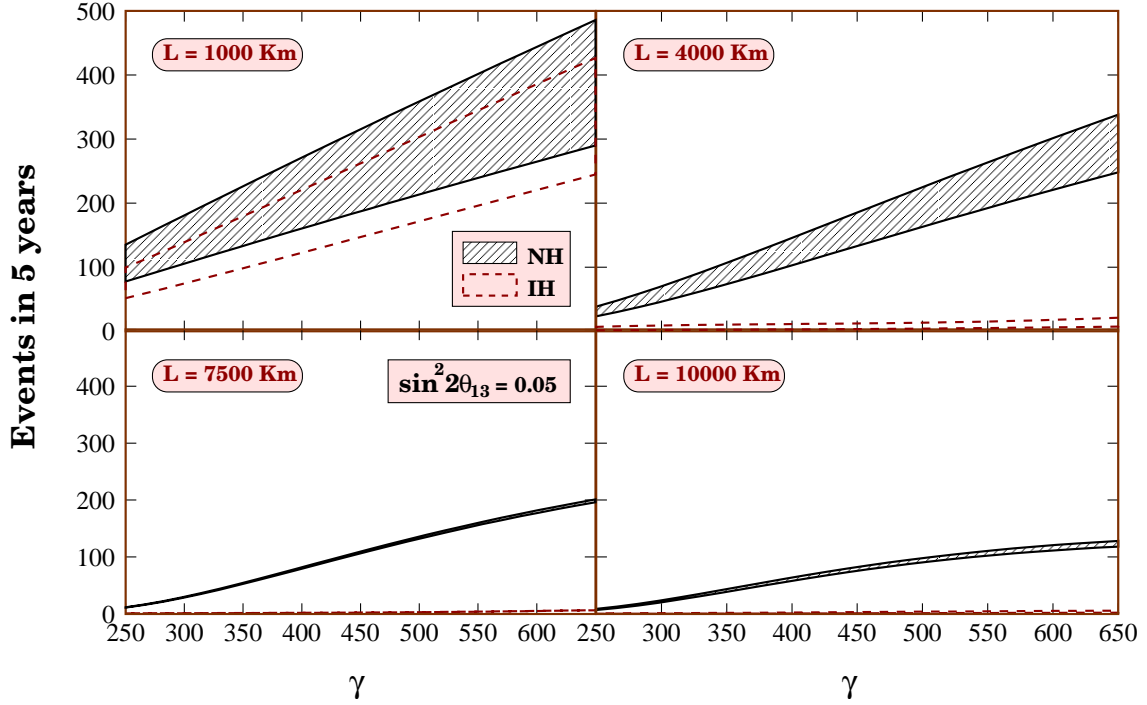}
\caption{\label{fig:eventshier}
Total number of events as a function of $\gamma$ for the
$^8$B source, for different values of $L$ are shown in
the four panels. The black hatched area shows the uncertainty
range due to the CP phase when NH is true,
while the area between the maroon dashed lines
shows the corresponding
uncertainty when IH is true.
For all cases we assume $\stch=0.05$.
}
\end{center}
\end{figure}

\begin{figure}[t]
\begin{center}
\includegraphics[width=15.0cm]{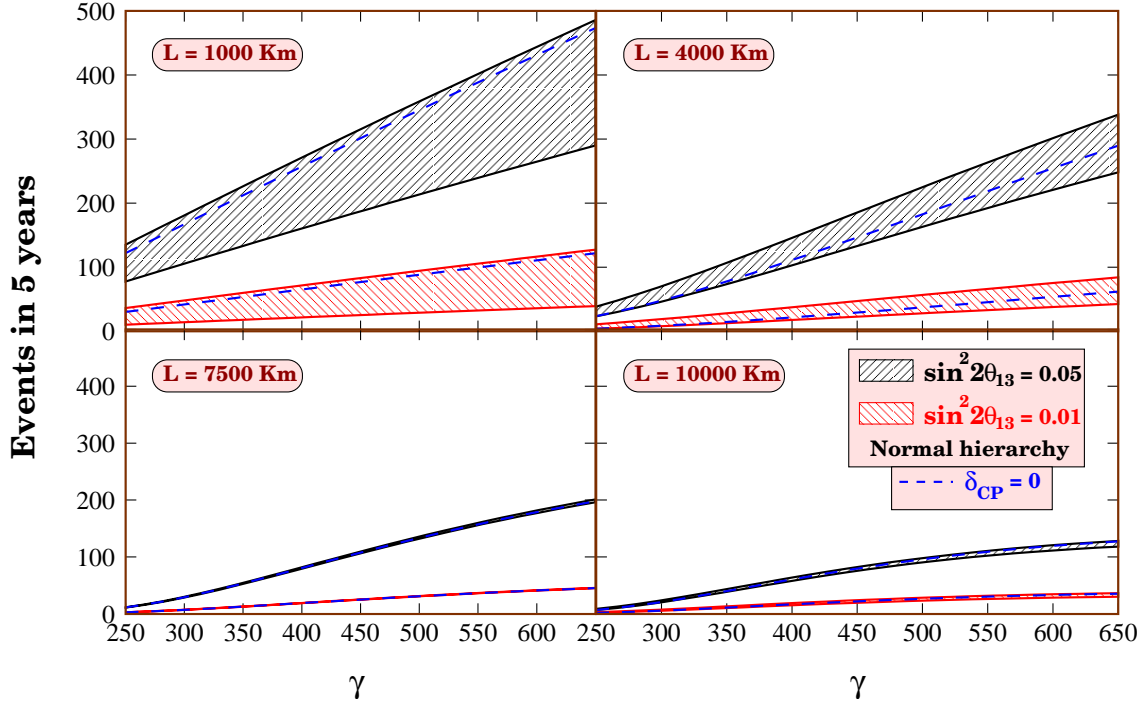}
\caption{\label{fig:eventsth13}
Total number of events as a function of $\gamma$ for the
$^8$B source, for different values of $L$ are shown in
the four panels.
The black hatched area shows the uncertainty
range in the events due to CP phase when $\stch=0.05$,
while the
red hatched area
shows the corresponding
uncertainty when $\stch=0.01$.
For all cases we assume NH to be true.
}
\end{center}
\end{figure}

\begin{figure}[t]
\begin{center}
\includegraphics[width=10.0cm]{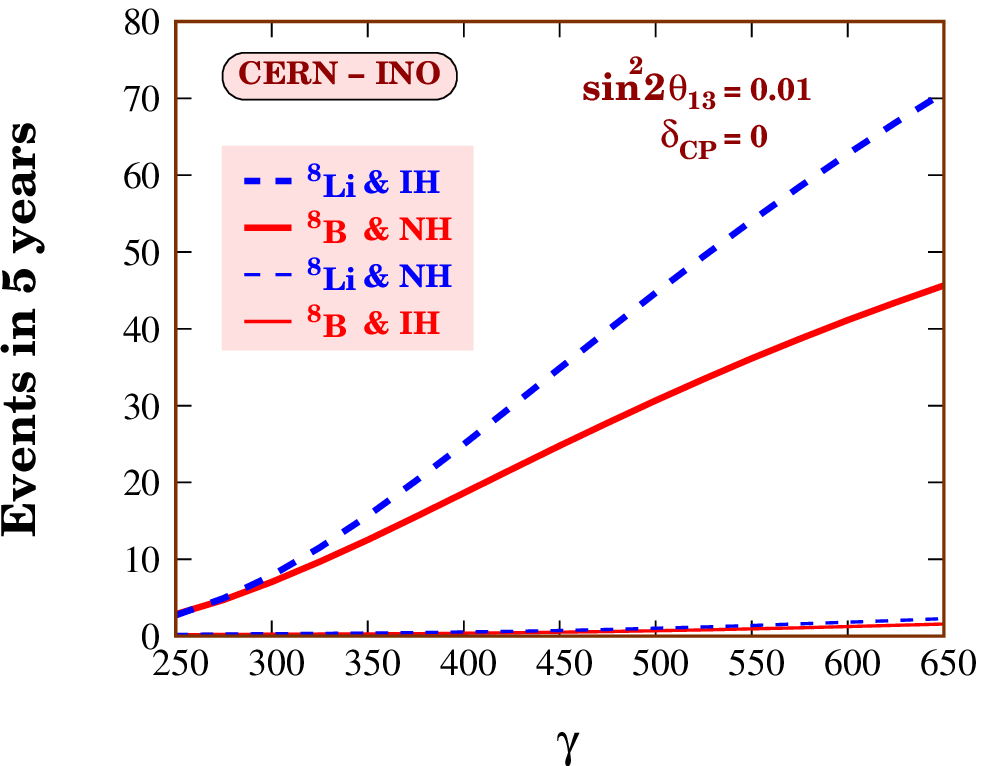}
\caption{\label{fig:eventsbetaino}
Total number of events as a function of $\gamma$ for the
$^8$B (solid lines) and
the $^8$Li (dashed lines)
sources. Results for both normal and inverted hierarchies are shown.
}
\end{center}
\end{figure}

Fig. \ref{fig:magic} depicts the number of events expected
in five years as a function of the baseline $L$,
if we run the experiment in the neutrino mode with
$\gamma=500$. A similar figure is expected for the antineutrino
beam. The upper black hatched area shows the
events for $\stch=0.05$ and
the lower red hatched area corresponds to $\stch=0.01$.
For each baseline $L$, the range
covered by the hatched area shows the uncertainty in the
expected value of the number of events due to the
completely unknown $\delta_{CP}$, which could
take any value from 0 to 2$\pi$. The baseline $L$ where
the width of this band reduces to (almost) zero is
the magic baseline. We see from the figure that the magic
baseline appears at about $L\simeq 7500$ km. Note that while
for $\stch=0.01$ the magic baseline is very clearly defined
with the CP dependence going completely to zero,
for the higher value of $\stch$ of 0.05, the
``magic'' is not complete.
The reason for this anomaly can be traced to the fact that
Eq. (\ref{eq:pemu}) was derived for only very small values of
$\theta_{13}$. For larger values of this angle,
higher order terms become important.  These terms might
depend on $\delta_{CP}$ and remain
non-zero even at the magic baseline.

\item
\fbox{\bf Event Rates vs. $\gamma$}

Figs. \ref{fig:eventshier} and \ref{fig:eventsth13}
demonstrate the impact of the magic baseline on the mass
hierarchy and $\theta_{13}$ sensitivity respectively.
In each of the four panels of both the figures the
expected events in five years as a function of $\gamma$ 
have been shown.
Each panel is for a certain fixed value of $L$, shown
in the corresponding panel. In Fig. \ref{fig:eventshier}
the black hatched area shows the band for NH
while the open band delimited by the dashed red lines
are for the IH. As in Fig. \ref{fig:magic},
the band corresponds to the uncertainty in the event rate
due to the unknown $\delta_{CP}$. The effect of the
uncertainty of $\delta_{CP}$ almost vanishes for $L=7500$ km
which is very close to the magic baseline.
It can be seen that for the smaller baseline $L=1000$ km, NH and IH
predictions are largely overlapping, making it almost
impossible for these experiments to give sensitivity
to the mass hierarchy unless $\stch$ turns out to be
extremely large and $\delta_{CP}$ favorable.
The hierarchy sensitivity is expected to improve as we
go to larger baselines and this is reflected from the
two bands for NH and IH separating out. It turns out that
because the matter effects are very large for the magic
baseline and effect of CP uncertainty is zero, this baseline
gives the best sensitivity to the mass hierarchy. For $L$
larger than magic, matter effects are higher but the
flux is lower, while for $L$ lower than magic,
flux is higher but the matter effects are lower. For
both above and below the magic baseline, the effect
of $\delta_{CP}$ is expected to further reduce the
sensitivity. This is particularly true for the lower $L$
baselines.
Fig. \ref{fig:eventsth13} shows the bands for NH but with two
different choices for $\stch$. Here the effect of the magic
baseline is seen even more clearly.

Fig. \ref{fig:eventsbetaino} shows
as a function of $\gamma$,
the number of events
expected in five years in the CERN-INO beta-beam set-up.
The solid (dashed)
lines are for neutrino
(antineutrino) events, with the thick line
showing the event rate for NH (IH) while the thin
line is for the IH (NH). We have assumed $\stch=0.01$ and
$\delta_{CP}=0$.
One point which is transparent from this figure and which will
be very relevant in understanding the behavior of the CERN-INO
beta-beam set-up is the following: For a given value of $\theta_{13}$
and for NH (IH), we expect a large
number of events in the neutrino (antineutrino)
channel and almost negligible
events in the antineutrino (neutrino) channel.
This means that for NH (IH) it will be
the neutrino (antineutrino) channel which will
be statistically more important.

\end{enumerate}


\subsection{Background Rejection}

Some technical issues pertaining to rejection of the beam related
and atmospheric neutrino backgrounds will be discussed here.
   
\subsubsection{\fbox{Beam Related Backgrounds}}

The possible backgrounds in a $\nue$ beta-beam experiment\footnote{The
discussion concerning the $\anue$ beta-beam is similar
and hence is not repeated
here.} using $\mu^-$ as an oscillation signal come from
NC events such as
\be
\nu_x + d(u) &\rightarrow& \nu_x + d(u)
\\
\nu_x + d(u) &\rightarrow& \nu_x + d(u) + q\bar{q}
\ee
and $\nue$ charged current events
\be
\nue + d &\rightarrow& e^- + u \; {\rm or} \; c \;({\rm Cabibbo
\; suppressed})
\\
\nue + d &\rightarrow& e^- + u + q\bar{q}
\ee
The quarks in the final state could produce mesons
as a part of the hadronic junk. These mesons might   
constitute backgrounds in the following way~:

\begin{enumerate}

\item
Some energetic mesons can give long tracks inside the detector
which can mimic the muon tracks.

\item
These mesons can decay in flight producing secondary muons, 
a possible source of background.

\end{enumerate}

As discussed earlier, ICAL@INO will have 6 cm thick iron plates.
Such a dense tracking detector will have excellent muon/pion and 
muon/electron separation capability in the energy range 
we are working with. The simulations carried out by the INO 
collaboration have shown that after the standard kinematical 
cuts are imposed, the electrons do not give any signal at all, 
while in 99\% of the cases, the pions and kaons for the energy range 
of interest to us get absorbed very quickly in the iron via 
strong interactions and therefore do not hit enough RPCs to give a signal. 
At the energies of a beta-beam considered here, production cross section of $D$
mesons (also Cabibbo suppressed) is small and they do not constitute a problem 
for the experiment. The associated strange or charm production is also highly 
suppressed at these energies. In addition, the fact that the detector 
will have a charge identification capability means that secondary 
$\mu^+$ produced can be safely discarded, reducing the background even further.
Therefore, in the analysis we estimate the NC backgrounds\footnote{Mesons 
produced in NC processes are degraded in energy.
Note that backgrounds from these mesons are very important in the
case of the neutrino factory. However, since our relevant energy
range is lower, the mesons produced in each event are much lower in
energy and hence can be easily rejected by putting suitable cuts.}
by assuming an energy independent background suppression factor of
$\sim 10^{-4}$ which agrees with \cite{mind}. We have noted that after 
five years of running of the CERN-INO beta-beam experiment with $\gamma=650$, 
we expect only about 0.1 NC background events. Nevertheless we take this
background into account in our numerical analysis.
The background is assumed to have the same shape as the signal. 
But one should keep in mind the fact that this shape is not much of 
an issue since anyway the background is very small.
we do not consider any backgrounds coming from the CC events of $\nue$.

Since the oscillation probability $\nue \rightarrow \nutau$ 
(silver channel) is about the same as that for 
$\nue \rightarrow \numu$ (golden channel), we expect
almost as many $\nutau$ arriving at the detector as $\numu$. The
$\tau^-$ produced through CC interaction may decay further inside 
the detector producing secondary $\mu^-$ with a branching ratio of 
17.36\%. But, the $\tau$ threshold (3.5 GeV) is high and the production
cross section suppressed compared to $\mu$. So we do not expect
any significant background from this source either. We have
estimated the number of secondary muons produced from the
$\nutau$ component of the beam. For $\stch=0.01$, we expect
0.3 muon events per year with $\gamma=500$. In addition, 
these secondary muons will be severely degraded in energy and 
therefore can be eliminated through energy cuts. We therefore 
neglect the backgrounds from this source.

\subsubsection{\fbox{Atmospheric Backgrounds}}

The atmospheric neutrino flux falls steeply with energy
and is expected to produce much fewer events
for the energy range that we are interested in\footnote{We
will show in the next section that even a threshold energy
of 4 GeV is easily admissible in our set-up, and above
4 GeV there are much fewer atmospheric events.}. The fact that
INO has charge identification capability further reduces the
atmospheric background. The most important handle on the
reduction of this background comes from the timing information
of the ion bunches inside the storage ring.
For 5T magnetic field and $\gamma=650$ for $^8B$ ions, the
total length of the storage ring turns out to be
19564 m. We have checked that with eight bunches inside this ring 
at any given time, a bunch size of about 40 ns would give an
atmospheric background to {\it signal} ratio of about $10^{-2}$,
even for a very low $\stch$ of $10^{-3}$. For a smaller bunch span, 
this will go down even further. In addition, atmospheric neutrinos will 
be measured in INO during deadtime and this can also be used to 
subtract them out. Hence we do not include this negligible 
background here.

\section{\fbox{Details of the Statistical Method}}
In order to quantify the sensitivity of this CERN-INO beta-beam
experimental set-up to the mixing angle $\theta_{13}$ and $sgn(\ma)$, 
we perform a statistical analysis of the ``data'' generated 
in ICAL@INO, assuming certain true values of the parameters. 
For our statistical analysis we employ a $\chi^2$ function defined as
\be
\chi^2_{total} = \chi^2_{\nu_e \rightarrow \nu_{\mu}}
               + \chi^2_{\bar\nu_e \rightarrow \bar\nu_{\mu}}
               + \chi^2_{prior}~,
\label{eq:tot_chisq}
\ee
where the first term is the contribution from the
neutrino channel, the second term comes from the antineutrino
channel, while the last term comes from imposing priors
on the oscillation parameters which we allow to vary freely
in our fit and which we expect will be determined better from
other experiments at the time when the data from the CERN-INO
beta-beam set-up would be finally available.
The $\chi^2$ for the neutrino channel is given by
\be
\chi^2_{\nu_e \rightarrow \nu_{\mu}} = min_{\xi_s, \xi_b}\left[2\sum^{n}_{i=1}
(\tilde{y}_{i}-x_{i} - x_{i} \ln \frac{\tilde{y}_{i}}{x_{i}}) +
\xi_s^2 + \xi_b^2\right ]~.
\label{eq:chipull}
\ee
where $n$ is the total number of bins,
\be
\tilde{y}_{i}(\{\omega\},\{\xi_s, \xi_b\}) = N^{th}_i(\{\omega\}) \left[
1+ \pi^s \xi_s \right] +
N^{b}_i \left[1+ \pi^b \xi_b \right] ~,
\label{eq:rth}
\ee
$N^{th}_i(\{\omega\})$ given by Eq. (\ref{eq:events})
being the predicted number of
events in the energy bin $i$ for a set of oscillation
parameters $\omega$ and $N_i^b$ are the number of
background events in bin $i$. The quantities
$\pi^s$ and $\pi^b$ in Eq. (\ref{eq:rth}) are the systematical
errors on signals and backgrounds respectively.
We have taken $\pi^s = 2.5\%$ and
$\pi^b = 5\%$ (see Table \ref{tab:detector}). The quantities
$\xi_s$ and $\xi_b$ are
the ``pulls'' due to the systematical error
on signal and background respectively.
The data in Eq. \ref{eq:chipull} enters through the
variables $x_i=N_i^{ex}+N_i^b$, where $ N_i^{ex}$
are the number of observed signal events in the detector and
$N_i^b$ is the background, as mentioned earlier. We simulate
the signal event spectrum using Eq. \ref{eq:events}
for our assumed true values for the set of oscillation
parameters which are given in
the first column of Table \ref{tab:bench}. 
Different options are used for $\stcht$, $\dcpt$ and the 
true hierarchy and these are mentioned wherever applicable.
In our $\chi^2$ fit we marginalize over {\it all}
oscillation parameters, the Earth matter density, as well
as the neutrino mass hierarchy, as applicable. We do this by allowing all
of these to vary freely in the fit and picking the
smallest value for the $\chi^2$ function. Of course, we
expect better determination of some of these parameters,
which are poorly constrained by this experimental set-up.
Therefore, we impose a ``prior'' on these parameters through the
$\chi^2_{prior}$ given by
\be
\chi^2_{prior} &=&
\left (\frac{|\Delta m^2_{31}|-
|\mat|}{\sigma(|\Delta m^2_{31}|)} \right )^2 +
\left (\frac{\sta-\stat}{\sigma(\sta)} \right )^2\nonumber \\
&+&
\left (\frac{\Delta m^2_{21}-
\mst}{\sigma(\Delta m^2_{21})} \right )^2 +
\left (\frac{\sss-\ssst}{\sigma(\sss)} \right )^2\nonumber \\
&+&
\left (\frac{\rho-1}{\sigma(\rho)} \right )^2 ~.
\label{eq:prior}
\ee
where the $1\sigma$ error on these that we use are
taken from \cite{huber10,solarprecision}
and are given in the right column of Table \ref{tab:bench}.
In our computation, we have used a matter
profile inside the Earth
with 24 layers. In Eq. \ref{eq:prior}, $\rho$ is
a constant number by
which the matter density of each layer has been scaled.
The external information
on $\rho$ is assumed to come from the study of the tomography of the earth
\cite{tomography}. In Eq. \ref{eq:prior}, $\rho$ varies from
0.95 to 1.05 i.e., 5$\%$ fluctuation around 1.

Note that in our definition of the $\chi^2$ function
given by Eqs. \ref{eq:tot_chisq} and \ref{eq:chipull},
we have assumed that the neutrino and antineutrino
channels are completely uncorrelated,
all the energy bins for a given channel
are fully correlated, and
$\xi_{s}$ and $\xi_{b}$ are fully uncorrelated.
We minimize the $\chi^2_{total}$ in two stages. First it
is minimized with respect to $\xi_{s}$ and $\xi_{b}$ to
get Eq. \ref{eq:chipull}, and then with respect to the
oscillation parameters ${\omega}$ to get the global best-fit.
For minima with respect to $\xi_{s}$ and $\xi_{b}$,
we require that
\be
\frac{\partial{\chi^2}}{\partial \xi_s} = 0 ~~{\rm and}~~
\frac{\partial{\chi^2}}{\partial \xi_b} = 0~.
\label{eq:minima}
\ee
From Eqs. \ref{eq:chipull}, \ref{eq:rth}, \ref{eq:minima}
we get,
\be
\left(
\begin{array}{cc}
a_{11} & a_{12} \\
a_{21} & a_{22} \end{array} \right)
\left(
\begin{array}{c}
\xi_{s} \\
\xi_{b} \end{array} \right)
=
\left(
\begin{array}{c}
c_{1} \\
c_{2} \end{array} \right)
\label{eq:matrix}
\ee
where,
\be
c_{1} = \sum^{n}_{i=1}
(\frac{x_{i} \pi^{s} N^{th}_{i}}{N^{th}_i + N^{b}_i} - \pi^{s} N^{th}_{i})~,
\nonumber \\
c_{2} = \sum^{n}_{i=1}
(\frac{x_{i} \pi^{b} N^{b}_{i}}{N^{th}_i + N^{b}_i} - \pi^{b} N^{b}_{i})~,
\nonumber \\
a_{11} = \sum^{n}_{i=1}
{\left[\frac{x_{i} (\pi^{s} N^{th}_{i})^{2}}{(N^{th}_i + N^{b}_i)^{2}} \right ]} + 1~,
\nonumber \\
a_{22} = \sum^{n}_{i=1}
{\left[\frac{x_{i} (\pi^{b} N^{b}_{i})^{2}}{(N^{th}_i + N^{b}_i)^{2}} \right ]} + 1~,
\nonumber \\
a_{12} = a_{21} = \sum^{n}_{i=1}
{\left[\frac{x_i N^{th}_{i} N^{b}_{i} \pi^{s} \pi^{b}}{(N^{th}_i + N^{b}_i)^{2}} \right ]}
\label{eq:coeffi}
\ee
Using Eqs. \ref{eq:matrix} and \ref{eq:coeffi},
we calculate the values of $\xi_{s}$ and $\xi_{b}$ and then we use
these values to calculate $\chi^2_{\nu_e \rightarrow \nu_{\mu}}$.
In a similar fashion, we estimate
$\chi^2_{\bar\nu_e \rightarrow \bar\nu_{\mu}}$
to obtain the $\chi^2_{total}$.

\section{\fbox{Measurement of the Neutrino Mass Ordering}}

\begin{figure}[t]
\includegraphics[width=8.0cm, height=7.0cm, angle=0]{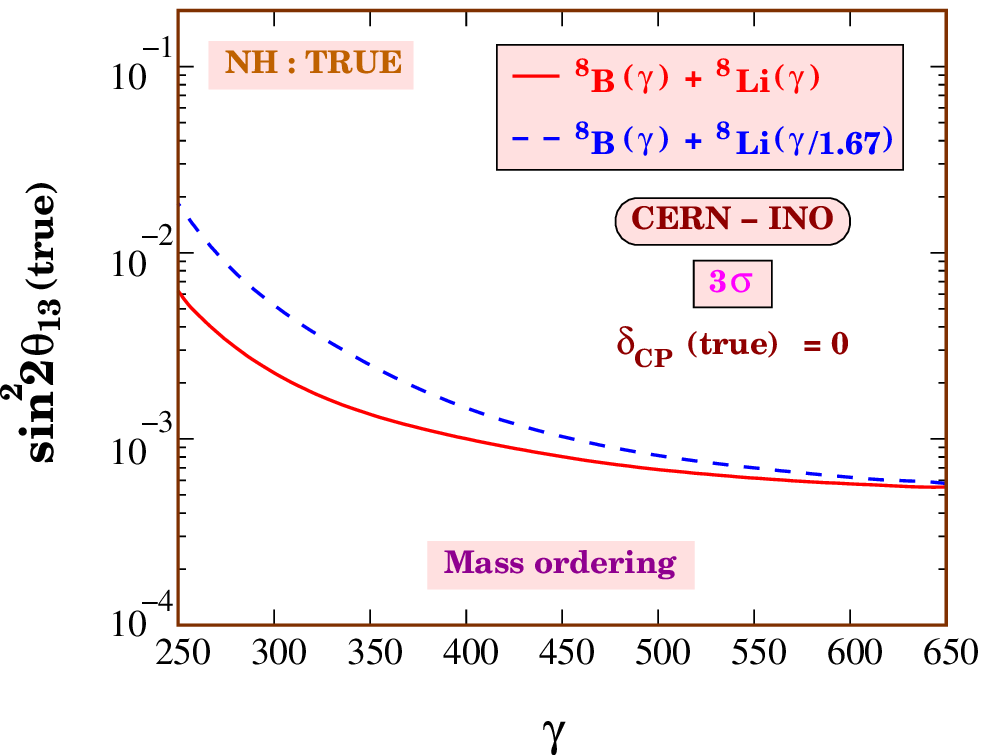}
\vglue -7.0cm \hglue 8.8cm
\includegraphics[width=8.0cm, height=7.0cm, angle=0]{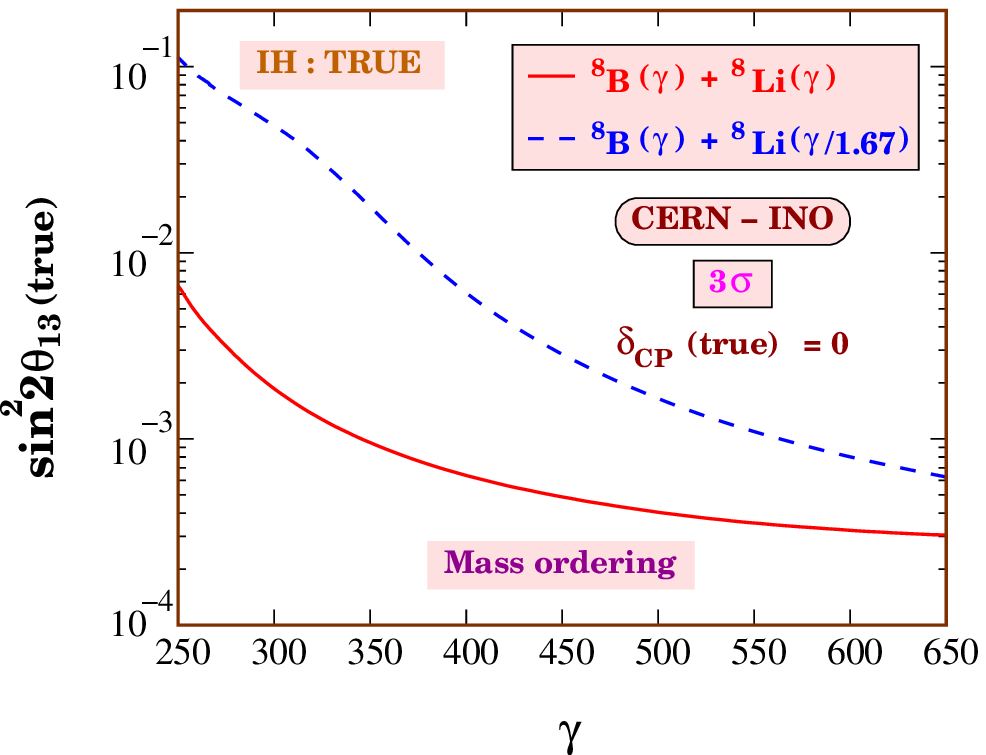}
\caption{\label{fig:senshier}
The range of $\stch$(true) for which the wrong
hierarchy can be ruled out at the $3\sigma$ C.L.,
as a function of $\gamma$. The left panel is for NH as true, while
the right panel is when IH is true. The red solid curves
show the sensitivity when the $\gamma$ is chosen to be the same
for both the neutrino and the antineutrino beams. The
blue dashed lines show the corresponding sensitivity when the 
$\gamma$ for the antineutrinos is scaled down by a factor of 1.67 
with respect to the $\gamma$ of the neutrino beam.
}
\end{figure}

\begin{figure}[p]
\includegraphics[width=8.0cm, height=7.0cm]{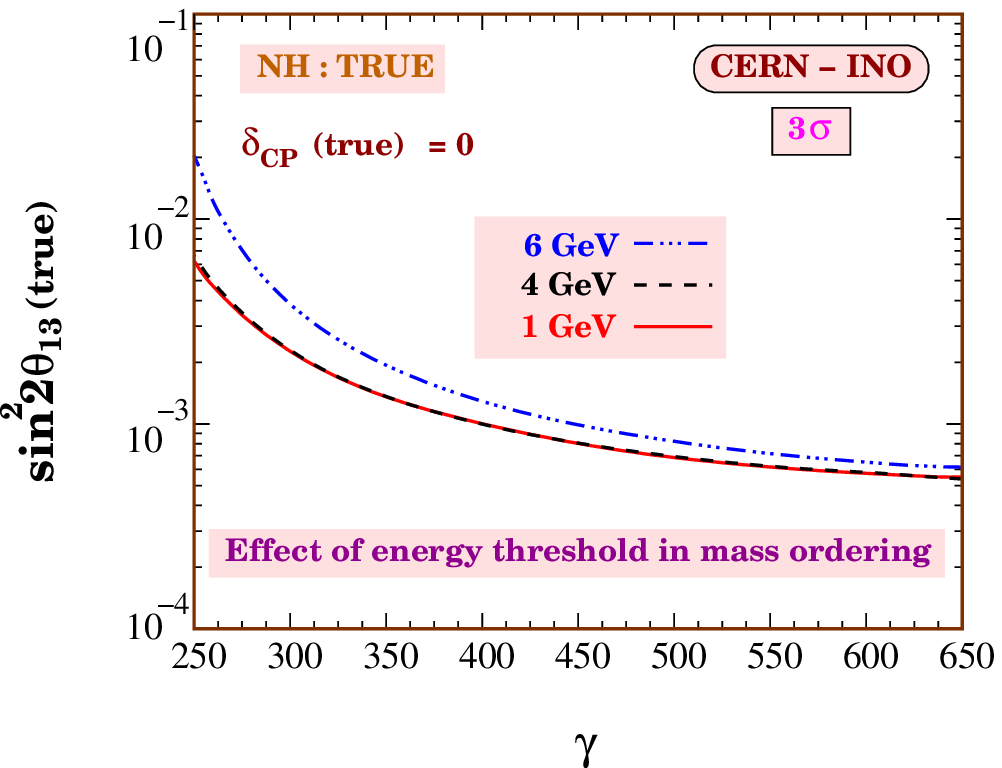}
\vglue -7.0cm \hglue 8.8cm
\includegraphics[width=8.0cm, height=7.0cm]{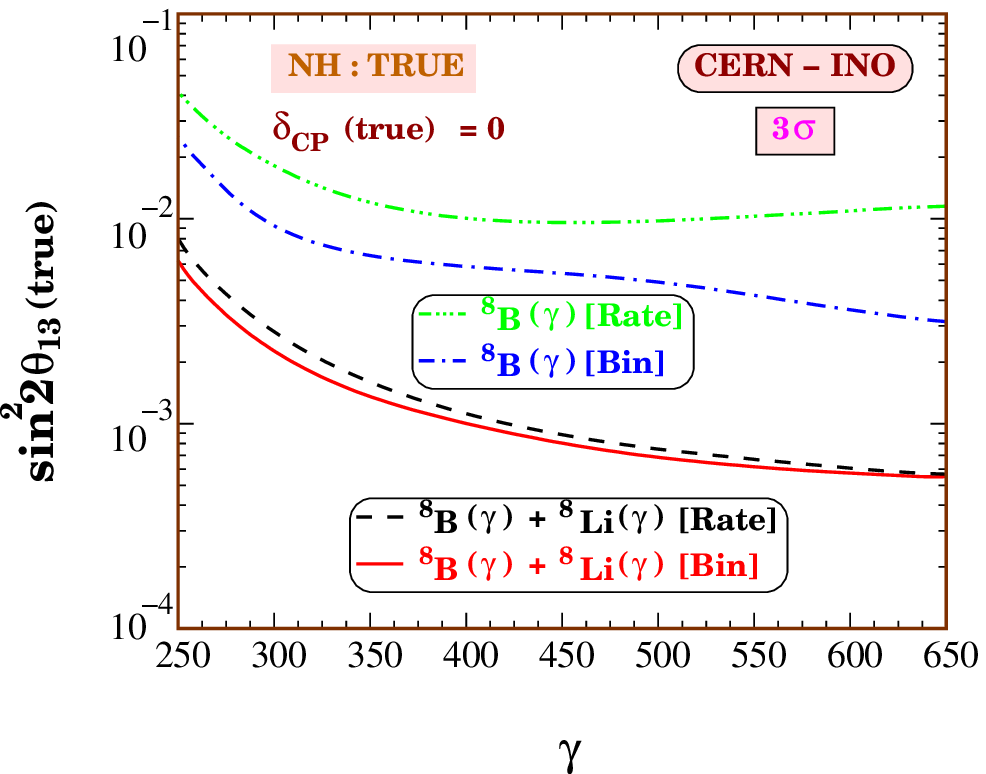}
\vglue -0.0cm \hglue 0.0cm
\includegraphics[width=8.0cm, height=7.0cm]{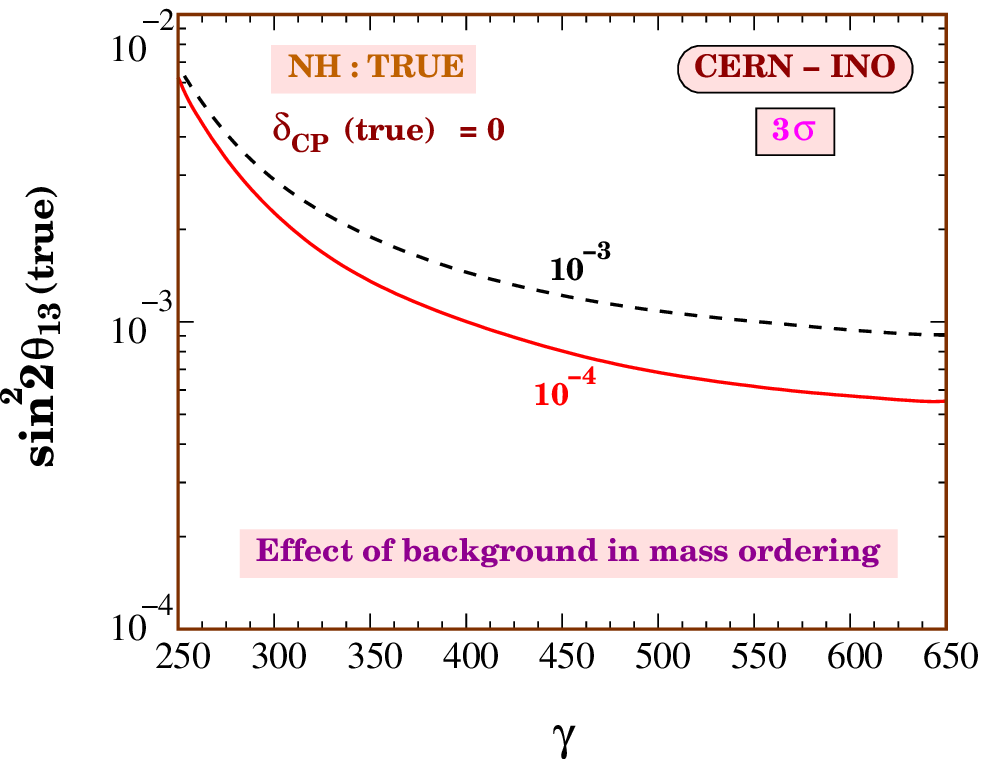}
\vglue -7.0cm \hglue 8.8cm
\includegraphics[width=8.0cm, height=7.0cm]{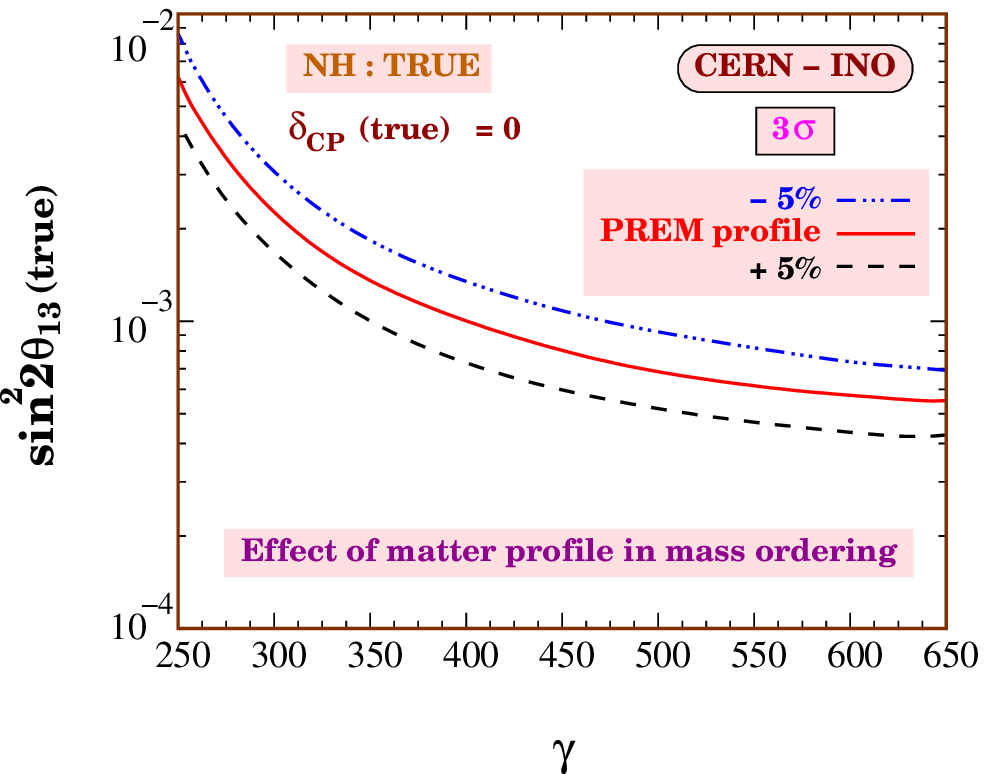}
\caption{\label{fig:senshiertest}
Plots showing the impact of various factors on the mass hierarchy
sensitivity of the CERN-INO beta-beam experiment. The top left panel
shows the impact of changing the detector threshold. The lower
left panel shows the effect of changing the background rejection
factor. The top right panel shows the difference in the sensitivity
between the rate and spectral analysis. The lower right panel
shows how the density profile would impact the hierarchy sensitivity.
}
\end{figure}

\begin{figure}[t]
\includegraphics[width=8.0cm, height=7.0cm]{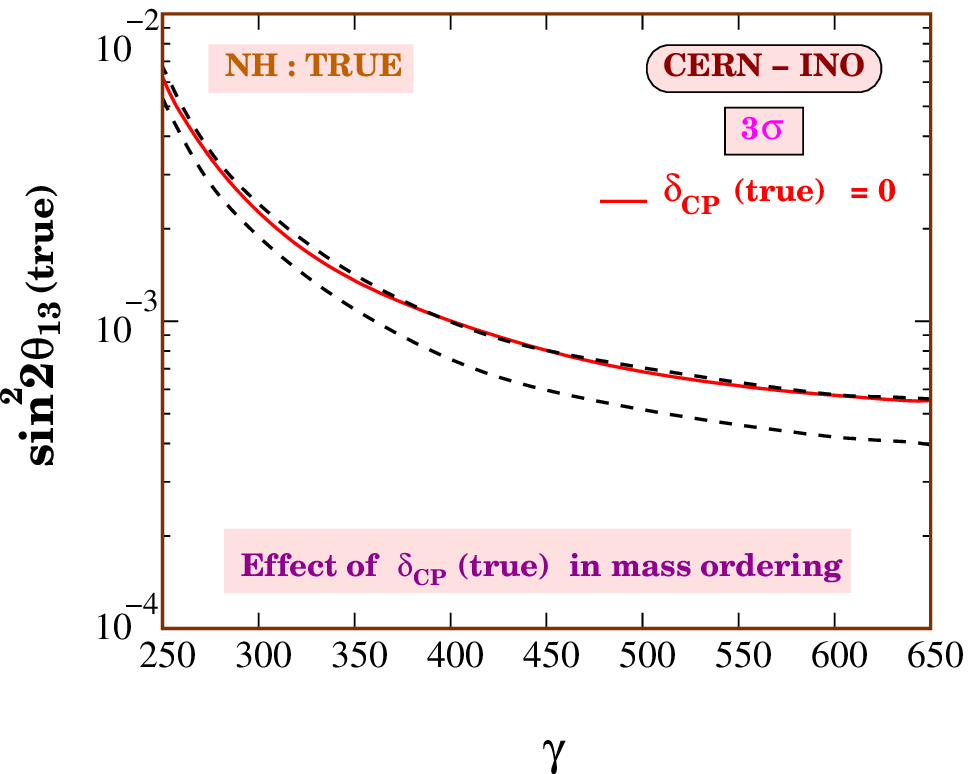}
\vglue -7.0cm \hglue 8.8cm
\includegraphics[width=8.0cm, height=7.0cm]{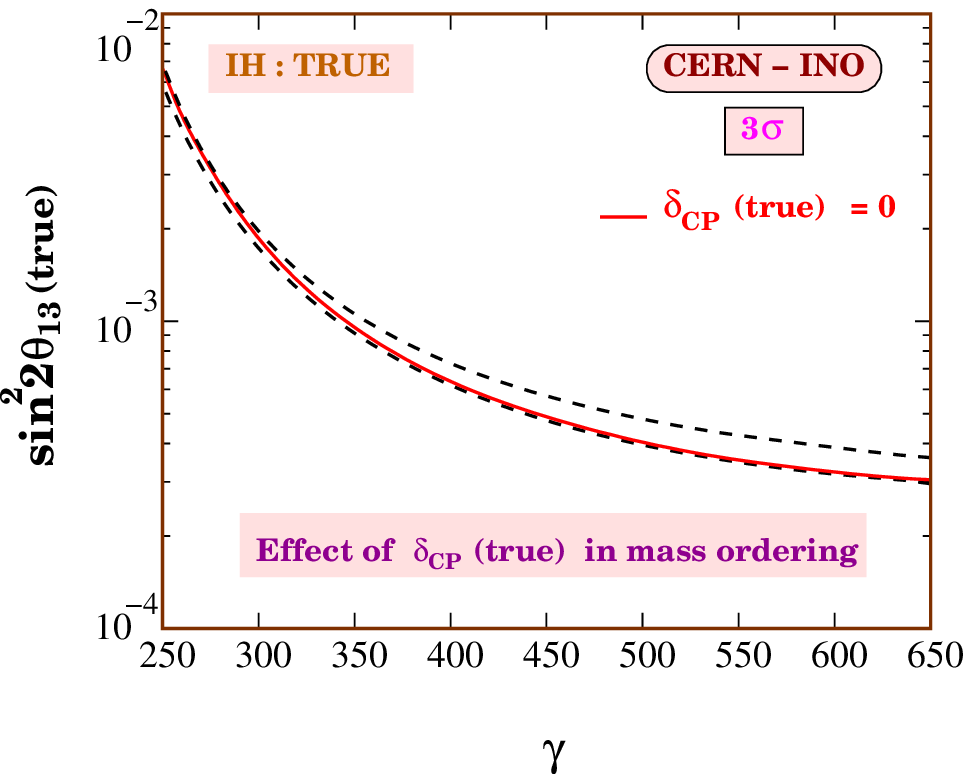}
\caption{\label{fig:senshiercploop}
Effect of $\dcpt$ on the hierarchy sensitivity.
The black dashed
lines show the worst and best
cases when we allow $\dcpt$ to take any
value between 0 and 2$\pi$. The red solid curve corresponds
to the reference case where $\dcpt=0$.
The left panel shows the
case for true NH while the right panel is for
true IH.
}
\end{figure}

\begin{figure}[t]
\includegraphics[width=8.0cm, height=7.0cm]{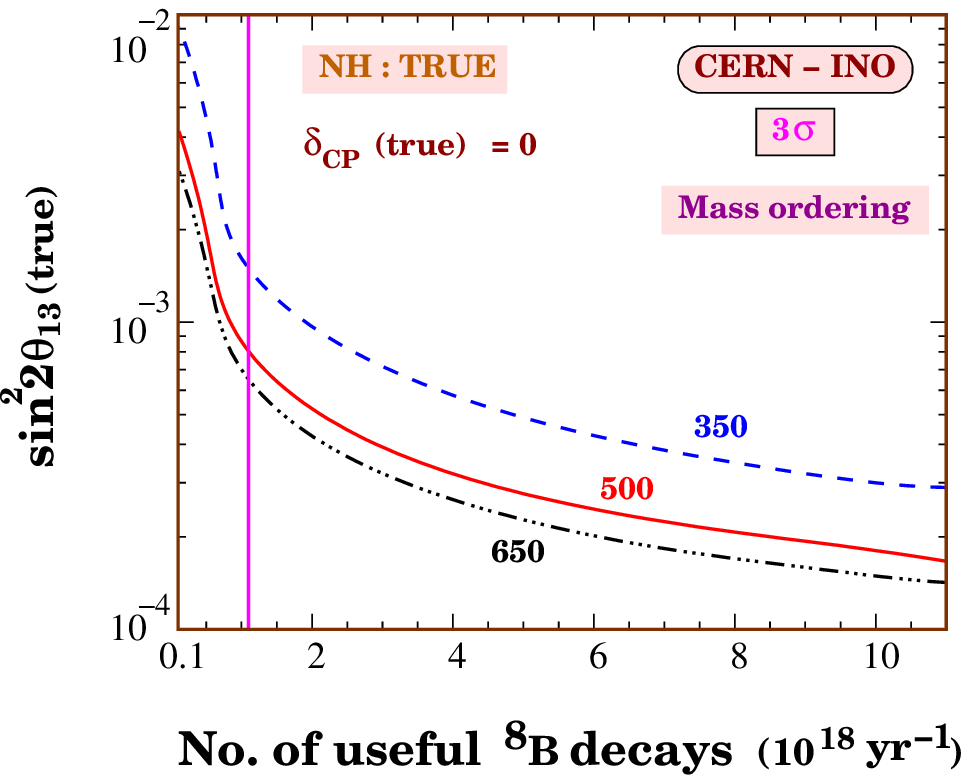}
\vglue -7.0cm \hglue 8.8cm
\includegraphics[width=8.0cm, height=7.0cm]{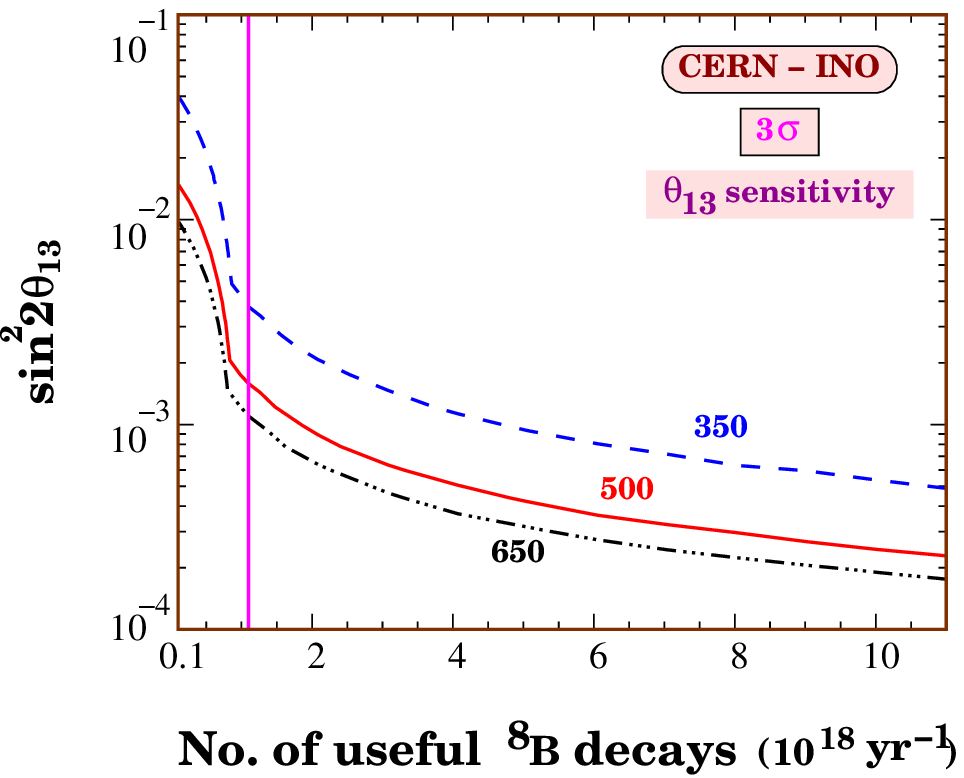}
\caption{\label{fig:senslum}
The variation of the experimental sensitivity on the
number of useful ion decays in the straight sections of the
storage ring. Left panel shows sensitivity to the
mass hierarchy assuming NH to be true. Right panel shows the
$\stch$ sensitivity reach. In both panels, the majenta solid
vertical line corresponds to the reference value used in the
rest of the analysis.
}
\end{figure}

Let us first focus on the issue of neutrino mass ordering.
If the true value of $\theta_{13}$ has indeed been chosen 
to be large by Nature, sizable matter effects can be expected 
in this magical CERN-INO beta-beam experimental set-up, giving 
us a handle on the mass hierarchy. The mass hierarchy sensitivity 
is defined as the range of $\stcht$ for which the wrong hierarchy 
can be excluded at the $3\sigma$ C.L. The left (right) panel of 
Fig. \ref{fig:senshier} shows the hierarchy sensitivity 
when the NH (IH) is true. 
We show this as a function of $\gamma$.
In both panels we show by the red solid lines the sensitivity
when we add neutrino and antineutrino data with the
same value of the Lorentz boost shown in the $x$-axis.
The blue dashed lines on the other hand correspond to the sensitivity
expected when the neutrino beam runs with the $\gamma$
shown in the $x$-axis while the $\gamma$ for the
antineutrino beam is scaled down by a factor of
1.67. We assume five\footnote{Here all sensitivity figures 
correspond to a five years run.} years of running of the beta-beam in both 
polarities and a full spectral analysis has been performed.
We note that using the combined neutrino and antineutrino beam
running at the same value of $\gamma$ the wrong
hierarchy could be ruled out at $3\sigma$ for
$\stch > 6.8 \times 10^{-4}$ ($\stch > 4.0 \times 10^{-4}$)
for $\gamma = 500$ if the NH (IH) is true.
Presence of both neutrino and antineutrino data simultaneously
in the analysis restricts the fitted value of $\theta_{13}$
to be in a range very close to the assumed true value.
For instance, for NH true, data corresponds to a large
number of events for neutrinos and a small number of
events for antineutrinos. When this is fitted with
IH, we have a small number of events predicted for the
neutrinos. In order to minimize the disparity between the
data and prediction for neutrinos, the fit tends to drive $\theta_{13}$
to its largest allowed value. However, larger values of
$\theta_{13}$ would give very large number of antineutrino
events for IH and this would be in clear conflict with the data.
Therefore, the net advantage of adding data from both neutrino and 
antineutrino channels is that one cannot artificially reduce the 
$\chi^2$ any longer by tinkering with $\theta_{13}$ in the fit. 
As a result, the sensitivity of the experiment to mass hierarchy 
witnesses a substantial improvement.

It can be noted from the plots that the hierarchy sensitivity falls when
the scaled $\gamma$ option for the antineutrino beam has been used. This is
particularly relevant when the true hierarchy is inverted and/or when
$\gamma$ is low. Since scaling the $\gamma$ reduces it by a factor of
1.67, the statistics for the antineutrinos fall by
nearly a factor of 1.67 for this case and this reflects
in the reduced hierarchy sensitivity of the experiment.
Its impact when true hierarchy is inverted is more because 
in that case, the data corresponds to larger events for the 
antineutrinos and very small events for the neutrinos. 
The antineutrino events are therefore the driving force and
an increase in their statistical uncertainty due to the scaled
down $\gamma$ accentuates the adverse effect on the
hierarchy sensitivity. In the case of NH, 
the events in the neutrino channel are the dominant factor and the
role of the antineutrinos is only to prevent the $\theta_{13}$
values in the fit to run to very large values, as discussed
before. As long as the antineutrino events have enough statistical power
to restrict $\theta_{13}$ to values close to the true value at which 
the data was generated, the hierarchy sensitivity
remains reasonably good. Therefore, for the NH only for very 
low values of $\gamma$ the hierarchy sensitivity
gets seriously affected by the Lorentz boost scaling.

Fig. \ref{fig:senshiertest} shows how the hierarchy
sensitivity depends on diverse input factors. As in
Fig. \ref{fig:senshier} we show the $3\sigma$
limit for $\stch$ as a function of $\gamma$ in all the
four panels and we assume that NH is true.
The reference curve (red solid line) in all panels
corresponds to the result obtained with a $\nue$ and
$\anue$ beam with a spectral analysis.
The upper left hand panel shows the effect of changing the 
threshold energy of the detector. The sensitivity of the experiment 
is seen to remain almost stable against the variation of the threshold
energy upto 4 GeV. Only for a threshold of 6 GeV and above 
the sensitivity falls, the lower $\gamma$ values getting more 
affected since they correspond to lower neutrino energies.
In the lower left hand panel of the figure the effect 
of the chosen background fraction on the hierarchy
sensitivity has been shown. The red solid line shows the sensitivity 
for our assumed background factor of $10^{-4}$ while the black dashed
line shows the corresponding sensitivity when the background rejection 
is poorer and we have a higher residual background fraction of $10^{-3}$.
The upper right hand panel shows how our sensitivity increases by taking 
into account the spectral information of the events. It also shows
how much improvement we get by combining the antineutrino
data with the neutrino data. The black dashed line shows how the sensitivity 
falls when we use the total event rates instead of the events spectrum. 
The blue dashed-dotted and green dashed-triple-dotted
lines show the sensitivity expected from the neutrino
data alone. The blue dashed-dotted line is for binned neutrino data while
the green dashed-triple-dotted lines shows the sensitivity for the
total event rate for neutrinos alone. It can be seen that the effect of using 
the spectral information is only marginal when both neutrino
and antineutrino are used, while the effect of combining
the antineutrino data with the neutrino data
on the sensitivity is huge.
For the neutrino data alone, the sensitivity improves
significantly when
one uses the spectral information.
In the lower right hand panel we show how the sensitivity of the
experiment to hierarchy would get affected if
we use a different profile for the
Earth matter density instead of PREM. The red solid line is for
earth density according to the PREM profile while
the blue dotted and
black dashed lines are when the matter density is 5\% lower
and 5\% higher respectively than the density predicted
by the PREM profile. When the density is higher (lower)
the matter effects are higher (lower) and therefore the
sensitivity improves (deteriorates).

One crucial point that we have not stressed so far concerns the dependence 
of the detector performance on the {\it true value} of $\delta_{CP}$.
All the earlier plots were presented assuming that $\delta_{CP}{\mbox {(true)}}=0$.
At exactly the magic baseline, it is expected that the sensitivity of
the experiment to be completely independent of $\delta_{CP}$.
The CERN-INO distance of 7152 km is almost magical,
but it is not the exact magic baseline.
Therefore, we do expect some remnant impact of $\delta_{CP}{\mbox {(true)}}$
on our results\footnote{Note that in all our results
presented in this work, we have
fully marginalized over all the oscillation parameters in the fit,
including $\delta_{CP}$.}.
To show how our results get affected by
$\dcpt$, we show in
Fig. \ref{fig:senshiercploop} the hierarchy sensitivity just as
in Fig. \ref{fig:senshier}, but here we show the full band corresponding
to all values of $\dcpt$ from 0 to $2\pi$. As before,
the left panel is for NH true while the the right panel is for IH true,
and we have taken in the analysis the full spectral data
for the neutrinos as well as the antineutrinos,
with the same $\gamma$. The lower edge of this band shows the best possible 
scenario where the experiment is most sensitive, while the
upper edge shows the worst possible sensitivity.
The red solid lines in both panels show for comparison
the hierarchy sensitivity
corresponding to $\dcpt=0$, which we had presented in
Fig. \ref{fig:senshier}. We note from the figure that
the hierarchy sensitivity is nearly the
best for $\dcpt=0$ when IH is true while if NH is
true then it would give us almost the worst sensitivity.
For NH (IH) as true the best possible sensitivity would be
$\stch > 3.96 \times 10^{-4}$ ($\stch > 2.96 \times 10^{-4}$)
for $\gamma=650$ to be compared with $\stch > 5.51 \times 10^{-4}$
($\stch > 3.05 \times 10^{-4}$) when $\dcpt=0$.
Therefore, we conclude that if NH is true then it would not be unfair
to expect an even better hierarchy sensitivity than what was reported in Fig.
\ref{fig:senshier}, while if IH is true then the best
sensitivity will be returned for $\dcpt\simeq 0$.

\begin{table}[t]
\begin{center}
\begin{tabular}{|c||c|c||c|c|} \hline
\multirow{2}{*}{$\gamma\backslash$N}
& \multicolumn{2}{|c||}{{\rule[0mm]{0mm}{6mm}Mass Hierarchy (\sig)}}
& \multicolumn{2}{|c|}{\rule[-3mm]{0mm}{6mm}{$\stch$ sensitivity (\sig)}}
\cr \cline{2-5} 
&  $1.1\times 10^{18}$ & $2.043\times 10^{18}$ &
 $1.1\times 10^{18}$ & $2.043\times 10^{18}$ \cr
\hline\hline
350 & $1.3\times 10^{-3}$ & $9.3\times 10^{-4}$ &
 $3.8\times 10^{-3}$ & $2.3\times 10^{-3}$ \cr
650 & $5.6\times 10^{-4}$ & $4.1\times 10^{-4}$ &
 $1.1\times 10^{-3}$ & $7.3\times 10^{-4}$ \cr
\hline
\end{tabular}
\caption{\label{tab:compare}
Comparison of the variation of the detector sensitivity
to mass hierarchy (columns 2 and 3) and $\stch$ sensitivity
(columns 4 and 5) with $\gamma$ and N, the
number of useful ion decays per
year.
}
\end{center}
\end{table}

It has been noted from Figs. \ref{fig:eventshier}
and \ref{fig:eventsth13} that
the total number of events
in the detector increases roughly linearly with $\gamma$,
except for extremely long baselines. Increasing the
number of ion decays per year will also bring about
a simple linear increase in the statistics.
It is therefore pertinent to make a
fair comparison between the dependence of the
mass hierarchy sensitivity to the Lorentz boost $\gamma$ and the
number of useful ion decays in the ring\footnote{Note that
this is also equivalent to increasing the total exposure time
of the experiment.
Both number of ion decays per year and exposure
appear as a normalization factor for the event rate and hence
increasing the number of ion decays by a factor $n$ keeping the
exposure same is equivalent to increasing the exposure by
a factor $n$ keeping the number of ion decays per year fixed.
}. In the left panel of Fig. \ref{fig:senslum} we show the effect 
of increasing the number of ion decays on the hierarchy 
sensitivity\footnote{We assume that the number of useful ion decays for 
both $^8$B and $^8$Li have been scaled by the same factor. 
In the figure along the $x$-axis only the $^8$B numbers are shown.}.
The plots exhibit the dependence of the sensitivity on
the number of useful ion decays per year for an exposure of five years,
for three different values of $\gamma$.
The same Lorentz boost for the neutrino and antineutrino beams
have been assumed. We present in Table \ref{tab:compare}
the relative increase in the hierarchy sensitivity when
we increase the $\gamma$ by a factor of 1.86 and compare it
against the increase in the sensitivity when the number of
ion decays are increased by the same factor.
It can be noted that while the hierarchy sensitivity improves by a factor of
2.54 in going from $\gamma=350$ to 650 keeping the
number of ion decays per year as $1.1\times 10^{18}$, it increases
1.5-fold when we raise the number of ion decays per year from
$1.1\times 10^{18}$ to $2.04\times 10^{18}$ keeping
$\gamma=350$. However, we would like to stress
that the improvement of the hierarchy sensitivity is not
linear with either $\gamma$ or number of ion decays per year.
The crucial thing is that the behavior of the
sensitivity dependence on both $\gamma$ and number of
ion decays per year is very similar.
It increases very fast initially and then
comparatively flattens out.

\section{\fbox{Measurement of $\stch$}}

The CERN-INO beta-beam set-up is also expected to give
very good sensitivity to the $\theta_{13}$ measurement.
In what follows, we will quantify our results in terms of
three ``performance indicators'',
\begin{enumerate}
\item  
\fbox{\bf $\stch$ sensitivity reach,}
\item 
\fbox{\bf $\stch$ discovery reach,}
\item 
\fbox{\bf $\stch$ precision.}
\end{enumerate}

Below a detailed description of our definitions of these
performance indicators has been given. All results 
in this section have been obtained by taking into account 
the full event spectrum and combining five years data
from both the neutrino and antineutrino channels.

\subsection{$\stch$ Sensitivity Reach}

\begin{figure}[t]
\includegraphics[width=8.0cm, height=7.0cm, angle=0]{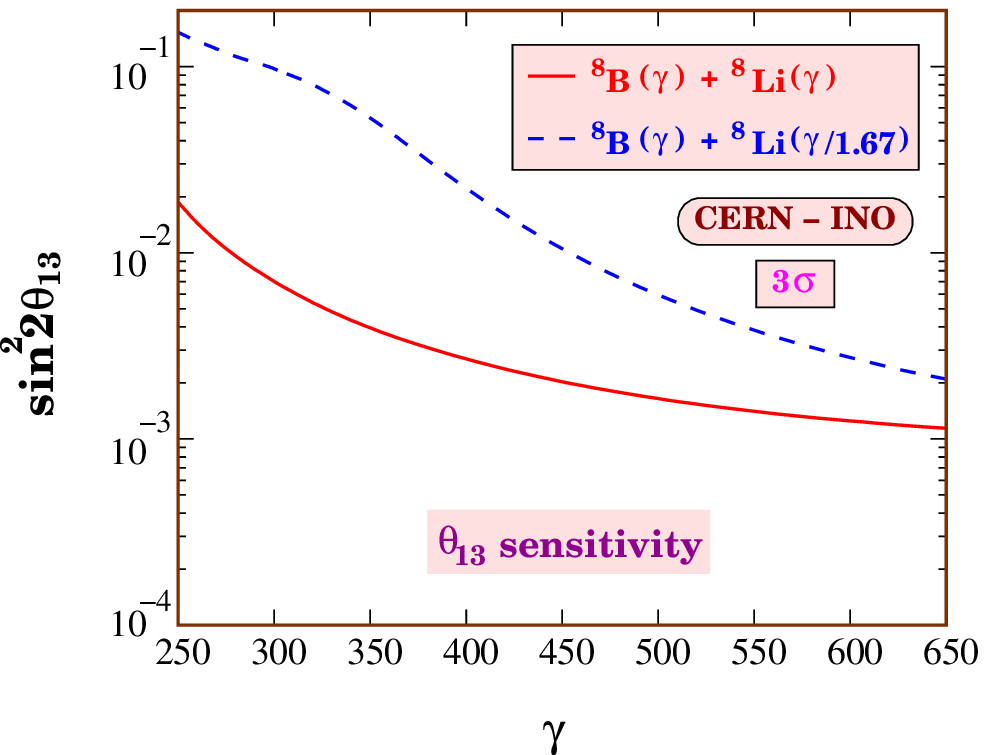}
\vglue -7.0cm \hglue 8.8cm
\includegraphics[width=8.0cm, height=7.0cm, angle=0]{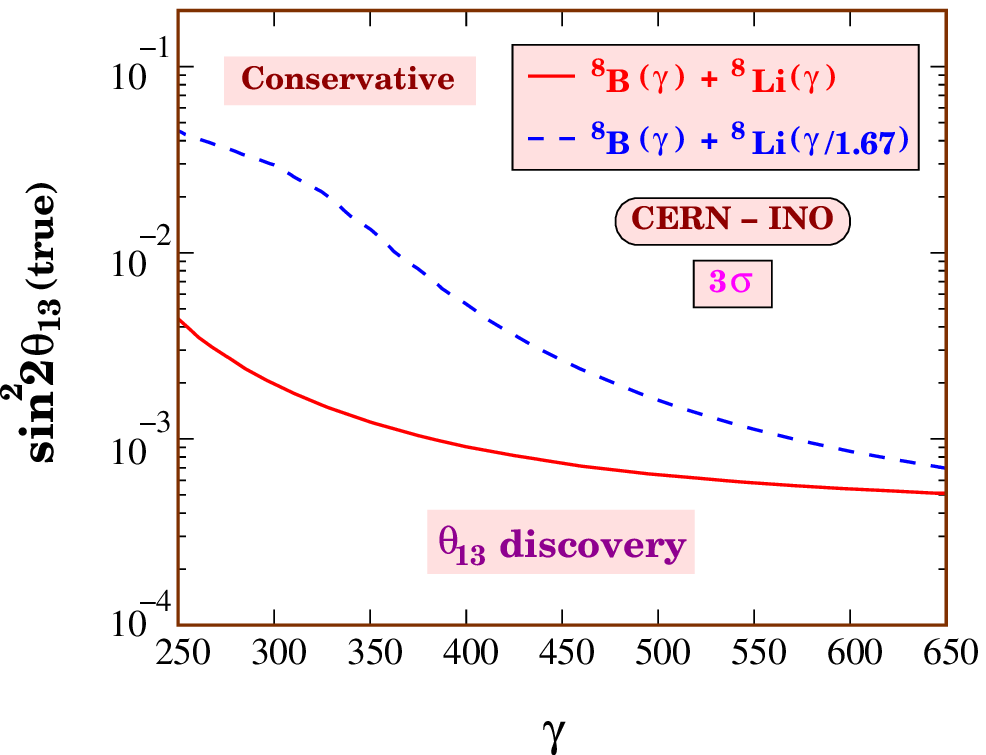}
\caption{\label{fig:sensth13}
Left panel shows
the $3\sigma$ sensitivity limit for $\stch$. Right
panel shows
the $3\sigma$ discovery reach for $\stcht$.
The red solid lines in the left and right panels
show the sensitivity reach and discovery potential respectively,
when the $\gamma$ is assumed to be the same
for both the neutrino and the antineutrino beams. The blue dashed lines
show the corresponding limits when the $\gamma$ for the
$^8$Li is scaled down by a factor of 1.67 with respect to the
$\gamma$ of the neutrino beam, which is plotted in the $x$-axis.
}
\end{figure}

The $\theta_{13}$ sensitivity reach is defined as the range of $\stch$
which is incompatible with the data generated for $\stcht=0$ at the 
$3\sigma$ C.L. This performance indicator
corresponds to the new $\stch$ limit if the experiment does not
see a signal for $\theta_{13}$-driven oscillations\footnote{Note
from Eq. \ref{eq:pemu}, while the first three terms go to
zero when $\theta_{13} \rightarrow 0$, the last term, which depends 
only on the solar parameters and $\theta_{23}$, remains non-vanishing.
Therefore, when the flux is high, \ie, for large $\gamma$ and/or enhanced
luminosity, we expect a sizable number of events even when $\stcht=0$.}.
In that case, we can exclude some allowed values of $\stch$, 
which we call our ``$\theta_{13}$ sensitivity reach''.
We simulate this situation in our analysis by generating the data 
at $\stcht=0$ and fitting it with some non-zero value of $\stch$
by means of the $\chi^2$ technique. In our fit we marginalize over
all the oscillation parameters including $\delta_{CP}$ and
the mass hierarchy\footnote{Note that
since $\stcht=0$, the data is independent of the $\dcpt$ and the true
mass hierarchy. However, since we allow for non-zero $\stch$ in the fit, the
predicted event rates in our ``theory'' depend on $\delta_{CP}$ and 
the mass hierarchy.} and choose the value of $\stch$ 
for which the fit yields $\chi^2 = 9$.
The result is shown in the left panel of Fig. \ref{fig:sensth13}, 
as a function of $\gamma$. The red solid line shows the $\stch$ sensitivity 
when $\gamma$ is assumed to be the same for both the neutrino and the 
antineutrino beams. The blue dashed lines shows the corresponding $3\sigma$
upper limit when $\gamma$ for the $^8$Li is scaled down by a factor of 1.67 
with respect to that for the neutrino beam.
In generating the data we have assumed that $\stcht=0$,
which means that we have negligible events in both the
neutrino as well as the antineutrino channels, irrespective
of the mass hierarchy. When this data is fitted allowing for
non-zero $\stch$, the neutrino (antineutrino) channel plays a 
dominating role when NH (IH) is assumed in the fit.
We reiterate that Fig. \ref{fig:sensth13} shows the $\stch$ sensitivity 
after marginalizing over hierarchy as well. In other words, the sensitivity 
shown in this figure corresponds to the statistically weaker channel.
For the case where the same $\gamma$ for $^8$B and $^8$Li has been used, 
the neutrino channel is weaker since the event rate is about 1.5 times less
than antineutrino events with the same $\gamma$.
On the other hand when we scale down the Lorentz boost for $^8$Li, the flux 
in the antineutrino channel goes down significantly and hence it becomes
the statistically weaker channel as can be seen from Fig. \ref{fig:eventsbetaino}
and therefore the marginalized $\chi^2$ corresponds mainly to that from
antineutrinos. Indeed one can check that the $\stch$ sensitivity that we 
exhibit by the blue dashed line for the scaled $\gamma$ case is comparable 
to what one can obtain for the antineutrino channel with IH and the 
corresponding lower $\gamma$. Similar feature is observed in addressing 
the issue of neutrino mass ordering taking IH as true hierarchy
(see right panel of Fig. \ref{fig:senshier}) and the $\gamma$ for the
antineutrinos is scaled down by a factor of 1.67 with respect to the
$\gamma$ of the neutrino beam.  

The dependence of the $\stch$ sensitivity on the number of
useful radioactive ion decays per year in the straight section of the
storage ring is shown in the right panel of Fig. \ref{fig:senslum}.
Here we have taken the same Lorentz boost for $^8$B and $^8$Li
and we have shown the results for three fixed values of $\gamma$.
The relative increase in the sensitivity by increasing $\gamma$ 
and/or the number of useful ion decays per year by the same
factor is quantified in the last two columns of Table \ref{tab:compare}.

\subsection{$\stch$ Discovery Reach}

How good are our chances of observing a positive signal
for oscillations and hence $\theta_{13}$ in the CERN-INO
beta-beam set-up? We answer this question in
terms of the parameter indicator which we call the ``discovery
reach'' of the experiment for $\stch$. 
This performance indicator is defined as the
range of $\stcht$ values which allow us to rule out 
$\stch=0$ at the $3\sigma$ C.L. To find this we simulate the
data at some non-zero value of $\stcht$ and fit it by assuming that
$\stch=0$, allowing all other oscillations parameters to take
any possible value in order to return back the smallest value
for the $\chi^2$. Note that since the fitted value of $\theta_{13}$
in this case always corresponds to 0, there is no need of any 
marginalizing over the hierarchy when fitting
the data. However, since the data here is generated at
a non-zero value of $\stcht$, it depends on the
true mass hierarchy. The discovery reach of the experiment
is therefore expected to be dependent on the true mass hierarchy.
Likewise, while the value of $\delta_{CP}$ in the fit is
inconsequential as $\stch=0$ in the fit, the data itself
and hence the discovery reach, would depend on $\dcpt$.
For each $\stcht$, we generate the data for all possible values of
$\dcpt$ and for both the true mass hierarchies. For each case,
the data is then fitted assuming $\stch=0$
and marginalizing over the other oscillation parameters,
returning a value of $\chi^2_{\rm min}$ for each data set.
We choose the minimum amongst these $\chi^2_{\rm min}$ and
find the value of $\stcht$ for which we could claim
a signal in the detector at the $3\sigma$ C.L.
In the right panel of Fig. \ref{fig:sensth13} we show this 
``most conservative''\footnote{This
is ``most conservative'' in the sense that no matter
what the choices of $\dcpt$ and the true neutrino mass ordering,
the $\theta_{13}$ discovery limit cannot be worse than the value presented.}
$\stch$ discovery reach
of our experiment as a function of $\gamma$.
We assume equal $\gamma$ for both the ions for the red solid curve.
One can see that for $\gamma=650$, the most conservative
discovery reach is $\stcht=5.11 \times 10^{-4}$ while if $\dcpt=0$ 
then we have checked that the reach is slightly better and 
this will be $\stcht=5.05 \times 10^{-4}$.
For the blue dashed line we assume that the $\gamma$ for
$^8$Li is scaled down by a factor of 1.67 compared to
that for $^8$B, plotted on the $x$-axis.
Since for same $\gamma$, neutrino is the statistically weaker channel,
the red line mainly corresponds to what we expect for the
true NH. For the scaled $\gamma$ case since the antineutrino
channel becomes statistically weaker, the lower $\chi^2$
comes from this channel and the blue dashed line corresponds to 
what we expect for the true IH.

\subsection{$\stch$ Precision}

\begin{figure}[t]
\includegraphics[width=8.0cm, height=7.0cm]{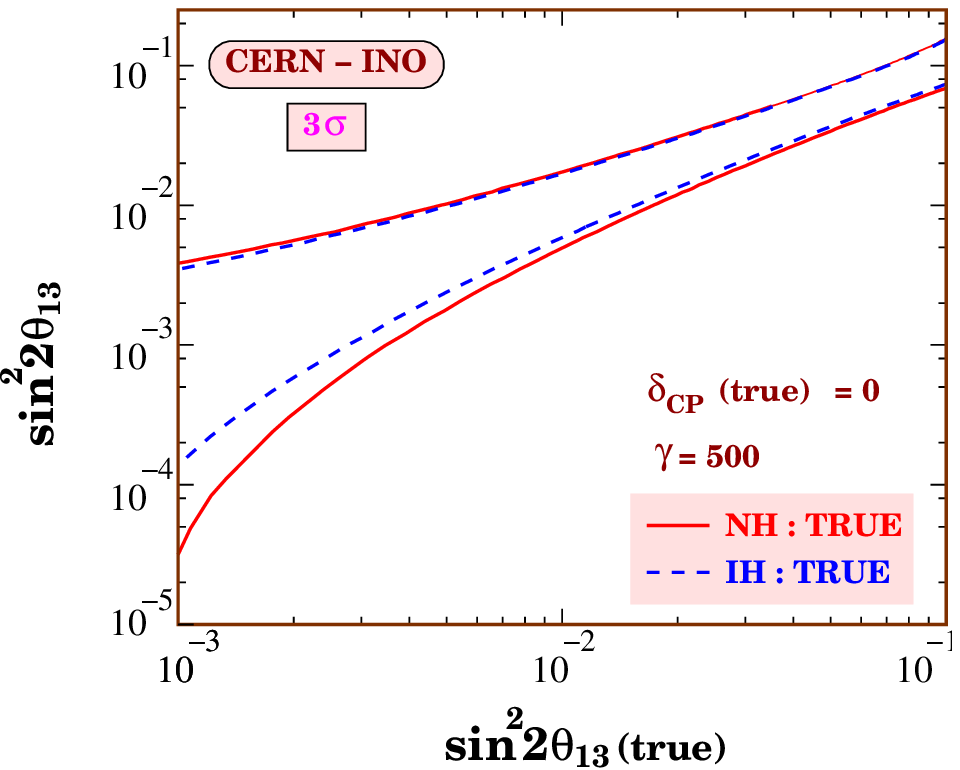}
\vglue -7.0cm \hglue 8.8cm
\includegraphics[width=8.0cm, height=7.0cm]{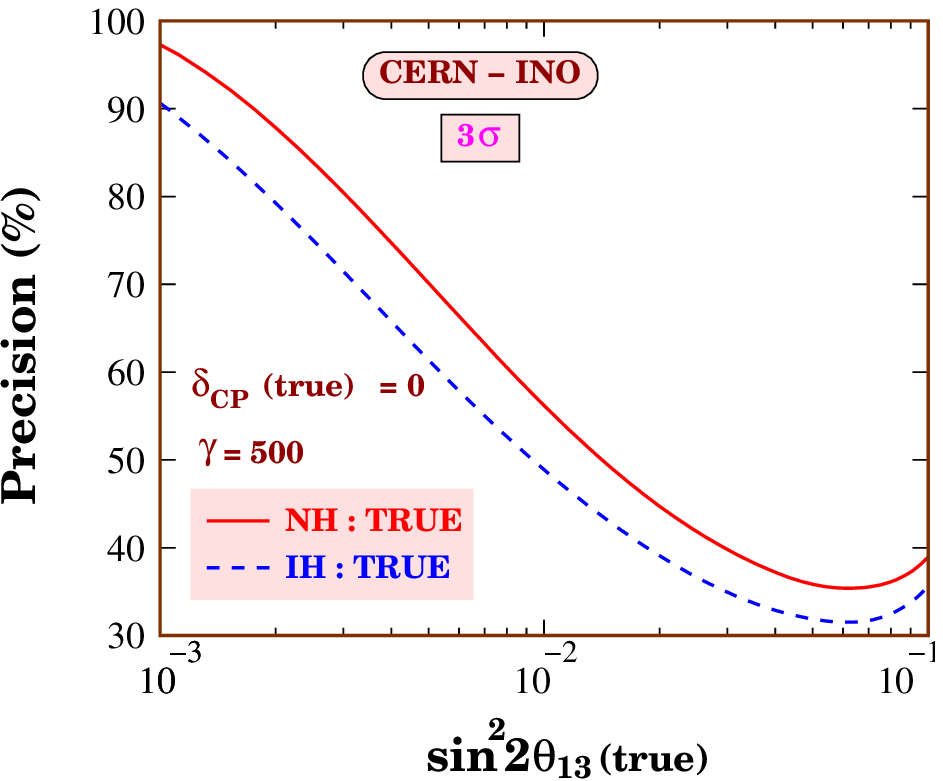}
\caption{\label{fig:precision}
The precision with which $\stch$ will be measured by
the CERN-INO beta-beam experiment as a function of
$\stcht$. Left panel shows the $3\sigma$ allowed range of $\stch$
while the right panel shows the precision defined in the text.
}
\end{figure}

In Fig. \ref{fig:precision} we show how {\it precisely} the mixing
angle $\stch$ will be measured, if we observe a $\theta_{13}$
driven signal at the detector. The left panel depicts as a function
of $\stcht$ the corresponding range of allowed values of
$\stch$ at the $3\sigma$ C.L. We have assumed $\gamma=500$ and $\dcpt=0$.
The solid line is assuming NH to be true, while the dashed line
is for IH true. Note that in the fit we always marginalize over the
hierarchy and $\delta_{CP}$. The right panel shows the variable
``precision'' which we define as
\be
{\rm precision} = \frac{{(\stch)}_{max}-{(\stch)}_{min}}
{{(\stch)}_{max}+{(\stch)}_{min}} \times 100\,\%~,
\ee
where ${(\stch)}_{max}$ and ${(\stch)}_{min}$
are the maximum and minimum allowed values of $\stch$
respectively at $3\sigma$.


\section{\fbox{Summary and Conclusions}}

Long baseline experiments which will use the golden $P_{e\mu}$ channel for
determining the neutrino oscillation parameters face a serious
threat from the menace of clone solutions due to the
so-called parameter degeneracies. These degeneracies
come in three forms: the $\delta_{CP}-\theta_{13}$ intrinsic
degeneracy, the $ \delta_{CP}-sgn(\ma)$ degeneracy and
the $\theta_{23}$ octant degeneracy, and necessarily result
in degrading the sensitivity of the experiment.
The CERN-INO near-magic distance of 7152 km offers the possibility of
setting up an experiment at a baseline where the $\delta_{CP}$ dependent 
terms almost drop out from the expression of the golden channel probability.
Thus two out of the three degeneracies are evaded, providing
a platform for clean measurement of $\theta_{13}$ and
$sgn(\ma)$, two major players in our understanding of the
origin of neutrino masses and mixing.
A large magnetized iron calorimeter with a total mass of at 
least 50 kton is expected to be built soon at INO. It will be ideal for
detecting multi-GeV $\numu$ and hence can be used as the far
detector for a high energy beta-beam.

In this chapter we studied in detail the physics reach of the
CERN-INO magical beta-beam set-up.
For $\gamma=650$, $\dcpt=0$ and true NH,
the sensitivity to hierarchy determination at $3\sigma$ 
is $\stcht=5.51 \times 10^{-4}$ when full spectral data from 
neutrino and antineutrino channels are combined.
Even though the effect of $\dcpt$ on the event rate of such an experiment 
is expected to be small, there is some residual
dependence on it because the CERN-INO distance does not
conform to the exact magic baseline. We studied the change in the
hierarchy sensitivity due to the uncertainty in $\delta_{CP}$.
It turns out that for $\gamma=650$ and with NH (IH) true, the best 
sensitivity to hierarchy determination corresponds to 
$\stcht =3.96 \times 10^{-4}$ ($\stcht =2.96 \times 10^{-4}$), while the 
worst case is $\stcht =5.58 \times 10^{-4}$ ($\stcht =3.59 \times 10^{-4}$).

A detailed analysis of the potential of probing $\theta_{13}$ 
at this experiment has been presented here.
We defined and studied the $\theta_{13}$ reach in terms of three 
performance indicators: the sensitivity reach, the discovery reach 
and the precision of $\stch$ measurement.
The $\theta_{13}$ sensitivity reach is defined as the range of $\stch$
which is incompatible with the data generated for $\stcht=0$ at the
$3\sigma$ C.L. The sensitivity reach corresponds to $\stch=1.14 \times 10^{-3}$
at $3\sigma$ C.L. for $\gamma=650$ and this is independent of the true hierarchy
and $\dcpt$. The discovery reach is defined as the range of true values of the 
mixing angle for which we have an unambiguous oscillation signal in the detector.
At $3\sigma$ C.L. the discovery reach corresponds to
$\stcht=5.05 \times 10^{-4}$ ($\stcht=2.96 \times 10^{-4}$)
for $\gamma=650$, $\dcpt=0$ and NH (IH) true while the most conservative 
limit irrespective of $\dcpt$ and the true neutrino mass ordering is 
$\stcht=5.11 \times 10^{-4}$. We also presented the expected precision 
with which $\stch$ would be determined in this experiment 
for $\stcht > 10^{-3}$.

Neutrino physics is in wait for the next great leap forward
in the decade ahead. The study of the golden channel probability 
$P_{e\mu}$ using a beta-beam neutrino source and an iron calorimeter
detector at a very long magical baseline may well turn out, as we have 
demonstrated in this chapter, to be very crucial in this endeavour.
In the next chapter, we will discuss the impact of matter effects 
in the $\nue$ survival probability at long baselines in extracting 
the information about the neutrino mass ordering and the 1-3 mixing
angle.

\newpage
\mbox{}
\chapter{Neutrino parameters from matter effects in $P_{ee}$ at
long baselines}
\section{\fbox{Introduction}}

Determination of the 1-3 neutrino mixing angle $\theta_{13}$,
\sgnma{}, the three CP phases and the absolute neutrino mass scale
are necessary for reconstruction of the neutrino mass matrix, which 
will have important consequences for nuclear and particle physics, 
astrophysics and cosmology. The golden channel $P_{e \mu}$ has the 
capability of measuring the Dirac phase $\delta_{CP}$, 
$sgn(\Delta m^2_{31})$ and $\theta_{13}$ in long baseline accelerator 
based experiments.
However, this strength of the golden channel also brings in
the well-known problem of parameter ``degeneracies'', where
one gets multiple fake solutions in addition to the true
one \cite{intrinsic,minadeg,th23octant,eight}.
Various ways to combat this vexing issue have been suggested 
in the literature, including combining the golden channel with the
``silver'' ($P_{e\tau}$) \cite{silver} and ``platinum'' ($P_{\mu e}$) 
channels. While each of them would have fake solutions, their combination 
helps in beating the degeneracies since each channel depends
differently on $\delta_{CP}$, $sgn(\Delta m^2_{31})$ and $\theta_{13}$.
In this chapter, we focus on the $\nu_e \to \nu_e$ survival channel, $P_{ee}$,
which is {\it independent} of $\delta_{CP}$ and the mixing angle $\theta_{23}$. 
It is therefore {\it completely} absolved of degeneracies
and hence provides a clean laboratory for the measurement
of $sgn(\ma)$ and $\theta_{13}$. This gives it an edge over the
conversion channels, which are infested with degenerate solutions.

The $P_{ee}$ survival channel has been extensively
considered for measuring $\theta_{13}$ with $\anue$ produced in
nuclear reactors \cite{white} and with detectors placed at a 
distance $\simeq 1$ km. Reducing systematic uncertainties 
to the sub-percent level is a prerequisite for this program 
and enormous R\&D is underway for this extremely challenging job.
For accelerator based experiments, the survival channel, $P_{ee}$, 
has been discussed with sub-GeV neutrinos from a beta-beam source 
at CERN and a megaton water detector in Fr\'{e}jus at a
baseline of 130 km \cite{bc,bc2,betaoptim,dissappear}.
However, no significant improvement on the $\theta_{13}$
limit was found in \cite{dissappear} for a systematic
error of $\gsim$ 5\%.
This stems mainly from the fact that in these experiments
one is trying to differentiate between two scenarios,
both of which predict a large number of events, differing from 
each other by a small number due to the small value of $\theta_{13}$.
Also, since $sgn(\ma)$, is ascertained using earth matter effects,
there is no hierarchy sensitivity in these survival channel
experiments due to the short baselines involved.

In this chapter, we emphasize on the existence of large matter
effects in the survival channel, $P_{ee}$, for an experiment with a
very long baseline. Recalling that $P_{ee}= 1 - P_{e \mu} - P_{e \tau}$ 
and since for a given $sgn(\ma)$ {\it both}
$P_{e \mu}$ and $P_{e \tau}$ will either increase 
or decrease in matter, the change  in $P_{ee}$ is almost twice that
in either of these channels. Using the multi-GeV $\nue$ flux from a 
beta-beam source, we show that this large matter effect
allows for significant, even maximal, deviation of $P_{ee}$
from unity. This, can thus be a convenient tool to explore $\theta_{13}$.
This is in contrast to the reactor option or the beta-beam experimental 
set-up in \cite{dissappear}, where increasing the
neutrino flux and reducing the systematic uncertainties
are the only ways of getting any improvement on
the current $\theta_{13}$ limit. We further show,
for the first time, that very good sensitivity to the
neutrino mass ordering can also be achieved
in the $P_{ee}$ survival channel owing to the large
matter effects. We discuss plausible experimental set-ups with 
the survival channel and show how the large matter effect propels 
this channel, transforming it into a very useful tool to probe $sgn(\ma)$
and $\theta_{13}$ even with relatively large room for systematic 
uncertainties.

\section{\fbox{The $\nu_e \to \nu_e$ Survival Probability}}

Under one mass scale dominance and the constant density approximation,
the matter conversion probabilites are,
\begin{equation}
P_{e x} =
Y_{23} \, \sin^2 2 \theta_{13}^{m}
\sin^2 \left[1.27 (\Delta m^2_{31})^{m} {L}/{E} \right],
\label{eq:conv_prob}
\end{equation}
where $Y_{23}=\sin^2 \theta_{23}$ for $x=\mu$ and
$Y_{23}=\cos^2 \theta_{23}$ for $x=\tau$.
Then we have
\begin{equation}
P_{e e} = 1 - P_{e \mu} - P_{e \tau}
= 1 - \sin^2 2 \theta_{13}^{m}
\sin^2 \left[1.27 (\Delta m^2_{31})^{m}
{L}/{E} \right].
\label{eq:peemat}
\end{equation}
The largest deviation of $P_{ee}$ from unity is obtained 
when the conditions \ref{eq:cond1} and \ref{eq:cond2}
are satisfied simultaneously. This ensures maximum matter 
effect in the conversion channels 
(see Eqs. \ref{eq:max_cond} and \ref{eq:eecondtn}).
However, there are suppression factors, $\sin^2\theta_{23}$ for 
\pemu and $\cos^2\theta_{23}$ for \petau, not present in $P_{ee}$. 
Moreover, since $P_{ee}$ does not contain $\theta_{23}$, the octant 
ambiguity as well as parameter correlations due to uncertainty in 
$\theta_{23}$ \cite{th23octant} are absent.
In addition, as mentioned earlier, the $P_{ee}$ channel does not
contain the CP phase, $\delta_{CP}$ and therefore this channel
does not suffer from the effect of the so-called
($\theta_{13}, \delta_{CP}$) intrinsic degeneracy \cite{intrinsic}
and the (\sgnma{}, $\delta_{CP}$) degeneracy \cite{minadeg}.
Both of these remain true in the presence of non-zero
$\Delta m^2_{21}$ \cite{akh}.

Using Eq. \ref{eq:eecondtn}, one can estimate the distance 
where $P_{ee}\simeq 0$ for a given value of $\theta_{13}$.
For instance for $p=0$ and $\sin^2 2 \theta_{13}$ = 0.2 and 0.1, 
these distances are 7600 km and 10200 km respectively which can 
be readily seen from Fig. \ref{fig:omsd}. For higher values of $p$ 
the distance exceeds the earth's diameter for $\theta_{13}$ in the 
current allowed range. Using $(\rho L)^{max}$ corresponding to the 
PREM profile, from Eq. \ref{eq:eecondtn} one can estimate that the
condition of maximal matter effects inside the earth's mantle
is satisfied only for $\stch \gsim 0.09$.
In our numerical work, we solve the full three flavour neutrino 
propagation equation assuming the PREM \cite{prem} profile and keep 
$\Delta m^2_{21}$ and $\sin^2\theta_{12}$ fixed at their present
best-fit values of $8.0 \times 10^{-5}$ eV$^2$ and 0.31 respectively 
\cite{limits} as mentioned in the first column of Table \ref{tab:bench}.
For the other oscillation parameters, the benchmark values are also 
taken from the same table. 

\begin{figure}[t]
\begin{center}
\includegraphics[width=16.0cm, height=9.5cm]{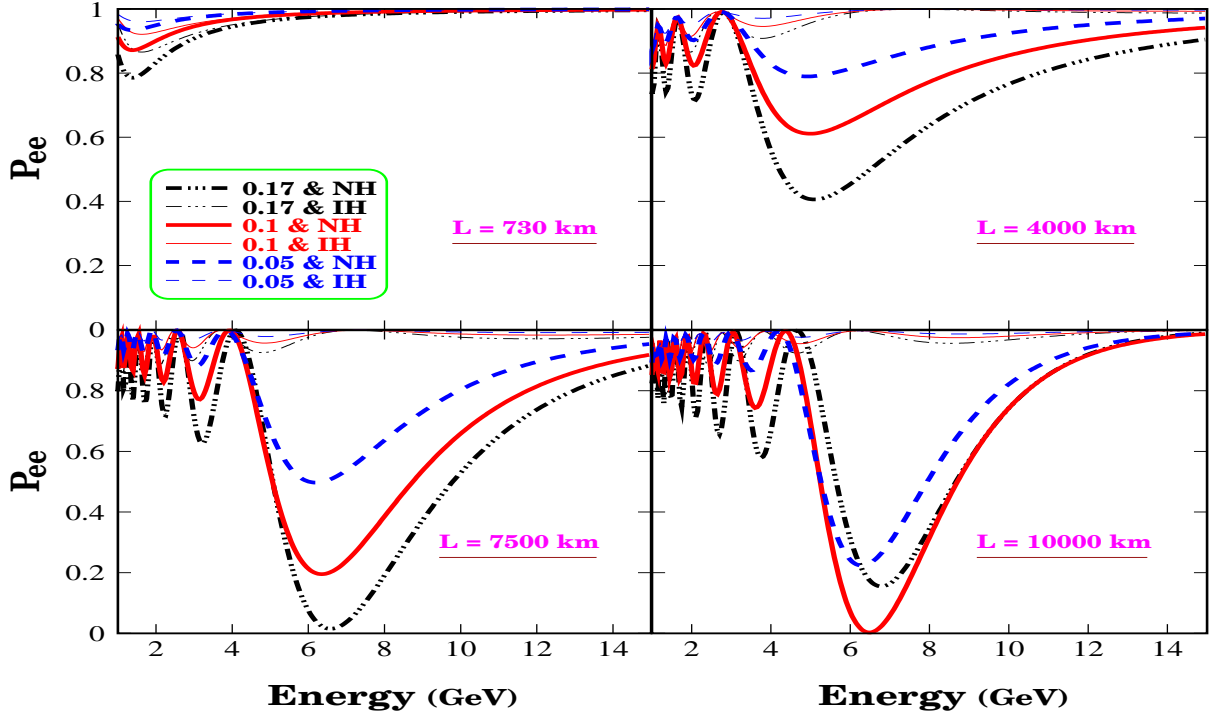}
\caption{\label{fig:fig1_pee}
The survival probability \pee in matter as a function of $E$ 
for four different values of the baseline $L$. For each $L$, 
the plots are given for three different values 
of $\stch$ (0.17, 0.1 and 0.05). Thick (thin) lines are for NH (IH).
}
\end{center}
\end{figure}

In Fig. \ref{fig:fig1_pee} we plot \pee as a function of energy,
at four different $L$ and for three values of $\sin^2 2\theta_{13}$.
The plots confirm that maximal matter effects
come at $L\simeq 10000$ km and $L\simeq 7500$ km for
$\stch=0.1$ and 0.17 respectively for the normal hierarchy (NH).
For the inverted hierarchy (IH) there is no significant matter effect
for $\nu_e$. This large difference in the probabilities for NH and IH
can be exploited for the determination of $sgn(\ma)$.
Further, since the matter effect is a sensitive function of
$\theta_{13}$ it may also be possible to obtain information on
this angle. We can also see that for a given value of $\stch$
($\gsim 0.09$) and $E$, the matter effect increases 
(almost linearly) with $L$, until the $L$ for maximal matter effect
is reached, beyond which matter effect falls.
For values of $\sin^2 2\theta_{13} <  0.09$ the condition
for maximum matter effect is not met inside the earth's mantle
and hence the matter effect and sensitivity to both hierarchy 
as well as $\theta_{13}$ increase with $L$.

In what follows, we will show how, in a plausible experiment, one
can use this near-resonant matter effect in the survival channel, 
$P_{ee}$, to constrain $\theta_{13}$ and $sgn(\ma)$.
Fig. \ref{fig:fig1_pee} shows that the requirements for such a 
program include 

\begin{enumerate}

\item
{\fbox{\bf a $\nu_e$ beam,}} 
\item
{\fbox{\bf a baseline of at least a few thousand km,}}
\item 
{\fbox{\bf average energies around 6 GeV,}} 
\item
{\fbox{\bf a detector capable of observing 
$e^-$ unambiguously at these energies.}}

\end{enumerate}

\section{\fbox{The Experimental Set-up}}

\subsection{Pure $\nu_e$ ($\bar\nue$) Source}

The various issues related to beta-beams have already been 
discussed in great detail in section 3.2 of the previous chapter.
The ions considered as possible sources for beta-beams are 
$^{18}$Ne and $^{8}$B for $\nu_e$ and $^{6}$He and $^{8}$Li for $\anue$.
The end point energies of $^{6}$He and $^{18}$Ne are $\sim 3.5$ MeV
while for $^{8}$B and $^{8}$Li this can be larger $\sim$ 13-14 MeV
(see Table \ref{tab:ions}). For the Lorentz boost factor $\gamma=250$ (500) 
the $^{8}$B and $^{8}$Li sources have peak energy around $\sim 4$ (7) GeV.
Since this is in the ball-park of the energy necessary for near-resonant 
matter effects as discussed above, we will work with $^8$B ($^8$Li) 
as the source ion for the $\nue$ ($\anue$) beta-beam and $\gamma=250$ and 500.

\subsection{Water \chr Detector}

Water \chr detectors have excellent capability of separating electron
from muon events. Since this technology is very well known, megaton water 
detectors are considered to be ideal for observing beta-beams.
Proposals for megaton water detectors include UNO \cite{uno} in USA, 
HyperKamiokande \cite{hk} in Japan and MEMPHYS \cite{memp} in Europe. 
If the beta-beam is produced at CERN, then baselines in the range 
7000-8600 km would be possible at any of the proposed locations for the UNO
detector. Likewise, if the beta-beam source be at FNAL, then the
far detector MEMPHYS would allow for $L=7313$ km.
HyperKamiokande could also be considered as the far detector and
in that case $L=10184$ ($9647$) km if the source be at FNAL (CERN). 
Such detectors do not have any charge identification capacity.
But in a beta-beam, the $\beta^{-}$ and $\beta^{+}$ emitters can be stacked 
in different bunches and the timing information at the detector can help 
to identify the $e^{-}$ and $e^{+}$ events \cite{doninibeta}.

\subsection{Backgrounds}

It is well known that there are no beam induced backgrounds for beta-beams.
In the previous chapter (subsection 3.4.3), these issues have been addressed 
in detail. In this experimental set-up, the process 
$\nu_e \to \nu_\tau \to \tau^- \to e^-$ could mimic the signal. 
We have checked that the background to signal ratio 
for these events in the relevant energy range is $\sim 10^{-2}$ and
can be neglected for the disappearance mode. The electron events from 
$K$ and $\pi^-$ decays are also negligible.
The atmospheric background can be estimated in the beam off mode
and reduced through directional, timing, and energy cuts.

\section{\fbox{Simulation Details and Event Rates}}

\begin{figure}[t]
\begin{center}
\includegraphics[width=10.0cm]{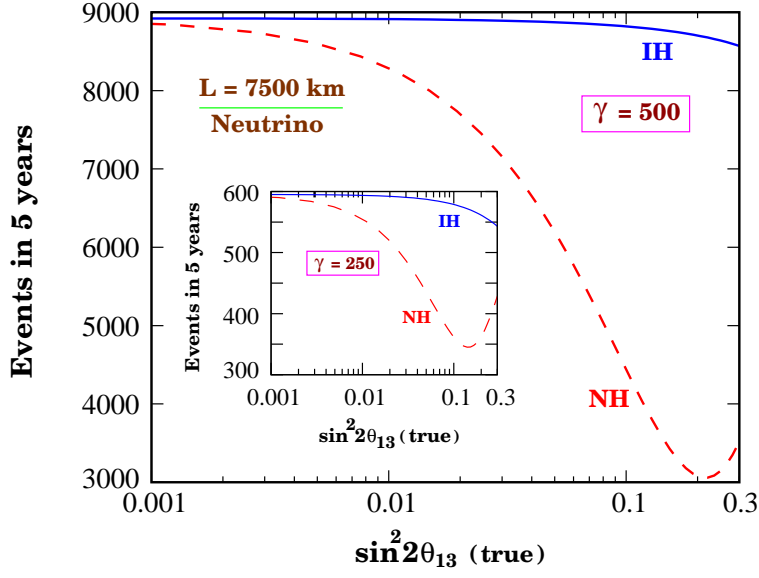}
\caption{\label{fig:fig2_pee}
Events in 5 years vs. $\stch$ for NH (dashed line) and IH (solid line) 
for $L=7500$ km and $\gamma=500$. The inset shows the same but for 
$\gamma=250$.
}
\end{center}
\end{figure}

For our numerical analysis we use the standard $\chi^2$ technique with
\begin{equation}
\chi^2_{total} = \chi^2_{\nue \to \nue} + \chi^2_{prior},
\label{eq:chisq_pee}
\end{equation}
where\footnote{No prior is taken on $\theta_{23}$ because
\pee is independent of $\theta_{23}$. Here we have kept 
$\Delta m^2_{21}$ and $\sin^2\theta_{12}$ fixed at their present 
best-fit values in the fit. No uncertainty has been 
taken in the PREM profile when we fit the simulated data.}
\begin{equation}
\chi^2_{prior}=\left (\frac{|\Delta m^2_{31}|-
|\mat|}{\sigma(|\Delta m^2_{31}|)} \right )^2.
\label{eq:prior_pee}
\end{equation}
$\chi^2_{\nue \to \nue}$ is calculated using the same expression that 
we have used for $\chi^2_{\nue \to \numu}$ (see Eqs. \ref{eq:chipull} \& 
\ref{eq:rth}) in the previous chapter. Here all the results have been 
presented using the data from only neutrino ($^8$B) run without any 
spectral information. We have not considered any 
background\footnote{Backgrounds are not important here because we are 
dealing with the survival channel (\pee) and the number of events is also quite large.} 
in our analysis. We have taken 3\% systematical error on signals.
The prospective ``data'' is generated at the ``true'' values of 
oscillation parameters as given in the Table~\ref{tab:bench},
assuming 440 kton of fiducial volume for the detector
with 90\% detector efficiency, threshold of 4 GeV and
energy smearing of width 15\%. For the $\nue$ beta-beam we have 
assumed $1.1 \times 10^{18}$ useful $^8$B decays per year and show 
results for 5 years of running of this beam. The number of events as a 
function of $\stch$ at $L=7500$ km with a $\gamma=500$ $\nue$ 
beta-beam is shown in Fig. \ref{fig:fig2_pee} for NH and
IH. The inset in Fig. \ref{fig:fig2_pee} shows the number of
events in 5 years expected from a lower $\gamma=250$. 
We have used the neutrino-nucleon interaction cross-sections from the 
Globes package \cite{globes} which are taken 
from \cite{Messier:1999kj,Paschos:2001np}.

\section{\fbox{Sensitivity to \sgnma{} and $\theta_{13}$}}

\begin{figure}[t]
\begin{center}
\includegraphics[width=10cm]{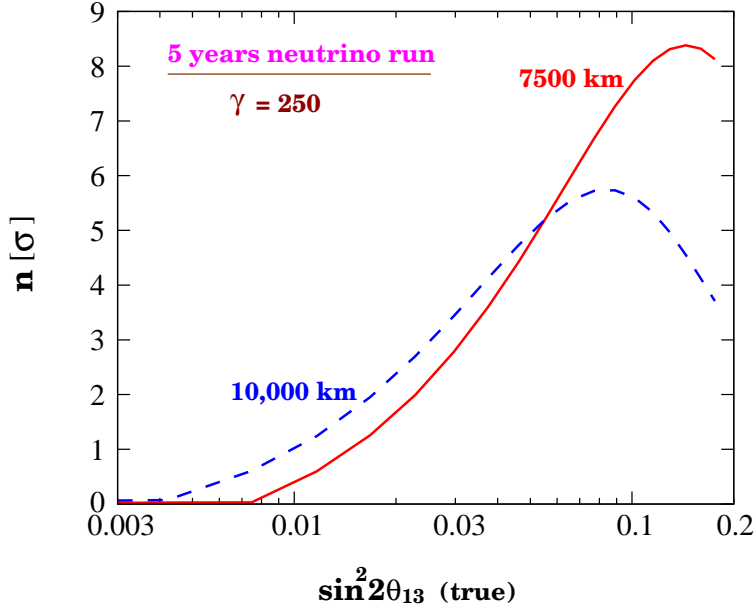}
\caption{\label{fig:fig3_pee}
Sensitivity to hierarchy for $L=7500$ (solid line) and 10000
km (dashed line) and $\gamma=250$,
as a function of $\stcht$.
}
\end{center}
\end{figure}

\begin{figure}[t]
\begin{center}
\includegraphics[width=12.0cm]{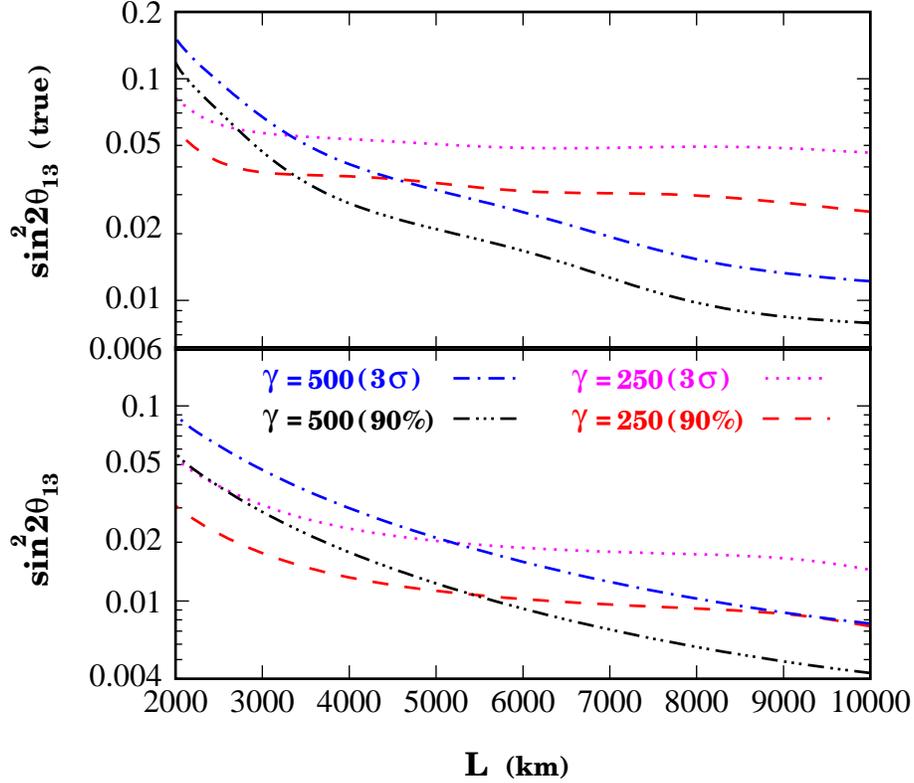}
\caption{\label{fig:fig4_pee}
The upper panel shows the range of $\stcht$ for which the 
wrong IH can be excluded at 90\% and 3$\sigma$ C.L. 
while the lower panel gives the sensitivity to 
$\sin^2 2\theta_{13}$ at various baselines at 90\% and 
3$\sigma$ C.L., for two values of $\gamma$.
}
\end{center}
\end{figure}

In Fig. \ref{fig:fig3_pee}, we show the sensitivity
($n \sigma,~n = \sqrt{\chi^2}$) of the survival channel 
to the neutrino mass ordering for $L$ = 7500 and
10000 km and $\gamma$ = 250. If the true value of $\stch$ = 0.05,
then one can rule out the IH at the 4.8$\sigma$ (5.0$\sigma$) C.L. 
with $L=$ 7500 (10000) km. For $L$ = 7500 (10000) km, the 
wrong IH can be disfavored at the 90\% C.L. if the true value of 
$\stch>$ 0.03 (0.025). The sensitivity improves significantly 
if we use $\gamma$ = 500 instead of 250, since (i) the flux at the 
detector increases, and (ii) the flux peaks at $E$ closer to 6 GeV, 
where we expect largest matter effects. For $\gamma$ = 500 and 
$L=$ 7500 (10000) km, the IH can be disfavored at 
2.6$\sigma$ (3.8$\sigma$) C.L. for a lower value of $\stcht$ = 0.015.
The range of $\stcht$ for which the IH can be ruled out at 
90\% and 3$\sigma$ C.L. for different values of $L$ are shown 
in the upper panel of Fig. \ref{fig:fig4_pee} for $\gamma$ = 250
and 500. From the figure one can see that for $\gamma$ = 500 and 
$L$ = 7500 (10000) km the wrong IH can be disfavored at the
90\% C.L. if $\stcht > 1.0\times 10^{-2}$
($8.0\times 10^{-3}$). If instead we use a total systematic error of 5\%
then we get the above sensitivity limits as $\stcht > 1.6\times 10^{-2}$
($1.2\times 10^{-2}$) at 90\% C.L.

If the $\stcht$ turns out to be smaller than the 
sensitivity reach shown in the upper panel of Fig. \ref{fig:fig4_pee} 
for a given $L$, then it would not be possible to determine the
hierarchy at the given C.L. However, we would still be able to put better
constraints on $\stch$ itself. The lower panel of Fig. \ref{fig:fig4_pee}
demonstrates as a function of $L$, the sensitivity to $\theta_{13}$, i.e.,
the range of $\stch$ which is incompatible with the data generated for 
$\stcht$ = 0 at 90\% and 3$\sigma$ C.L.
Both Figs. \ref{fig:fig3_pee} and \ref{fig:fig4_pee} show that
the sensitivity improves with $L$, even though the flux falls
as $1/L^2$. This results from matter effects increasing with $L$,
as noted before. For $L=7500$ (10000) km, we can constrain
$\stch < 6.3  \times 10^{-3}$ ($4.3 \times 10^{-3}$)
at the 90\% C.L. for $\gamma$ =500. For a 5\% systematic error 
the above numbers are changed to $\stch < 1.0  
\times 10^{-2}$ ($7.3 \times 10^{-3}$).

How does this compare with alternate possibilities?
If the energy can be reconstructed accurately, then the result 
can be improved further. For instance, for $L$=7500 km, if one could
preferentially select the energy in the range 5 to 7.5 GeV,
then the wrong IH can be excluded from true NH for 
$\stcht = 7.47 \times 10^{-3}$ at 90\% C.L. for $\gamma$ = 500.

We have presented our results using a $\nue$ beam and assuming
NH to be the true hierarchy. Similar results can also be obtained 
with a $\anue$ beam for IH. It is also possible to run both beams 
simultaneously.

\section{\fbox{Discussions and Conclusions}}

In conclusion, we propound the possibility of using large matter effects 
in the survival channel, $P_{ee}$, at long baselines for determination 
of the neutrino mass ordering ($sgn(\Delta m^2_{31})$) and the yet 
unknown leptonic mixing angle $\theta_{13}$. Matter effects in the 
transition probabilites $P_{e \mu}$ and $P_{e \tau}$ act in consonance
to give an almost two-fold effect in the survival channel. In addition,
the problem of spurious solutions due to the leptonic CP phase and the 
atmospheric mixing angle $\theta_{23}$ does not crop up.
The development of beta-beams as sources of pure $\nu_e/\bar\nu_e$ beams
enables one to exclusively study the $P_{ee}$ survival probability and adds 
a new direction to the prospects of a future beta-beam.
In the next chapter, we will study the performance of the 
magical CERN-INO beta-beam set-up in exploring the signals of 
new physics which are present in the R-parity violating supersymmetric 
models.

\chapter{Can R-parity violating supersymmetry be seen in 
long baseline Beta-beam experiments?}
\section{\fbox{Introduction}}

Long baseline neutrino oscillation experiments using beta-beam 
hold promise of refining our knowledge on the third mixing angle 
$\theta_{13}$, \sgnma{} and the CP phase $\delta_{CP}$, vital missing 
parameters of the neutrino mass matrix.
The phenomenon of neutrino oscillation firmly establishes the
evidence for physics beyond the Standard Model and the sub dominant
effects due to possible new physics can leave their imprints in the
future long baseline beta-beam experiments which are supposed to
give us precision below 1\%. In this chapter, we will try to address 
the following questions~:
\begin{enumerate}
\item
{\bf Can the proposed magical CERN-INO long baseline beta-beam 
neutrino experiment probe non-standard interactions (NSI) that are 
present in RPVSM?}
\item
{\bf Can these NSI become fatal in attempts to further sharpen 
our understanding of the neutrino properties?}
\item
{\bf Can new physics leave its imprints at the propagation 
stage of neutrinos?}
\end{enumerate}

Interaction of neutrinos with matter affect long baseline experiments 
and this becomes more prominent at higher values of $\theta_{13}$.
Various authors \cite{addatm,gandhi2,matter} have considered this effect 
for atmospheric neutrinos. Apart from the electroweak effects, 
there may well be NSI leading to flavour diagonal and flavour changing
neutral currents. Here we have in mind interactions with quarks and 
leptons involving an initial and a final neutrino. 
If there is no change in the 
neutrino flavour -- as, for example, 
in $Z^0$ exchange --  this is classified as an FDNC process, while it 
would be FCNC otherwise. RPVSM \cite{rparity}, which have such 
interactions already built-in\footnote{{\em e.g.} through squark 
($\lambda^\prime$-type couplings) or slepton ($\lambda$-type couplings) 
exchange.}, will be the main focus of our work. Recently a model in
which couplings associated with FCNC and FDNC can be quite a bit
higher than permitted in RPVSM has also been considered
\cite{david, fcnc1}. Naturally, here the matter effect will be further 
enhanced. However, as RPVSM is a well-studied, renormalizable model 
which can satisfy all phenomenological constraints currently available, 
we shall restrict our main analysis only to it and shall make
qualitative remarks about the other model, for which our results 
can be easily extended.

Consequences of FCNC and FDNC for solar and atmospheric neutrinos
\cite{lma_msw,solar1,atm1}, and neutrino factory experiments \cite{fac1}
have been looked into. Our focus is on long baseline beta-beam experiments.
Our analysis encompasses both NH and IH and we also incorporate all 
relevant trilinear R-parity violating couplings leading to FCNC and FDNC. 
Huber {\em et al} \cite{huber} have a somewhat similar analysis using
neutrino beams obtained from muon decays.

The very long baseline from CERN to INO will capture a
significant matter effect and offers a scope to signal NSI.
We examine whether the presence of R-parity violating ($\Rsl$) 
interactions will come in the way of constraining the mixing 
angle $\theta_{13}$ or unraveling the neutrino mass ordering. 
The possibility to obtain bounds on some $\Rsl$ couplings 
is also probed.  

\section{\fbox{$\Rsl$ Supersymmetry}}

In supersymmetric theories \cite{rparity}, gauge invariance does
not imply baryon number ($B$) and lepton number ($L$) conservation.
In the minimal supersymmetric Standard Model, $L$ and $B$ conservation 
is ensured by invoking `$R$-parity' (defined as $R = (-1)^{3B + L + 2S}$
where $S$ is the spin). It is a discrete $Z_2$ symmetry under
which the Standard Model particles are even and their superpartners are odd.
The imposition of such a symmetry, while it serves a purpose, is rather 
{\em ad hoc}. In general, from the na\"{i}ve theoretical point of view 
it is expected that $L$ and $B$ conservation does not hold in supersymmetric 
theories. However, as this leads to a very fast proton decay, we follow a 
common practice and assume that $B$ is conserved. This can be ensured by
replacing the $Z_2$ symmetry of $R$-parity by a $Z_3$ symmetry,
the so-called `baryon triality'~\cite{ross}. 
In such a scenario, in addition to the usual Yukawa interactions, 
the superpotential contains renormalizable $L$-violating trilinear 
$\lambda$- and $\lambda^\prime$-type couplings and bilinear 
$\mu_i$ couplings~:
\be
W_{\not L} =
\sum_{i,j,k} \frac{1}{2}\lambda_{ijk} {L}_i {L}_j {E}_k^c +
\lambda_{ijk}' {L}_i {Q}_j {D}_k^c + \mu_i {L}_i {H}_u ,
\label{super}
\ee
where $i,j,k = 1, 2, 3$ are generation indices and colour and $SU(2)$ 
indices are suppressed. Here ${L}_i$ and ${Q}_i$ are $SU(2)$-doublet lepton
and quark superfields respectively; ${E}_i$, ${D}_i$ denote the
right-handed $SU(2)$-singlet charged lepton and down-type quark
superfields respectively; ${H}_u$ is the Higgs superfield which
gives masses to up-type quarks. Particularly, $\lambda_{ijk}$ is
antisymmetric under the interchange of the first two generation
indices. The bilinear couplings, $\mu_i$, are severely constrained by the
small neutrino masses. So we will discuss the phenomenology of
$\lambda$- and $\lambda^\prime$-type couplings only. 
Then, expanding the above superpotential in standard four-component 
Dirac notation, we have for $\lambda^\prime$-type couplings
\be
{\cal {L_{\lambda^\prime}}} &=&  \lambda'_{ijk} ~\big[ ~\tilde d^j _L
\,\bar d ^k _R \nu^i _L
  + (\tilde d ^k_R)^\ast ( \bar \nu ^i_L)^c d^j _L +
   \tilde \nu ^i _L \bar d^k _R d ^j _L  \nonumber \\
& & ~~~~~ -\tilde e^i _L \bar d ^k _R u^j _L - \tilde u^j _L
\,\bar d ^k _R e^i _L
-(\tilde d^k _R)^\ast (\bar e ^i _L)^c u^j _L \big] + \mbox{h.c.} ,
\label{lag_1}
\ee
where only the first two terms and their hermitian conjugates are
relevant for the quark-neutrino interactions inside the matter
{\em via} squark exchange. Above, the sfermion fields are characterized 
by the tilde sign. 
For $\lambda$-type couplings, one can write
\be
{\cal {L_{\lambda}}} &=& \frac{1}{2}\lambda_{ijk} ~\big[ ~\tilde e^j _L \,
\bar e^k_R \nu^i_L
  + (\tilde e ^k_R)^\ast (\bar \nu ^i _L)^c e^j _L +
   \tilde \nu ^i _L \bar e^k _R e ^j _L  - (i \leftrightarrow j)
\big] + \mbox{h.c.} , 
\label{lag_2}
\ee
where only the first two terms with $i \leftrightarrow j$ and their hermitian 
conjugates are responsible for the interactions of neutrinos with the charged
leptons inside the matter {\em via} slepton exchange. The interaction terms given by
Eq. \ref{lag_1} and \ref{lag_2} violate $L$, as well as lepton flavour number. 
Suitable combinations of two such terms can lead to processes which are 
lepton flavour violating but $L$-conserving. In what follows, all
these new physics couplings are assumed to be real but will entertain 
both positive and negative values. The interactions of neutrinos with 
electrons and $d$-quarks in matter induce transitions    
(i) $\n_i + d \rightarrow \n_j + d $ 
and (ii)
 $\n_i + e \rightarrow \n_j
+e $. (i) is possible through $\lambda^{\prime}$ couplings {\em via} 
squark exchange for all $i, j$ and through $Z$ exchange for $ i = j$ 
while (ii) can proceed {\em via} $W$ and $Z$ exchange for $i = j$,
as well as through $\lambda$ couplings {\em via} slepton
exchange for all $i, j$. 

\section{\fbox{Golden Channel Oscillations including NSI}}
 
The golden channel probability $P_{e\mu}$ in the presence of
standard Earth matter effects has been discussed in great detail
in section 3.3. Now let us see what are the changes that we would 
have in the presence of NSI inside the Earth matter.
In the mass basis of neutrinos 
\begin{equation}
     M^2 = {\rm{diag}}(m_1^2,m_2^2,m_3^2) = U^\dagger M_\nu^+ M_\nu U,
\label{e:m1}
\end{equation}
where $M_\nu$ is the neutrino mass matrix in the flavour basis and 
$m_1, m_2$, and $m_3$ correspond to masses of three neutrinos.
$U$ is the mixing matrix defined by Eq. \ref{eq:upmns}.
Eq. \ref{eq:evo_matter_3flav} depicts the time evolution of 
flavour eigenstates (in the presence of three neutrinos) 
in matter which can be rewritten in the form
\be
i\frac{d}{d t}\left(
\begin{array}{c}
\n_e\\
\n_{\mu}\\
\n_{\tau} \end{array} \right)  
=  
H \left(
\begin{array}{c}
\n_e\\
\n_{\mu}\\
\n_{\tau} \end{array} \right),
\ee
where 
\begin{equation}
H = U \left( { M^2 \over 2 E} \right) 
   U^\dagger 
+  R = {{\tilde M}^2 \over 2E}.
\label{e:m2}
\end{equation}
$\tilde M^2 \over 2E$ is the effective mass squared matrix.  
$R$ is a $3 \times 3$ matrix reflecting the matter effect in the form
\begin{equation}
R_{ij} = R_{ij}(\mbox{Standard Model}) + R_{ij}(\lambda^{\prime}) + R_{ij}(\lambda).
\end{equation}
Specifically, 
\begin{equation}
R_{ij}(\mbox{Standard Model})   = \sqrt{2}  G_F N_e \delta_{ij} (i,j=1),
\label{e:rsm}
\end{equation}
\begin{equation}
R_{ij}(\lambda^{\prime})  =  \sum_m \left( {\lambda_{im1}^\prime 
\lambda_{jm1}^{\prime} \over 4 m^2 ({\tilde d_m})} N_d + 
{\lambda_{i1m}^{\prime  } \lambda_{j1m}^{\prime} 
\over 4 m^2({\tilde d_m})  } N_d \right) ,
\label{e:rlp}
\end{equation}
\begin{equation} 
R_{ij}( \lambda ) = 
\sum_{k \neq i, j}  {    \lambda_{ik1} \lambda_{jk1} \over
 4 m^2 ({\tilde { l^{\pm}_k}}) }  N_e
 + \sum_n  {    \lambda_{i1n} \lambda_{j1n} \over
 4 m^2 ({\tilde { l^{\pm}_n}}) }  N_e ,
\label{e:rl}
 \end{equation}  
where $N_d$ is the down-quark density in Earth matter. 
Note that $R$ is a symmetric matrix and also that antineutrinos 
will have an overall opposite sign for $R_{ij}$.
Assuming Earth matter to be neutral and isoscalar, $N_e = N_p = N_n$ 
and $N_d = 3 N_e$. 

The current bounds on the $\Rsl$ couplings
\cite{rparity} imply that the $\lambda^\prime$ induced contributions
to $R_{11}$, $R_{12}$ and $R_{13}$ are several orders less than
$\sqrt{2} G_F N_e$. We neglect those terms in our analysis. The
upper bounds on all couplings in $R_{ij}(\lambda)$ are also very
tight \cite{rparity} in comparison to $\sqrt{2} G_F N_e$ and
their effect will be discussed later. So, first we consider, in addition
to the Standard Model contribution, only
\be
R_{23} &=& R_{32}= {N_d \over 4 m^2 ( {\tilde d_m} )}
 \left( \lambda_{2m1}^{\prime }
  \lambda_{3m1}^{\prime} +
  \lambda_{21m}^{\prime}
  \lambda_{31m}^{\prime} \right) ,  \nonumber \\
R_{22} &=& {N_d \over 4 m^2 ( {\tilde d_m} )}
 \left( \lambda_{2m1}^{\prime \; 2}
 + \lambda_{21m}^{\prime \; 2} \right)   \; ,\;\; 
R_{33} = {N_d \over 4 m^2 ( {\tilde d_m})}
 \left( \lambda_{3m1}^{\prime \; 2}
 + \lambda_{31m}^{\prime \; 2} \right) ,  
\label{e:R1}
\ee
which are comparable to $\sqrt{2} G_F N_e$. One can see from
Eq. \ref{e:R1} that $R_{23} \neq 0$ implies both $R_{22}$ and
$R_{33}$ are non-zero\footnote{However, in other models 
\cite{david} this may not be the case.}.

The current bounds on the relevant couplings are 
as follows \cite{rparity}~:
\begin{equation}
|\lambda_{221}^{\prime}, \lambda_{231}^{\prime}|  < 0.18 ;  \;
|\lambda_{21m}^{\prime}|  < 0.06 ; \; |\lambda_{331}^{\prime}|  <
0.58 ; \; |\lambda_{321}^{\prime}|  <  0.52 ; \;
|\lambda_{31m}^{\prime}| < 0.12 ,
\label{e:lim}
\end{equation}
for down squark mass  $m_{\tilde d} = 100$ GeV.  The chosen
limits on $\lambda_{21m}^{\prime}$ and $\lambda_{31m}^{\prime}$
do not conflict with the ratio 
$R_{\tau \pi}=\Gamma (\tau \rightarrow \pi \nu_{\tau})
/\Gamma (\pi \rightarrow \mu \bar{\nu}_{\mu} )$ \cite{rparity}. 
However, the recently published BELLE bound on the mode 
$\tau \rightarrow \mu \pi^0$ \cite{belltau} tightly constrains
precisely those products of the $\lambda^\prime$ couplings which
enter in $R_{23} = R_{32}$ in Eq.
(\ref{e:R1}). It has been shown that $|{\lambda'_{21m}
\lambda'^{\ast} _{31m}}|$ and  $|{\lambda'_{2m1}
\lambda'^{\ast} _{3m1}}|$ both must be $< 1.8 \times 10^{-3}
(\frac{\tilde m}{100~{\rm GeV}})^2$ \cite{rpartau}. This 
effectively makes $R_{23}$ negligible for our purposes.

In general, it is cumbersome to write an analytical form of the
oscillation probability in the three-flavour scenario
with matter effects. However, under certain reasonable
approximations it is somewhat tractable. For the energies 
and baselines under consideration, $\Delta m_{21}^2 L/E << 1$
and we can use the one mass scale dominance approximation. 
Under OMSD, the $V_{12}$ part of $U$ 
(see Eq. \ref{eq:euler_rotations}) drops out from Eq. \ref{e:m2}. 
In the special case where $R_{22} = R_{33}$, if one uses 
the best-fit value of the vacuum mixing angle $\theta_{23} = \pi/4$ 
then the neutrino mass squared eigenvalues are:
\begin{equation}
\left( {\tilde M_2}^2 \over 2E\right)  = R_{22} - R_{23}, \;\; 
\left( {\tilde M_{1,3}}^2 \over 2E\right)  = {1 \over 2} 
\left(  {\Delta m_{31}^2 \over 2E} +R_{11} + R_{22} + R_{23} \mp
B \right) ,
\label{e:Msq}
\end{equation}
where
\begin{equation} B = {\left[ {\left( {\Delta m_{31}^2 \over 2E} \right) }^2
+ {\left(  -R_{11} + R_{22} + R_{23} \right)}^2  - 2 {\Delta
m_{31}^2 \over 2E}  \cos 2 \theta_{13}\left( R_{11} - R_{22}
-R_{23} \right)  \right]}^{1/2} .
\end{equation}
The matter induced neutrino mixing matrix is given by
\begin{equation}
U^m = \left( 
\begin{array}{ccc}
U_{11}^m & 0 & U_{13}^m\\
N_1 & -{1 \over \sqrt{2}} & N_3 \\
N_1 & {1 \over \sqrt{2}} & N_3 
\end{array}
\right).
\label{e:U}
\end{equation}
Here
\begin{equation}
N_{1,3} =  {\left[  {2  {\left ( 
-R_{11} + R_{22} + R_{23} +  {\Delta m_{31}^2 \over 2E} \cos 2\theta_{13}
\pm B \right)}^2 
\over {\left( {\Delta m_{31}^2 \over 2E} \right) }^2 \sin^2 2 \theta_{13} } +
2 \right]}^{-1/2} ,
\end{equation}
where  $N_1$ ($N_3$) corresponds  to +(--) sign in the above expression.
Neglecting the CP phase $\delta_{CP}$ in
the standard parameterization of $U^m$, one may write $U_{13}^m =
\sin \theta_{13}^m$ and $U_{23}^m =
\sin \theta_{23}^m \cos \theta_{13}^m$. From eq. (\ref{e:U}) it follows
that $\theta_{23}^m = \theta_{23}$, the vacuum mixing angle.
$\theta_{13}^m$, on the other hand, changes from its vacuum value
and it is $\pi/4$ for
\begin{equation}
R_{11} - R_{22} -R_{23} = {\Delta m_{31}^2 \over 2E} \cos 2 \theta_{13}.
\label{e:mmax}
\end{equation}
In the absence of non-standard interactions, 
$R_{22} = R_{23} = R_{33} =0$ and $R_{11} = \sqrt{2} G_F N_e$,  
this is the well-known condition for matter induced 
maximal mixing (see Eq. \ref{eq:resonance}).  
Since in Eq. \ref{e:U}, $U_{12}^m = 0$, in the $\nue$ to $\numu$ 
oscillation probability the terms involving 
$({\tilde M_2}^2 - {\tilde M_1}^2)$ and $({\tilde M_3}^2 - {\tilde M_2}^2)$ 
will not survive and we get~:
\begin{equation}
P_{\n_e \rightarrow \n_{\mu}}  = 4 \;\;{\left(U_{13}^m\right)}^2\;\; 
{\left(U_{23}^m\right)}^2 \;\;\sin^2 \left(  2.54 \;B \;L \right),
\end{equation}
where $E$, $\Delta m_{31}^2$ and $L$ are expressed in GeV, eV$^2$, 
and km, respectively. This expression is also valid for antineutrinos.  
Using Eqs. \ref{e:Msq} and \ref{e:U} one can easily obtain the oscillation 
probabilities for other channels. 

\begin{figure}[t]
\includegraphics[width=8.0cm, height=7.0cm, angle=0]{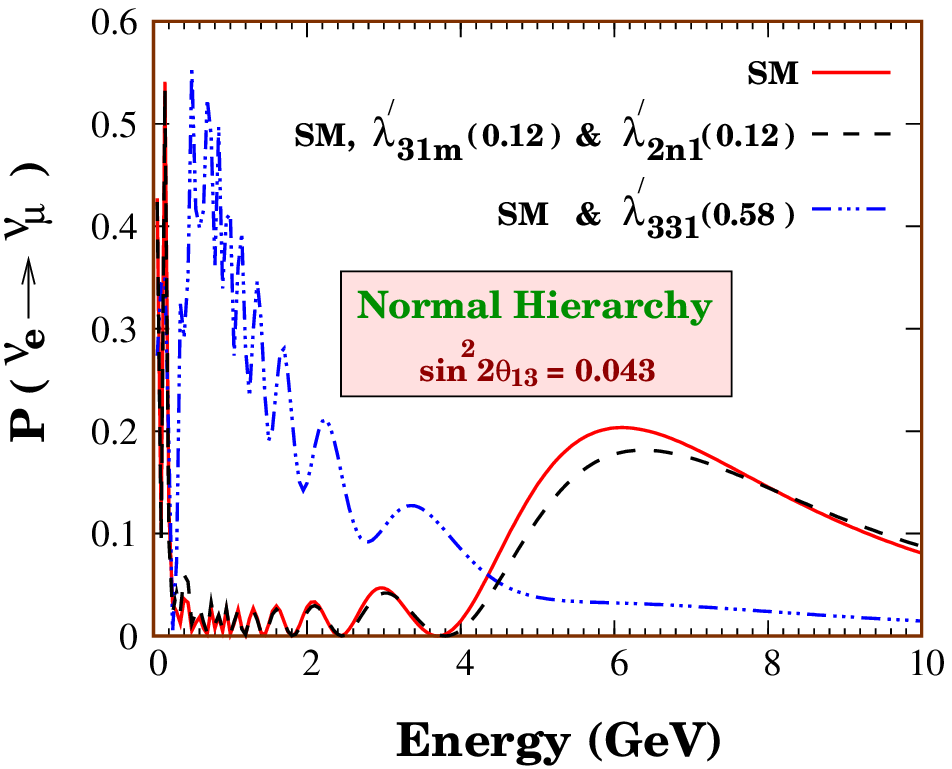}
\vglue -7.0cm \hglue 8.8cm
\includegraphics[width=8.0cm, height=7.0cm, angle=0]{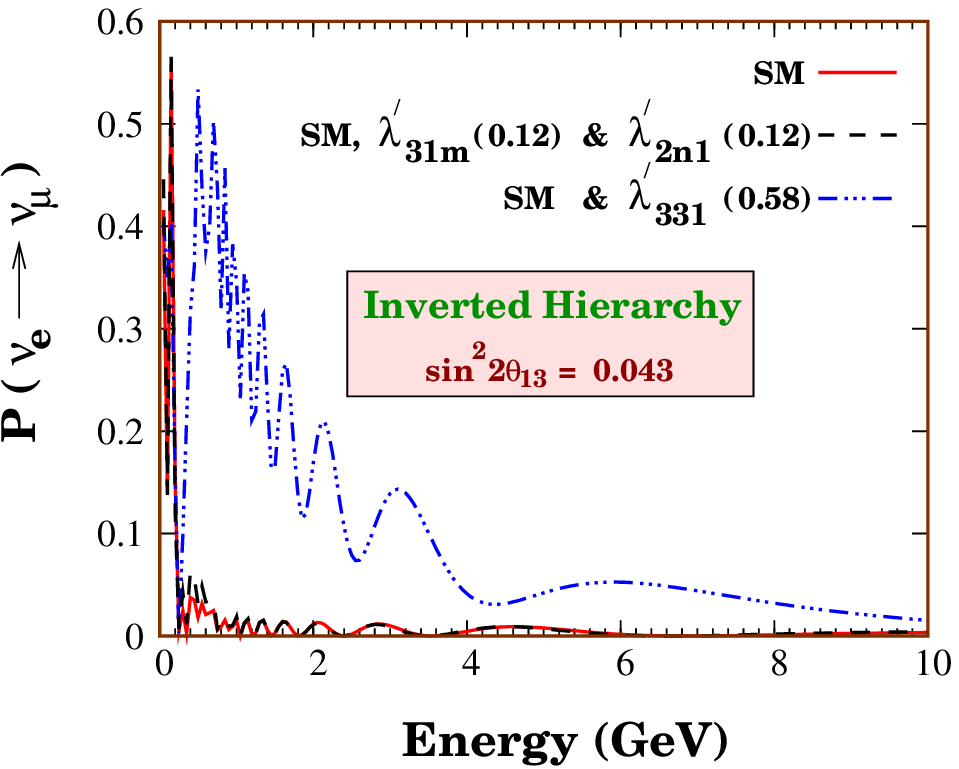}
\caption{\label{f:osc}
$P_{\nue \rightarrow \numu}$ for the NH and IH. 
SM corresponds to only standard electroweak interactions. The values of
$\lambda^{\prime}$ are given in parentheses. $m$ can take any
value, $n =$ 2 or 3.}
\end{figure}

The above analytical formulation has been used as a 
cross-check on our numerical results.
For example, Fig. \ref{f:osc}, which shows the variation
of $P_{\n_e \rightarrow \n_{\mu} }$ as a function of the energy,
is obtained by solving the full three flavour neutrino
propagation equation assuming the PREM \cite{prem} profile 
and including NSI. The range of energy is chosen in line with 
the discussions in the rest of this work. The probability falls 
with decreasing $\theta_{13}$ and, for illustration, we have chosen 
$\stch=0.043$ and for all other oscillation parameters we assume the
benchmark values given in Table \ref{tab:bench}.
Apart from some qualitative remarks, all the results have been 
presented considering $\delta_{CP}$ = 0 corresponding to the 
CP conserving case. The purpose of Fig. \ref{f:osc} is twofold: 
(a) to show how the distinguishability between the NH and IH
may get blurred by the RPVSM interactions, and (b) how
irrespective of the hierarchy chosen by Nature the results may be
completely altered by the presence of these interactions. Each
panel of Fig. \ref{f:osc} has three curves: the solid line (only electroweak
interactions), dot-dashed line (in addition, $R_{33}$ gets a
non-zero RPVSM contribution), and dashed line 
($R_{22} = R_{33}$  are nonzero, in addition to the
electroweak contribution). Only in the
last case is the analytical formula we have presented above
applicable. We find excellent agreement. Two aspects of the results
are worth pointing out.

First, in the absence of NSI, for the IH with neutrinos,  
the resonance condition Eq. \ref{e:mmax} is not satisfied and the 
oscillation probability is negligible (right panel solid line).  
This could be altered prominently by the RPVSM interactions 
(dot-dashed curve) so that the distinguishability between the two 
hierarchy scenarios may well get marred by $\Rsl$ supersymmetric
interactions.

Secondly, for the NH, it is seen that the peak in
the probability may shift to a different energy in the presence of the
RPVSM interactions. This is because the condition for maximal
mixing in Eq. \ref{e:mmax} is affected by the $\Rsl$ interactions.  
For the IH, the oscillation probability is considerably enhanced for 
some energies. Thus, physics expectations for both hierarchies will 
get affected by RPVSM.

In the following section we dwell on the full impact of this
physics on a long baseline CERN-INO beta-beam experiment.

\section{\fbox{Results}}

In this work the same magical CERN-INO long baseline beta-beam 
experimental set-up has been considered with which we dealt
earlier in chapter 3. Here all the results will be presented 
taking $^8$B as the candidate ion for the $\nue$ 
beta-beam\footnote{Broadly speaking, the results obtained
with $^8$B neutrino beta-beam taking NH (IH) are similar
to that with $^8$Li antineutrino beta-beam source
considering IH (NH) but details do differ.} 
source and with $\gamma = 350$. All the presented results  
are based on a five years of ICAL@INO\footnote{A detailed 
description of the ICAL@INO detector has already been given 
in section 3.4 and we use the same detector characteristics 
as depicted in Table \ref{tab:detector}.} data sample.
The event rates have been estimated using Eq. \ref{eq:events}. 
Details about the backgrounds and neutrino-nucleon cross sections 
may be found in section 3.4.

At the production and detection levels, FCNC and
FDNC effects can change the spectrum and detection cross sections
by a small (\raisebox{-.6ex}{\rlap{$\sim$}}\raisebox{.6ex}{$<$}
0.1\%) amount but this would not alter the conclusions. At the
source and detector, they may also mimic the oscillation signal
itself, but these effects are tiny\footnote{This is due to the
very tight constraints from $\mu \rightarrow e$ transition limits
in atoms \cite{mu2e}.} ($\sim {\cal O}(10^{-14})$). Here we
discuss how FCNC and FDNC may significantly modify the
propagation of neutrinos through matter over large distances.

\subsection{Extraction of $\theta_{13}$ and determination of hierarchy}

\begin{figure}[t]
\includegraphics[width=8.0cm, height=7.0cm, angle=0]{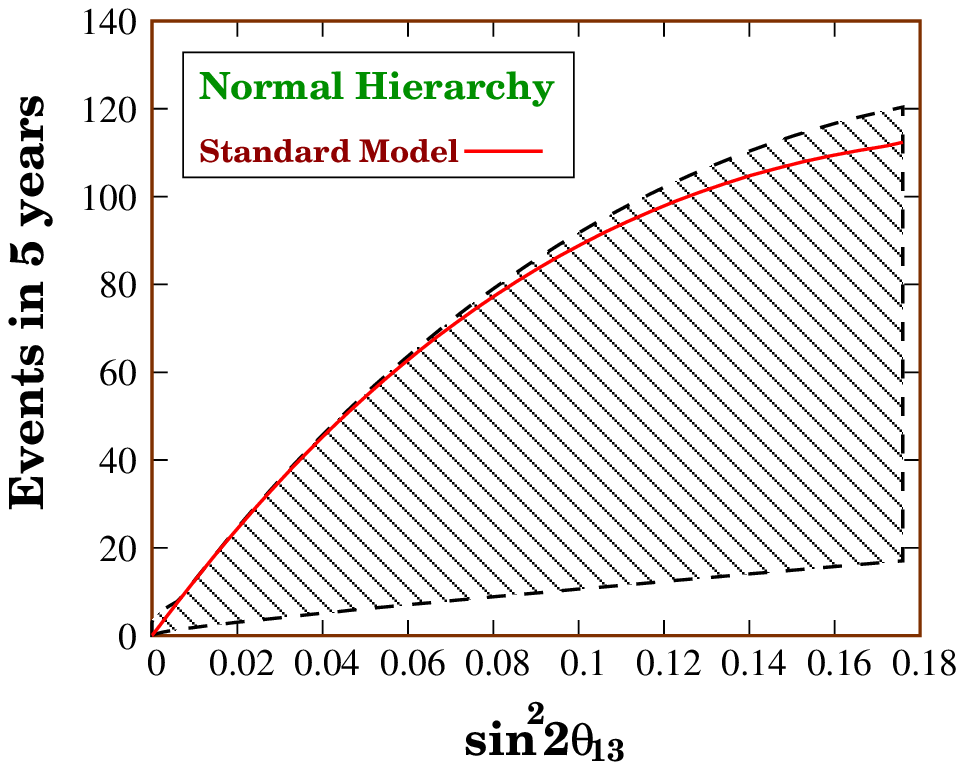}
\vglue -7.0cm \hglue 8.8cm
\includegraphics[width=8.0cm, height=7.0cm, angle=0]{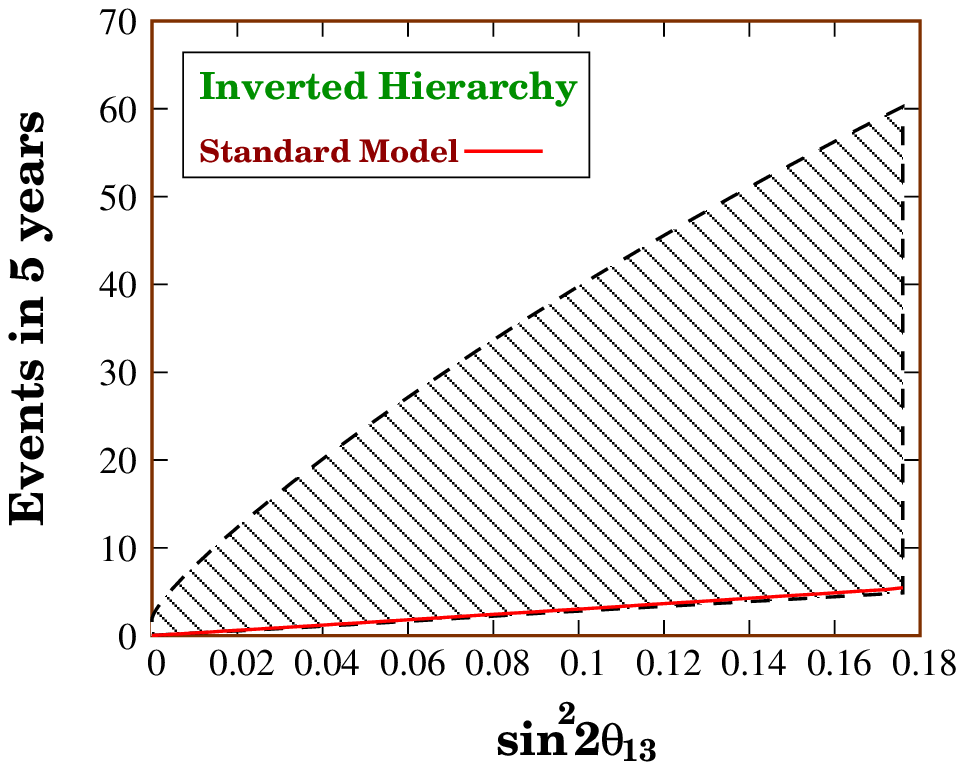}
\caption{\label{f:lpno}
Number of muon events for the NH (left panel) and IH (right panel) 
as a function of $\sin^2 2\theta_{13}$ for a five years of ICAL@INO 
run. The solid lines correspond to the absence of any NSI. 
The hatched area is covered if the $\lambda^{\prime}$ couplings are 
varied over their entire allowed range.}
\end{figure}

If neutrinos have only Standard Model interactions then the expected 
number of muon events is fixed\footnote{Recall we assume that, 
but for $\theta_{13}$ and the mass hierarchy, the
other neutrino mass and mixing parameters are known.} for a
particular value of $\theta_{13}$ with either NH or IH
as may be seen from the solid lines in Fig. \ref{f:lpno}. 
The vast difference for the alternate hierarchies
picks out such long baseline experiments as good laboratories for
addressing this open question of the neutrino mass spectrum.

If non-standard interactions are present then, depending on their
coupling strength, the picture can change dramatically. In Fig.
\ref{f:lpno}, the shaded region corresponds to the allowed values
when supersymmetric FCNC and FDNC interactions are at play. It is obtained
by letting the $\lambda^{\prime}$ couplings\footnote{In fact, we have
chosen the subscript $m$ in the $\lambda^{\prime}$ couplings in 
Eq. \ref{e:R1} to be any one of 1, 2 or 3.} vary over their entire
allowed range -- both positive and negative -- given in 
Eq. \ref{e:lim}, subject to the further constraints on particular
products. 

It is seen that to a significant extent the distinguishability of
the two hierarchies is obstructed by the $\Rsl$ interactions
unless the number of events is more than about 60. Also, the
one-to-one correspondence is lost between $\theta_{13}$ and the
number of events and, at best, a lower bound can now be placed on
$\theta_{13}$ from the observed number. Of course, if the
neutrino mass ordering is known from other experiments, then
this lower bound can be strengthened, especially for the inverted
hierarchy.

It is also noteworthy that for some values of
$\lambda^{\prime}$-couplings there may be more events than can be
expected from the Standard Model interactions, no matter what the
value of $\theta_{13}$. Thus, observation of more than 112 (5)
events for the NH (IH) would be a clear signal of new physics.

\subsection{Constraining $\lambda^{\prime}$}

If $\theta_{13}$ is determined from other experiments then it
will be easier to look for non-standard signals from this
beta-beam experiment. However, even if the precise value
remains unknown at the time, considering the upper bound on
$\theta_{13}$ one may tighten the constraints on the
$\lambda^{\prime}$ couplings. Fig. \ref{f:lpno} reflects the
overall sensitivity of the event rate to the $\Rsl$ interactions
obtained by letting all RPVSM couplings vary over their entire
allowed ranges. In this subsection, we want to be more specific
and ask how the event rate depends on any chosen
$\lambda^{\prime}$ coupling.

At the outset, it may be worth recalling that the BELLE bound on
$\tau \rightarrow \mu \pi^0$ \cite{belltau} severely limits the
products $\lambda^{\prime}_{2m1} \lambda^{\prime}_{3m1}$ and
$\lambda^{\prime}_{21m} \lambda^{\prime}_{31m}$. Thus, $R_{23}$
can be dropped in the effective neutrino mass matrix 
(see Eq. \ref{e:m2}). If only $\lambda^{\prime}_{2m1}$ and/or 
$\lambda^{\prime}_{21m}$ ($\lambda^{\prime}_{3m1}$ and/or 
$\lambda^{\prime}_{31m}$) is non-zero, then $R_{22}$ ($R_{33}$) 
alone receives an RPVSM contribution. Both $R_{22}$ and $R_{33}$
can be simultaneously non-zero if $\lambda^{\prime}_{21m}$ and
$\lambda^{\prime}_{3m1}$ (or $\lambda^{\prime}_{2m1}$ and
$\lambda^{\prime}_{31m}$) are non-zero at the same time.

\begin{figure}[t]
\includegraphics[width=8.0cm, height=7.0cm, angle=0]{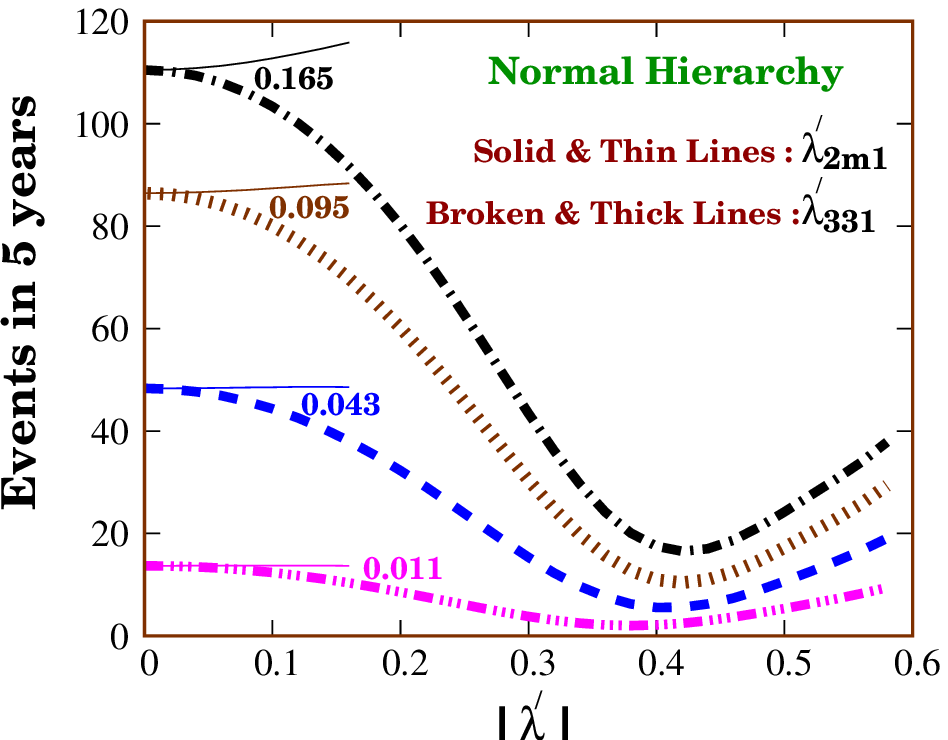}
\vglue -7.0cm \hglue 8.8cm
\includegraphics[width=8.0cm, height=7.0cm, angle=0]{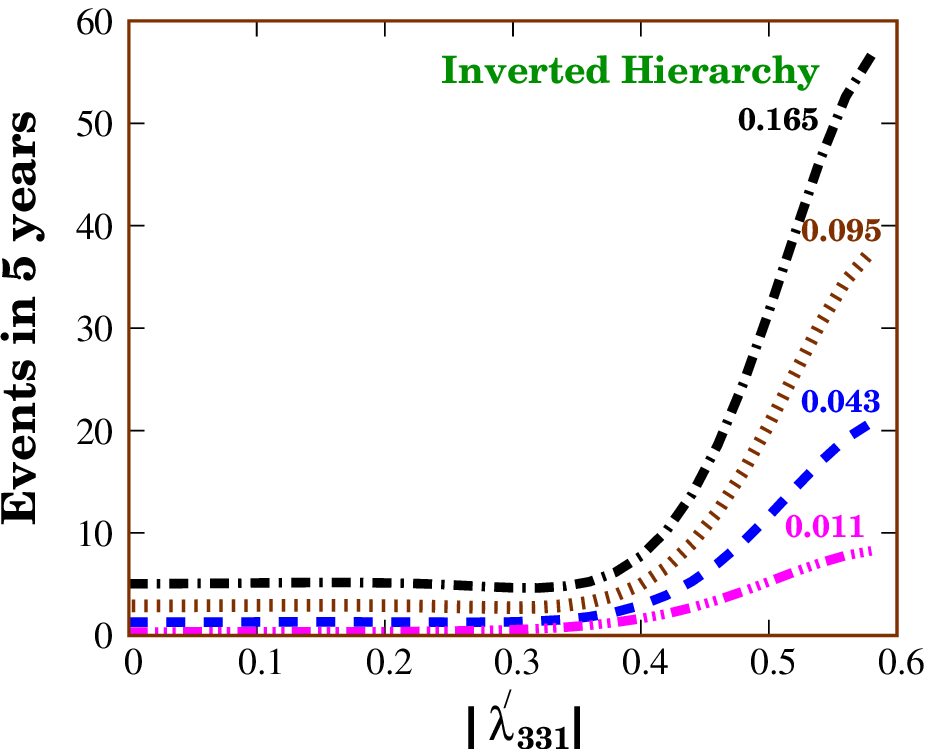}
\caption{\label{f:bndl3}
The number of events as a function of a coupling $|\lambda^{\prime}|$,
present singly, for the NH (left panel) and IH (right panel).
The thick (thin) lines are for $|\lambda^{\prime}_{331}|$ 
($|\lambda^{\prime}_{2m1}|$, $m=$ 2, 3). The chosen
$\stch$ are indicated next to the curves.}
\end{figure}

In the light of this, we consider the situation where only 
one of the above $\Rsl$ coupling is non-zero. In such an event, 
only one of $R_{22}$, $R_{33}$ is non-zero. The dependence of 
the number of events on a non-zero $\lambda^{\prime}_{331}$ or
$\lambda^{\prime}_{2m1}$, for a chosen $\stch$, can be seen 
from Fig. \ref{f:bndl3}. In this figure, we use the fact that 
if only one of these $\Rsl$ couplings is non-zero, it appears 
in the results through $|\lambda^{\prime}|$. For the NH, 
the curves for $\lambda^{\prime}_{2m1}$, for $m =$ 2,3, are
terminated at the maximum allowed value of 0.18. 
Fig. \ref{f:bndl3} can also be used for $\lambda^{\prime}_{321}$, 
$\lambda^{\prime}_{31m}$ and $\lambda^{\prime}_{21m}$, bearing in mind 
their different upper bounds. For the IH, the number of events is 
small for $\lambda^{\prime}_{2m1}$ and $\lambda^{\prime}_{21m}$ and 
insensitive to the magnitude of the coupling. These are not shown. 
It is seen that for the NH there is a good chance to determine the $\Rsl$
couplings from the number of events. In fact, if the number of
events is less than about 38 there is a disallowed region for
$|\lambda^{\prime}|$, while for larger numbers there is only an upper
bound. For the IH, more than about five events
will set a lower bound on the coupling.

\subsection{Effect of $\lambda$} 

\begin{figure}[t]
\includegraphics[width=8.0cm, height=7.0cm, angle=0]{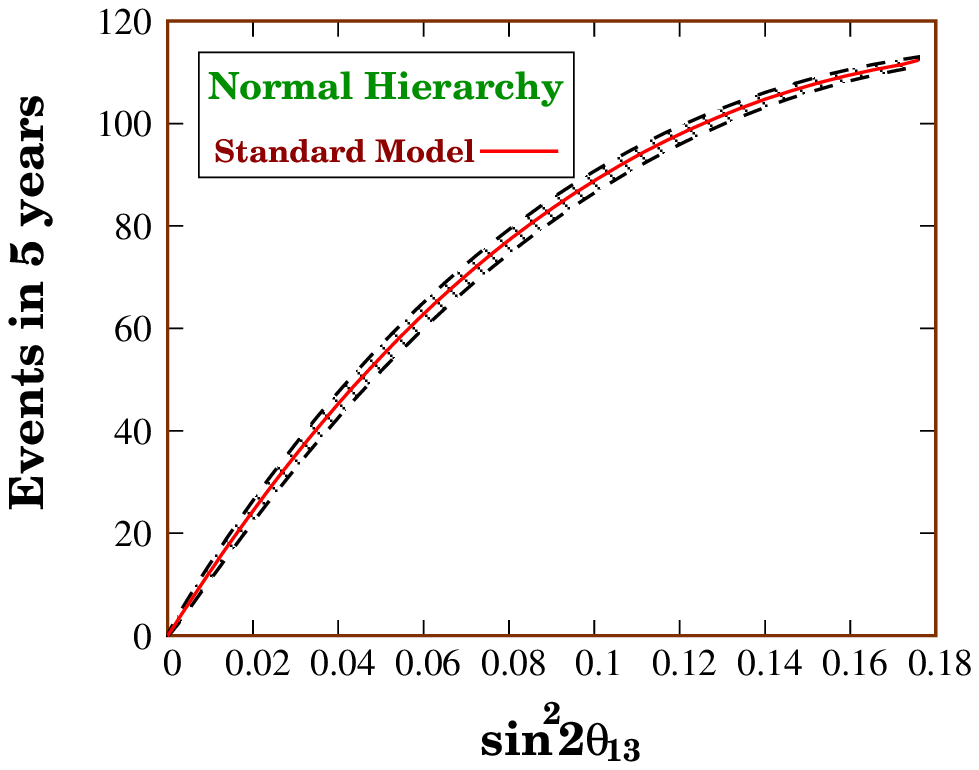}
\vglue -7.0cm \hglue 8.8cm
\includegraphics[width=8.0cm, height=7.0cm, angle=0]{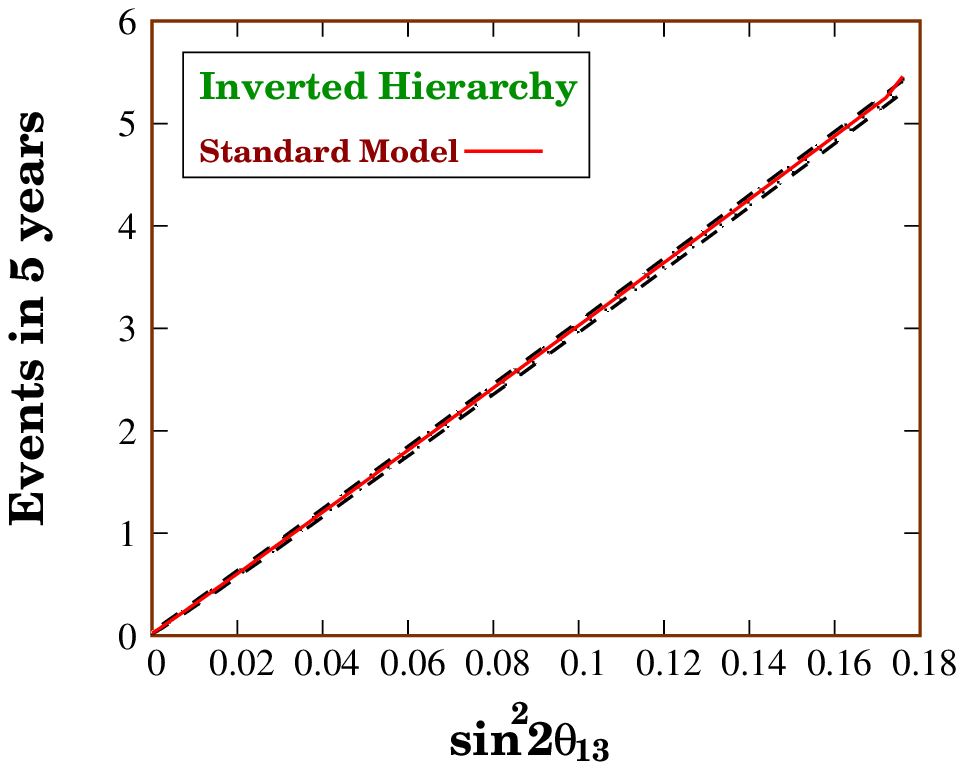}
\caption{\label{f:lno}
Number of muon events for the NH (left panel) and IH (right panel)
as a function of $\stch$ for a five years of ICAL@INO run. The solid lines 
correspond to the absence of any NSI. The hatched area is covered if
the $\lambda$ couplings are varied over their entire allowed range.}
\end{figure}

The $\lambda$ couplings which can contribute in 
Eq. \ref{e:rl} have strong existing bounds \cite{rparity} 
and their contribution to $R$ is rather small in comparison 
to $\sqrt{2} G_F N_e$. Among them, the bounds
$\lambda_{121} < 0.05 $ and $\lambda_{321} < 0.07 $ for
$m_{\tilde l} = 100$ GeV are relatively less stringent
\cite{rparity}. We show their very modest impact in Fig.
\ref{f:lno}. It is clear from this figure that (a) the
$\lambda$-type couplings cannot seriously deter the extraction of
$\theta_{13}$ or the determination of the neutrino mass
ordering, and (b) when $\theta_{13}$ is known in future it will
still not be possible to constrain these couplings through long
baseline experiments.

\section{\fbox{Conclusions}}

R-parity violating supersymmetry is among several extensions 
of the Standard Model crying out for experimental verification. 
The model has flavour diagonal and flavour changing neutral currents 
which can affect neutrino masses and mixing and
can leave their imprints in long baseline experiments. 
This is the focus of this work.

We consider a beta-beam experiment with the source at CERN and
the detector at INO. We find that the $\Rsl$ interactions may
obstruct a clean extraction of the mixing angle $\theta_{13}$ or
determination of the mass ordering unless the bounds on the
$\lambda^{\prime}$ couplings are tightened. 
On the other hand, one might be able to see a clean signal of 
new physics. Here, the long baseline comes as a boon over experiments 
like MINOS which cover shorter distances. Two experiments of these 
contrasting types, taken together, can expose the presence of NSI 
like RPVSM.

There are other non-standard models \cite{david} where four-fermion 
neutrino couplings with greater strength have been invoked. The signals 
we consider will be much enhanced in such cases.

Our results are presented for the CP conserving case. 
As $\theta_{13}$ is small, the CP violating effect is 
expected to be suppressed. We have checked this for the Standard Model, 
where the `magic' nature of the baseline \cite{magic} also plays a role.

Finally, in this work we have restricted ourselves to a beta-beam 
neutrino source. Much the same could be done for antineutrinos as well; 
then the signs of all terms in $R$ (see Eq. \ref{e:m2}) will be reversed. 
It follows from Eq. \ref{e:mmax} that $\theta_{13}^m$ can then be maximal 
only for the IH and as such more events are expected here than
in the NH. Broadly, results similar to the ones presented here with neutrinos 
can be obtained with antineutrinos if NH is replaced by IH and {\em vice-versa}.
In the next chapter, we will show that a small detector placed near a 
beta-beam storage ring can probe $L$ violating interactions, as predicted 
by supersymmetric theories with R-parity non-conservation at the 
production and detection stages of neutrinos.

\chapter{Probing Lepton Number Violating Interactions with Beta-beam 
using a Near-Detector}

\section{\fbox{Introduction}}\label{sec1}

In the Standard Model, lepton number conservation is only
accidental; the particle content and the requirement of
renormalizability ensure that each lepton flavour number 
is conserved separately. However, non-zero neutrino masses, 
as indicated by recent neutrino oscillation experiments, 
have proved that the success of the Standard Model should be 
viewed as that of a low energy effective theory. It is
not unreasonable to expect that in some extensions of the 
Standard Model, $L$ conservation may not hold. Indeed, a 
Majorana mass term for the neutrinos violates total lepton number.  
The non-observation of direct $L$ violation in the past experiments 
have put stringent constraints on some of these interactions. 
In this chapter we show that beta-beams and a nearby detector can be 
a good further probe of such interactions. For this work, 
the advantage of a `near' detector is twofold. 
\begin{enumerate}

\item
{\bf Firstly, due to the short base-length, neutrinos do not get
much scope to oscillate before being detected, which could otherwise
mimic signals of the $L$-violating ($\not{\! L}$) interactions.}  
\item
{\bf The other obvious advantage is that a larger part of the beam 
can be picked up with a smaller detector.}

\end{enumerate}

We consider placing a 5 kton cylindrical detector, aligned with the
beam axis, within 1 km from the beta-beam storage ring. The
$\not{\! L}$ interactions can lead to tau leptons in
near-detector experiments in two ways. A $\nutau$ can be
produced due to such interactions during beta-decay, yielding
a $\tau$ through weak CC interactions in the detector. Alternatively, 
the electron neutrinos in the beam, produced through usual beta-decay, 
can undergo $\not{\! L}$ interactions with the detector, leading to tau leptons. 
The taus promptly decay, part of the time in a muonic channel. Iron
calorimeters with active detector elements serve well for
identifying these muons, which leave long tracks in the detector,
and for filtering out backgrounds. We will also briefly comment
on water \u{C}erenkov and other detectors.

In the following section a brief account of the experimental 
set-up is presented. In section \ref{sec3}, $\not{\! L}$ interactions
are discussed in the context of R-parity violating supersymmetric models
\cite{rparity,couplings}. We stress how beta-decay can be affected in the 
presence of such interactions, yielding $\nutau$ in a few cases in place 
of the standard $\nue$. The processes via which $\nue$ produce tau leptons
in the detector are also described. The expected number of muon events 
from tau decay and the constraints ensuing in the event of their 
non-observation will be presented in section \ref{sec4}.

\section{\fbox{Beta-beam Flux at a Near-Detector}}\label{sec2}

The choice of ions for a beta-beam is predicated by the intended
physics. The low end-point energies (see Table \ref{tab:ions}) of 
the $^{6}$He and $^{18}$Ne ions restrict the energy reach of the beam; 
a threshold energy of $3.5$ GeV is necessary to produce a $\tau$-lepton 
from an incoming neutrino. Therefore, we will work with $^{8}$B and 
$^{8}$Li ions, offering higher end-point energies. We pick only the 
neutrino beam for our discussion.

The geometry of the beta-beam storage ring determines the neutrino
flux at a near-detector. For a low-$\gamma$ design, a 6880 m decay
ring with straight sections of length ($\equiv S$) $2500$ m each 
($36\%$ useful length for ion decays) has been proposed. In such a 
configuration, $N_{\beta} = 1.1\times 10^{18}$ useful
decays (decays in one of the straight sections) per year can be
obtained with $^{18}$Ne ions. We have used this same luminosity for $^{8}$B 
and higher $\gamma$. To settle these issues a dedicated study 
is on at CERN.

\begin{figure}[t]
\begin{center}
\includegraphics[width=12.0cm]{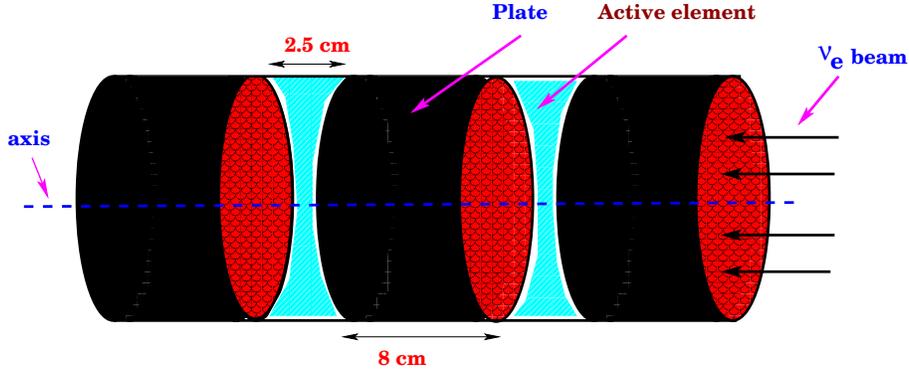}
\caption{\label{fig1_subhendu} 
A schematic diagram of the proposed
detector (a part only). The incoming $\nue$ beam may have a very small
contamination of neutrinos of other flavours in the presence of lepton
flavour violating interactions.}
\end{center}
\end{figure}

\subsection{Detector Simulation Study}

We consider a cylindrical 5 kton detector 
(as in Fig. \ref{fig1_subhendu}) aligned with one of the straight 
sections of the storage ring. The detector is made of iron slabs 
(thickness 8 cm) with interleaved active detector elements 
(thickness 2.5 cm) such as resistive plate chambers (RPCs). 
The readouts from these RPCs will be concentric
annular strips of small width with further segmentation to improve the
position resolution. In this proposal, iron is the main content of the
detector\footnote{Lead may be an interesting alternative material to
enhance the event rate.}. The thickness of the slabs ensures that
electrons do not propagate in the detector.  The signal muons are of
sufficient energy to give rise to long tracks. To eliminate possible
beam-induced backgrounds (see below) from pions produced in CC and NC
processes, typically 6 to 13 hits (depending on the boost $\gamma$ 
ranging from 250 to 450) are required of a putative muon track.

As noted earlier, the signature of new physics we consider is the
appearance of prompt tau leptons which decay into muons with a
branching fraction of $17.36\%$ \cite{hagiwara}. The tau production 
threshold is around $3.5$ GeV. This is what necessitates the higher 
boost $\gamma$.

Backgrounds, other than those of the beam-induced variety
discussed below, are controllable, as we now point out. A
beam-off run will help make a first estimate of these
backgrounds. Further, an important aspect of the beta-beam
source is its capability of eliminating backgrounds through
timing information. The beam itself will consist of bunches of
typically 10ns size and the number of bunches will be chosen so
as to ensure that the ratio of the active- to the dead-time is
$\cal{O}$($10^{-3}$). Backgrounds from other sources, namely, 
atmospheric neutrinos, spallation neutrons, cosmic rays, etc. can 
thus be largely rejected from the time-stamp of a recorded event.  
Even further reductions of the backgrounds of external origin can be 
envisioned through fiducial and directionality cuts.

Now let us turn our attention to the issue of beam-induced backgrounds
caused by NC and CC interactions of unoscillated $\nue$. Electrons produced 
through weak interactions by the incoming $\nue$ are quickly absorbed and 
do not leave any track. Formation of prompt muons through $\Rsl$ 
supersymmetric interactions is suppressed by strong bounds on the relevant 
couplings arising from limits on $\mu-e$ transitions in atoms \cite{mu2e}. 
However, the beta-beam neutrinos can produce pions along with other hadrons 
at the detector via CC and NC processes. They undergo strong interactions with the
detector material and are quickly absorbed before they can decay. But
as numerous pions are produced, it needs to be checked whether some of
them can fake the signal.

We have checked our na\"{i}ve expectations with a detector simulation
study using GEANT \cite{geant3} aided by NUANCE \cite{nuance}. 
It is observed that for neutrino-nucleon interactions at energies 
interesting for our study, the produced lepton preferentially carries 
most of the energy of the incident neutrino. Moreover, pions are usually 
produced with multiplicity more than unity. Hence it is not unreasonable to
expect that the pions will be less energetic compared to the taus
produced via $\not{\! L}$ interactions and hence in detectors of this
genre, it is possible to distinguish hadronic showers from a muon track.

However, a conservative approach is followed in pion background estimation. 
Although pions do not leave behind a straight track
like a muon, we still count the number of hits as a measure of
the distance traversed by a pion. We impose a criterion of
minimum number of hits to identify a track to be a muon one. We
find from a simulation that, for $\gamma=250/350/450$, imposing a
cut of 6/10/13 hits will reduce the pion background at least to
the $10^{-3}$ level.

\begin{figure}[t]
\begin{center}
\includegraphics[width=12.0cm]{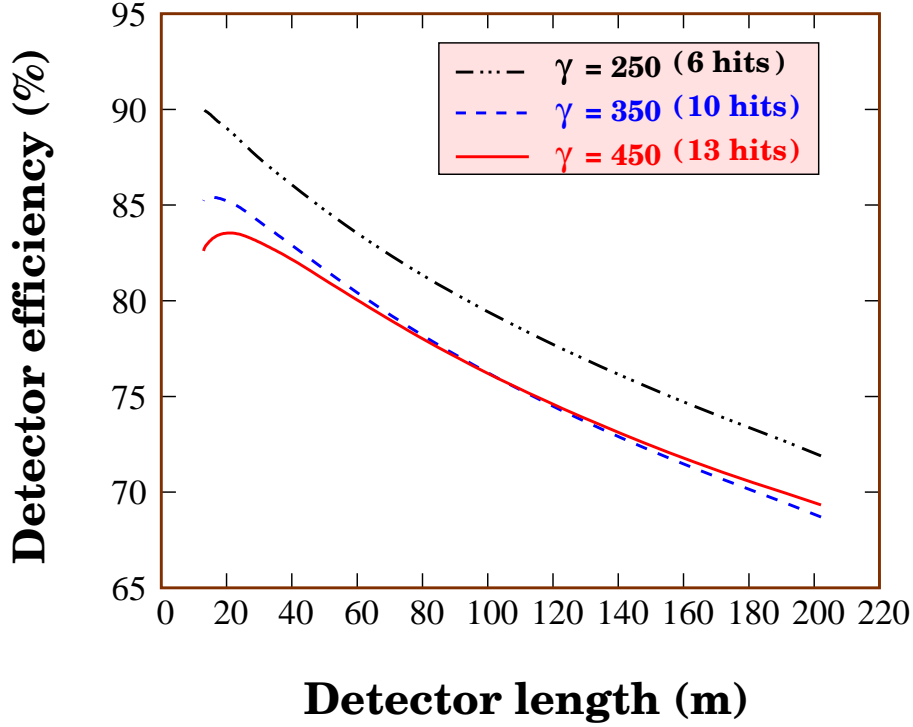}
\caption{\label{efficiency}
Detector efficiency for $\gamma=250,350,450$. The corresponding cuts 
on muon hits used are 6, 10 and 13 respectively.}
\end{center}
\end{figure}

The detector geometry plays a role in determining the signal
efficiency after imposition of these cuts. Since the muons produced
from boosted tau lepton decay carry transverse momentum, some of them
may exit the detector through its sides, failing to satisfy the cuts.
For a fixed detector mass (5 kton), a longer detector has a smaller
cross-sectional area, resulting in a drop in the detector efficiency
for the above reason.  As the detector length increases from 20 m to
200 m, with our set of cuts, the efficiency factor reduces from 
85\% to 70\% approximately, showing little dependence on $\gamma$ 
(see Fig. \ref{efficiency}).

\subsection{Neutrino Fluxes}

Neglecting small Coulomb interactions, the lab frame neutrino
beta-beam flux (per unit solid angle per unit energy bin per unit
time per unit length of the straight section) emitted at an angle
$\theta$ with the beam axis is described by 
(see also Eq. \ref{eq:flux})
\begin{equation}
\phi^{Near}(E,\theta)
 =\frac{1}{4\pi}\frac {g} {m_e^5 \,f}
\frac{1}{\gamma(1-\beta \cos\theta)}
 (E_0 - E^*) E^{*2} \sqrt{ (E_0-E^*)^2-m_e^2},
\label{eq:near_flux}
\end{equation}
where $g \equiv N_{\beta}/S$ is the number of useful decays
per unit time per unit length of the straight section.

To calculate the resulting number of events at a cylindrical
near-detector of radius $R$ and length $D$ aligned with the beam
axis it is necessary to integrate over the length $S$ of the
straight section of the storage ring and the volume of the
detector. The event rate at a detector placed at a distance $L$
from the storage ring is given by \cite{low_volpe}
\begin{equation} 
\frac{dN_{\not L}}{dt}
=n \varepsilon \int_0^S \: dx \int_0^D \: d\ell
\int_0^{\theta'} d\theta \:{2\pi} \sin\theta
\int_{E_{min}}^{E^{\prime}} \: dE \:\phi^{Near}(E,\theta)\:
\sigma(E), 
\end{equation}
where
\begin{equation}
\tan\theta^{'}(x,\ell)=\frac{R}{L+x+\ell}\,~~{\rm
and}~~E^{\prime}=\frac{E_0-m_e}{\gamma(1-\beta\cos\theta)}.
\end{equation}
Here $n$ represents the number of target nucleons per unit
detector volume, $\varepsilon$ is the detector efficiency as
presented in Fig. \ref{efficiency}, $E_{min}$ denotes the tau
production threshold, and $\sigma(E)$ stands for the neutrino-nucleon 
cross section. Note that the source of $L$-violation may lie either in 
$\phi^{Near}(E,\theta)$ (in case of $\Rsl$ beta-decay) 
or in $\sigma(E)$ (in case of $\Rsl$ tau production).

To help subsequent discussion, following \cite{low_volpe}, we
rewrite the above formula isolating the geometry integrated total
flux $\Phi(E;S,D,R,L)$ (per unit time per unit energy bin)
falling on the detector and emitted from the whole length of the
straight section as follows~: 
\begin{equation} 
\frac{dN_{\not L}}{dt} =n \varepsilon
\int_{E_{min}}^{E_{max}} \:\: dE\:\Phi(E;S,D,R,L)\:
\sigma(E), 
\end{equation}
where
\begin{equation} 
\Phi(E;S,D,R,L)=\int_0^S \: dx \int_0^D \: d\ell
\int_0^{\theta'} d\theta \:{2\pi} \sin\theta \:
\phi^{Near}(E,\theta)
\label{PHIeqn}
\end{equation}
and
\begin{equation} 
E_{max}=\frac{E_0-m_e}{\gamma(1-\beta)}. 
\end{equation}

The beta-beam also involves a few small uncertainties which we neglect
in our analysis. However for completeness, these are listed here~:
\begin{itemize}
\item There exist different excited states of the daughter nuclei
of the decaying ion, which additionally lead to small
contributions to the spectra with different end-point energies.

\item The ion beam has a finite transverse size. However, as this
size varies \cite{adrian} between only $3.0$ cm to $5.1$ cm, with
an average of $4$ cm ($3\sigma$), in both transverse directions
inside the ring, the variation in flux at the detector due to this
is negligible.

\item The decaying ions may have small transverse momentum due to
thermal fluctuations ($k_{B}T \sim 2.6\times 10^{-3}$ eV), but this
can be safely ignored in comparison with the end-point energy of 
the beta-decay.

\end{itemize}

\section{\fbox{$\Rsl$ Processes}}\label{sec3}

A detailed description of the RPVSM \cite{rparity,couplings} 
containing renormalizable $\not{\! L}$ trilinear $\lambda$- and
$\lambda'$-type couplings is given in section 5.2. To impose 
conservative upper bounds, we work in a minimal $\Rsl$ framework 
where only a pair of such couplings are assumed to be non-zero 
at a time. For a near-detector, $\Rsl$ interactions can come into 
effect in two ways as described in the following subsections.

\subsection{$\nutau$ Production in Beta-decay via $\Rsl$ Interactions}

$\Rsl$ interactions can drive beta-decay producing $\nutau$ 
instead of $\nue$. $\nutau$ so produced give rise to $\tau$ 
leptons in the detector which may decay in the leptonic channel 
producing muons.

\begin{figure}[t]
\begin{center}
\includegraphics[width=13.0cm]{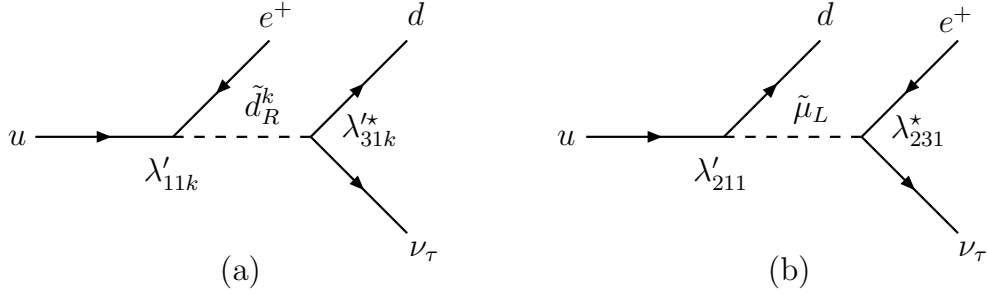}
\caption{\label{Feyn} 
Feynman diagrams for $\Rsl$ interactions during beta-decay through 
(a) $\lambda'\lambda'$ and (b) $\lambda\lambda'$
type trilinear product couplings. 
Substantial event rates are obtained in (a) when $k = 2, 3$.}
\end{center}
\end{figure}

Simultaneous presence of ${\lambda}'_{31k}$ and ${\lambda}'_{11k}$
couplings (see Fig.~\ref{Feyn}(a)) can be responsible for producing a 
$\nutau$ in beta-decay. Of these, $\lambda'_{111}$ is tightly constrained from
neutrinoless double beta-decay \cite{double_beta}. But the upper bound
on the combination $|{\lambda'^{\star}_{31k}
\lambda'_{11k}}|, k = 2, 3$ is rather relaxed; a limit of
$2.4\times 10^{-3} ({\tilde m}/{100~{\rm GeV}})^2$, $\tilde m$ being a
common sfermion mass, follows from $\tau^{-} \rightarrow e^{-}
\rho^{0}$ \cite{mu2e,hagiwara}. The corresponding decay amplitude 
can be written as,
\begin{eqnarray}
{M}_{\not L}(u \longrightarrow d e^+ \nu_\tau) &=&
\frac{\lambda'^{\star}_{31k}\lambda'_{11k}}
{2 (\hat s - {\tilde m}^2)}\; \big[\bar u_{\nu_\tau} \gamma_{\mu} P_L
u_{e}\big]\, \big[\bar u_{d} \gamma^{\mu} P_L u_u \big].
\label{process01}
\end{eqnarray}
\noindent
Alternatively, $\nutau$ can be produced in beta-decay if another
combination of $\Rsl$ couplings ${\lambda^\star_{i31}\lambda'_{i11}}$
($i$ = 1, 2) is non-zero (see Fig.~\ref{Feyn}(b)). As mentioned
earlier, $\lp_{111}$ is severely constrained. The combination
$|{\lambda^\star_{231} \lambda'_{211}}|$ is bounded from above by 
$1.6\times 10^{-3} ({\tilde m}/{100~{\rm GeV}})^2$ arising from the 
decay channel $\tau^{-} \rightarrow e^{-} \eta^{0}$ \cite{mu2e,hagiwara}, 
which is not too small to produce an observable effect. The
corresponding decay amplitude is given by, 
\begin{eqnarray}
{M}_{\not L}(u \longrightarrow d e^+ \nu_\tau) &=&
\frac{\lambda^\star_{231}\lambda'_{211}}
{(\hat t - {\tilde m}^2)}\;
 \big[\bar
u_{\nu_\tau} P_R \,u_{e}\big]\, \big[\bar u_{d} P_L u_u \big].
\label{process02}
\end{eqnarray}
\noindent

\subsection{$\tau$ production from $\nu_e$ via $\Rsl$ Interactions}

$\nu_e$ produced through ordinary beta-decay driven by weak
interactions can undergo $\Rsl$ interactions with the detector producing
$\tau$ which subsequently decay into muons.

\begin{figure}[t]
\begin{center}
\includegraphics[width=13.0cm]{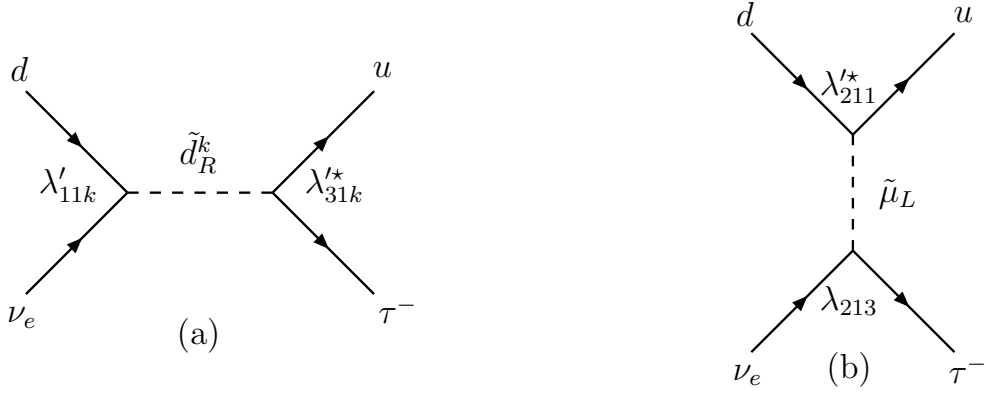}
\caption{\label{Feyn2}
Feynman diagrams for tau production from an incoming $\nue$ beta-beam 
through (a) $\lambda'\lambda'$ and (b) $\lambda\lambda'$ type trilinear
product couplings. Substantial event rates are obtained in (a)
when $k = 2, 3$.}
\end{center}
\end{figure}

Simultaneous presence of ${\lambda}'_{31k}$ and ${\lambda}'_{11k}$
couplings can give rise to $\tau^{-}$ in the final state from an
incoming $\nue$ beta-beam (see Fig.~\ref{Feyn2}(a)). The amplitude 
for the corresponding $s$-channel diagram can be written,
after a Fierz transformation, as
\begin{eqnarray}
{M}_{\not L}(\nu_e \;d \longrightarrow \tau^- \; u) &=&
\frac{\lambda'^{\star}_{31k}\lambda'_{11k}}
{2 (\hat s - {\tilde m}^2)}\; \big[\bar u_{\tau} \gamma_{\mu} P_L
u_{\nu_e}\big]\, \big[\bar u_{u} \gamma^{\mu} P_L u_d \big].
\label{process1}
\end{eqnarray}
\noindent

An alternative channel of tau production from an incoming $\nue$ beam
exists (see Fig.~\ref{Feyn2}(b)) if a particular combination of the
$\lambda$ and $\lambda'$ couplings ${\lambda_{i13}\lambda'^\star_{i11}}$ 
($i$ = 2, 3) is non-zero. Here again, $\lambda_{313}$ is severely constrained 
from neutrinoless double beta-decay experiments \cite{Bhattacharyya}. 
An upper bound of $1.6\times 10^{-3} ({\tilde m}/{100~{\rm GeV}})^2$ applies to the
combination $|{\lambda_{213} \lambda'^\star_{211}}|$, from the decay
channel $\tau^{-} \rightarrow e^{-} \eta^{0}$ \cite{mu2e,hagiwara}. 
The amplitude for this $t$-channel process is
\begin{eqnarray}
{M}_{\not L}(\nu_e \;d \longrightarrow \tau^- \; u) &=&
\frac{\lambda_{213}\lambda'^\star_{211}} {(\hat t - {\tilde
m}^2)}\;
 \big[\bar
u_{\tau} P_L u_{\nu_e}\big]\, \big[\bar u_{u} P_R u_d \big].
\label{process2}
\end{eqnarray}
\noindent

In what follows, we categorize the above two kinds of diagrams (a) and
(b) in Figs. 6.3 and 6.4 as $\lambda'\lambda'$ and $\lambda \lambda'$
processes, respectively.

Note that, if $|{\lambda'^{\star}_{31k} \lambda'_{11k}}|,~k = 2, 3$ is
non-zero, tau leptons can be produced at the detector either due to
$\Rsl$ interactions affecting beta-decay or due to $\Rsl$ interactions 
of a $\nue$ with the detector material. These two equal contributions
add in the total rate of tau production.

However, for the $\lambda\lp$ process, we see that the $\Rsl$
combinations $|{\lambda^\star_{231} \lambda'_{211}}|$ 
(which drive the $\Rsl$ beta-decay) and 
$|{\lambda_{213} \lambda'^\star_{211}}|$ 
(which is responsible for producing a tau from an incoming $\nue$ 
in the detector) are different. As we are following the strategy of 
taking only two $\Rsl$ couplings non-zero at a time, these contributions, 
which are of the same magnitude, cannot be present at the same time.

In passing, a few comments are in order~:
\begin{itemize}
\item In both the diagrams of Fig. \ref{Feyn2}, the incoming $\nue$
can interact with a $\bar u$ quark from the sea to
produce a tau. Due to the smallness of the corresponding parton
distribution function, this contribution is suppressed
but we do include it in the numerical evaluations.

\item Here we should mention that the FCNC process, 
$K^{+} \rightarrow \pi^{+} \nu \bar\nu$ \cite{kpipi} 
puts stringent bounds on all the $\lambda'$ couplings. 
However, these are basis dependent and hence can be evaded.

\item As already noted, the non-observation of the process $\mu
\rightarrow e~(Ti)$ severely restricts \cite{rparity,mu2e}
the possibility of emitting a $\numu$ in beta-decay and
direct production of muons from an incoming $\nue$ beam.

\item Since the beta-beam energy is $\sim$ a few GeV, the expected
event rate will be essentially independent of the sfermion mass as
the bounds on $\lambda, \lambda^\prime$ scale with 
$(\tilde m/100~{\rm GeV})^2$.

\end{itemize}

At this energy range it is important to consider contributions from
deep-inelastic, quasi-elastic, and single-pion production channels. To
estimate the $\Rsl$ deep-inelastic scattering cross section, we have used
CTEQ4LQ parton distributions \cite{parton}. $\Rsl$ quasi-elastic
scattering and single-pion production cross sections have been
evaluated from the corresponding Standard Model 
cross sections\footnote{These cross sections include all nuclear effects 
for an iron target.} \cite{Paschos:2001np,paschos} by a rescaling 
of the couplings. It is noticed that, as Eq. \ref{process2} is not reducible 
to a Standard Model-like $(V-A) \otimes (V-A)$ Lorentz structure, in calculating 
deep-inelastic cross section a factor $\sim 1/3$ appears from polar integration 
compared to that for Eq. \ref{process1}. For the $\lambda\lambda'$ process 
we have adopted the same suppression factor for the Standard Model quasi-elastic 
and single-pion production cross sections as well.  
Conservatively, we assume that a similar suppression also applies 
to the case of $\Rsl$ beta-decay. It bears stressing that the effect of the
tau mass is felt on the neutrino-nucleus cross section throughout the
energy range beyond the $\tau$-threshold and this is included in the
analysis.

\section{\fbox{Results}}\label{sec4}

A near-detector set-up is qualitatively different from a far-detector
as in the former case the storage ring and the detector really `sees'
each other and relative geometric considerations are of much
relevance. The observed number of events in a given period of time
depends on the choice of the radioactive ion, the boost factor
$\gamma$ and the details of the set-up (which include storage ring
parameters, detector configuration and the short base-length between
them). As alluded to earlier, the maximum $\gamma$ available is
limited by the storage ring configuration. With a view to optimizing
the set-up, the essential inputs are summarized as follows~:

\begin{itemize}

\item
\fbox{\bf Storage Ring Parameters ~:}

Total length 6880 m, length of a straight section, 
$S=2500$ m, number of beta-decays in the
straight section, $N_{\beta} = 1.1\times 10^{18}$ per year.

\item
\fbox{\bf Detector Configuration ~:}

The detector material is\footnote{Brief comments are made about a water \u{C}erenkov
detector in section 6.4.2.} iron ($\rho = 7.87~\mbox{gm}/\mbox{cm}^3$). 
A detector of mass 5 kton is considered. For a given material, this fixes
the length of the detector as the radius is changed. It varies from
202.13 m to 12.63 m as the radius ranges over 1 m to 4 m.

\item
\fbox{\bf Base-length ~:}

Results are presented for three representative
values of the distance of the detector from the storage ring,
$L=200$ m, 500 m, 1 km.

\item
\fbox{\bf Boost factor $\gamma$ ~:}

The tau production threshold ($3.5$ GeV) calls for a high $\gamma$. 
We consider $\gamma=250, 350, 450$ for $^8$B and as large as $800$ for $^{18}$Ne
with fixed $N_{\beta}$.

\end{itemize}

The high collimation achievable in the beta-beams encourages the
choice of a detector of cylindrical shape coaxial with the storage
ring straight section. As $\gamma$ increases, the $\Rsl$ event rates
increase for the following reasons~:
\begin{enumerate}
\item
an increasingly larger part of the beam falls onto the detector,
\item
more neutrinos have enough energy to produce a tau lepton,
\item
with the more energetic neutrinos the cross section is larger.
\end{enumerate}

\begin{figure}[t]
\begin{center}
\includegraphics[width=13.0cm]{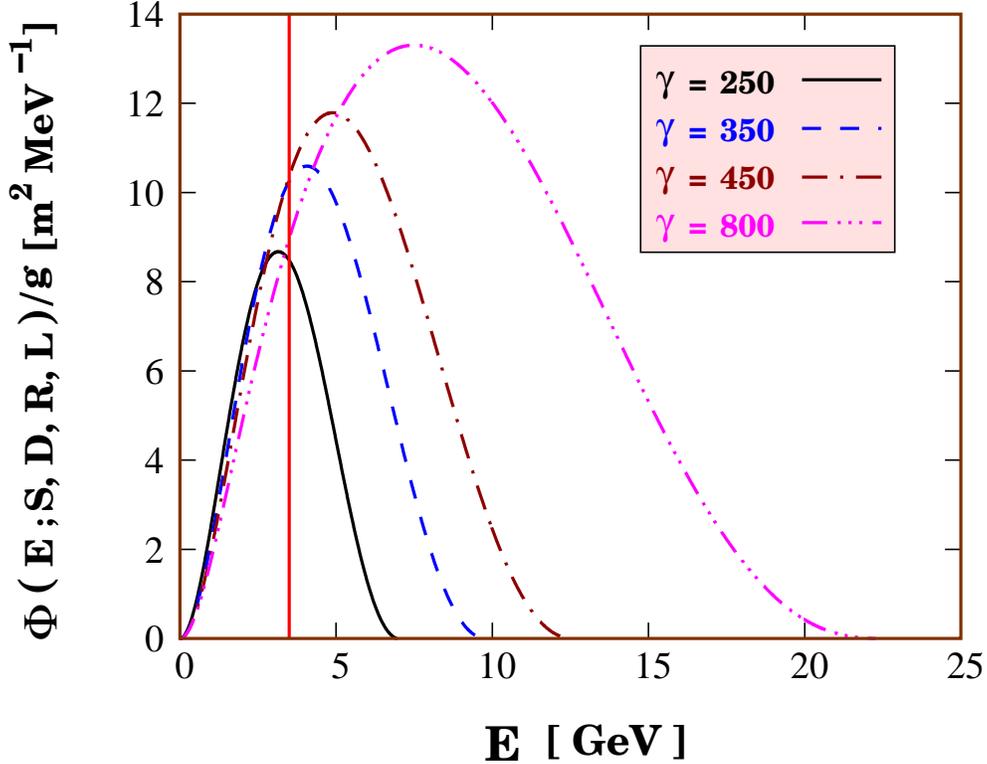}
\caption{\label{capphi}
Geometry integrated flux $\Phi(E;S,D,R,L)/g$ taking $^8$B as the decaying 
ion is plotted against neutrino energy $E$ for different $\gamma$ for
$S=2500$ m, $D=202.13$ m, $R=1$ m, and $L=200$ m. The vertical
line at 3.5 GeV indicates the tau production threshold energy.}
\end{center}
\end{figure}

The first two effects are demonstrated by Fig. \ref{capphi}.
The geometry integrated flux, $\Phi$, as defined in Eq. \ref{PHIeqn},
represents the beta-beam neutrino flux spectrum falling onto the
detector per unit time. It is seen that as $\gamma$ increases, the
total area under the curve also increases, illustrating the first
effect. The area under the curves on the right side of the vertical
line (the threshold) also increases with $\gamma$, in conformity with
the second expectation. For a high $\gamma$ the beam should
saturate. However, with the $\gamma$ used in Fig. \ref{capphi} this is
not evident due to the enormous length of the straight section of the
storage ring. To collimate the flux emanating from the rear part of
the ring a very high $\gamma$ will be needed.

As geometry plays a crucial role in optimizing the near-detector
set-up, we study the detector length dependence of the expected number
of $\Rsl$ events for different base-lengths and different $\gamma$. We
consider the contribution coming from the two options -- the
$\lambda'\lambda'$ and $\lambda\lambda'$ processes -- in different
panels for every figure, assuming the $\Rsl$ coupling constants saturate
present experimental upper limits.

\subsection{Choice of Ion Source and Detector}

The choice of $^8$B as the ion source provides the most attractive
option due to its high end-point energy. Iron calorimeters are
preferred for the smaller size and significant background removal.

To get a glimpse of the number of events one might expect in such
a set-up, let us present the following estimate. A 5 kton Fe detector
of radius 1 m (length 202.13 m) placed at a distance 200 m from
the decay ring can give rise to 92 (24) muon events via the
$\lambda'\lambda'$ ($\lambda\lambda'$) process in 5 years
for\footnote{The corresponding
numbers for $\gamma = 350$ are 421 (103).} $\gamma=250$.

\begin{figure}[t]
\includegraphics[width=8.0cm, height=7.0cm, angle=0]{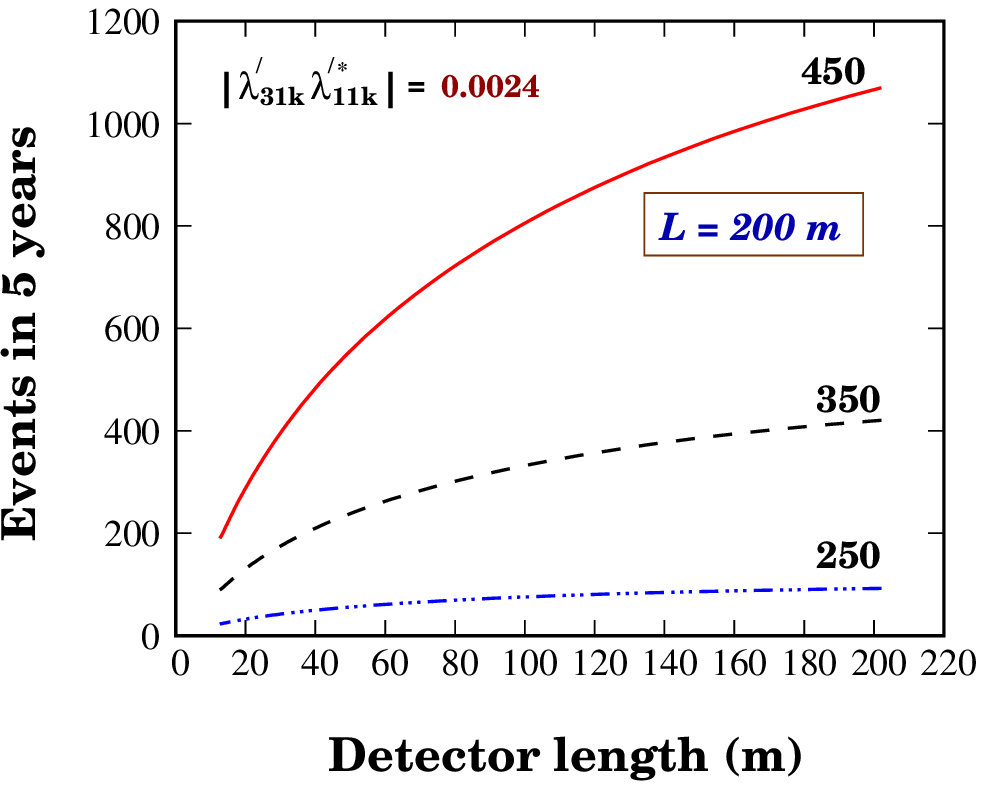}
\vglue -7.0cm \hglue 8.8cm
\includegraphics[width=8.0cm, height=7.0cm, angle=0]{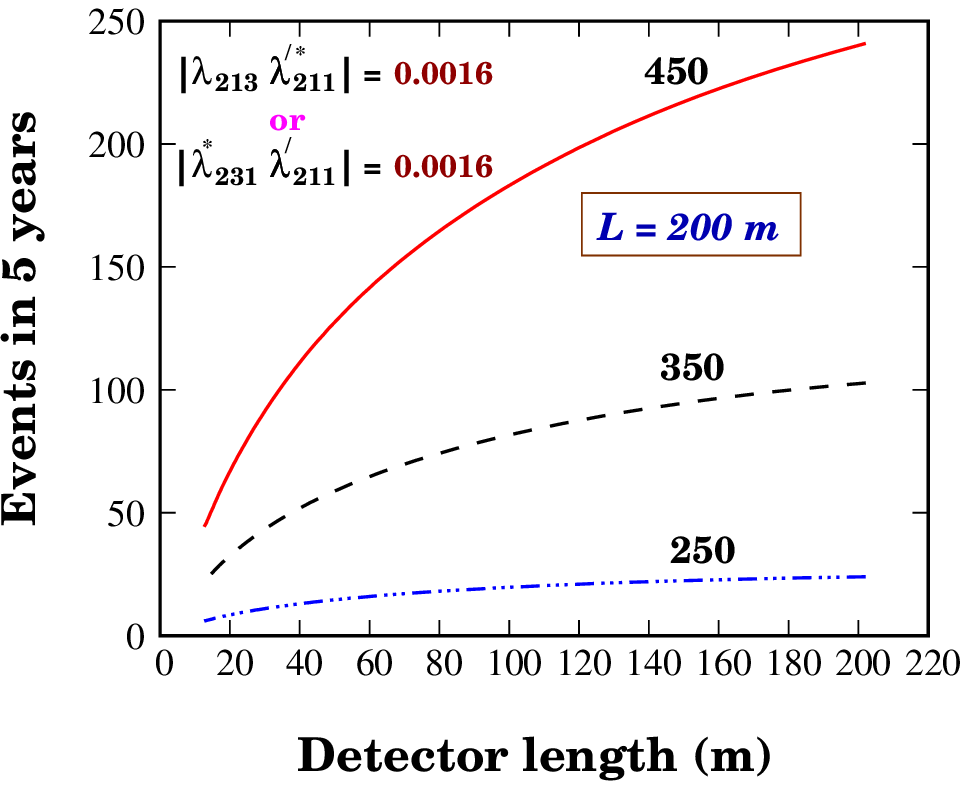}
\caption{\label{figgam}
Expected number of $\Rsl$ muon events in five years for a 5 kton 
iron detector vs. the detector length for
$\gamma$ = 250, 350, and 450 for $^8$B beta-beam flux. The
left~(right) panel is for the $\lp\lp$~($\lambda\lp$) driven
process. $k = 2, 3$.}
\end{figure}

In Fig. \ref{figgam} we exhibit the $\gamma$ dependence of the
expected number of muon events over a five-year period for a fixed
base-length of 200 m. Collimation plays a role as is demonstrated
by the increase in the number of events for higher $\gamma$. As
expected, a long detector serves better as it provides more
opportunity for a neutrino interaction to occur. However, this
increase with the length is not linear; a part of the beam is lost
due to the concomitant decrease in the radius (to keep the total
mass fixed at 5 kton). In addition, with the increase in detector
length as the detector efficiency decreases, the increase
in the rates is also somewhat restricted.

\begin{figure}[t]
\includegraphics[width=8.0cm, height=7.0cm, angle=0]{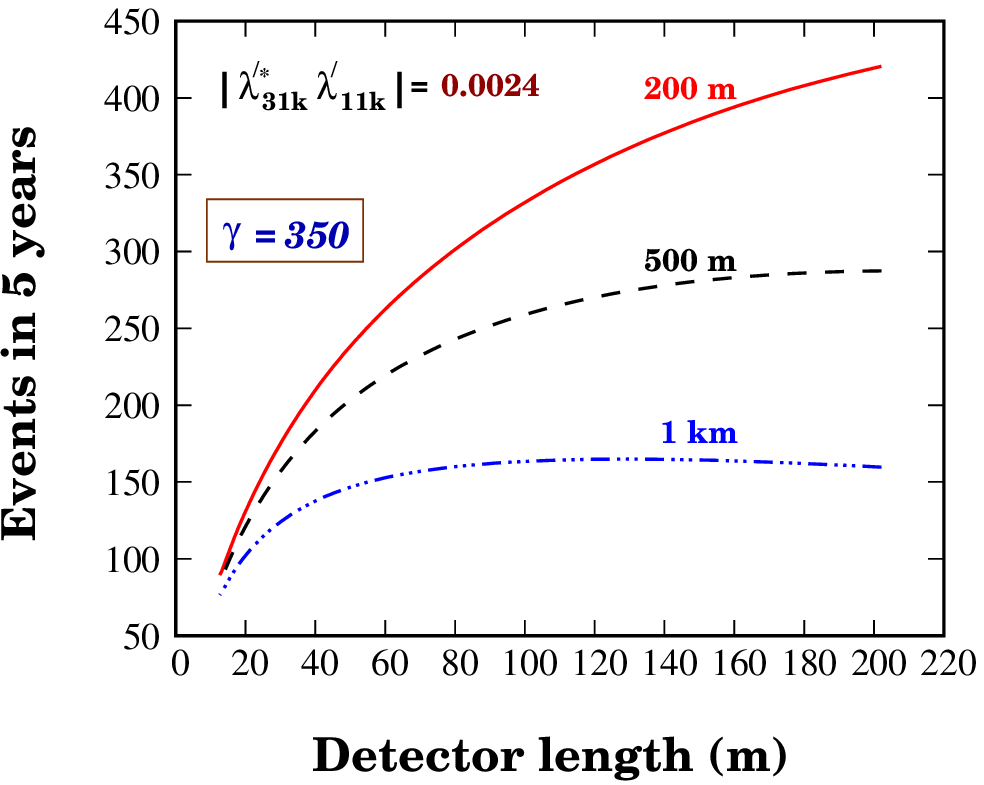}
\vglue -7.0cm \hglue 8.8cm
\includegraphics[width=8.0cm, height=7.0cm, angle=0]{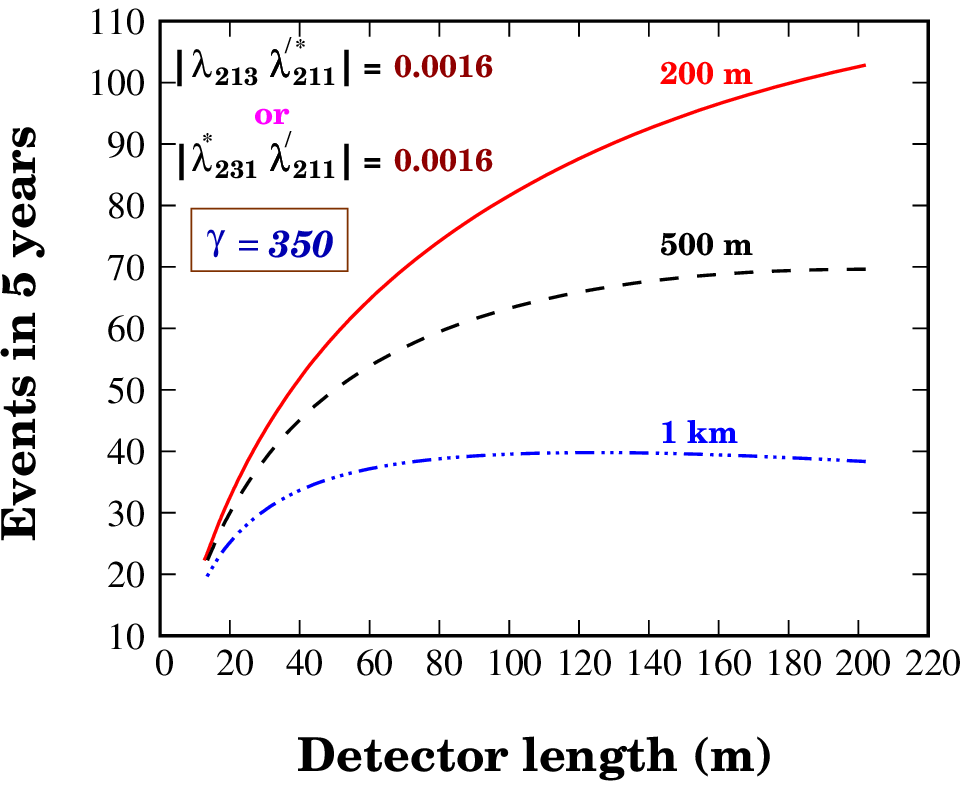}
\caption{\label{figL}
Muon signal event rate in 5 years as a function of the detector (Fe) 
length for three different choices of base-length have been shown for 
$^8$B beta-beam flux. The left~(right) panel corresponds to the 
$\lp\lp$~($\lambda\lp$) driven process. $k = 2, 3$.}
\end{figure}

It is also of interest to study the base-length dependence of the
number of events. The beam spreads with an increase in the
base-length, reducing the effective flux hitting the detector. This
causes a fall in the number of events (other parameters remaining the
same) as shown in Fig. \ref{figL}. It is interesting to note that the
increase in the number of events with increase in the length of the
detector gets severely diluted at larger base-lengths.

\begin{figure}[t]
\includegraphics[width=8.0cm, height=7.0cm, angle=0]{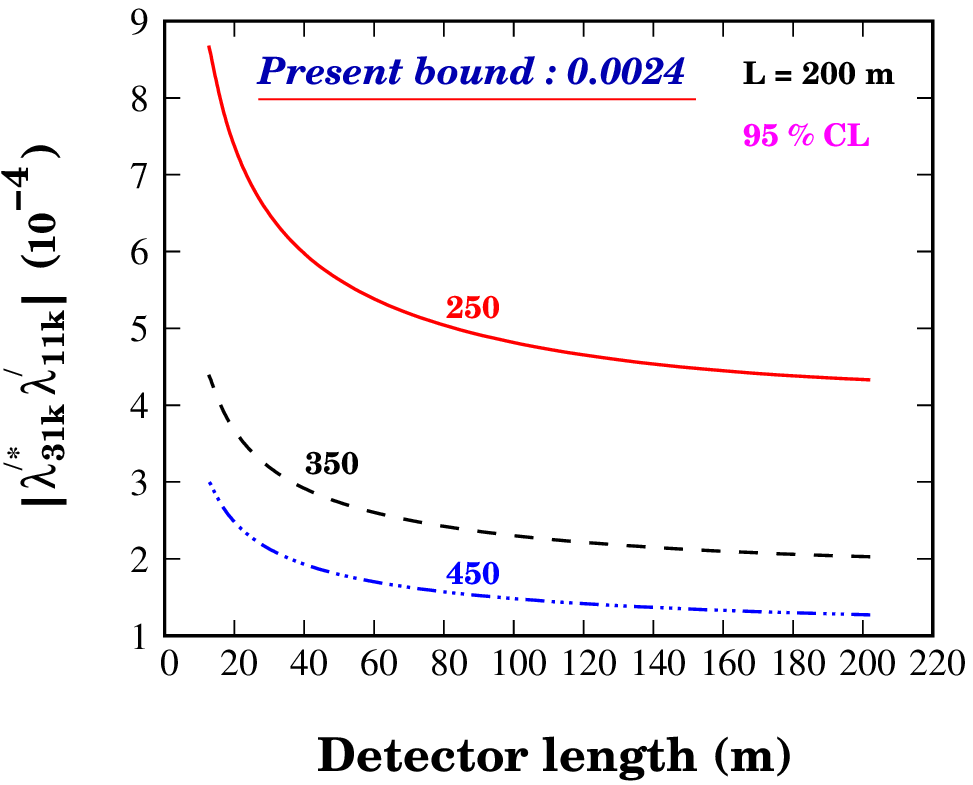}
\vglue -7.0cm \hglue 8.8cm
\includegraphics[width=8.0cm, height=7.0cm, angle=0]{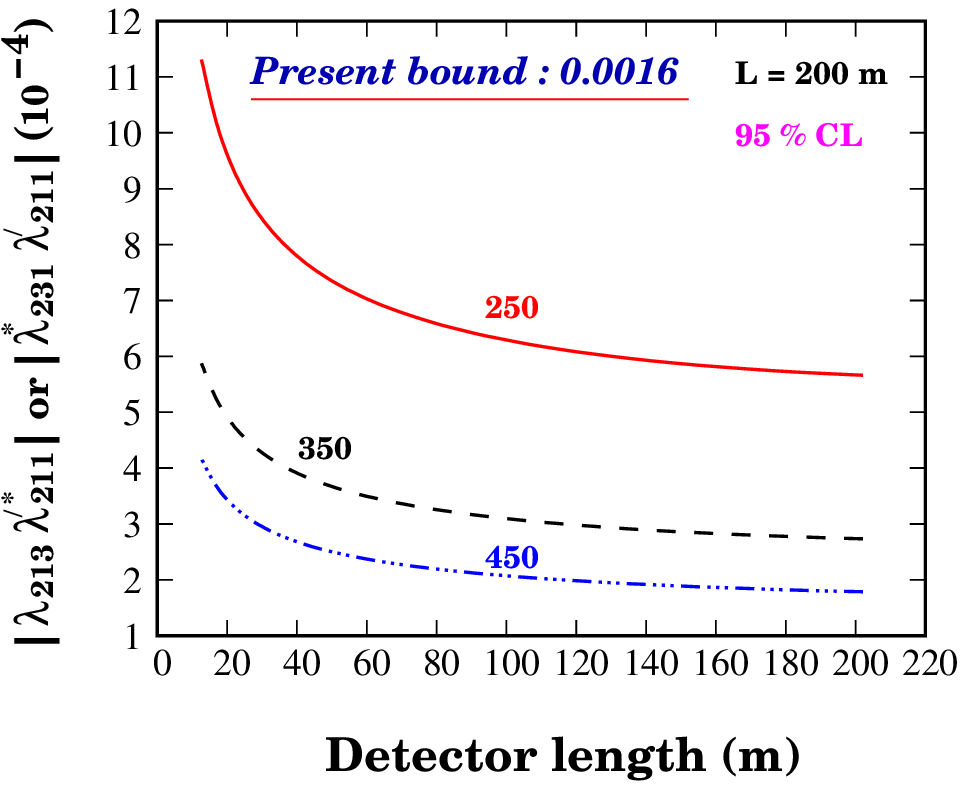}
\caption{\label{figbound} 
Bounds on $|\lambda'^{\star}_{31k}\lambda'_{11k}|,$ $k = 2, 3$
($|\lambda^\star_{231}\lambda'_{211}|$ or
$|\lambda_{213}\lambda'^{\star}_{211}|$) versus detector size at
$95\%$ C.L. for zero observed events is depicted in left (right) panel
for $\gamma=250,350,450$. The bounds scale as $(\tilde m/100~{\rm
GeV})^2$. The results are for a five-year run for a 5 kton Fe detector
placed at a distance of 200 m from the front end of the storage ring
for $^8$B beta-beam flux.}
\end{figure}

While presenting the expected number of events we assumed the $\Rsl$
couplings saturate the present experimental upper bounds. In case less 
or even no events are seen, the existing limits on the combinations 
$|\lambda'^{\star}_{31k}\lambda'_{11k}|, k = 2, 3$,
$|\lambda^{\star}_{231}\lambda'_{211}|$ and
$|\lambda_{213}\lambda'^{\star}_{211}|$ will be improved. Choosing the
minimum number of non-zero $\Rsl$ couplings, one can put conservative
upper bounds. In Fig. \ref{figbound} we show the bounds -- the region
above the curves are disallowed -- achievable in the case of
`no-show'\footnote{At 95\% C.L. this corresponds to not more than 3
events.}. It is seen that to put stringent bounds it is necessary to
go for a higher $\gamma$ and a longer detector.

\subsection{Alternative Set-ups}

\begin{figure}[t]
\includegraphics[width=8.0cm, height=7.0cm, angle=0]{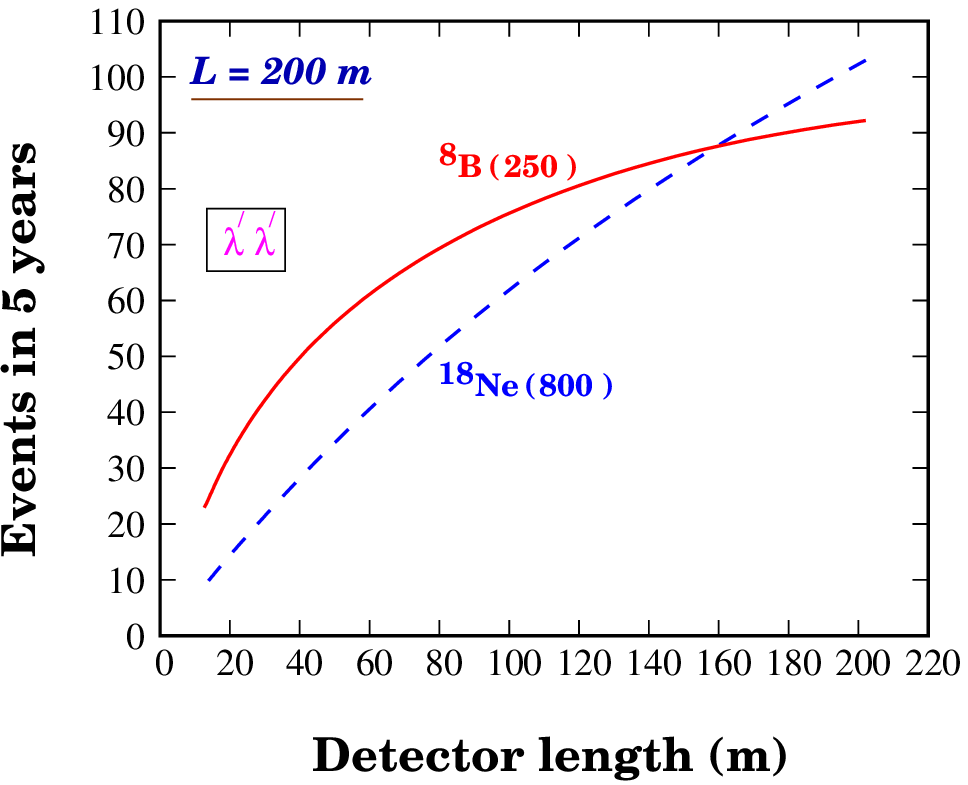}
\vglue -7.0cm \hglue 8.8cm
\includegraphics[width=8.0cm, height=7.0cm, angle=0]{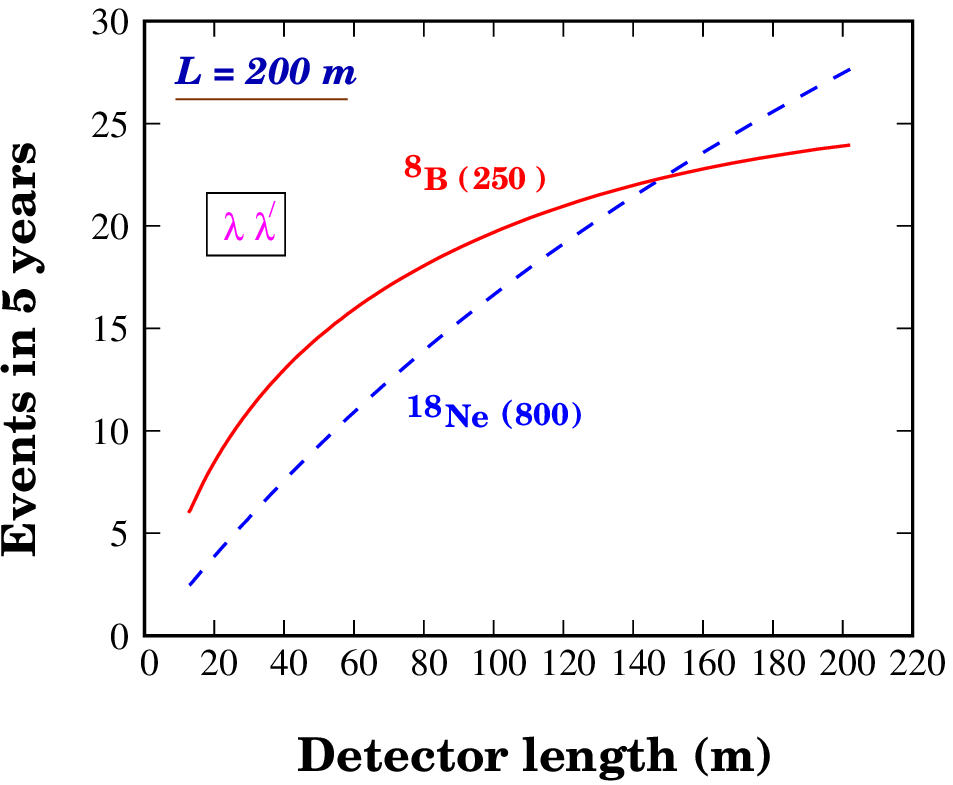}
\caption{\label{figNe}
Comparison of the muon signal event rates as a function of the detector 
length for a 5 kton iron calorimeter placed at a distance of 200 m from 
the storage ring for $\gamma=800$~$(250)$ with $^{18}$Ne ($^8$B).
The left and right panels correspond to $\lp\lp$
and $\lambda\lp$ driven processes, respectively.}
\end{figure}

Although so far we have presented results with $^8$B as the
beta-beam source, $^{18}$Ne is the most discussed decaying ion in
the literature. As mentioned earlier, due to the smaller
end-point energy of $^{18}$Ne, a high $\gamma$ is required to
cross the $\tau$ threshold. Fig. \ref{figNe} depicts the
variation in the expected event rate with detector length for
$^{18}$Ne with $\gamma=800$ using a 5 kton iron calorimeter. 
We see that due to high $\gamma$ for $^{18}$Ne,
the beam is so collimated that the event rates increase almost
linearly with increasing detector length in contrast to the $^8$B
case we have presented. However even in such an extreme scenario,
where we use the same storage ring configuration to reach such a
high $\gamma$, the expected event rates are comparable to that in
the $^8$B case. Hence we conclude that $^8$B is preferred to
$^{18}$Ne in exploring lepton number violating signatures with
beta-beams.

The use of water \u{C}erenkov detectors with good capability of
muon-electron separation and moderate efficiency of neutral current
rejection may be an interesting option to see the signals of new
physics and to normalize the incoming flux. The disadvantage of this
set-up turns out to be the huge background. Consider a 5 kton water
\u{C}erenkov detector with radius 2.5 m at a distance 200 m from the
decay ring. In five years, this will lead to 45 (12) muon events from
$\tau$-lepton decay for $\lp\lp$ ($\lambda\lp$) driven processes from
an incoming $^8$B $\nu_{e}$ beam accelerated with a $\gamma$ of 250
and with a muon detection threshold of 200 MeV. For the same
configuration and duration, one expects roughly $10^8$ pions produced
from charged and neutral current interactions of the $\nu_e$
beam. Muons produced from $\pi$ decay will thus completely swamp the
signal.

The number of signal events may be increased by designing a very long
water detector with small radius though this could be technologically
challenging. In any case, the background events will continue to be
very high. So, this option also does not hold much promise. The basic
problem of high backgrounds, avoided in the Fe detector, will also
plague totally active scintillator based detectors.

\section{\fbox{Discussion and Conclusion}}

Beta-beam experiments may be sensitive to the $\not{\! L}$ interactions. 
In the previous chapter it was shown that $\Rsl$ interactions can interfere 
with pure oscillation signals in long-baseline beta-beam experiments. 
In this chapter we explore a complementary scenario. It has been proposed 
that to probe such interactions, an iron calorimeter detector placed close 
to the storage ring holds promise as it provides essentially a neutrino oscillation 
free environment. In particular, the combination of a 5 kton cylindrical iron
detector placed within a distance of 200 m to 1 km from the decay ring
and a neutrino beam from an $^{8}$B ion source with $\gamma$ in the
range 250 to 450, running for 5 years is well-suited in this regard. 
We have examined the impact of non-trivial design details of
such a near-detector setup.

At production, low energy beta-decay experiments may get
contaminated by tau neutrinos through $\Rsl$ interactions.  
We show that, this contamination, though small, can be probed using the above
set-up. $\Rsl$ interactions can also play a role in such an experiment
during the interactions of the beta-beam electron neutrinos with the
detector.

It is interesting to explore if $\Rsl$ interactions can affect beta-beam
experiments in other ways. For example, we have checked that the
impact of these interactions on the $\mu$ detection cross section is
insignificant.  As mentioned earlier, $\nu_\mu$ may be produced in
beta-decay through $\Rsl$ interactions but this also is severely
suppressed as the corresponding couplings have stringent upper
limits.

Results are presented for a neutrino beam. Antineutrino beams can
also be produced using $^8$Li or $^6$He as sources. In fact, a
storage ring design may allow both beams to be present
simultaneously. The expected event rates for antineutrinos are of
similar order as for the neutrinos.

In conclusion, we find a near-detector setup can be useful for
exploring lepton number violating interactions with beta-beams.  
It may allow us to put stringent bounds on some of these couplings.
In the next chapter, we will present a summary of this doctoral work.

\chapter{Summary and Conclusions}
Neutrino physics is traversing through an exciting phase with
lots of new data making this field more interesting. There is 
an excellent progress in our understanding of the neutrinos over 
the last ten years or so, thanks to the experiments on neutrino 
oscillations which confirm the fact that neutrinos have a tiny, 
but non-zero, mass quite against the expectations of our best 
theory - the Standard Model. We have now a rough picture of the parameters 
governing three-flavour oscillations and we are all set to move into
the precision regime. There is no doubt that the use of artificial 
neutrino sources is mandatory in the era of high precision experiments. 
In this direction the beta-beam is a recently proposed technique of 
producing a pure, intense and collimated beam of $\nue$ or $\anue$ 
through the beta-decay of completely ionized radioactive ions. 
My thesis sheds light on the importance of beta-beams in probing 
some unknown territories in the field of neutrino physics.

We started with a brief description of neutrino properties in the 
first half of chapter 1 and then we took a glance at the fascinating journey 
of discovery into one of Nature's most impalpable particles - the 
neutrino which is the main theme of this thesis. In the second 
half of chapter 1 we focused on the main sources of neutrinos.

The first part of chapter 2 dealt with a basic introduction to the
quantum mechanics of neutrino oscillation in vacuum under both two and
three flavour frameworks. Then the importance of matter effects
in neutrino oscillations was discussed. In the second half of chapter 2 
we took a look at our present understanding of neutrino parameters and 
we identified the major unknowns in the neutrino sector. Finally we 
closed chapter 2 by giving a brief note on the future neutrino road-map 
based on long baseline experiments.

In chapter 3 we expounded in detail the physics reach of an 
experimental set-up in which the proposed large magnetized 
iron detector at the INO would serve as the far detector for 
a beta-beam. If this pure $\nue$ and/or $\anue$ beam is shot 
from some source location like CERN such that the source-detector
distance $L \simeq 7500$ km, the impact of the CP phase
$\delta_{CP}$ on the oscillation probability and associated
parameter correlation and degeneracies are almost negligible. 
This ``magical'' long baseline beta-beam experiment would have
unprecedented sensitivity to the neutrino mass hierarchy and
$\theta_{13}$, two of the missing ingredients needed for
our understanding of the neutrino sector. With Lorentz boost 
$\gamma=650$ and irrespective of the true value of $\delta_{CP}$,
the neutrino mass hierarchy could be determined at $3\sigma$ C.L.
if $\sin^22\theta_{13}{\rm {(true)}} > 5.6 \times 10^{-4}$
and we can expect an unambiguous signal for $\theta_{13}$
at $3\sigma$ C.L. if $\sin^22\theta_{13}{\rm {(true)}} > 5.1 \times 10^{-4}$
independent of the true neutrino mass ordering.

In chapter 4 we showed that the earth matter effects in the
${\rm {\nu_e \to \nu_e}}$ survival probability can be used 
to cleanly determine the third leptonic mixing angle 
$\theta_{13}$ and the sign of the atmospheric neutrino 
mass squared difference, $\Delta m^2_{31}$, using a 
beta-beam as a $\nue$ source.

In chapter 5 we discussed possible new physics scenarios with 
a beta-beam neutrino source and made a strong physics case for
these high-precision experiments. Long baseline neutrino oscillation 
experiments may well emerge as test beds for neutrino interactions 
as are present in R-parity violating supersymmetry. We showed that 
flavour diagonal and flavour changing neutral currents arising 
therefrom prominently impact a neutrino beta-beam experiment with the
source at CERN and the detector at the proposed India-based
Neutrino Observatory. These interactions may preclude any
improvement of the present limit on $\theta_{13}$ and cloud the
hierarchy determination unless the upper bounds on $\Rsl$
couplings, particularly $\lambda^{\prime}$, become significantly 
tighter. If $\Rsl$ interactions are independently established then 
from the event rate a lower bound on $\theta_{13}$ may be set. It has
been shown that there is scope to see a clear signal of non-standard FCNC 
and FDNC interactions, particularly in the IH scenario and also
sometimes for the NH. In favourable cases, it may be possible to set 
lower and upper bounds on $\lambda^{\prime}$ couplings. 
FCNC and FDNC interactions due to $\lambda$ type $\Rsl$ couplings 
are unimportant.

A detector placed near a beta-beam storage ring can probe $\not{\! L}$ 
interactions, as predicted by supersymmetric theories with R-parity 
non-conservation. This issue has been discussed in great detail in 
chapter 6. In the presence of $\Rsl$ interactions, $\nutau$ can be 
produced during beta-decay leading to tau leptons through weak interactions.
Alternatively, electron neutrinos from beta-decay of radioactive ions can 
produce tau leptons in a nearby detector through these interactions. 
The muons from the decay of these tau leptons can be readily identified 
in a small iron calorimeter detector and will signal violation of R-parity.

The study of neutrinos has always landed up with lots of surprises.
So we can expect further surprises in store and, obviously, beta-beams can
play a leading role in this direction. This doctoral work is an effort 
to judge the expected performance of a beta-beam neutrino source in the
future progress of neutrino physics and also for the hunt for signals of   
non-standard new physics. We hope that this work
will provide a boost to the detailed and thorough R\&D of the novel
beta-beam neutrino source in future.


\newpage
\addcontentsline{toc}{chapter}{~~~~List of Publications}
\begin{center}
{\Large\bf\underline{ List of Publications}}
\end{center}
\vskip 20pt
\begin{enumerate}

\item
{\bf{Exploration prospects of a long baseline beta beam
neutrino experiment with an iron calorimeter detector}},\\
Sanjib Kumar Agarwalla, Amitava Raychaudhuri and Abhijit Samanta, \\
Phys.\ Lett.\ B {\bf 629}, 33 (2005); hep-ph/0505015.

\vskip1.0cm

\item
{\bf{Can R-parity violating supersymmetry be seen in long baseline
beta-beam experiments?}},\\
Rathin Adhikari, Sanjib Kumar Agarwalla and Amitava Raychaudhuri, \\
Phys.\ Lett.\ B {\bf 642}, 111 (2006); hep-ph/0608034.

\vskip1.0cm

\item
{\bf{Probing Lepton Number Violating Interactions with Beta-beams}},\\
Sanjib Kumar Agarwalla, Subhendu Rakshit, and Amitava Raychaudhuri, \\
Phys.\ Lett.\ B {\bf 647}, 380 (2007); hep-ph/0609252.

\vskip1.0cm

\item
{\bf{Neutrino mass hierarchy and $\theta_{13}$ with a magic
baseline beta-beam experiment}},\\
Sanjib Kumar Agarwalla, Sandhya Choubey, and Amitava Raychaudhuri, \\
Nucl.\ Phys.\ B {\bf 771}, 1 (2007); hep-ph/0610333.

\vskip1.0cm

\item
{\bf{Neutrino Mixings and Leptonic CP Violation from
CKM Matrix and Majorana Phases}},\\
S.K. Agarwalla, M.K. Parida, R.N. Mohapatra, and G. Rajasekaran, \\
Phys.\ Rev.\ D {\bf 75}, 033007 (2007); hep-ph/0610333.

\vskip1.0cm

\item
{\bf{Neutrino parameters from matter effects in the
$\nu_e$ survival probability at long baselines}},\\
Sanjib Kumar Agarwalla, Sandhya Choubey, Srubabati Goswami and Amitava Raychaudhuri, \\
Phys.\ Rev.\ D {\bf 75}, 097302 (2007); hep-ph/0611233.

\vskip1.0cm

\item
{\bf{Unraveling neutrino parameters with a magical beta-beam experiment
at INO}},\\
Sanjib Kumar Agarwalla, Sandhya Choubey, and Amitava Raychaudhuri, \\
Nucl.\ Phys.\ B {\bf 798}, 124 (2008); arXiv:0711.1459 [hep-ph].

\vskip1.0cm

\item
{\bf{Optimizing the greenfield Beta-beam}},\\
Sanjib Kumar Agarwalla, Sandhya Choubey, Amitava Raychaudhuri and Walter Winter, \\
JHEP {\bf 0806}, 090 (2008); arXiv:0802.3621 [hep-ex].

\vskip1.0cm

\item
{\bf{ Exceptional Sensitivity to Neutrino Parameters with a Two Baseline 
Beta-Beam Set-up}},\\
Sanjib Kumar Agarwalla, Sandhya Choubey and Amitava Raychaudhuri, \\
Accepted in Nucl.\ Phys.\ B; arXiv:0804.3007 [hep-ph]. 

\end{enumerate}


\newpage
\addcontentsline{toc}{chapter}{~~~~Bibliography}
\begin{thebibliography}{999}

\bibitem{pauli}
W. Pauli, {\it letter sent to the Tubingen conference}, 4th December, 1930.

\bibitem{cowan_reines} 
  F.~Reines and C.~L.~Cowan,
  Phys.\ Rev.\  {\bf 113} (1959) 273.

\bibitem{sup1}
  K.~Hirata {\it et al.}  [KAMIOKANDE-II Collaboration],
  Phys.\ Rev.\ Lett.\  {\bf 58}, 1490 (1987).

\bibitem{sup2}
  R.~M.~Bionta {\it et al.},
  Phys.\ Rev.\ Lett.\  {\bf 58}, 1494 (1987).

\bibitem{sup3}
  G.~G.~Raffelt,
  Ann.\ Rev.\ Nucl.\ Part.\ Sci.\  {\bf 49}, 163 (1999).

\bibitem{sk1}
  Y.~Fukuda {\it et al.}  [Super-Kamiokande Collaboration],
  Phys.\ Rev.\ Lett.\  {\bf 81}, 1562 (1998).

\bibitem{sno1}
  Q.~R.~Ahmad {\it et al.}  [SNO Collaboration],
  Phys.\ Rev.\ Lett.\  {\bf 89}, 011301 (2002).

\bibitem{kl}
K.~Eguchi {\it et al.},
  [KamLAND Collaboration],
Phys.\ Rev.\ Lett.\  {\bf 90}, 021802 (2003);
%
T.~Araki {\it et al.}, [KamLAND Collaboration],
Phys.\ Rev.\ Lett.\  {\bf 94}, 081801 (2005);
%
S.~Abe {\it et al.}  [KamLAND Collaboration],
  arXiv:0801.4589 [hep-ex].

\bibitem{kltalk}
I. Shimizu, talk at 10th International Conference on
Topics in Astroparticle and Underground Physics, TAUP 2007.

\bibitem{lma_msw}
  L.~Wolfenstein,
  Phys.\ Rev.\  D {\bf 17}, 2369 (1978);
  L.~Wolfenstein,
  Phys.\ Rev.\  D {\bf 20}, 2634 (1979);
  S.~P.~Mikheev and A.~Y.~Smirnov,
  Sov.\ J.\ Nucl.\ Phys.\  {\bf 42}, 913 (1985)
  [Yad.\ Fiz.\  {\bf 42}, 1441 (1985)];
  S.~P.~Mikheev and A.~Y.~Smirnov,
  Nuovo Cim.\  C {\bf 9}, 17 (1986).

\bibitem{k2k_ahn}
  M.~H.~Ahn {\it et al.}  [K2K Collaboration],
  Phys.\ Rev.\ Lett.\  {\bf 90}, 041801 (2003).

\bibitem{p1} 
  B.~Pontecorvo,
  Sov.\ Phys.\ JETP {\bf 26}, 984 (1968)
  [Zh.\ Eksp.\ Teor.\ Fiz.\  {\bf 53}, 1717 (1967)].

\bibitem{p2} 
  V.~N.~Gribov and B.~Pontecorvo,
  Phys.\ Lett.\  B {\bf 28}, 493 (1969).

\bibitem{maltoni_garcia}
  M.~C.~Gonzalez-Garcia and M.~Maltoni,
  Phys.\ Rept.\  {\bf 460}, 1 (2008).

\bibitem{strumia_vissani}
  A.~Strumia and F.~Vissani,
  arXiv:hep-ph/0606054.

\bibitem{bahcall} 
  J.~N.~Bahcall, A.~M.~Serenelli and S.~Basu,
  Astrophys.\ J.\  {\bf 621}, L85 (2005).

\bibitem{solar}
B.~T.~Cleveland {\it et al.},
Astrophys.\ J.\  {\bf 496}, 505 (1998);
%
J.~N.~Abdurashitov {\it et al.}  [SAGE Collaboration],
J.\ Exp.\ Theor.\ Phys.\  {\bf 95}, 181 (2002)
[Zh.\ Eksp.\ Teor.\ Fiz.\  {\bf 122}, 211 (2002)];
%
W.~Hampel {\it et al.}  [GALLEX Collaboration],
Phys.\ Lett.\ B {\bf 447}, 127 (1999).
%
%
\bibitem{borex}
C.~Arpesella {\it et al.}
  [Borexino~Collaboration],
  Phys.\ Lett.\  B {\bf 658}, 101 (2008).

\bibitem{sk2}
  S.~Fukuda {\it et al.}  [Super-Kamiokande Collaboration],
  Phys.\ Lett.\ B {\bf 539}, 179 (2002);

\bibitem{sno2}
B.~Aharmim {\it et al.}  [SNO Collaboration],
Phys.\ Rev.\ C {\bf 72}, 055502 (2005);

\bibitem{kgf} 
  C.~V.~Achar {\it et al.},
  Phys.\ Lett.\ {\bf 18}, 196 (1965).

\bibitem{africa} 
  F.~Reines {\it et al.},
  Phys.\ Rev.\ Lett.\  {\bf 15}, 429 (1965).

\bibitem{soudan} 
  M.~C.~Sanchez {\it et al.}  [Soudan 2 Collaboration],
  Phys.\ Rev.\  D {\bf 68}, 113004 (2003).

\bibitem{macro} 
  M.~Ambrosio {\it et al.}  [MACRO Collaboration],
  Phys.\ Lett.\  B {\bf 517}, 59 (2001).

\bibitem{ino}
  M.~S.~Athar {\it et al.}  [INO Collaboration],
 A Report of the INO Feasibility Study,\\
{http://www.imsc.res.in/$\sim$ino/OpenReports/INOReport.pdf}.
%

\bibitem{geoneutrino}
  T.~Araki {\it et al.},
  Nature {\bf 436}, 499 (2005).

\bibitem{tele}
  F.~Halzen,
  arXiv:astro-ph/0506248.

\bibitem{gosgen} 
  G.~Zacek {\it et al.}  [CALTECH-SIN-TUM Collaboration],
  Phys.\ Rev.\  D {\bf 34}, 2621 (1986).

\bibitem{krasnoyarsk} 
  G.~S.~Vidyakin {\it et al.},
  JETP Lett.\  {\bf 59}, 390 (1994)
  [Pisma Zh.\ Eksp.\ Teor.\ Fiz.\  {\bf 59}, 364 (1994)].

\bibitem{bugey} 
  Y.~Declais {\it et al.},
  Nucl.\ Phys.\  B {\bf 434}, 503 (1995).

\bibitem{CHOOZ} 
  M.~Apollonio {\it et al.}  [CHOOZ Collaboration],
  Phys.\ Lett.\  B {\bf 466}, 415 (1999).

\bibitem{paloverde} 
  A.~Piepke  [Palo Verde Collaboration],
  Prog.\ Part.\ Nucl.\ Phys.\  {\bf 48}, 113 (2002).

\bibitem{report1}
M.~Apollonio {\it et al.},
arXiv:hep-ph/0210192.

\bibitem{report2}
  A.~Guglielmi, M.~Mezzetto, P.~Migliozzi and F.~Terranova,
  arXiv:hep-ph/0508034.

\bibitem{offaxis}
  A.~Para and M.~Szleper,
  arXiv:hep-ex/0110032.

\bibitem{minos}
D. G. Michael {\it et al.}, [MINOS Collaboration],
  arXiv:hep-ex/0607088.

\bibitem{opera}
  M.~Guler {\it et al.}  [OPERA Collaboration],
CERN-SPSC-2000-028 (2000).

\bibitem{icarus}
  P.~Aprili {\it et al.}  [ICARUS Collaboration],
CERN-SPSC-2002-027 (2002).

\bibitem{t2k}
  Y.~Itow {\it et al.},
  arXiv:hep-ex/0106019.

\bibitem{nova}
  D.~S.~Ayres {\it et al.}  [NOvA Collaboration],
  arXiv:hep-ex/0503053.

\bibitem{geer}
  S.~Geer,
  Phys.\ Rev.\  D {\bf 57}, 6989 (1998)
  [Erratum-ibid.\  D {\bf 59}, 039903 (1999)].

\bibitem{mice}
  R.~Edgecock,
  J.\ Phys.\ G {\bf 29}, 1601 (2003).

\bibitem{others}
  M.~M.~Alsharoa {\it et al.}  [Muon Collider/Neutrino Factory
                  Collaboration],
  Phys.\ Rev.\ ST Accel.\ Beams {\bf 6}, 081001 (2003);
N. Holtkamp and S. Geer, Feasibility study of neutrino source based on
muon storage ring, ICFA Beam Dyn. Newslett. {\bf 21}, 37 (2000);
S. Ozaki {\it et al.}, Feasibility study-II of a muon-based neutrino 
source, (2001), BNL-52623; P. Gruber (ed.) {\it et al.}, The study of a 
European neutrino factory complex, (2002), CERN-PS-2002-080-PP.

\bibitem{silver_nufact}
A.~Donini, D.~Meloni and P.~Migliozzi,
  Nucl.\ Phys.\ B {\bf 646}, 321 (2002);
D.~Autiero {\it et al.},
  Eur.\ Phys.\ J.\ C {\bf 33}, 243 (2004).

\bibitem{zucc}
P.~Zucchelli,
Phys.\ Lett.\ B {\bf 532}, 166 (2002).

\bibitem{volpe}
For a recent review see
  C.~Volpe,
  J.\ Phys.\ G {\bf 34}, R1 (2007).

\bibitem{cernmemphys}
 J.~E.~Campagne {\it et al.},
 JHEP {\bf 0704}, 003 (2007).

\bibitem{paper1}
  S.~K.~Agarwalla, A.~Raychaudhuri and A.~Samanta,
  Phys.\ Lett.\ B {\bf 629}, 33 (2005).

\bibitem{betaino1}
  S.~K.~Agarwalla, S.~Choubey and A.~Raychaudhuri,
  Nucl.\ Phys.\  B {\bf 771}, 1 (2007).

\bibitem{betaino2}
  S.~K.~Agarwalla, S.~Choubey and A.~Raychaudhuri,
  Nucl.\ Phys.\  B {\bf 798}, 124 (2008).

\bibitem{pee}
  S.~K.~Agarwalla, S.~Choubey, S.~Goswami and A.~Raychaudhuri,
  Phys.\ Rev.\  D {\bf 75}, 097302 (2007).

\bibitem{bboptim}
  S.~K.~Agarwalla, S.~Choubey, A.~Raychaudhuri and W.~Winter,
  JHEP {\bf 0806}, 090 (2008).

\bibitem{twobaseline}
  S.~K.~Agarwalla, S.~Choubey and A.~Raychaudhuri,
  arXiv:0804.3007 [hep-ph].

\bibitem{rparity1}
  R.~Adhikari, S.~K.~Agarwalla and A.~Raychaudhuri,
  Phys.\ Lett.\ B {\bf 642}, 111 (2006).

\bibitem{rparity2}
  S.~K.~Agarwalla, S.~Rakshit and A.~Raychaudhuri,
  Phys.\ Lett.\  B {\bf 647}, 380 (2007).

\bibitem{conf_talks}
  S.~K.~Agarwalla, S.~Choubey and A.~Raychaudhuri,
  AIP Conf.\ Proc.\  {\bf 981}, 84 (2008);
%

\bibitem{sanjib_talks}
  S.~K.~Agarwalla, S.~Choubey and A.~Raychaudhuri,
  AIP Conf.\ Proc.\  {\bf 939}, 265 (2007);
S. K. Agarwalla, Talk at NuFact08, Valencia-Spain, 
June 30 - July 5, 2008, http://ific.uv.es/nufact08/

\bibitem{oldpapers}
  M.~Mezzetto,
  J.\ Phys.\ G {\bf 29}, 1771 (2003)
  [arXiv:hep-ex/0302007].
%
  M.~Mezzetto,
  Nucl.\ Phys.\ Proc.\ Suppl.\  {\bf 143}, 309 (2005).
%
  M.~Mezzetto,
  Nucl.\ Phys.\ Proc.\ Suppl.\  {\bf 155}, 214 (2006);
%

\bibitem{donini130}
  A.~Donini, E.~Fernandez-Martinez, P.~Migliozzi, S.~Rigolin and
L.~Scotto Lavina,
  Nucl.\ Phys.\  B {\bf 710}, 402 (2005).


\bibitem{doninibeta}
  A.~Donini, E.~Fernandez, P.~Migliozzi, S.~Rigolin, L.~Scotto Lavina,
  T.~Tabarelli de Fatis and F.~Terranova,
  arXiv:hep-ph/0511134;
%
  A.~Donini, E.~Fernandez-Martinez, P.~Migliozzi, S.~Rigolin,
  L.~Scotto Lavina, T.~Tabarelli de Fatis and F.~Terranova,
  Eur.\ Phys.\ J.\  C {\bf 48}, 787 (2006).

\bibitem{newdonini}
  P.~Coloma, A.~Donini, E.~Fernandez-Martinez and J.~Lopez-Pavon,
  JHEP {\bf 0805}, 050 (2008).

\bibitem{bc}
  J.~Burguet-Castell, D.~Casper, E.~Couce, J.~J.~G\'{o}mez-Cadenas and P.~Hernandez,
  Nucl.\ Phys.\  B {\bf 725}, 306 (2005).
%

\bibitem{bc2}
  J.~Burguet-Castell, D.~Casper, J.~J.~G\'{o}mez-Cadenas, P.~Hernandez and F.~Sanchez,
  Nucl.\ Phys.\  B {\bf 695}, 217 (2004).

\bibitem{fnal}
  A.~Jansson, O.~Mena, S.~Parke and N.~Saoulidou,
  arXiv:0711.1075 [hep-ph].

\bibitem{betaoptim}
  P.~Huber, M.~Lindner, M.~Rolinec and W.~Winter,
  Phys.\ Rev.\  D {\bf 73}, 053002 (2006).

\bibitem{doninialter}
  A.~Donini and E.~Fernandez-Martinez,
  Phys.\ Lett.\ B {\bf 641}, 432 (2006).

\bibitem{boulby}
  D.~Meloni, O.~Mena, C.~Orme, S.~Palomares-Ruiz and S.~Pascoli,
  arXiv:0802.0255 [hep-ph].

\bibitem{lindroos}
 M.~Lindroos,
  arXiv:physics/0312042;
%
  M.~Lindroos,
  Nucl.\ Phys.\ Proc.\ Suppl.\  {\bf 155}, 48 (2006).

\bibitem{betabeampage}
http://beta-beam.web.cern.ch/beta\%2Dbeam/

\bibitem{jacques}
J.~Bouchez, M.~Lindroos and M.~Mezzetto,
AIP Conf.\ Proc.\  {\bf 721}, 37 (2004).

\bibitem{iss}
The ISS Physics Working Group,
  arXiv:0710.4947 [hep-ph]. http://www.hep.ph.ic.ac.uk/iss/

\bibitem{rubbia}
  C.~Rubbia, A.~Ferrari, Y.~Kadi and V.~Vlachoudis,
  Nucl.\ Instrum.\ Meth.\ A {\bf 568}, 475 (2006);
  C.~Rubbia,
  arXiv:hep-ph/0609235.

\bibitem{mori}
  Y.~Mori,
  Nucl.\ Instrum.\ Meth.\  A {\bf 562}, 591 (2006).

\bibitem{beta}

L. P. Ekstrom and R. B. Firestone, WWW Table of Radioactive Isotopes, \\
database version 2/28/99 from URL http://ie.lbl.gov/toi/

\bibitem{beamnorm}
  B.~Autin, R.~C.~Fernow, S.~Machida and D.~A.~Harris,
  J.\ Phys.\ G {\bf 29}, 1637 (2003);
%
  F.~Terranova, A.~Marotta, P.~Migliozzi and M.~Spinetti,
  Eur.\ Phys.\ J.\  C {\bf 38}, 69 (2004).

\bibitem{recent_chooz}
M.~Apollonio {\it et al.},
Eur.\ Phys.\ J.\ C {\bf 27}, 331 (2003).

\bibitem{k2k}
E.~Aliu {\it et al.}  [K2K Collaboration],
  Phys.\ Rev.\ Lett.\  {\bf 94}, 081802 (2005).

\bibitem{limits}
  M.~C.~Gonzalez-Garcia and M.~Maltoni,
  arXiv:0704.1800 [hep-ph];
%
M.~Maltoni,
T.~Schwetz, M.~A.~Tortola and J.~W.~F.~Valle,
New J.\ Phys.\  {\bf 6}, 122 (2004), hep-ph/0405172 v6;
%
  S.~Choubey,
  arXiv:hep-ph/0509217;
%
  S.~Goswami,
  Int.\ J.\ Mod.\ Phys.\ A {\bf 21}, 1901 (2006);
%
  A.~Bandyopadhyay,
S.~Choubey, S.~Goswami, S.~T.~Petcov and D.~P.~Roy,
  Phys.\ Lett.\ B {\bf 608}, 115 (2005);
%
  G.~L.~Fogli {\it et al.},
  Prog.\ Part.\ Nucl.\ Phys.\  {\bf 57}, 742 (2006);
  G.~L.~Fogli {\it et al.},
  Phys.\ Rev.\  D {\bf 75}, 053001 (2007).

\bibitem{thomas}
The plenary talk of Thomas Schwetz at Nufact07, Okayama, Japan.

\bibitem{rparity}
  R.~Barbier {\it et al.},
  Phys.\ Rept.\  {\bf 420}, 1 (2005);
  M.~Chemtob,
  Prog.\ Part.\ Nucl.\ Phys.\  {\bf 54}, 71 (2005).

\bibitem{pontecorvo}
  B.~Pontecorvo,
  Sov.\ Phys.\ JETP {\bf 6}, 429 (1957)
  [Zh.\ Eksp.\ Teor.\ Fiz.\  {\bf 33}, 549 (1957)];
  Sov.\ Phys.\ JETP {\bf 7}, 172 (1958)
  [Zh.\ Eksp.\ Teor.\ Fiz.\  {\bf 34}, 247 (1957)].

\bibitem{pmns}
  Z.~Maki, M.~Nakagawa and S.~Sakata,
  Prog.\ Theor.\ Phys.\  {\bf 28}, 870 (1962).

\bibitem{jarlskog}
  L.~L.~Chau and W.~Y.~Keung,
  Phys.\ Rev.\ Lett.\  {\bf 53}, 1802 (1984);
  C.~Jarlskog,
  Z.\ Phys.\  C {\bf 29}, 491 (1985);
  C.~Jarlskog,
  Phys.\ Rev.\  D {\bf 35}, 1685 (1987).

\bibitem{sk3}
  Y.~Ashie {\it et al.}  [Super-Kamiokande Collaboration],
  Phys.\ Rev.\ D {\bf 71}, 112005 (2005).

\bibitem{mass}
  J.~Dunkley {\it et al.}  [WMAP Collaboration],
  arXiv:0803.0586 [astro-ph];
  E.~Komatsu {\it et al.}  [WMAP Collaboration],
  arXiv:0803.0547 [astro-ph];
  C.~L.~Reichardt {\it et al.},
  arXiv:0801.1491 [astro-ph];
  C.~Dickinson {\it et al.},
  Mon.\ Not.\ Roy.\ Astron.\ Soc.\  {\bf 353}, 732 (2004);
  A.~C.~S.~Readhead {\it et al.},
  Astrophys.\ J.\  {\bf 609}, 498 (2004);
  C.~J.~MacTavish {\it et al.},
  Astrophys.\ J.\  {\bf 647}, 799 (2006);
  W.~L.~Freedman {\it et al.}  [HST Collaboration],
  Astrophys.\ J.\  {\bf 553}, 47 (2001);
  P.~Astier {\it et al.}  [The SNLS Collaboration],
  Astron.\ Astrophys.\  {\bf 447}, 31 (2006);
  D.~J.~Eisenstein {\it et al.}  [SDSS Collaboration],
  Astrophys.\ J.\  {\bf 633}, 560 (2005);
  U.~Seljak, A.~Slosar and P.~McDonald,
  JCAP {\bf 0610}, 014 (2006);
  P.~McDonald {\it et al.}  [SDSS Collaboration],
  Astrophys.\ J.\  {\bf 635}, 761 (2005).
  G.~L.~Fogli {\it et al.},
  arXiv:0805.2517 [hep-ph].

\bibitem{0vbbus}
  S.~Pascoli, S.~T.~Petcov and T.~Schwetz,
  Nucl.\ Phys.\  B {\bf 734}, 24 (2006);
%
  S.~Choubey and W.~Rodejohann,
  Phys.\ Rev.\  D {\bf 72}, 033016 (2005).

\bibitem{barger_msw}
  V.~D.~Barger, K.~Whisnant, S.~Pakvasa and R.~J.~N.~Phillips,
  Phys.\ Rev.\ D {\bf 22}, 2718 (1980).

\bibitem{golden}
  A.~Cervera, A.~Donini, M.~B.~Gavela, J.~J.~G\'{o}mez-Cadenas, P.~Hernandez, O.~Mena and S.~Rigolin,
  Nucl.\ Phys.\ B {\bf 579}, 17 (2000)
  [Erratum-ibid.\ B {\bf 593}, 731 (2001)].

\bibitem{chooz2}
  F.~Ardellier {\it et al.},
  arXiv:hep-ex/0405032;
%
  F.~Ardellier {\it et al.}  [Double Chooz Collaboration],
  arXiv:hep-ex/0606025.

\bibitem{huber10}
  P.~Huber,
M.~Lindner, M.~Rolinec, T.~Schwetz and W.~Winter,
  Phys.\ Rev.\ D {\bf 70}, 073014 (2004)
and references therein.

\bibitem{crossthomas}
  P.~Huber, M.~Mezzetto and T.~Schwetz,
  JHEP {\bf 0803}, 021 (2008).

\bibitem{volpelow}
  C.~Volpe,
  J.\ Phys.\ G {\bf 30}, L1 (2004).

\bibitem{intrinsic}
  J.~Burguet-Castell,
M.~B.~Gavela, J.~J.~G\'{o}mez-Cadenas, P.~Hernandez and O.~Mena,
  Nucl.\ Phys.\ B {\bf 608}, 301 (2001).

\bibitem{minadeg}
  H.~Minakata and H.~Nunokawa,
  JHEP {\bf 0110}, 001 (2001).

\bibitem{th23octant}
  G.~L.~Fogli and E.~Lisi,
  Phys.\ Rev.\ D {\bf 54}, 3667 (1996).

\bibitem{eight}
  V.~Barger, D.~Marfatia and K.~Whisnant,
  Phys.\ Rev.\ D {\bf 65}, 073023 (2002).

\bibitem{magic}
  P.~Huber and W.~Winter,
  Phys.\ Rev.\ D {\bf 68}, 037301 (2003).

\bibitem{magic2}
  A.~Y.~Smirnov,
  arXiv:hep-ph/0610198.

\bibitem{petcov}
  M.~Freund, M.~Lindner, S.~T.~Petcov and A.~Romanino,
  Nucl.\ Phys.\  B {\bf 578}, 27 (2000).

\bibitem{nufactoptim}
  P.~Huber, M.~Lindner, M.~Rolinec and W.~Winter,
  Phys.\ Rev.\  D {\bf 74}, 073003 (2006).

\bibitem{autin}
  B.~Autin {\it et al.},
  J.\ Phys.\ G {\bf 29}, 1785 (2003)
  [arXiv:physics/0306106].

\bibitem{freund}
  M.~Freund, P.~Huber and M.~Lindner,
  Nucl.\ Phys.\  B {\bf 615}, 331 (2001).

\bibitem{d21msw}
  W.~Winter,
  Phys.\ Lett.\ B {\bf 613}, 67 (2005).

\bibitem{diffLnE}
  H.~Minakata and H.~Nunokawa,
  Phys.\ Lett.\  B {\bf 413}, 369 (1997);
%
  V.~Barger, D.~Marfatia and K.~Whisnant,
  Phys.\ Rev.\  D {\bf 66}, 053007 (2002);
%
  V.~Barger, D.~Marfatia and K.~Whisnant,
  Phys.\ Lett.\  B {\bf 560}, 75 (2003);
%
  O.~Mena and S.~J.~Parke,
  Phys.\ Rev.\  D {\bf 70}, 093011 (2004);
%
  O.~Mena Requejo, S.~Palomares-Ruiz and S.~Pascoli,
  Phys.\ Rev.\  D {\bf 72}, 053002 (2005);
%
  M.~Ishitsuka, T.~Kajita, H.~Minakata and H.~Nunokawa,
  Phys.\ Rev.\  D {\bf 72}, 033003 (2005);
%
  K.~Hagiwara, N.~Okamura and K.~i.~Senda,
  Phys.\ Rev.\  D {\bf 76}, 093002 (2007).

\bibitem{t2ksimulation}
  P.~Huber, M.~Lindner and W.~Winter,
  Nucl.\ Phys.\  B {\bf 645}, 3 (2002);
%
  P.~Huber, M.~Lindner and W.~Winter,
  Nucl.\ Phys.\  B {\bf 654}, 3 (2003).

\bibitem{silver}
  A.~Donini, D.~Meloni and P.~Migliozzi,
  Nucl.\ Phys.\  B {\bf 646}, 321 (2002);
%
  D.~Autiero {\it et al.},
  Eur.\ Phys.\ J.\  C {\bf 33}, 243 (2004).

\bibitem{dissappear}
  A.~Donini, E.~Fernandez-Martinez and S.~Rigolin,
  Phys.\ Lett.\  B {\bf 621}, 276 (2005);
%
  A.~Donini, E.~Fernandez-Martinez, D.~Meloni and S.~Rigolin,
  Nucl.\ Phys.\  B {\bf 743}, 41 (2006).

\bibitem{addatm}
  P.~Huber, M.~Maltoni and T.~Schwetz,
  Phys.\ Rev.\  D {\bf 71}, 053006 (2005);

\bibitem{addreact}
  P.~Huber, M.~Lindner, T.~Schwetz and W.~Winter,
  Nucl.\ Phys.\  B {\bf 665}, 487 (2003).

\bibitem{prem}
  A.~M.~Dziewonski and D.~L.~Anderson,
  Phys.\ Earth Planet.\ Interiors {\bf 25}, 297 (1981);
\\
S.~V.~Panasyuk, Reference Earth Model (REM) webpage,\\
 http://cfauves5.harvrd.edu/lana/rem/index.html.

\bibitem{Takamura:2005df}
  A.~Takamura and K.~Kimura,
  JHEP {\bf 0601}, 053 (2006).

\bibitem{solarprecision}

See for example,
A.~Bandyopadhyay, S.~Choubey, S.~Goswami and S.~T.~Petcov,
  Phys.\ Rev.\ D {\bf 72}, 033013 (2005);
%
  J.~N.~Bahcall and C.~Pena-Garay,
  JHEP {\bf 0311}, 004 (2003).

\bibitem{tomography}
R.~J.~Geller and T.~Hara,
Nucl.\ Instrum.\ Meth.\  A {\bf 503}, 187 (2001).

\bibitem{anuls:2001zn}
  M.~C.~Banuls, G.~Barenboim and J.~Bernabeu,
  Phys.\ Lett.\ B {\bf 513}, 391 (2001).

\bibitem{gandhi1}
  R.~Gandhi, P.~Ghoshal, S.~Goswami, P.~Mehta and S.~Uma Sankar,
  Phys.\ Rev.\ Lett.\  {\bf 94}, 051801 (2005).

\bibitem{gandhi2}
  R.~Gandhi, P.~Ghoshal, S.~Goswami, P.~Mehta and S.~Uma Sankar,
  Phys.\ Rev.\ D {\bf 73}, 053001 (2006).

\bibitem{mind}
A. Cervera, talk at NuFact07.

\bibitem{globes}
  P.~Huber, J.~Kopp, M.~Lindner, M.~Rolinec and W.~Winter,
  Comput.\ Phys.\ Commun.\  {\bf 177}, 432 (2007);
%
  P.~Huber, M.~Lindner and W.~Winter,
  Comput.\ Phys.\ Commun.\  {\bf 167}, 195 (2005).

\bibitem{Messier:1999kj}
  M.~D.~Messier,
PhD thesis, UMI-99-23965.

\bibitem{Paschos:2001np}
  E.~A.~Paschos and J.~Y.~Yu,
  Phys.\ Rev.\ D {\bf 65}, 033002 (2002).

\bibitem{white}
  K.~Anderson {\it et al.},
  arXiv:hep-ex/0402041.

\bibitem{akh}
  E.~K.~Akhmedov {\it et al.},
  JHEP {\bf 0404}, 078 (2004).

\bibitem{uno}
  C.~K.~Jung,
  AIP Conf.\ Proc.\  {\bf 533}, 29 (2000).

\bibitem{hk}
  Y.~Itow {\it et al.},
  arXiv:hep-ex/0106019.

\bibitem{memp}
  A.~de Bellefon {\it et al.},
  arXiv:hep-ex/0607026.

\bibitem{matter}
  G.~L.~Fogli, E.~Lisi, D.~Montanino and G.~Scioscia,
  Phys.\ Rev.\  D {\bf 55}, 4385 (1997);
  E.~K.~Akhmedov, A.~Dighe, P.~Lipari and A.~Y.~Smirnov,
  Nucl.\ Phys.\  B {\bf 542}, 3 (1999);
  D.~Choudhury and A.~Datta,
  JHEP {\bf 0507}, 058 (2005);
  D.~Indumathi and M.~V.~N.~Murthy,
  Phys.\ Rev.\  D {\bf 71}, 013001 (2005);
  S.~Choubey and P.~Roy,
  Phys.\ Rev.\  D {\bf 73}, 013006 (2006).

\bibitem{david}
  S.~Davidson, C.~Pena-Garay, N.~Rius and A.~Santamaria,
  JHEP {\bf 0303}, 011 (2003).

\bibitem{fcnc1}
  N.~Kitazawa, H.~Sugiyama and O.~Yasuda,
  arXiv:hep-ph/0606013;
  A.~Friedland and C.~Lunardini,
  Phys.\ Rev.\  D {\bf 74}, 033012 (2006).

\bibitem{solar1}
  J.~W.~F.~Valle,
  Phys.\ Lett.\  B {\bf 199}, 432 (1987);
  M.~M.~Guzzo, A.~Masiero and S.~T.~Petcov,
  Phys.\ Lett.\  B {\bf 260}, 154 (1991);
  E.~Roulet,
  Phys.\ Rev.\  D {\bf 44}, 935 (1991);
  V.~D.~Barger, R.~J.~N.~Phillips and K.~Whisnant,
  Phys.\ Rev.\  D {\bf 44}, 1629 (1991);
  S.~Bergmann,
  Nucl.\ Phys.\  B {\bf 515}, 363 (1998);
  E.~Ma and P.~Roy,
  Phys.\ Rev.\ Lett.\  {\bf 80}, 4637 (1998);
  S.~Bergmann, M.~M.~Guzzo, P.~C.~de Holanda, P.~I.~Krastev and H.~Nunokawa,
  Phys.\ Rev.\  D {\bf 62}, 073001 (2000);
  M.~Guzzo, P.~C.~de Holanda, M.~Maltoni, H.~Nunokawa, M.~A.~Tortola 
and J.~W.~F.~Valle,
Nucl.\ Phys.\  B {\bf 629}, 479 (2002);
  R.~Adhikari, A.~Sil and A.~Raychaudhuri,
  Eur.\ Phys.\ J.\  C {\bf 25}, 125 (2002).

\bibitem{atm1} 
  M.~C.~Gonzalez-Garcia {\it et al.},
  Phys.\ Rev.\ Lett.\  {\bf 82}, 3202 (1999);
  G.~L.~Fogli, E.~Lisi, A.~Marrone and G.~Scioscia,
  Phys.\ Rev.\  D {\bf 60}, 053006 (1999);
  S.~Bergmann, Y.~Grossman and D.~M.~Pierce,
  Phys.\ Rev.\  D {\bf 61}, 053005 (2000);
  N.~Fornengo, M.~Maltoni, R.~T.~Bayo and J.~W.~F.~Valle,
  Phys.\ Rev.\  D {\bf 65}, 013010 (2002).

\bibitem{fac1} 
  P.~Huber and J.~W.~F.~Valle,
  Phys.\ Lett.\  B {\bf 523}, 151 (2001);
  A.~Datta, R.~Gandhi, B.~Mukhopadhyaya and P.~Mehta,
  Phys.\ Rev.\  D {\bf 64}, 015011 (2001);
  A.~Bueno, M.~Campanelli, M.~Laveder, J.~Rico and A.~Rubbia,
  JHEP {\bf 0106}, 032 (2001).

\bibitem{huber}
  P.~Huber, T.~Schwetz and J.~W.~F.~Valle,
  Phys.\ Rev.\  D {\bf 66}, 013006 (2002).

\bibitem{ross}
  L.~E.~Ib\'{a}\~{n}ez and G.~G.~Ross,
  Nucl.\ Phys.\ B {\bf 368}, 3 (1992);
  H.~K.~Dreiner, C.~Luhn and M.~Thormeier,
  Phys.\ Rev.\ D {\bf 73}, 075007 (2006).

\bibitem{belltau}
  Y.~Enari {\it et al.}  [Belle Collaboration],
  Phys.\ Lett.\  B {\bf 622}, 218 (2005).

\bibitem{rpartau}
  W.~j.~Li, Y.~d.~Yang and X.~d.~Zhang,
  Phys.\ Rev.\  D {\bf 73}, 073005 (2006).

\bibitem{mu2e}
P. Wintz (on behalf of SINDRUM II Collaboration),
Proc. of the 14th Intl. Conf. on Particles and Nuclei (PANIC96),
(World Scientific 1997, Eds. C. E. Carlson and J. J. Domingo) 458;
  J.~E.~Kim, P.~Ko and D.~G.~Lee,
  Phys.\ Rev.\  D {\bf 56}, 100 (1997).

\bibitem{couplings}
P.~Fayet,
Phys.\ Lett.\ B {\bf 69}, 489 (1977);
G.~R.~Farrar and P.~Fayet,
Phys.\ Lett.\ B {\bf 76}, 575 (1978).

\bibitem{hagiwara}
K.~Hagiwara {\it et al.} [Particle Data Group],
Phys.\ Rev.\ D {\bf 66}, 010001 (2002).

\bibitem{geant3}
GEANT - Detector Description and Simulation Tool CERN Program Library
Long Writeup W5013.

\bibitem{nuance}
  D.~Casper,
  Nucl.\ Phys.\ Proc.\ Suppl.\ {\bf 112}, 161 (2002).

\bibitem{low_volpe}
  J.~Serreau and C.~Volpe,
  Phys.\ Rev.\  C {\bf 70}, 055502 (2004).

\bibitem{adrian} Adrian Fabich, private communication.

\bibitem{double_beta}
M.~Hirsch, H.~V.~Klapdor-Kleingrothaus and S.~G.~Kovalenko,
Nucl.\ Phys.\ Proc.\ Suppl.\ {\bf 62}, 224 (1998).

\bibitem{Bhattacharyya}
G.~Bhattacharyya, H.~V.~Klapdor-Kleingrothaus and H.~P\"{a}s,
Phys.\ Lett.\ B {\bf 463}, 77 (1999).

\bibitem{kpipi}
K.~Agashe and M.~Graesser,
Phys.\ Rev.\ D {\bf 54}, 4445 (1996). 

\bibitem{parton}
H.~L.~Lai {\it et al.},
Phys.\ Rev.\ D {\bf 55}, 1280 (1997). 

\bibitem{paschos}
  E.~A.~Paschos, L.~Pasquali and J.~Y.~Yu,
Nucl.\ Phys.\ B {\bf 588}, 263 (2000).

\end{thebibliography}
\end{document}